\documentclass[a4paper,11pt,twoside]{article}
\usepackage[T1]{fontenc}
\usepackage[utf8]{inputenc}
\usepackage[english]{babel}
\usepackage[titletoc,title]{appendix}
\usepackage{titlesec, layout, setspace, afterpage}
\usepackage{fancyhdr}
\usepackage{amsmath, amssymb, amsfonts, calrsfs, calligra, fixmath, mathtools, cases, dsfont}
\counterwithin{equation}{section}
\usepackage{graphicx, subfig, float, tikz, xcolor}
\usetikzlibrary{decorations.pathmorphing, arrows, shapes.geometric}
\tikzset{snake it/.style={decorate, decoration=snake}}
\tikzstyle{startstop} = [rectangle, rounded corners, minimum width=3cm, minimum height=1cm, text centered, draw=black, fill=white]
\tikzstyle{process} = [diamond, aspect=2, text centered, draw=black, fill=white]
\tikzstyle{arrow} = [thick,->,>=stealth]
\graphicspath{ {./Figures/Thesis-Logo} {./Figures/Plots/CTCs}{./Figures/Plots/Plot-RN}{./Figures/Plots/Plot-KN}}
\usepackage[framemethod=tikz]{mdframed}
\mdfdefinestyle{mystyle}{innerleftmargin=0.5cm, innerrightmargin=0.5cm,
innertopmargin=0.5cm, innerbottommargin=0.5cm, roundcorner=2pt, backgroundcolor = gray!10}
\usepackage{url, float, enumitem}
\definecolor{dark-red}{RGB}{175, 0, 0}
\definecolor{dark-blue}{RGB}{0, 0, 175}
\definecolor{ergo}{RGB}{0, 0, 75}
\definecolor{dark-green-div}{RGB}{25, 125, 25}
\usepackage[
top   = 4cm,
bottom = 3.5cm,
left   = 2.8cm,
right  = 2.8cm,
footskip = 1cm,
headsep = 0.925cm]{geometry}
\newcommand{\pms}[1]{{#1} ' \mkern +1mu ^2}
\DeclareMathAlphabet{\pazocal}{OMS}{zplm}{m}{n}
\DeclareMathAlphabet{\mathbfcal}{OMS}{cmsy}{b}{n}

\newcommand{\Ia}{\pazocal{I}}

\newcommand{\R}{\mathbb{R}}
\newcommand{\E}{\mathbb{E}}

\newcommand{\nb}{\nonumber}

\newcommand{\myparagraph}[1]{\paragraph{#1}\mbox{}\\ \vspace{-0.35cm} \\}
\DeclareMathSymbol{\Gamma}{\mathalpha}{operators}{0}
\DeclareMathSymbol{\Delta}{\mathalpha}{operators}{1}
\DeclareMathSymbol{\Theta}{\mathalpha}{operators}{2}
\DeclareMathSymbol{\Lambda}{\mathalpha}{operators}{3}
\DeclareMathSymbol{\Xi}{\mathalpha}{operators}{4}
\DeclareMathSymbol{\Pi}{\mathalpha}{operators}{5}
\DeclareMathSymbol{\Sigma}{\mathalpha}{operators}{6}
\DeclareMathSymbol{\Upsilon}{\mathalpha}{operators}{7}
\DeclareMathSymbol{\Phi}{\mathalpha}{operators}{8}
\DeclareMathSymbol{\Psi}{\mathalpha}{operators}{9}
\DeclareMathSymbol{\Omega}{\mathalpha}{operators}{10}
\renewcommand{\Re}{\operatorname{Re}}
\renewcommand{\Im}{\operatorname{Im}}
\newcommand{\ernst}{\pazocal{E}}

\usepackage{comment}
\newcommand{\fonte}[1]{{\color{red}[fonte mancante]}}

\titleformat{\section}[display]
  {\centering\normalfont\Large\bfseries}
  {\textit{}}
  {0pt}  {\titlerule[0.8pt]\vspace{3mm}\Huge}
  [{\vspace{3mm}\titlerule[0.8pt]\vspace{15mm}}]
\numberwithin{equation}{section}
\numberwithin{figure}{section}
\numberwithin{table}{section}
\allowdisplaybreaks
\newcommand{\beq}{\begin{equation}}
\newcommand{\eeq}{\end{equation}}
\newcommand{\bea}{\begin{eqnarray}}
\newcommand{\eea}{\end{eqnarray}}
\newcommand{\ba}{\begin{array}}
\newcommand{\ea}{\end{array}}
\newcommand{\bal}{\begin{align}}
\newcommand{\bit}{\begin{itemize}}
\newcommand{\eit}{\end{itemize}}

\newcommand{\beqNN}{\begin{equation*}}
\newcommand{\eeqNN}{\end{equation*}}
\usepackage{epigraph}

\let\originalepigraph\epigraph 
\renewcommand\epigraph[2]{\originalepigraph{\textit{#1}}{\textsc{#2}}}
\makeatletter
\renewcommand{\@dotsep}{10000} 
\makeatother
\usepackage{hyperref}
\hypersetup{
    colorlinks=true,
    linkcolor=dark-blue,
    filecolor=magenta,      
    urlcolor=blue,
    citecolor=dark-green-div
}
\begin{document}
\thispagestyle{empty}
\pagenumbering{Alph}
\thispagestyle{empty}
{\begingroup
\fontsize{12}{12}\selectfont
\newgeometry{
top       = 2.5cm,
bottom = 1.5cm,
left       = 2.5cm,
right     = 2.5cm}
\begin{figure}[htp]
\includegraphics[scale=0.1]{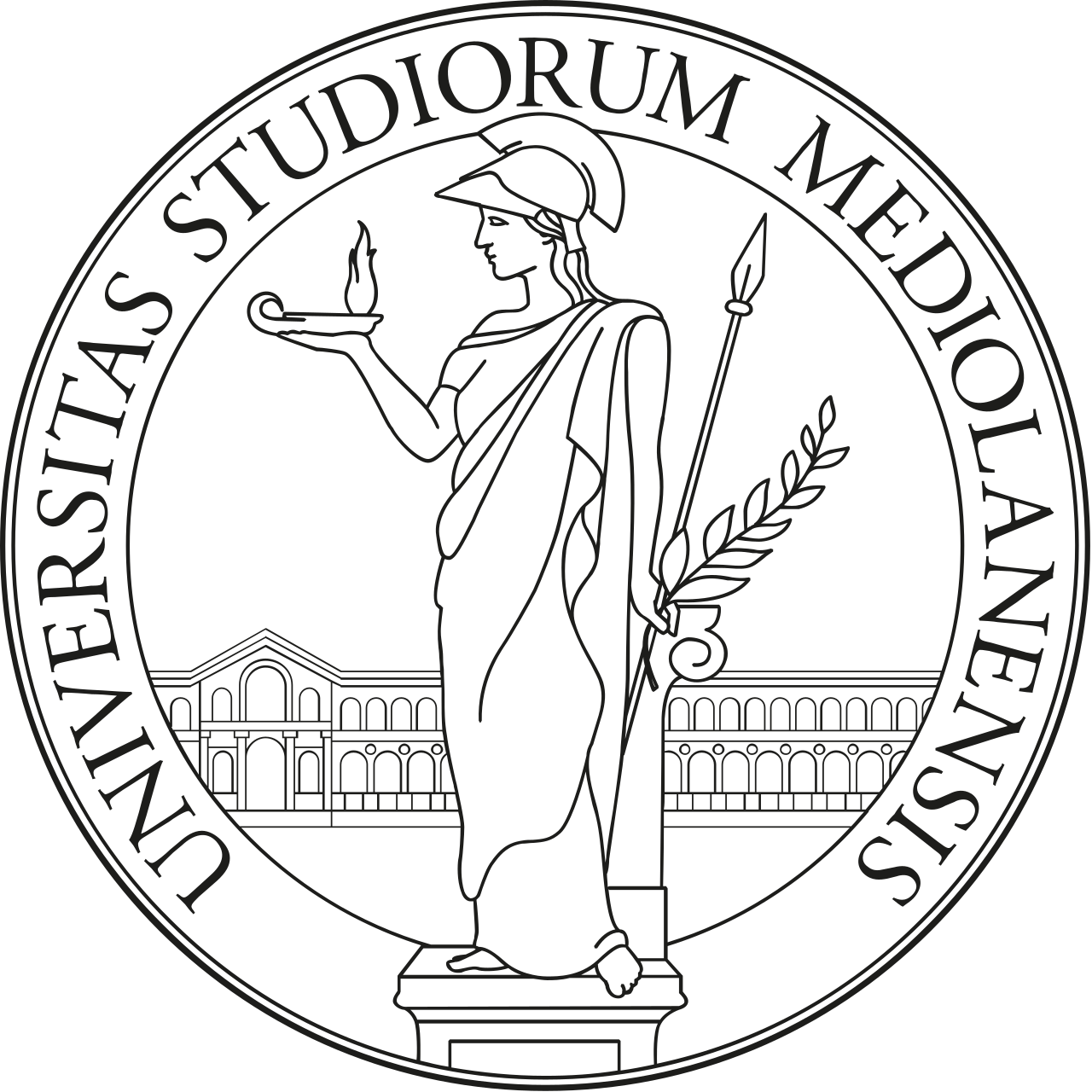}
\centering
\end{figure}
\begin{center}
{\Huge
$\copyright$ Università degli Studi di Milano
}
\end{center}
\vspace{1cm}
\LARGE
\setstretch{1.5}
\renewcommand{\thefootnote}{\alph{footnote}}
\noindent The following thesis was deposited on 25 June 2024 and defended on 09 July 2024 as part of the requirements for the Master's degree in Physics at the Università degli Studi di Milano. \\
\vspace{-0.05cm} \\
\noindent {\bfseries Title:} \\Charged and Rotating Black Holes in a Melvin-swirling Universe \\
\vspace{-0.05cm} \\
\noindent {\bfseries Candidate:} Andrea Di Pinto\footnote{andrea.dipinto@studenti.unimi.it, andreadipinto.physics@gmail.com} \\
\noindent{\bfseries Supervisor:} Prof.~Silke Klemm \\
\noindent{\bfseries Cosupervisor:} Dr.~Adriano Viganò
\renewcommand{\thefootnote}{\arabic{footnote}}
\setcounter{footnote}{0}
\restoregeometry
\endgroup}

\clearpage
\thispagestyle{empty}
\thispagestyle{empty}
{\begingroup
\fontsize{12}{12}\selectfont
\newgeometry{
top       = 2.5cm,
bottom = 1.25cm,
left       = 2.5cm,
right     = 2.5cm}
\begin{figure}[htp]
\includegraphics[scale=0.1]{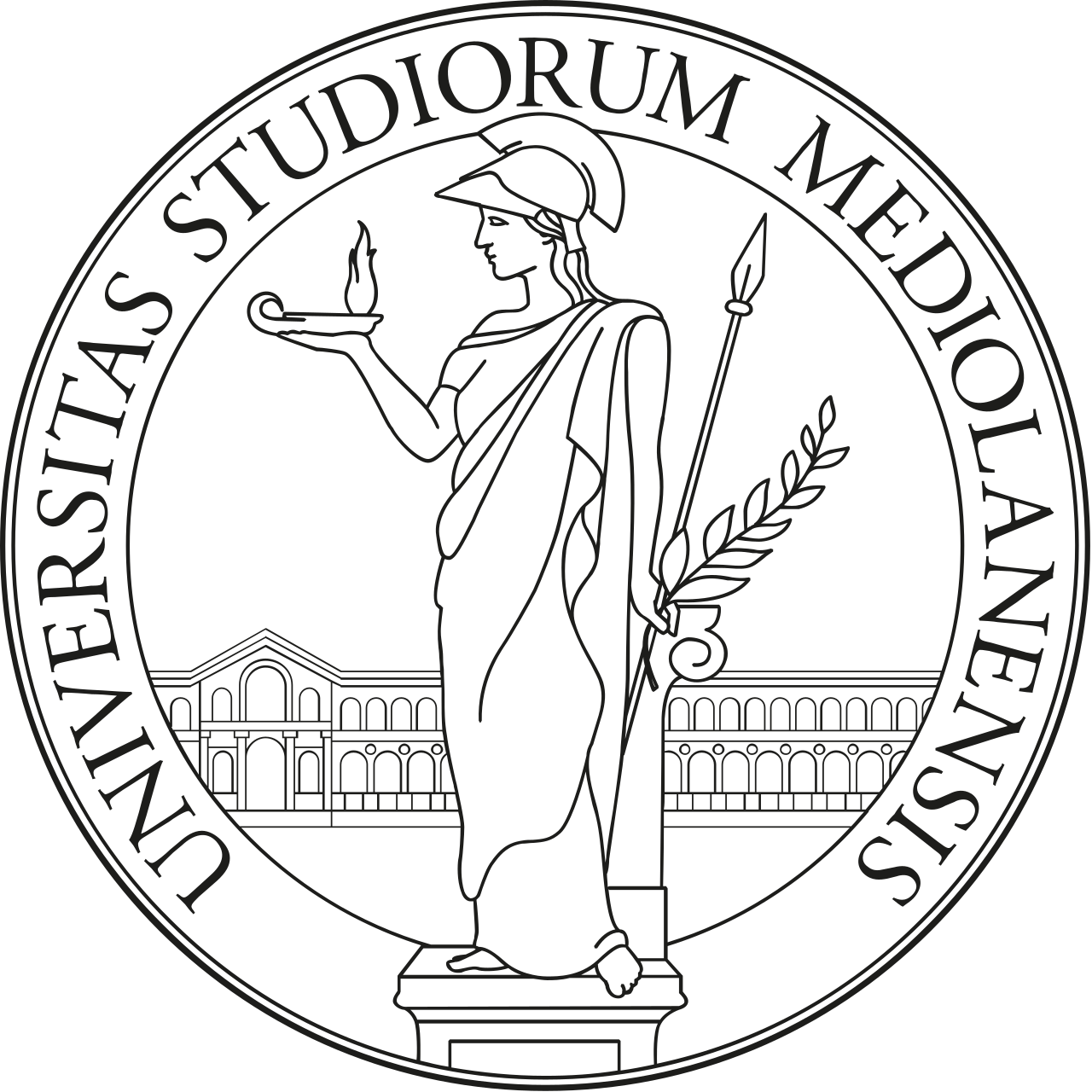}
\centering
\end{figure}
\huge
\vspace{-0.8cm}
\begin{center}
\setstretch{1.4}
\textbf{UNIVERSITÀ DEGLI STUDI DI MILANO} \\
FACOLTÀ DI SCIENZE E TECNOLOGIE \\
Corso di Laurea Magistrale in Fisica \\
\vspace{2cm}
\textbf{
CHARGED AND ROTATING} \\
\setstretch{1.25}
\textbf{
BLACK HOLES \\ IN A MELVIN-SWIRLING UNIVERSE
}
\end{center}
\vspace{2.5cm}
\LARGE
\textbf{Relatrice:} \hfill  \textbf{Candidato:}\\
Prof.ssa Silke Klemm \hfill  Andrea Di Pinto \\
\null \hfill Matricola 981352\\
\textbf{Correlatore:}  \\
Dott. \!Adriano Viganò
\vspace{2cm}
\begin{center}
\noindent \makebox[\linewidth]{\rule{\textwidth}{0.75pt}}
\vfill Anno Accademico 2023/2024
\end{center}
\restoregeometry
\endgroup}
\clearpage
\pagestyle{empty}
\null
\null\vfil
\begin{LARGE}
\begin{flushright}
{ \it To my parents} 
\end{flushright}
\end{LARGE}
\vfil
\newcommand{\conventionindent}{\vspace{0.5cm} \null \hspace{0.5 cm}}
\newcommand{\conventionspacing}{\vspace{1cm} \newline}
\newcommand{\conventionhalfspacing}{\vspace{0.6cm} \newline}
\clearpage
\thispagestyle{empty}
\section*{Notations and conventions}
\label{Notations}
 {\bfseries \large Planck Natural Units:} \vspace{0.5cm} \\
\conventionindent {\it Speed of light:} $ c = 1$ \\
\conventionindent {\it Universal gravitational constant:} $G = 1$ \\
\conventionindent {\it Reduced Planck constant:} $ \hbar = 1 $ \\
\conventionindent {\it Boltzmann constant:} $ k_B = 1 $
\vspace{0.125cm}
\newline
{\bfseries \large Metric Signature:} $\bigl(-,+,+,+\bigr)$
\conventionhalfspacing
{\bfseries \large World Indices:} Greek Letters ($\mu, \nu, \rho, \dots$)
\conventionhalfspacing
{\bfseries \large Tangent Space Indices:} Latin Letters ($a, b, c, \dots$)
\conventionhalfspacing
{\bfseries \large Einstein Summation Convention:} \vspace{0.25cm} \\
\conventionindent Repeated indices are implicitly summed over (e.g. $\displaystyle V^{\mu}V_{\mu}  \coloneqq \sum_{\mu = 0}^{\mu =3} V^{\mu} V_{\mu} $)
\vspace{-0.25cm}
\newline
{\bfseries \large Symmetrization:} \vspace{0.25cm} \\
\conventionindent ${T^{\sigma}}_{(\mu_{1} \dots \mu_{n}) \rho}  \coloneqq \frac{1}{n!} \bigl( {T^{\sigma}}_{\mu_{1} \dots \mu_{n} \rho} + \text{sum over permutations of indices } \mu_{1} \dots \mu_{n} \bigr)$
\newline
{\bfseries \large Antisymmetrization:} \vspace{0.25cm} \\
\conventionindent ${T^{\sigma}}_{[\mu_{1} \dots \mu_{n}] \rho}  \coloneqq \frac{1}{n!} \bigl( {T^{\sigma}}_{\mu_{1} \dots \mu_{n} \rho} + \text{alternating sum over permutations of indices } \mu_{1} \dots \mu_{n} \bigr)$
\newline
{\bfseries \large Gamma Matrices:} \vspace{0.25cm} \\
\conventionindent $\{ \gamma_{a}, \gamma_{b} \}  = 2 \eta_{a b}\, \mathds{1}_{4} $ \\
\conventionindent $\gamma_{a_{1} \dots a_{n}} \coloneqq \gamma_{[a_{1}} \dots \gamma_{a_{n}]}$
\clearpage
\thispagestyle{empty}
{\hypersetup{linkcolor=black}
\tableofcontents
}
\clearpage
\pagenumbering{arabic}
\setcounter{page}{1}
\pagestyle{fancy}
\fancyhf{}
\fancyhead[LE, RO]{\bfseries \large \thepage}
\fancyhead[RE, LO]{\nouppercase{\it Introduction}}
\addcontentsline{toc}{section}{Introduction}
\section*{Introduction}
\label{Introduction}
\vspace{0.25cm}
\setlength{\epigraphwidth}{0.5\textwidth}
\epigraph{When I ask myself what are the great things, looking back in history, that we got from the era of the Renaissance, it’s the great art, the great music, the science insights of Leonardo da Vinci. \\
Two hundred years from now, when you ask what are the great things that came from this era, I think it’s going to be an understanding of the universe around us.}%
{Kip Thorne}

\vspace{1.5cm}

\noindent In 1915, Albert Einstein published the theory of General Relativity~\cite{Einstein:GR, Einstein:GR2}, which represents one of the most profound intellectual breakthroughs in the history of physics and our current best description of gravity.

Before Einstein, the leading theory of gravity was Isaac Newton's law of universal gravitation. In this framework gravity is seen as a long-range attractive force, acting between any two massive objects, directly proportional to the product of their masses and inversely proportional to the square of their distance. 

On the other hand, in General Relativity, thanks to the sophisticated mathematical formalism of differential geometry, developed by Bernhard Riemann and other mathematicians in the 19th century~\cite{Riemann}, Einstein elevated the concept of spacetime given in Special Relativity~\cite{Einstein:SR}, in which it is seen as a \emph{flat} and \emph{immutable} object, to a \emph{dynamic} and possibly \emph{curved} description, where gravity emerges not as a force but rather as a manifestation of the curvature of spacetime, due to the presence of density and flux of energy and momentum. Therefore, the main essence of this theory can be eloquently captured in the words of the renowned physicist John Archibald Wheeler: “\emph{Spacetime tells matter how to move; matter tells spacetime how to curve}”~\cite{Wheeler1, Wheeler2}. In other words, matter (or energy, or mass) moves according to the geometry of spacetime where it is located, while simultaneously, the presence of that matter (or energy, or mass) alters the curvature of spacetime everywhere.

\noindent General Relativity has undergone extensive testing across various scales, consistently demonstrating an excellent agreement with experimental observations. From its first successes in the so-called “\emph{Three Classical Tests of General Relativity}”, namely reproducing the perihelion precession of Mercury's orbit~\cite{Clemence}, predicting the deflection of light by the Sun~\cite{Eddington}, and the gravitational redshift of light~\cite{Pound-Rebka}; to the remarkable foresight of the recently detected gravitational waves~\cite{LIGOScientific-grav-waves}. Among all predictions, black holes~\cite{Schwarzschild} are arguably the most intriguing: they are regions of spacetime where the gravitational field is so intense that nothing can escape from it, not even light.
\newline

\noindent The primary object of this thesis is to analytically construct a new particular type of black hole, which can be called the \emph{dyonic Kerr-Newman black hole in a Melvin-swirling universe}, that corresponds to a rotating and electromagnetically charged black hole, i.e. a (dyonic) Kerr-Newman black hole~\cite{Kerr-Newman1, Kerr-Newman2}, embedded in a universe that itself is both rotating (\emph{swirling}~\cite{Swirling}) and permeated by a uniform magnetic field (\emph{Melvin}~\cite{Melvin}). 

Starting from this new solution, we will also obtain other new black holes, corresponding to its non-charged and non-rotating sub-cases, namely the \emph{Kerr black hole in a Melvin-swirling universe} and the \emph{dyonic Reissner-Nordstr\"om black hole in a Melvin-swirling universe}.
\newline

\noindent More precisely, we will begin by introducing the basic concepts of General Relativity in Sec.~\ref{sec:GR}, and some advanced ones in Sec.~\ref{sec:GR-Advanced}, following the most common textbooks~\cite{Wheeler2, Carroll, Wald}.

Subsequently, Sec.~\ref{sec:Ernst} will consist in explaining the \emph{Ernst formalism}, a solution-generating technique necessary to construct the desired black hole, where we will follow the original Ernst's papers~\cite{ErnstI, ErnstII}, as done in~\cite{Vigano-Thesis, Martelli-Thesis}. This Ernst formalism is then applied in Sec.~\ref{sec:Kerr-Newman-Melvin-Swirling} in order to obtain the aforementioned new solutions. 

In Sec.~\ref{sec:Melvin-Swirling-Universe} we will discuss the physical interpretation of the Melvin-swirling universe, where we will also add a cosmological constant to this background. Following that, in Sec.~\ref{sec:Properties} we will study the physical and mathematical properties of the new black hole and of all its sub-cases. Finally, Sec.~\ref{sec:Sugra} is dedicated to discussing a possible supersymmetric extension of these spacetimes in Supergravity, using~\cite{DallAgata, Romans, Klemm-SusyPD, Klemm-aDs, Ortin} as references.

\vfill
\renewcommand{\sectionmark}[1]{\markright{\it \thesection \hspace{0.25cm} \ #1}}
\renewcommand{\subsectionmark}[1]{\markright{\it \thesubsection \hspace{0.25cm} \ #1}}
\fancyhead[LE,RO]{\bfseries \large \thepage}
\fancyhead[RE,LO]{\nouppercase{\rightmark}}
\titleformat{\section}[display]
  {\normalfont}
  {\hspace{0.25 em} \Large \it CHAPTER \textsl{\thesection}}
  {0pt}  {\bfseries \centering \titlerule[0.8pt]\vspace{3mm}\Huge }
  [{\vspace{3mm}\titlerule[0.8pt]\vspace{10mm}}]
\titleformat{\subsection}{\Large\bfseries}{\thesubsection}{0.5em}{}[\vspace{2.5mm}]
\titleformat{\subsubsection}{\large\bfseries}{\thesubsubsection}{0.5em}{}[\vspace{2.5mm}]
\clearpage
\section{Road to General Relativity}
\label{sec:GR}
\vspace{0.5cm}
\subsection{Basic Notions of (pseudo-)Riemannian Geometry} \label{subsec:Riemannian-geometry}
The fundamental object in General Relativity is the {\bfseries metric tensor} $g_{\mu \nu}$, a $(0,2)$ symmetric tensor ($g_{\mu \nu} = g_{\nu \mu}$), taken to be non-degenerate (the determinant $g \coloneqq |g_{\mu \nu}|$ never vanishes), that represents a generalization of the dot product of ordinary Euclidean space.

The metric captures all the causal and geometric structure of spacetime; some of its main purposes consist of being utilized to lower and raise (with the inverse metric $g^{\mu \nu}$) indices on tensors
\begin{subequations}
\begin{align}
g_{\mu \rho_{i}} T^{\rho_{1} \dots \rho_{i} \dots \rho_{n}}{}_{\sigma_{1} \dots \sigma_{m}} & = T^{\rho_{1} \dots}{}_{\mu}{}^{\dots \rho_{n}}{}_{\sigma_{1} \dots \sigma_{m}} \,, \\
g^{\mu \sigma_{i}} T^{\rho_{1} \dots \rho_{n}}{}_{\sigma_{1} \dots \sigma_{i} \dots \sigma_{m}} & = T^{\rho_{1} \dots \rho_{n}}{}_{\sigma_{1} \dots}{}^{\mu}{}_{\dots \sigma_{m}} \,,
\end{align}
\end{subequations}
from which it is possible to define the norm of any vector field
\beq
|V^{\mu}|^2=V^{\mu}V_{\mu}=g_{\mu \nu} V^{\mu}V^{\nu} \,,
\eeq
and thus to determine the {\bfseries line element}, which is the squared distance $ds^2$ between two points infinitesimally close to each other
\beq
\label{basic-line-element}
ds^2=g_{\mu \nu} dx^{\mu}dx^{\nu} \,,
\eeq
where $dx^{\mu}$ represents the dual basis of the coordinates $x^{\mu}$.

If the metric is not positive-definite, such as in General Relativity, the norm of a vector may not be positive either, depending on the sign of this number, vectors are:
\begin{subequations}
\begin{align}
\normalfont{timelike: \quad} &|V^{\mu}|^2<0 \,,\\
\normalfont{lightlike: \quad} &|V^{\mu}|^2=0 \,,\\
\normalfont{spacelike: \quad} &|V^{\mu}|^2>0 \,.
\end{align}
\end{subequations}
\clearpage
\noindent The partial derivative $\partial$ is a good tensor operator only if the spacetime is flat; in contrast, on curved spacetimes, it is necessary to use the so-called {\bfseries covariant derivative} (or {\bfseries connection}) $\tilde{\nabla}$, whose action on a generic $(n,m)$ tensor $T$ is:
\begin{align}
\begin{split}
\label{cov-der}
\tilde{\nabla}_{\mu}{T^{\rho_{1}\dots \rho_{n}}}_{\sigma_{1} \dots \sigma_{m}} = \partial_{\mu}{T^{\rho_{1}\dots \rho_{n}}}_{\sigma_{1} \dots \sigma_{m}} & + \tilde{\Gamma}{}^{\rho_{1}}{}_{\mu \lambda}\, {T^{\lambda \dots \rho_{n}}}_{\sigma_{1} \dots \sigma_{m}} + \, \dots \, + \tilde{\Gamma}{}^{\rho_{n}}{}_{\mu \lambda}\, {T^{\rho_{1} \dots \lambda}}_{\sigma_{1} \dots \sigma_{m}} \\ 
& -  \tilde{\Gamma}{}^{\lambda}{}_{\mu \sigma_{1}} {T^{\rho_{1} \dots \rho_{n}}}_{\lambda \dots \sigma_{m}} \, - \, \dots \, -  \tilde{\Gamma}{}^{\lambda}{}_{\mu \sigma_{m}} {T^{\rho_{1} \dots \rho_{n}}}_{\sigma_{1} \dots \lambda} \,,
\end{split}
\end{align}
where $ \tilde{\Gamma}{}^{\lambda}{}_{\mu \nu}$ are the {\bfseries connection coefficients}, which are \emph{not} the components of a tensor, but serve to correct the non-tensorial behavior of the partial derivative in such a way that the result becomes covariant.

Starting from the connection coefficients it is possible to construct two important tensors: the {\bfseries torsion tensor}
\beq
\label{torsion-tensor}
T_{\mu \nu}{}^{\lambda}  =  \tilde{\Gamma}{}^{\lambda}{}_{\mu \nu} -  \tilde{\Gamma}{}^{\lambda}{}_{\nu \mu} = 2 \,  \tilde{\Gamma}{}^{\lambda}{}_{[\mu \nu]} \,,
\eeq
and the {\bfseries curvature tensor}, also called the {\bfseries Riemann tensor},
\beq
\label{Riemann-tensor}
{R^{\rho}}_{\sigma \mu \nu}=\partial_{\mu}  \tilde{\Gamma}{}^{\rho}{}_{\nu \sigma} - \partial_{\nu}  \tilde{\Gamma}{}^{\rho}{}_{\mu \sigma}  +  \tilde{\Gamma}{}^{\rho}{}_{\mu \lambda} \tilde{\Gamma}{}^{\lambda}{}_{\nu \sigma} - \tilde{\Gamma}{}^{\rho}{}_{\nu \lambda}  \tilde{\Gamma}{}^{\lambda}{}_{\mu \sigma} \,,
\eeq
from which it can be verified that the commutator of two covariant derivatives satisfies:
\beq
\label{commutator-cov-dev}
\bigl[\tilde{\nabla}_{\mu},\tilde{\nabla}_{\nu}\bigr]V^{\rho}={R^{\rho}}_{\sigma \mu \nu}V^{\sigma} + {T_{\mu \nu}}^{\lambda}\tilde{\nabla}_{\lambda}V^{\rho} \,.
\eeq
The covariant derivative has another important application, indeed it can be used to define the {\bfseries parallel transport}, which is the generalization on a curved spacetime of the concept \emph{“keeping a vector constant as we move it along a path”}.
Given a $(m,n)$ tensor $T$ and a curve $x^{\mu}(\lambda)$, the definition of the parallel transport of $T$ along $x^{\mu}(\lambda)$ is then the requirement:
\beq
\label{equation-of-parallel-transport}
\frac{dx^{\nu}}{d\lambda}\tilde{\nabla}_{\nu}{T^{\rho_{1}\dots \rho_{n}}}_{\sigma_{1} \dots \sigma_{m}}=0 \,,
\eeq
also known as the {\bfseries equation of parallel transport}.
In particular, if $T$ is the tangent vector $\frac{dx^{\mu}}{d\lambda}$ to the path $x^{\mu}(\lambda)$, then Eq.~\eqref{equation-of-parallel-transport} becomes the {\bfseries geodesic equation}
\beq
\label{geodesic-equation}
\frac{d^2x^{\mu}}{d^2\lambda} + \tilde{\Gamma}{}^{\mu}{}_{\rho \sigma}\frac{dx^{\rho}}{d\lambda}\frac{dx^{\sigma}}{d\lambda}=0 \,,
\eeq
which describes the paths of the shortest distance between two points, called {\bfseries geodesics}.

Alternatively, by denoting $u^{\mu} = \frac{dx^{\mu}}{d\lambda}$ as the tangent vector to the geodesic $x^{\mu}(\lambda)$, this equation is also typically expressed as
\beq
\label{geodesic-equation-velocity}
\tilde{\nabla}_{u}u^{\mu}=u^{\nu}\tilde{\nabla}_{\nu}u^{\mu}=0 \,.
\eeq
\clearpage
\noindent With the concept of parallel transport, torsion can be interpreted as a description of the twisting of tangent spaces upon parallel transport along a curve, while curvature describes how tangent spaces roll along a path. Finally, the commutator of two covariant derivatives~\eqref{commutator-cov-dev} measures the difference between parallel transporting a vector (tensor) first in one direction and then the other, compared to the reverse transport order.

Moreover, it is always possible to define a unique connection on a manifold with a metric. Indeed, the {\bfseries Fundamental Theorem of Riemannian Geometry} states:

\begin{mdframed}[style=mystyle]
{\it Let $\mathcal{M}$ be a differentiable manifold with a metric tensor $g_{\mu \nu}$. Then there always exists a unique connection $\nabla$, called the {\bfseries Levi-Civita connection}, which satisfies the following conditions:
\begin{align}
& \bullet \quad \text{Metric compatibility: } \nabla_{\rho}g_{\mu \nu}  = 0 \,, \label{metric-compatibility}  \\
& \bullet \quad  \text{Torsion free: } {T_{\mu \nu}}^{\lambda} = 0 \,. \label{torsion-free}
\end{align}}
\end{mdframed}

\noindent The metric compatibility condition~\eqref{metric-compatibility} ensures the preservation, during parallel transport, of the norm of vectors, the sense of orthogonality, and so on. It also implies that a metric-compatible covariant derivative commutes with the raising and lowering of indices
\beq
V^{\mu}\nabla_{\rho}T_{\sigma \mu}=V_{\mu}\nabla_{\rho}{T_{\sigma}}^{\mu} \,.
\eeq
From the definition of the covariant derivative of a tensor~\eqref{cov-der} and the metric compatibility condition~\eqref{metric-compatibility}, it is straightforward to verify that the associated connection coefficients $\Gamma^{\nu}{}_{\mu \lambda}$ of the Levi-Civita connection, called the {\bfseries Christoffel symbols}, are entirely determined by the metric tensor via the relation:
\beq
\label{Christoffel-simbols}
\Gamma^{\nu}{}_{\mu \lambda}=\frac{1}{2} g^{\nu \rho} \Bigl(\partial_{\mu}g_{\rho \lambda}+\partial_{\lambda}g_{\rho \mu}-\partial_{\rho}g_{\mu \lambda}\Bigr) \,,
\eeq
whose symmetry $\bigl(\Gamma^{\lambda}{}_{\mu \nu} = \Gamma^{\lambda}{}_{(\mu \nu)}\bigr)$ is a direct consequence of the definition of the torsion tensor~\eqref{torsion-tensor} and the torsion-free requirement~\eqref{torsion-free} of the connection.

Starting from the Riemann tensor~\eqref{Riemann-tensor} it is possible to define several other significant objects. The most notable among these are: the {\bfseries Ricci tensor}
\beq
\label{Ricci-tensor}
R_{\mu \nu}={R^{\lambda}}_{\mu \lambda \nu} \,,
\eeq
the {\bfseries Ricci scalar} (also know as the {\bfseries scalar curvature})
\beq
\label{Ricci-scalar}
R=g^{\mu \nu} R_{\mu \nu} \,,
\eeq
the {\bfseries Einstein tensor}
\beq
\label{Einstein-tensor}
G_{\mu\nu}= R_{\mu \nu}-\frac{1}{2}R\, g_{\mu \nu} \,,
\eeq
and lastly, the {\bfseries Weyl tensor}, which, for a spacetime of dimension $d$, takes the form:
\beq
\label{weyl-tensor}
C_{\mu \nu \rho \sigma} = R_{\mu \nu \rho \sigma} - \frac{2}{(d-2)}\bigl(g_{\mu [\rho}R_{\sigma]\nu} - g_{\nu [\rho}R_{\sigma] \mu}\bigr) + \frac{2}{(d-2)(d-1)}R\,g_{\mu [\rho}g_{\sigma] \nu} \,.
\eeq
The principal properties of these entities can be summarized as follows:
\begin{subequations}
\begin{align}
R_{\mu [\nu \rho \sigma]} & = 0 \,, \\
\nabla_{[\lambda} R_{\mu \nu] \rho \sigma} & = 0 \,, \\
R_{\mu \nu \rho \sigma} = R_{\rho \sigma \mu \nu} & = - R_{\nu \mu \rho \sigma} = - R_{\mu \nu \sigma \rho} \,, \\
C_{\mu [\nu \rho \sigma]} & = 0 \,, \\
C^{\mu}{}_{ \nu \mu \sigma} & = 0 \,, \\
C_{\mu \nu \rho \sigma} = C_{\rho \sigma \mu \nu} & = - C_{\nu \mu \rho \sigma} = - C_{\mu \nu \sigma \rho} \,, \\
R_{\mu \nu} & = R_{\nu \mu} \,, \\
\nabla_{\mu}R^{\mu}{}_{ \nu} & = \frac{1}{2} \nabla_{\nu}R \,, \\
G_{\mu \nu} & = G_{\nu \mu} \,, \\
\nabla_{\mu}G^{\mu \nu} & = 0 \label{Einstein-tensor-divergence},
\end{align}
\end{subequations}
of these, the first two are respectively referred to as the {\bfseries first} and {\bfseries second Bianchi identity}. 

Moreover, the same name is sometimes also used for those that are derived starting from a Bianchi identity; in particular, the one involving the four-divergence of the Einstein tensor~\eqref{Einstein-tensor-divergence} is also often referred to as the second Bianchi identity.


\subsection{Einstein Field Equations}  \label{subsec:Einstein-equations}
The most important equation of General Relativity is represented by the {\bfseries Einstein field equations}, or just {\bfseries Einstein equations}, for which, given the importance, the gravitation constant $G$ and the speed of light $c$ will just be temporarily restored:
\beq
\label{Einstein-equations-complete}
G_{\mu\nu}+\Lambda g_{\mu\nu}=\frac{8\pi G}{c^4} T_{\mu\nu} \,,
\eeq
wherein $G_{\mu\nu}$ is Einstein tensor~\eqref{Einstein-tensor}, $\Lambda$ is the {\bfseries cosmological constant}, and $T_{\mu\nu}$ is the {\bfseries energy-momentum tensor}, which describes the density and flux of energy and momentum within the spacetime.

\clearpage
\noindent Therefore, the core concept of General Relativity lies in the physical interpretation of the Einstein field equations: the curvature of spacetime is fundamentally tied to the distribution of energy and momentum.

These equations can be considered as second-order partial differential equations (PDEs) for the metric $g_{\mu \nu}$. Indeed, the Ricci tensor~\eqref{Ricci-tensor} and the scalar curvature~\eqref{Ricci-scalar} result from contractions of the Riemann tensor~\eqref{Riemann-tensor}, which involves derivatives and products of the Christoffel symbols~\eqref{Christoffel-simbols}, which in turn are a combination of the inverse and derivatives of the metric. Moreover, the energy-momentum tensor $T_{\mu \nu}$ typically also involves the metric. 

The Einstein equations are also non-linear, meaning that two known solutions cannot be superposed to form a third.  Nevertheless, there is a physical reason for this, which is that, in General Relativity, the gravitational field couples to itself. 

Moreover, being second-order non-linear PDEs, finding a solution without simplifying assumptions is usually quite a difficult task. 

Since both sides of these equations are symmetric two-index tensors, there are apparently only ten independent equations, which appear to correspond precisely to the ten unknown components of the metric. However, the Bianchi identity~\eqref{Einstein-tensor-divergence} imposes four additional constraints, thus leaving only six truly independent equations. This aligns well with the notion of covariance: if a metric is a solution to the Einstein equations in one coordinate system then it should also be a solution in any other coordinate system. Hence, there exist four unphysical degrees of freedom in $g_{\mu \nu}$, corresponding to the freedom to choose any coordinate system, and Einstein equations solely constrain the six coordinate-independent degrees of freedom.

Furthermore, the Einstein equations automatically imply the energy-momentum conservation $\nabla_{\mu}T^{\mu\nu}=0$, this is straightforwardly derived from the requirement of metric compatibility~\eqref{metric-compatibility} and the fact that the four-diverge of the Einstein tensor is always zero~\eqref{Einstein-tensor-divergence}.

\subsubsection{Trace-reversed Einstein Field Equations}  \label{subsubsec:Einstein-equations-B}
There is another useful way of expressing the Einstein equations~\eqref{Einstein-equations-complete}. Indeed, using the definition of the Einstein tensor~\eqref{Einstein-tensor} and taking the trace with respect to the metric of both sides of these equations, one obtains:
\beq
\label{Einstein-equations-trace-d}
\frac{\bigl(2 - d \bigr)}{2} R + d \, \Lambda = 8 \pi T \,,
\eeq
where it has been used that the trace of the metric tensor is equal to the spacetime dimension, $g^{\mu \nu} g_{\mu \nu} = d$, and where the trace of the stress-energy tensor as has been defined as $T \coloneqq T^{\mu}{}_{\mu}$. 
\clearpage
\noindent Solving for the Ricci scalar $R$ and substituting into the Einstein equation, the following equivalent form of Einstein equations is obtained:
\beq
\label{Ricci-Einstein-equations-d}
R_{\mu \nu} - \frac{2 \, \Lambda}{\bigl(d - 2 \bigr)} g_{\mu \nu} = 8 \pi \biggl[\, T_{\mu \nu} - \frac {T}{\bigl(d - 2\bigr)} g_{\mu \nu}\biggr] \,,
\eeq
which are sometimes called the {\bfseries trace-reversed Einstein (field) equations}.

In particular, for $d=4$, these reduce to:
\beq
\label{Ricci-Einstein-equations-d4}
R_{\mu \nu}  - \Lambda g_{\mu \nu} = 8\pi \Bigl(T_{\mu \nu} - \frac{1}{2} T g_{\mu \nu}\Bigr) \,.
\eeq

\subsection{General Relativity} \label{subsec:General-Relativity}
With all the concepts introduced in this chapter, it is possible to ultimately summarize the entire content of General Relativity as follows:
\begin{mdframed}[style=mystyle]
{\emph{{\bfseries Spacetime is a four-dimensional differentiable manifold equipped with a Lorentzian metric $\boldsymbol{g_{\mu \nu}}$. \\
\newline
The curvature of spacetime is fundamentally tied to the distribution of \, matter by the Einstein field equations~\eqref{Einstein-equations-complete}.} \\
\null \hspace{0.5cm} (Matter tells spacetime how to curve) \\
\newline
{\bfseries In the curved manifold, free particles move along geodesics~\eqref{geodesic-equation}.} \\
\null \hspace{0.5cm}  (Spacetime tells matter how to move)}}
\end{mdframed}
 
\clearpage
\section{Advanced Topics in General Relativity}
\label{sec:GR-Advanced}
\vspace{-0.5cm}
\subsection{Einstein-Hilbert Action} \label{subsec:Einstein-Hilbert-Action}
\vspace{-0.2cm}
As understood by Hilbert~\cite{Hilbert:Action}, Einstein equations~\eqref{Einstein-equations-complete} can be completely derived by means of an action principle, using the {\bfseries Einstein-Hilbert action}
\beq
\label{EH-action}
\Ia_{E.H.} = \Ia_{matter}+\frac{1}{16 \pi}\int d^4 x \sqrt{-g} \Bigl(R - 2 \Lambda\Bigr) \,.
\eeq
Indeed, to recover a physical law through the stationary-action principle, the variation of this action with respect to the inverse metric must be zero, thus yielding
\beq
0 = \delta \Ia_{E.H.}=\int d^4x \frac{\delta \Ia_{matter}}{\delta g^{\mu \nu}} \delta g^{\mu \nu} + \frac{1}{16 \pi}\int d^4 x \sqrt{-g} \Bigl(G_{\mu \nu} + \Lambda g_{\mu \nu}\Bigr) \delta g^{\mu \nu} \,.
\eeq
Since this equation should hold for any variation $\delta g^{\mu \nu}$, it is straightforward that it reduces precisely to the Einstein equations~\eqref{Einstein-equations-complete} if the energy-momentum tensor is defined as:
\beq
\label{energy-momentum-tensor}
T_{\mu\nu} \coloneq - \frac{2}{\sqrt{-g}}\frac{\delta \Ia_{matter}}{\delta g^{\mu \nu}} \,.
\eeq
Therefore, in principle, given the action that describes the theory, it is always possible to recover the energy-momentum tensor~\eqref{energy-momentum-tensor}. Thus, through the integration of the Einstein equations, the metric tensor $g_{\mu \nu}$.

To be more precise, the matter action is also dependent on the matter fields $\Ia_{matter}[g_{\mu \nu}, \psi^{i}]$. This means that there are additional field equations of motion, resulting from the variation of this action with respect to the matter fields $\psi^{i}$.
\vspace{-0.2cm}
\subsubsection{Einstein-Maxwell Equations} \label{subsec:Einstein-Maxwell equations}
In relativity, the electric scalar potential and the magnetic vector potential are combined into a single differential $1-$form, known as the {\bfseries electromagnetic four-potential} $A_{\mu}$, from which it is possible to construct a differential $2-$form, and thus a totally antisymmetric $(0,2)$ tensor
\beq
\label{Faraday-tensor}
F_{\mu \nu}=\partial_{\mu}A_{\nu}-\partial_{\nu}A_{\mu} \,,
\eeq 
referred to as the {\bfseries Faraday Tensor}.

\noindent If no external currents are considered, the action of the electromagnetic field is given by
\beq
\label{EM-action}
\Ia_{EM}=-\frac{1}{16 \pi}\int d^4 x \sqrt{-g} F_{\mu \nu}F^{\mu \nu} \,,
\eeq
therefore, by means of Eq.~\eqref{energy-momentum-tensor}, the electromagnetic energy-momentum tensor is
\beq
\label{EM-energy-momentum-tensor}
T_{EM}^{\mu\nu} = \frac{1}{4 \pi}\Bigl(F^{\mu \lambda}{F^{\nu}}_{\lambda}-\frac{1}{4}g^{\mu \nu}F_{\rho \sigma}F^{\rho \sigma}\Bigr) \,,
\eeq
and the Einstein equations~\eqref{Einstein-equations-complete} become
\beq
\label{Einstein-equations-Maxwell}
G^{\mu\nu} + \Lambda g^{\mu\nu} = 8\pi T_{EM}^{\mu\nu} \,,
\eeq
or equivalently, being the electromagnetic energy-momentum tensor~\eqref{EM-energy-momentum-tensor} traceless, \mbox{$T_{EM}\hspace{-1pt} = 0$}, in the trace-reversed~\eqref{Ricci-Einstein-equations-d4} form:
\beq
\label{Ricci-Einstein-equations-Maxwell}
R^{\mu \nu}  - \Lambda g^{\mu \nu} = 8\pi T_{EM}^{\mu \nu} \,.
\eeq
Analogously, the variation of the electromagnetic action~\eqref{EM-action}, with respect to the four-potential $A_{\mu}$, leads to the inhomogeneous {\bfseries Maxwell equations}
\beq
\label{Maxwell-equations}
\nabla_{\mu}F^{\mu \nu}=0 \,.
\eeq
A priori one should also consider the homogeneous Maxwell equations, which take the form:
\beq
\label{homogeneus-Maxwell-equations}
\nabla_{[\mu}F_{\rho \sigma]}=0 \,,
\eeq
however, these are always verified, given the definition of the Faraday tensor~\eqref{Faraday-tensor} as a differential $2-$form.
\clearpage
\subsection{Killing Vectors} \label{subsec:Killing Vectors}
Under an infinitesimal diffeomorphism $x'^{\mu} = x^{\mu} + \xi^{\mu}(x)$, it can be verified that the variation of the metric is given by
\beq
\label{metric-variaton}
\delta g_{\mu \nu} = -\bigl(\nabla_{\mu} \xi_{\nu} + \nabla_{\nu} \xi_{\mu} \bigr) \,.
\eeq
A four-vector $\xi^{\mu}$ that satisfies the condition $\delta g_{\mu \nu} = 0$, i.e.
\beq
\label{Killing-equation}
\nabla_{\mu} \xi_{\nu} + \nabla_{\nu} \xi_{\mu} = 0 \,,
\eeq
is called a {\bfseries Killing vector}, while this last equation is known as the {\bfseries Killing equation}.

Thus, in general, a Killing vector generates a one-parameter family of diffeomorphism, under which the metric remains invariant. Such a diffeomorphism is called an {\bfseries isometry}.

It can be proved that if a spacetime possesses a Killing vector $\xi^{\mu}$, it is always possible to find a coordinate system whose metric is independent of one of the coordinates $\tilde{x}$,  corresponding to the integral curve coordinate of that Killing vector, and for which $\xi^{\mu}=\partial_{\tilde{x}}$.

Moreover, Killing vectors imply conserved quantities associated with the motion of free particles. 
Indeed, if $x^{\mu}(\lambda)$ is a geodesic~\eqref{geodesic-equation} with tangent vector $u^{\mu} = \frac{d x^{\mu}}{d \lambda}$, and $\xi^{\mu}$ is a Killing vector, then
\beq
\label{cons-Killing}
u^{\nu}\nabla_{\nu}(\xi_{\mu}u^{\mu}) = u^{\nu}u^{\mu}\nabla_{\nu}\xi_{\mu} + \xi_{\mu}u^{\nu}\nabla_{\nu}u^{\mu} = 0 \,,
\eeq
where the first term is zero due to the antisymmetrical nature of the Killing equation~\eqref{Killing-equation}, while the second vanishes because of the assumption that $x^{\mu}(\lambda)$ is a geodesic~\eqref{geodesic-equation-velocity}. Thus, the quantity $\xi_{\mu}u^{\mu}$ remains conserved along the particle’s worldline. Since, by definition, the metric remains unchanged along the direction of a Killing vector, the physical interpretation of this conservation is that a free particle will not experience any “forces” in the direction given by a Killing vector, and therefore the component of the particle's momentum in that direction will be conserved.

Furthermore, it can be proved that, for any Killing vector $\xi$, the following useful relations hold:
\begin{subequations}
\begin{align}
\nabla_{\rho}\nabla_{\mu} \xi^{\nu} & =  R^{\nu}{}_{\mu \rho \sigma} \xi^{\sigma} \,, \label{Killing-prop-1}  \\
\nabla^{\nu}\nabla_{\nu} \xi^{\mu} & =  R^{\mu \nu}\xi_{\nu} \,, \label{Killing-prop-2} \\
\nabla_{\nu}\nabla_{\mu}\nabla^{\nu}\xi^{\mu}& = 0 \,. \label{Killing-prop-3}
\end{align}
\end{subequations}
\clearpage

\subsection{Spacetimes with Symmetries} \label{subsec:Stationary-Axisymmetric}

Since the presence of Killing vectors and symmetries simplifies the form of the metric it is useful to state the following definitions:

\begin{mdframed}[style=mystyle]
{\it $\bullet$\quad A spacetime is said to be {\bfseries stationary} if there exists a one-parameter group of isometries $\sigma_{t}$ whose orbits are timelike curves. Equivalently, a stationary spacetime is one which possesses a timelike Killing vector $k^{\mu}$.

\it \noindent $\bullet$\quad A spacetime is said to be {\bfseries axisymmetric} (axially symmetric) if there exists a one-parameter group of isometries $\chi_{\phi}$ whose orbits are closed spacelike curves, which implies the existence of a spacelike Killing vector $m^{\mu}$ whose integral curves are closed.}

\it \noindent $\bullet$\quad A spacetime is said to be stationary and axisymmetric if it possesses both these symmetries and if, in addition, the action of $\sigma_{t}$ and $\chi_{\phi}$ commutes:
\beq
\label{Commuting-Killing-isometries}
\sigma_{t}\circ \chi_{\phi} = \chi_{\phi}\circ \sigma_{t} \,.
\eeq
\it Equivalently, a stationary and axisymmetric is one which possesses both these symmetries and if, in addition, the Killing vectors $k^{\mu}$ and $m^{\mu}$ commute:
\beq
\label{Commuting-Killing-stationary-axisymmetric}
[k,m] = 0 \,.
\eeq

\it \noindent $\bullet$\quad A spacetime is said to be {\bfseries static} if it is stationary and if, in addition, there exists a spacelike hypersurface $\Sigma$ that is orthogonal to the orbits of the isometry.

\it \noindent $\bullet$\quad A spacetime is said to be {\bfseries spherically symmetric} if its isometry group contains a subgroup isomorphic to the group \normalfont{SO}$(3)$, and the orbits of this subgroup are two-dimensional spheres.
\end{mdframed}
\vspace{0.5cm}
\noindent It can be proved that, for a static spacetime, there always exists a coordinate transformation such that, in the new coordinates, the timelike Killing vector is $k=\partial_{t}$, with $t=x^{0}$ timelike coordinate, and where
\beq
\label{line-element-static}
ds^2_{static} = -f\bigr(x^{1}, x^{2}, x^{3}\bigr) dt^2 + \sum_{i, j =1}^{3} h_{i j}\bigr(x^{1}, x^{2}, x^{3}\bigr)dx^{i} dx^{j} \,.
\eeq
\clearpage
\noindent For a spherically symmetric spacetime the metric induces a metric on each orbit $2-$sphere, which, because of the rotational symmetry, must be a multiple of the metric of a unit $2-$sphere
\beq
\label{line-element-2-sphere}
ds^2_{unit-2-sphere} = \bigl(d \theta^2 + \sin^2\theta d\phi^2\bigr) \,,
\eeq
thus, if a spacetime is both static and spherically symmetric, and if the static Killing vector $k=\partial_{t}$ is unique, then $k^{\mu}$ must be orthogonal to the orbit $2-$spheres. Consequently, it is always possible to choose a set of coordinates $x^{\mu}=(t, r, \theta, \phi)$ for which it holds
\beq
\label{line-element-static-spherically}
ds^2_{static-spherically-symmetric} = -f\bigr(r) dt^2 + h(r)dr^2 + r^2 \bigl(d \theta^2 + \sin^2\theta d\phi^2\bigr) \,.
\eeq
Similarly, due to the definition of a stationary and axisymmetric spacetime, and in particular for the commutativity of the two Killing vectors~\eqref{Commuting-Killing-stationary-axisymmetric}, it holds that, for a spacetime of this type, it is always possible to choose a set of coordinates $x^{\mu}=(t, x^{1}, x^{2}, \phi)$, in such a way that $k = \partial_{t}$, $m=\partial_{\phi}$, and for which the metric has no dependence on the two coordinates $t$ and $\phi$. Thus, in these coordinates, the line element of  a stationary and axisymmetric spacetime takes the form
\beq
\label{line-element-stationary-axisymmetric}
ds^2_{stationary-axisymmetric} = g_{\mu \nu}(x^2, x^3)dx^{\mu}dx^{\nu} \,.
\eeq
Moreover, due to Eq.~\eqref{cons-Killing}, for a stationary and axisymmetric spacetime, it holds that along a geodesic with tangent vector $u^{\mu}$ are both constant the energy (per unit mass) $E$, associated with the Killing vector $k=\partial_{t}$, and the angular momentum (per unit mass) $L$, associated with $m=\partial_{\phi}$:
\begin{subequations}
\begin{align}
E &= g_{\mu \nu} u^{\mu} k^{\nu} \,, \label{Killing-energy}\\
L &= g_{\mu \nu} u^{\mu} m^{\nu} \,.\label{Killing-momentum}
\end{align}
\end{subequations}

\subsubsection{Special Stationary and Axisymmetric Spacetimes} \label{subsubsec:Special-Stationary-Axisymmetric}

The line element of a stationary and axisymmetric spacetime~\eqref{line-element-stationary-axisymmetric} can be further simplified if the following theorem is satisfied:

\begin{mdframed}[style=mystyle]
{\bfseries Theorem $\star$:} \\
{\it Let $k^{\mu}$ \it and $m^{\mu}$ \it be two commuting Killing fields such that: \\\normalfont{(}$\ast$\normalfont{)} $k_{[\mu}m_{\nu}\nabla_{\sigma}k_{\lambda]}$ \it and $k_{[\mu}m_{\nu}\nabla_{\sigma}m_{\lambda]}$ \it each vanishes at, at least, one point of the spacetime. \\
\normalfont{(}$\ast\ast$\normalfont{)} $k^{\mu}R_{\mu [\nu} k_{\sigma} m_{\lambda]} = m^{\mu}R_{\mu [\nu} k_{\sigma} m_{\lambda]} = 0$. \\
\it Then the $2-$planes orthogonal to the Killing fields are integrable.}
\end{mdframed}
\clearpage
\noindent In particular, ($\ast$) is satisfied whenever the spacetime is asymptotically flat since there must be a “rotation axis” on which $m^{\mu}$ vanishes; similarly ($\ast \ast$) is trivially satisfied for vacuum spacetimes without a cosmological constant, due to the trace-reversed Einstein equations~\eqref{Ricci-Einstein-equations-d4}, which yield $R_{\mu \nu} = T_{\mu \nu} = 0$. Moreover, ($\ast \ast$) can be proved to hold for spacetimes with specific energy-momentum tensors $T_{\mu \nu}$, such as that of a stationary and axisymmetric electromagnetic field~\cite{Carter}, or that of a perfect fluid with the four-velocity lying in the plane spanned by the Killing vectors $k^{\mu}$ and $m^{\mu}$.

The meaning of Theorem $\star$ is that, at each point, the two-dimensional subspaces of the tangent space spanned by the vectors orthogonal to $k^{\mu}$ and $m^{\mu}$ are integrable, thus tangent to two-dimensional surfaces. Hence, it is possible to choose coordinates $(x^1, x^2)$ in one of the orthogonal two-dimensional surfaces and then extend these coordinates to the rest of the spacetime by “transporting” them along the integral curves of $k^{\mu}$ and $m^{\mu}$.

Therefore, for a stationary and axisymmetric spacetime with commuting Killing vectors $k=\partial_{t}$ and $m=\partial_{\phi}$, it is always possible to choose a set of coordinates $(t, x^1, x^2, \phi)$ such that
\beq
\label{metric-stationary-axisymmetric}
g_{\mu \nu}=\begin{pmatrix} -V & 0 & 0 & W \\ 0 & g_{11} & g_{12} & 0 \\ 0 & g_{12} & g_{22} & 0 \\ W & 0 & 0 & X \end{pmatrix} \,,
\eeq
where $V = - k_{\mu}k^{\mu} = - g_{t t}$, $W = k_{\mu}m^{\mu} = g_{t \phi}$, and $X = m_{\mu}m^{\mu} = g_{\phi \phi}$, thus reducing the number of the non-vanishing independent metric components from ten to six.

In order to define one of the two-dimensional surface coordinates it is useful to introduce the quantity
\beq
\label{LWP-contidion}
\rho^2 = VX + W^2 \,,
\eeq
which corresponds to (minus) the determinant of the $t-\phi$ part of the metric. Under the assumption $\nabla_{\mu} \rho \neq 0$ it is possible to choose $x^1$ as $\rho$. 
Similarly, $x^2$ is chosen as a certain $z$ such that $\bigl(\nabla_{\mu} z \bigr)\bigl( \nabla^{\mu} \rho\bigr) = 0$, where this requirement of orthogonality is accomplished by setting $z$ as constant along the integral curves of $\nabla^{\mu} \rho$, requirement which uniquely determines $z$ up to $ z \rightarrow z’= f(z)$. 

These coordinates completely determine the metric of a stationary and axisymmetric spacetime that satisfies the hypothesis of Theorem $\star$, whose line element is then 
\beq
\label{LWP-0}
ds^2 = \frac{\rho^2}{V}d\phi^2 + \Omega^2 \bigl(d\rho^2 + \Sigma\, dz^2\bigr) - V \biggl[dt - \frac{W}{V} d \phi\biggr]^2 \,,
\eeq
and for which there are only four unknown components of the metric, corresponding to $V$, $W$, $\Sigma$, and $\Omega$, that are functions only of the two variables $\rho$ and $z$. 

\subsubsection{The Electric LWP Metric} \label{subsubsec:Electric-LWP}

The form of the metric given by Eq.~\eqref{LWP-0} can be furthermore simplified for spacetimes in a vacuum, $R_{\mu \nu} = 0$. Indeed, using Eq.~\eqref{Killing-prop-2} it is possible to compute the components of $R_{\mu \nu}$ in the plane spanned by the Killing vectors $k^{\mu}$ and $m^{\mu}$. Therefore, the equation
\beq
\label{eqn-D-D-rho-0}
0 = R^{tt} + R^{\phi \phi} = R^{\mu \nu} k_{\nu} \nabla_{\mu}t  + R^{\mu \nu} m_{\nu} \nabla_{\mu}\phi \,,
\eeq
yields
\beq
\label{eqn-D-D-rho}
D_{\mu}D^{\mu}\rho = 0 \,,
\eeq
where $D$ is the covariant derivative on the two-dimensional surface spanned by $\rho$ and $z$. 

This equation has the consequence that $\Sigma$ is a function of only the coordinate $z$. Hence, given the remaining degree of freedom on $z$ that was discussed above, it is possible to transform $z \rightarrow \int \Sigma^{\frac{1}{2}} dz$ and thereby set $\Sigma=1$. Finally, defining $\gamma = \frac{1}{2}\log\bigl(V\, \Omega^2\bigr)$, $f = V$ and $\omega = \frac{W}{V}$, the metric takes the remarkably simpler form
\beq
\label{Electric-LWP}
ds^2_{electric-LWP} = \frac{1}{f}\Bigl[\rho^2 d\phi^2 + e^{2\gamma} \bigl(d\rho^2 + dz^2\bigr)\Bigr] - f \bigl(dt - \omega d \phi\bigr)^2 \,,
\eeq
known as the {\bfseries electric Lewis-Weyl-Papapetrou (LWP) metric}~\cite{LWP1,LWP2,LWP3,LWP4}. Thus, the electric LWP metric~\eqref{Electric-LWP} represents the most general metric for stationary and axisymmetric spacetimes in a vacuum without a cosmological constant.

Actually, the condition $R_{\mu \nu} = 0$ is sufficient but not necessary for the electric LWP metric to be the most general for a stationary and axisymmetric spacetime. Indeed, this metric is also the most general for a spacetime of this type equipped with a stationary and axisymmetric electromagnetic field~\cite{Heusler}.

It can be noted that a flat spacetime corresponds to $V = 1$, $\omega = \gamma = 0$, for which the electric LWP metric reduces to the Minkowski metric, where the coordinates $(\rho,z,\phi)$ are the ordinary cylindrical coordinates. For this reason, the coordinates $(t,\rho,z,\phi)$ are called Weyl, or cylindrical, coordinates.

\subsubsection{The Magnetic LWP Metric}  \label{subsubsec:Magnetic-LWP}
Historically the electric LWP metric~\eqref{Electric-LWP} has always been called just as the LWP metric, without the \emph{electric} adjective. On the other hand, in recent years, it has been widely used another rewriting of this metric, dubbed as \emph{magnetic}~\cite{Vigano-Thesis, Swirling}, which is particularly useful in the Ernst formalism, that will be explained in Sec.~\ref{sec:Ernst} where it will also be given another interpretation on the use of the terms “magnetic” and “electric” for these metrics. 

\noindent At this stage, it is still possible to explain the terminology by drawing a parallel with the dualism between electric and magnetic charges. Indeed, it is possible to apply a “duality” transformation on the electric LWP metric~\eqref{Electric-LWP} by an analytical continuation, or discrete double-Wick rotation:
\beq
\label{double-Wick}
DW \coloneqq
\Bigl\{
t \to i\psi  \,, \quad \quad \\
\phi \to i \tau
\Bigr\} \,,
\eeq
which leads to the {\bfseries magnetic Lewis-Weyl-Papapetrou (LWP) metric}
\beq
\label{Magnetic-LWP}
ds^2_{magnetic-LWP} = \frac{1}{f}\Bigl[-\rho^2 d \tau^2 + e^{2\gamma} \bigl(d\rho^2 + dz^2\bigr)\Bigr] + f \bigl(d\psi - \omega d \tau\bigr)^2 \,,
\eeq
that, given the construction of this metric, is another independent solution of the Einstein equations~\eqref{Einstein-equations-complete} if the same double-Wick rotation~\eqref{double-Wick} is also applied to all the other possible fields of the theory.

\subsubsection{LWP Conjugation}  \label{subsubsec:LWP-Conjugation}
Moreover, without using any transformation, it is still always possible to rewrite the \emph{same} solution from an electric to a magnetic form, and vice-versa. Indeed, the electric metric~\eqref{Electric-LWP} can be rewritten as
\beq
\label{LWP-electric-to-magnetic}
ds^2_{e\rightarrow m} = - \frac{1}{x f}\Bigl[-\rho^2 d t^2 - x e^{2\gamma} \bigl(d\rho^2 + dz^2\bigr)\Bigr] - x f \Bigl[d\phi - \frac{\omega}{x} d t\Bigr]^2 \,,
\eeq
where $\displaystyle  x  = \frac{f^2 \omega^2 - \rho^2}{f^2}$ , which leads to the magnetic form~\eqref{Magnetic-LWP} if the following redefinitions are applied:
\beq
\label{LWP-W0}
\tilde{f} = - x f \,, \quad
\tilde{\omega} = \frac{\omega}{x} \,, \quad
\tilde{\gamma} = \gamma + \frac{1}{2} \log \bigl( - x \bigr) \,.
\eeq
In the same way, the magnetic metric~\eqref{Magnetic-LWP} can be rewritten as
\beq
\label{LWP-magnetic-to-electric}
ds^2_{m \rightarrow e} = - \frac{1}{x f}\Bigl[\rho^2 d\psi^2 -  x e^{2\gamma} \bigl(d\rho^2 + dz^2\bigr)\Bigr] + x f \Bigl[d\tau -  \frac{\omega}{x} d \psi\Bigr]^2 \,,
\eeq
which results in the electric form~\eqref{Electric-LWP} by using the same redefinitions~\eqref{LWP-W0}.

The reason why the same redefinitions~\eqref{LWP-W0} are applied in both cases relies upon the fact that the double-Wick rotation~\eqref{double-Wick} can be seen as the transformation
\beq
\label{LWP-W}
W \coloneqq \biggl\{ f \to - \biggl[\frac{f^2 \omega^2 - \rho^2}{f^2}\biggr] f  \,, \,\,\,
\omega \to \biggl[\frac{f^2}{f^2 \omega^2 - \rho^2}\biggr]\omega \,, \,\,\,
e^{2\gamma} \to - \biggl[\frac{f^2 \omega^2 - \rho^2}{f^2}\biggr]e^{2\gamma} \biggr\} \,,
\eeq
which is commonly referred to as {\bfseries conjugation}~\cite{Chandra-book}, and that acts as an involution operator, i.e.~$W \circ W = \mathds{1}$. A widely known example of the use of the conjugation is that it maps the Schwarzschild black hole~\eqref{Schwarzschild-metric} into the Witten bubble of nothing~\cite{Horowitz}.

\subsection{Black Holes} \label{subsec:Black-Holes}

The spacetime geometry in the region surrounding an object of mass $M$, in a vacuum and without a cosmological constant, is described by the static and spherically symmetric {\bfseries Schwarzschild metric}~\cite{Schwarzschild}
\beq
\label{Schwarzschild-metric}
ds^2=-\biggl(1 - \frac{2 M}{r}\biggr) dt^2 + \frac{dr^2}{\Bigl(1 - \frac{2 M}{r}\Bigr)} + r^2\bigl(d\theta^2 + \sin^2\theta\, d\phi^2\bigr)\,.
\eeq
It is straightforward to verify that this spacetime is singular whenever $r=0$ or $r=2M$.

Spacetime's singularities can be classified as {\bfseries coordinate singularities}, if at that point the coordinate system breaks down while maintaining a finite curvature, thus being removable by choosing a different coordinate system, or as {\bfseries curvature singularities}, if the curvature at that point diverges. 

To distinguish between curvature and coordinate singularities one could na\"ively use the components of the Riemann tensor $R^{\mu}{}_{ \nu \rho \sigma}$. On the other hand, since these are the components of a tensor, they are actually coordinate-dependent. Thus, the correct choice relies on constructing scalar quantities starting from $R^{\mu}{}_{ \nu \rho \sigma}$, which are always coordinate-independent. The scalar commonly used for this purpose is the {\bfseries Kretschmann scalar}
\beq
\label{Kretschmann-scalar}
K=R_{\mu \nu \rho \sigma}R^{\mu \nu \rho \sigma}\,.
\eeq
In the case that the radius of the object described by the Schwarzschild metric~\eqref{Schwarzschild-metric} becomes less than two times the mass $r<2M$, the object undergoes gravitational collapse, forming what is known as a {\bfseries black hole}, a region of spacetime where gravity is so strong that nothing, not even light, can escape from it. This region is delimited by a sphere of radius $r=2M$, which is called an {\bfseries event horizon}, thus representing a surface that, once crossed, can never be traveled back. The singularity at $r=2M$, i.e.~where the event horizon is located, is actually just a \emph{coordinate} singularity, which can thus be transformed away using alternative coordinate systems, such as the Kruskal coordinates. Instead, at the center of the black hole, $r = 0$, lies a \emph{curvature} singularity, representing the point where all the mass of the starting object concentrates after the gravitational collapse.
\newpage
\noindent The generalization to an electromagnetically charged and rotating black hole is given by the {\bfseries dyonic Kerr-Newman black hole}~\cite{Kerr-Newman1, Kerr-Newman2}, where the term \emph{dyonic} signifies the presence of both the electric and magnetic charge. The spacetime of this black hole is stationary and axisymmetric, and can be expressed in the canonical (ADM) form as:
\begin{subequations}
\label{kn-magnetic}
\begin{align}
ds^2 & = R^2 \biggl[ -\frac{\Delta}{\Sigma}\,dt^2 + \frac{dr^2}{\Delta} + d\theta^2\biggr] + \frac{\Sigma\sin^2\theta}{R^2}\biggl[ d\phi - \frac{a\, \lambda}{\Sigma} \,dt \biggr]^2 \,, \\
A & = \frac{A_0}{\Sigma} \,dt + \frac{A_{3}}{R^2}\biggl[ d\phi - \frac{a\, \lambda}{\Sigma}\,dt \biggr] \,,
\end{align}
\end{subequations}
where
\vspace{-0.25cm}
\begin{subequations}
\begin{align}
Z^2 & = Q^2 + H^2 \,, \\
\Delta & = r^2 - 2Mr + Z^2 + a^2 \,, \\
\lambda &= r^2 + a^2 - \Delta = 2Mr - Z^2 \,, \\
\Sigma & = \bigl(r^2 + a^2\bigr)^2 - \Delta a^2 \sin^2\theta \,, \\
R^2 & = r^2 + a^2\cos^2\theta \,, \\
A_0 & = a H\Delta \cos \theta - Q r \bigl(r^2+a^2\bigr) \,, \\
A_3 & = a Q r \sin^2\theta - H \bigl( r^2 + a^2 \bigr) \cos\theta \,,
\end{align}
\end{subequations}
$M$ is the mass, $a$ is the angular momentum per unit mass, while $Q$ and $H$ are respectively the electric and the magnetic charge.

It is straightforward to verify that the only curvature singularity of this spacetime is located at the “center” $R^2 = 0$, i.e.~for $r=\cos\theta=0$, which is called a \emph{ring singularity} due to its topology. Conversely, the other singularities at $\theta = 0$, $\Delta = 0$, and $\Sigma = 0$, are all coordinate singularities. 

In particular, and in contrast to the Schwarzschild case, the presence of at least the rotation or one of the charges gives rise to two horizons, located at:
\beq
\label{KN-horizon}
\Delta = 0 \quad \Rightarrow \quad r_{\pm} = M\pm\sqrt{M^2-a^2-Z^2}\,,
\eeq
that may return to a single event horizon at $r=M$, resulting in what is called an {\bfseries extremal black hole}, if the mass satisfies the condition:
\beq
\label{KN-extremality}
M^2 = a^2 + Z^2  \quad \Rightarrow \quad r_{+}=r_{-} \,.
\eeq
Additionally, Eq.~\eqref{KN-horizon} also implies an upper limit for the charges and the angular momentum, indeed, if $M^2 < a^2 + Z^2$ the black hole becomes what is known as a {\bfseries naked singularity}, i.e.~a curvature singularity not hidden behind an event horizon.

\subsubsection{Frame-dragging}  \label{subsubsec:Frame-dragging}

An important property of rotating black holes is the {\bfseries frame-dragging}, which is an effect where observers with zero angular momentum actually rotate with respect to infinity. Indeed, for observers with four-velocity 
\beq
\label{zamo}
u=\sqrt{\frac{g_{\phi \phi}}{g_{t \phi}^2\,-\,g_{tt}g_{\phi\phi}}}\biggl[\partial_{t} - \frac{g_{t\phi}}{g_{\phi \phi}} \partial_{\phi}\biggr] \,,
\eeq
it is straightforward to verify that $u \cdot \partial_{r} = u \cdot \partial_{\theta} =0$, meaning that these observers do not move with respect to the surfaces of constant time. Furthermore, since $u \cdot \partial_{\phi} = 0$, they also have zero angular momentum~\eqref{Killing-momentum}.

On the other hand, they \emph{do} rotate with respect to infinity, with angular velocity:
\beq
\frac{d \phi}{d t} = \frac{u^{\phi}}{u^{t}} = - \frac{g_{t\phi}}{g_{\phi \phi}} \,, \label{frame-dragging-eq}
\eeq
which implies that, despite being locally non-rotating observers due to their zero angular momentum, they actually rotate with respect to infinity with angular velocity $\omega = - \frac{g_{t\phi}}{g_{\phi \phi}}$. 

In this sense, the quantity~\eqref{frame-dragging-eq} represents the frame-dragging, or gravitational dragging, of the whole spacetime. 

This angular velocity increases as the observer approaches the black hole, and it goes in the same direction as the black hole rotation. In particular, this implies that the angular velocity of the event horizons is
\beq
\omega^{H}_{\pm} = -\biggr(\frac{g_{t\phi}}{g_{\phi \phi}}\biggr) \bigg\rvert_{r = r_{\pm}} \! = \frac{a}{r_{\pm}^2 + a^2} \label{angular-velocity-hn-kn}\\ \,.
\eeq

\subsubsection{Ergoregions}  \label{subsubsec:Ergoregions}

Another peculiar property of rotating black holes is that the Killing vector $k=\partial_{t}$ is not always timelike outside the outer event horizon. Taking, for simplicity, the non-charged sub-case $Q=H=0$ of the Kerr-Newman black hole~\eqref{kn-magnetic}, which is called the {\bfseries Kerr black hole} and is described by the line element
\begin{subequations}
\begin{align}
ds^2 & = R^2 \biggl[ -\frac{\Delta}{\Sigma}\,dt^2 + \frac{dr^2}{\Delta} + d\theta^2\biggr] + \frac{\Sigma\sin^2\theta}{R^2}\biggl[ d\varphi - \frac{2\,a M r}{\Sigma} \,dt \biggr]^2 \label{kerr-magnetic} \,, \\
\Delta & = r^2 - 2Mr + a^2 \,, \\
\Sigma & = \bigl(r^2 + a^2\bigr)^2 - \Delta a^2 \sin^2\theta \,, \\
R^2 & = r^2 + a^2\cos^2\theta \,.
\end{align}
\end{subequations}

\noindent It is straightforward to verify that this black hole has the horizons~\eqref{KN-horizon} located at:
\beq
r_\pm=M\pm\sqrt{M^2-a^2} \,.
\eeq
On the other hand,
\beq
k^2=g_{tt} = - \frac{\Delta - a^2 \sin^2\theta}{R^2}=-\biggl(1- \frac{2 M r}{r^2 + a^2 \cos^2\theta} \biggr) \,,
\eeq
is not always negative for $r > r_+$. In fact, $k$ becomes timelike for $r > r_+$ only outside the surface described by
\beq
\label{ergosphere}
r_{erg}(\theta)=m+\sqrt{m^2-a^2\cos^2\theta} \,,
\eeq
which is known as the {\bfseries ergosphere}, while the region between the outer event horizon $r_{+}$ and the ergosphere $r_{erg}(\theta)$ is called the {\bfseries ergoregion}. It is interesting to note that the ergosphere intersects the outer event horizon only for $\theta = 0, \pi$. Thus, except for these two angles, observers can freely enter and exit the ergoregion, as doing so does not require crossing the event horizon.
\newline
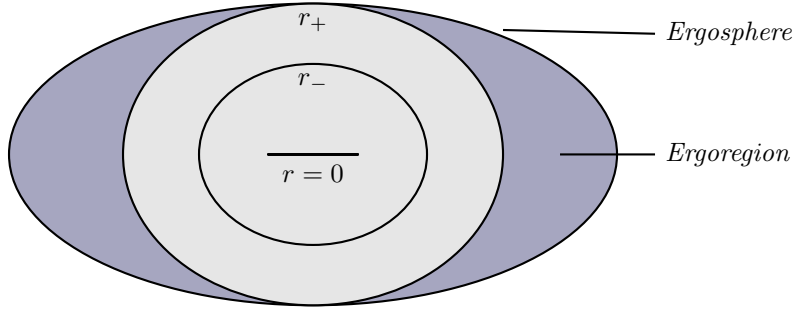
\begin{figure}[H]
\begin{center}
\begin{tikzpicture}[trim right=5.57cm]
\filldraw[color = black, thick, fill = ergo!35] (0,0) ellipse (4cm and 2cm);
\filldraw[color = black, thick, fill = black!10] (0,0) ellipse (2.5cm and 2cm);
\filldraw[color = black, thick, fill = black!10] (0,0) ellipse (1.5cm and 1.2cm);
\filldraw[color = black, thick, fill = black!10] (0,0) ellipse (0.6cm and 0cm);
\path [draw = black, thick,] (2.5,1.65) -- (4.5,1.6);
\path [draw = black, thick,] (3.25,0) -- (4.5,0);
\draw (0,0) node[anchor=north] {\small $r = 0$};
\draw (0,1.2) node[anchor=north] {\small $r_{-}$};
\draw (0,2) node[anchor=north] {\small $r_{+}$};
\draw (4.5,0) node[anchor=west] {\small \it Ergoregion};
\draw (4.5,1.6) node[anchor=west] {\small \it Ergosphere};
\end{tikzpicture}
\label{fig:ergo}
\caption{\small Graphical representation of the Kerr black hole. $r_{\pm}$ are the inner and outer horizons, $r=0$ is the ring singularity, while the ergoregion is the region between the ergosphere~\eqref{ergosphere} and the outer event horizon $r_{+}$.}
\end{center}
\end{figure}
\vspace{-0.25cm}
\noindent An interesting theoretical phenomenon due to ergoregions is the Penrose process, which offers a method for extracting energy from a rotating black hole. Indeed, if a particle with four-momentum $p^{\mu}$ enters the ergoregion along a geodesic~\eqref{geodesic-equation-velocity}, from Eq.~\eqref{Killing-energy}, there is a constant of motion associated with the Killing vector $k=\partial_{t}$, which corresponds to the energy, $E = - g_{\mu \nu} p^{\mu} k^{\nu}$. If this particle decays into two other particles, one of which falls into the black hole with energy $E_1 = - p_{1} \cdot k$, while the other exits the ergoregion and escapes to infinity, then, conservation of energy dictates that the escaping particle will have energy $E_{2} = E - E_{1}$. On the other hand, since $k$ is spacelike inside the ergoregion, it is possible to obtain $E_2 > E$, thereby extracting energy from a rotating black hole.

\noindent Furthermore, the ergosphere is also referred to as the {\bfseries static limit surface}. This is because once observers enter the region inside the ergosphere, it is impossible for them to remain \emph{static}, in the sense that they cannot maintain a four-velocity equal to $u=\sqrt{-\frac{1}{g_{tt}}}\partial_{t}$.

Similarly, the outer event horizon $r_{+}$ is also the {\bfseries stationary limit surface}, because inside this horizon it is impossible for observers to remain \emph{stationary}, in the sense that a stationary observer is one that does not perceive any time variation in the black hole’s gravitational field, due to having the four-velocity equal to $u=\sqrt{\frac{g_{\phi \phi}}{g_{t \phi}^2\,-\,g_{tt}g_{\phi\phi}}}\biggl[\partial_{t} - \frac{g_{t\phi}}{g_{\phi \phi}} \partial_{\phi}\biggr]$~\eqref{zamo}.

Finally, it can be proved that the ergoregion is also the region where the black hole's frame dragging becomes so extreme that, even if an arbitrarily large force is applied, all observers must orbit with a non-null angular velocity whose direction matches that of the black hole's rotation.

\subsubsection{Black Holes Thermodynamics} \label{subsubsec:Black-Holes-Thermodynamics}

A hypersurface $\pazocal{N}$ whose normal vectorial fields are lightlike is called a {\bfseries Killing horizon} of a Killing vector $\xi$ if, on $\pazocal{N}$, this Killing vector $\xi$ is normal to $\pazocal{N}$, and thus lightlike. 

As an example, the two horizons $r_{\pm}$~\eqref{KN-horizon} of the Kerr-Newman black hole are indeed Killing horizons of the Killing vectors
\beq
\xi^{KN}_{\pm} = \partial_{t} + \omega^{H}_{\pm} \partial_{\phi} \,, \label{killing-horizon-zero}\\
\eeq
where $\omega^{H}_{\pm}$ are the angular velocities of the horizons~\eqref{angular-velocity-hn-kn}.

Additionally, for a Killing horizon $\pazocal{N}$ of the Killing vector $\xi$, it is possible to define the {\bfseries surface gravity}
\beq
\kappa =  \sqrt{-\frac{1}{2}\nabla_{\mu}\xi_{\nu}\nabla^{\mu}\xi^{\nu}}\bigg\rvert_{\pazocal{N}}\,. \label{surface-gravity-zero}
\eeq
Classically, a black hole should have zero temperature, however, in 1974, Hawking~\cite{HawkingRadiation} proposed that due to the quantum effect known as {\bfseries Hawking radiation}, an event horizon should radiate like a black body with a temperature
\beq
T=\frac{\kappa}{2\pi}\,, \label{temperature}
\eeq
in this sense, the surface gravity~\eqref{surface-gravity-zero} actually represents the black hole temperature.

Moreover, as proposed by Bekenstein and Hawking~\cite{Bekenstein, BekesteinHawking}, black holes should also have an entropy, proportional to the area of the event horizon,
\beq
\label{entropy-zero}
S=\frac{A}{4} \,.
\eeq
These considerations have solidified the idea that black holes are thermodynamic objects, since a few years earlier it had already been formulated what are called {\bfseries The Four Laws of Black Hole Mechanics}~\cite{FourLawsBH}, which already bore a strong resemblance to the four laws of thermodynamics:
{\begin{table}[H]
\begin{center}
\doublespacing
\begin{tabular}{ |p{1.125cm}||p{6.2cm}|p{6.3cm}|  }
\hline
\multicolumn{3}{|c|}{The Four Laws of Black Hole Mechanics} \\
\hline
& Thermodynamics                          & Black Holes                                  \\
\hline
Zeroth                        
& The temperature $T$ is constant throughout a system in thermal equilibrium.                 
& The surface gravity $\kappa$ is constant over the entire horizon of a stationary black hole.                                 \\
\hline
First                            
& $dE = T \,dS \,\, + $\, work terms, where $E$ is the energy, $T$ is the temperature and $S$ is the entropy.                 
& $dM = \frac{\kappa}{8 \pi} \,dA \,+\, \Omega_{H} \,dJ \,+\, \Phi \,dQ$, where $M$ is the black hole mass, $\kappa$ is the surface gravity, $A$ is the horizon area, $\Omega_{H}$ is the angular velocity of the horizon, $J$ is the black hole angular momentum, $\Phi$ is the electrostatic potential and $Q$ is the black hole electric charge.     \\
\hline
Second                       
& In any process, the variation of the entropy $S$ is always greater than or equal to zero $\delta S\geq0$.                
& In any process, the variation of the black hole area $A$ is always greater than or equal to zero $\delta A\geq0$.           \\
\hline
Third                           
& It is impossible to achieve zero temperature $T=0$ in any physical process.                 
& It is impossible to achieve zero surface gravity $\kappa=0$ in any physical process.                      \\
\hline
\end{tabular}
\caption{\small The Four Laws of Black Hole Mechanics, expressed in Planck's units, alongside their counterparts in Classical Thermodynamics.}
\end{center}
\label{table:Four-Laws-BH}
\end{table}}
\vspace{-0.25cm}
\noindent However, it must be noted that the second law fails in the presence of magnetic charges, such as in the case of the dyonic Kerr-Newman black hole. Additionally, the third law excludes the possibility of extremal black holes~\eqref{KN-extremality}, since for these black holes the surface gravity is always zero.
Furthermore, one might argue that Hawking's discovery that black holes radiate supersedes the second law, as this radiation leads, over time, to a decrease in the black hole mass, and consequently a reduction in the area of the event horizon.

\noindent In particular, the surface gravity and the area of the Kerr-Newman black hole are respectively:
\begin{align}
\kappa^{KN}_{\pm} & = \sqrt{-\frac{1}{2}\nabla_{\mu}\xi_{\nu}\nabla^{\mu}\xi^{\nu}} \,\bigg\rvert_{r=r_\pm} \!=\frac{r_{\pm}-r_{\mp}}{ 2 \bigl( r_{\pm}^2 + a^2 \bigr)} \,, \label{surface-gravity-kn}\\
\pazocal{A}^{KN}_{\pm} & =  \int_{0}^{2 \pi}\!d\phi \int_{0}^{\pi} \!d\theta\, \sqrt{g_{\theta \theta} g_{\varphi \varphi}} \,\bigg\rvert_{r=r_\pm} \!= 4 \pi \bigl( r_{\pm}^2 + a^{2} \bigr) \,, \label{area-kn}
\end{align}
from which the temperature and the entropy of this black hole, i.e.~those of the outer horizon, are given by:
\begin{align}
T^{KN} & = \frac{\kappa_{+}}{2 \pi}  = \frac{r_{+}-r_{-}}{ 4\pi \bigl( r_{+}^2 + a^2 \bigr)} \,, \label{temperature-kn}\\
S^{KN} & = \frac{\pazocal{A}_{+}}{4} = \pi \bigl( r_{+}^2 + a^{2} \bigr) \,. \label{entropy-kn}
\end{align}

\subsection{Tetrad Formulation of General Relativity}   \label{subsubsec:Tetrad}
Up until now, at each point in the manifold, it has been used a natural basis for tangent spaces given by the partial derivatives with respect to the coordinates at that point, $e_{\mu} = \partial_{\mu}$, which led to a basis for the cotangent spaces given by the gradients of the coordinate functions, $\theta^{\mu} = dx^{\mu}$.
However, a coordinate basis $\partial_{\mu}$ is in general \emph{not} orthonormal, except of course for the trivial case of the flat spacetime in Cartesian coordinates. Therefore, at each point in the manifold, it can be introduced a basis $e_{a}$ not related to any coordinate system\footnote{The index is a Latin letter, instead of a Greek one, as a reminder that this is a basis not related to any coordinate system.} as a linear combination of the coordinate basis:
\beq
e_{a} = e^{\mu}_{a}e_{\mu} \,,
\eeq
and also required to be orthonormal, which traduces in the following conditions:
\beq
\label{orto-tetrad}
g_{\mu \nu}e^{\mu}_{a}e^{\nu}_{b}=\eta_{a b} \,,
\eeq
where $\eta_{a b} = \text{diag}(-1, +1, +1, +1)$ is the Minkowski metric. 

Thus, $e^{\mu}_{a}$ represents the coefficients matrix to pass from a non-orthonormal coordinate system to an orthonormal non-coordinate one. 
Denoting their inverse as $e^{a}_{\mu}$, which then satisfy the conditions
\begin{subequations}
\begin{align}
e^{\mu}_{a} e^{a}_{\nu} = \delta^{\mu}_{\nu} \,,\\
e^{a}_{\mu} e^{\mu}_{b} = \delta^{a}_{b} \,,
\end{align}
\end{subequations}
it is possible to define an orthonormal basis for the cotangent space as
\beq 
e^{a} = e_{\mu}^{a}\theta^{\mu} \,,
\eeq
which is compatible with the basis vector $e_{a}$, in the sense that
\beq
e^{a}(e_{b}) = \delta^{a}_{b} \,.
\eeq
The set of vectors $e_{a}$ comprising the orthonormal basis is known as a {\bfseries tetrad}\footnote{From the Greek \emph{tetras}, “a group of four”.}, or sometimes as a {\bfseries vielbein}\footnote{From the German for “many legs”. However, in different numbers of dimensions, it occasionally becomes a zweibein (two), dreibein (three), vierbein (four), and so on.}.

Furthermore, the requirement of ortogonality~\eqref{orto-tetrad} also results in the following useful relation between the determinant of the tetrad $e \coloneqq |e^{a}_{\mu}|$ and the determinant of the metric $g$:
\beq
e = \sqrt{-g} \,.
\eeq
With these definitions, for any tensor, it is then possible to convert \emph{world} Greek indices into \emph{tangent space} Latin indices, and vice-versa, as:
\begin{subequations}
\begin{align}
e^{a}_{\rho_{i}} \, T^{\rho_{1} \dots \rho_{i} \dots \rho_{n}}{}_{\sigma_{1} \dots \sigma_{m}} & =  T^{\rho_{1} \dots a \dots \rho_{n}}{}_{\sigma_{1} \dots \sigma_{m}} \,, \\
e^{\sigma_{i}}_{a} \, T^{\rho_{1} \dots \rho_{n}}{}_{\sigma_{1} \dots \sigma_{i} \dots \sigma_{m}} & =  T^{\rho_{1} \dots \rho_{n}}{}_{\sigma_{1} \dots a \dots \sigma_{m}} \,, \\
e^{\mu}_{c_{i}} \, T^{c_{1} \dots c_{i} \dots c_{n}}{}_{d_{1} \dots d_{m}} & =  T^{c_{1} \dots 	\mu \dots c_{n}}{}_{d_{1} \dots d_{m}}  \,, \\
e^{d_{i}}_{\mu} \, T^{c_{1} \dots c_{n}}{}_{d_{1} \dots d_{i} \dots d_{m}} & =  T^{c_{1} \dots c_{n}}{}_{d_{1} \dots \mu \dots d_{m}} \,.
\end{align}
\end{subequations}
Being a non-coordinate basis, the tetrad can be changed independently of the coordinates, with the only restriction of preserving the orthonormality~\eqref{orto-tetrad}. Therefore, since the transformations that preserve the flat metric $\eta_{ab}$ are the Lorentz transformations, it holds that, at any point in spacetime $x$, the tetrad can be transformed as
\beq
e'_{a} = \Lambda_{a}{}^{b}(x) e_{b} \,,
\eeq  
where $\Lambda(x)$ are now {\bfseries \emph{local} Lorentz transformations}, in the sense that $\Lambda(x)$ represents a position-dependent Lorentz transformation. 

Moreover, there is still the usual freedom to perform a {\bfseries general coordinate transformations}, i.e.~a diffeomorphism. Thus, the transformation law for tensors with mixed indices is in general given by:
\beq
T'^{a \mu}{}_{c \nu} = \Lambda^{a}{}_{b}\, \frac{\partial x'^{\mu}}{\partial x^{\rho} }\, \Lambda_{d}{}^{c} \,\frac{\partial x^{\sigma}}{\partial x'^{\nu} } \,T^{b \rho}{}_{d \sigma} \,.
\eeq
For this reason, it is necessary to define the action of the covariant derivative upon tensors with Latin indices, in such a way that it has the correct transformation law under local Lorentz transformations. This requires new connection coefficients, known as {\bfseries the spin connection coefficients} $\omega_{\mu}{}^{a}{}_{b}$, by means of which the covariant derivative on a tensor with Latin indices is defined as:
\begin{align}
\begin{split}
\label{cov-der-spin}
\nabla_{\mu}{T^{c_{1}\dots c_{n}}}_{d_{1} \dots d_{m}} = \partial_{\mu}{T^{c_{1}\dots d_{n}}}_{c_{1} \dots d_{m}} & + \omega_{\mu}{}^{c_{1}}{}_{a} \,{T^{a \dots c_{n}}}_{d_{1} \dots d_{m}} + \, \dots \, + \omega_{\mu}{}^{c_{n}}{}_{a}  \, {T^{c_{1} \dots a}}_{d_{1} \dots d_{m}} \\ 
& -  \omega_{\mu}{}^{a}{}_{d_{1}} {T^{c_{1} \dots c_{n}}}_{a \dots d_{m}} \, - \, \dots \, - \omega_{\mu}{}^{a}{}_{d_{m}}  {T^{c_{1} \dots c_{n}}}_{d_{1} \dots a} \,,
\end{split}
\end{align}
while the action of the covariant derivative on Greek indices is the same as before, using the Christoffel symbols~\eqref{cov-der} as the connection coefficients.
These new spin connection coefficients are related to the Christoffel symbols via the so-called {\bfseries tetrad postulate}
\beq
\nabla_{\mu} e^{a}_{\nu} = 0\,,
\eeq
which is \emph{always} true, and not only for the Levi-Civita connection, and gives:
\beq
\omega_{\mu}{}^{a}{}_{b} = e^{a}_{\nu}e^{\lambda}_{b}\Gamma^{\nu}{}_{\mu \lambda} - e^{\lambda}_{b}\partial_{\mu}e^{a}_{\lambda}\,.
\eeq
On the other hand, for the Levi-Civita connection~\eqref{metric-compatibility}, the metric compatibility requirement~\eqref{metric-compatibility}, which now reads $\nabla \eta_{a b} = 0$, yields that the spin connection coefficients are antisymmetric in the Latin indices:
\beq
\omega_{\mu a b} = -\omega_{\mu b a} \,.
\eeq
\subsubsection{Spinors in Curved Spacetime}
The connection $\omega_{\mu}{}^{a}{}_{b}$ is dubbed with the adjective \emph{spin} because it can be used to construct the {\bfseries Lorentz-covariant derivate on spinors} $\pazocal{D}_{\mu}$, which on a spinor field $\psi$ act as:
\beq
\label{lorentz-cov-der-spin}
\pazocal{D}_{\mu} \psi = \Bigl(\partial_{\mu} + \frac{1}{4} \omega_{\mu}{}^{ab}\gamma_{ab} \Bigr) \psi \,.
\eeq
Indeed, as the name suggests, under a local Lorentz transformation $\Lambda(x)$ for which spinors transform as
\beq
\psi' = \rho\bigl(\Lambda (x) \bigr) \psi \,,
\eeq
it holds that $\pazocal{D}_{\mu} \psi$ transforms covariantly:
\beq
\pazocal{D}_{\mu} \psi' =  \pazocal{D}_{\mu}\Bigl[\rho\bigl(\Lambda (x) \bigr) \psi\Bigr] = \rho\bigl(\Lambda (x) \bigr) \pazocal{D}_{\mu} \psi \,.
\eeq
\vfill
\clearpage
\noindent Moreover, for a spin $3/2$ field $\psi_{\nu}$, the Lorentz-covariant derivative satisfies the following relation with the Levi-Civita connection:
\beq
\nabla_{[\mu}\psi_{\nu]} = \pazocal{D}_{[\mu}\psi_{\nu]} \,,
\eeq
which is an equality that holds because the extra term in the Levi-Civita connection, $\Gamma^{\lambda}{}_{[\mu \nu]} \psi_{\lambda}$, is zero due to the requirement of the connection being torsion free~\eqref{torsion-free}.

It is interesting to note that the commutator of two Lorentz-covariant derivatives is related to the curvature, indeed it can be proved that
\beq
\bigl[\pazocal{D}_{\mu}, \pazocal{D}_{\nu}]\psi = \frac{1}{4} \Bigl(R_{\mu \nu}{}^{a b}\gamma_{a b}\Bigr) \psi \,,
\eeq
which, in a certain sense, is similar to what happens for a connection with no torsion~\eqref{commutator-cov-dev}.

Finally, it is possible to define the gamma matrices with a curved index as
\beq
\gamma_{\mu} \coloneqq e^{a}_{\mu} \gamma_{a} \,.
\eeq
\clearpage
\section{The Ernst Formalism}
\label{sec:Ernst}
As pointed out in Sec.~\ref{subsec:Einstein-equations}, Einstein equations are a set of highly non-linear PDEs of the second order. For this reason, finding new exact analytical solutions just resting on ans\"atze is usually extremely challenging, a problem that led to the development of numerous methods and techniques with the aim of constructing new exact solutions \emph{without} the need to perform a direct integration of the equations of motion. The fundamental idea behind these methods consists of starting from a known solution, often referred to as a seed, and then applying a transformation or a map to generate a new, usually non-equivalent, solution of the field equations, thereby circumventing the necessity of directly integrating the equations of motion.

One of these methods is due to Ernst~\cite{ErnstI, ErnstII}, which is applicable to stationary and axisymmetric spacetimes whose metric can be expressed in the electric LWP form~\eqref{Electric-LWP}. The core concept of this method involves utilizing what are called the complex Ernst potentials, which are defined from the metric and the electromagnetic potential, in order to rewrite the Einstein-Maxwell equations as a pair of new complex equations, known as the Ernst equations, due to the fact these new equations unveil certain symmetries whose associated transformations can map one solution of the field equations to another.

In general, not all symmetry maps of the Ernst equations actually generate new independent solutions; in fact, some are merely gauge transformations. However, there are two maps, namely the \emph{Ehlers transformations}~\cite{Ehlers} and the \emph{Harrison transformations}~\cite{Harrison}, which always act non-trivially on a seed black hole solution. It is worth noting that these maps actually encompass four distinct transformations. Indeed, as noted in Sec.~\ref{subsubsec:LWP-Conjugation}, the electric LWP metric~\eqref{Electric-LWP} can be always double-Wick rotated~\eqref{double-Wick} in the magnetic LWP~\eqref{Magnetic-LWP} and subsequently reshuffled again in the electric form. This means that the same transformation can generate another independent solution when applied to the same seed expressed in the magnetic form, provided that the same double-Wick rotation is also applied to the potentials of the Ernst method.
\vfill
\clearpage
 \subsection{The Electric Ansatz} \label{subsec:electric-ansatz}
The action of the electromagnetic theory without a cosmological constant is given by the {\bfseries Einstein-Maxwell action}
\beq
\Ia_{E.M.} =\frac{1}{16 \pi} \int d^4x \sqrt{-g} \biggl( R - \frac{1}{4} F_{\mu\nu} F^{\mu\nu} \biggr) \,,
\eeq
which, as seen in Sec.~\ref{subsec:Einstein-Maxwell equations}, leads to the following Einstein-Maxwell field equations
\begin{subequations}
\begin{align}
\nabla_{\mu}F^{\mu \nu} & = 0 \,, \label{em-motion-faraday} \\
G_{\mu\nu} & = 8\pi T_{\mu\nu} \,, \label{em-motion-einstein}
\end{align}
\end{subequations}
with the electromagnetic energy-momentum tensor $T_{\mu \nu}$ given by Eq.~\eqref{EM-energy-momentum-tensor}.

As discussed in Sec.~\ref{subsubsec:Electric-LWP}, the most general metric for a stationary and axisymmetric spacetime, eventually in the presence of a stationary and axisymmetric electromagnetic field, is the electric Lewis-Weyl-Papapetrou metric~\eqref{Electric-LWP}. Thus, the first step in the Ernst formalism consists of choosing an ansatz, in cylindrical coordinates, for the metric and the electromagnetic field, which is called the {\bfseries electric ansatz} and given by\footnote{Actually, as explained in~\cite{Heusler}, this is not the most general form for a stationary and axisymmetric four-potential. Nonetheless, this is the electromagnetic potential that is used for the Ernst formalism.}
\begin{subequations}
\label{electric-ansatz}
\begin{align}
ds_{e}^2 & = \frac{1}{f}\Bigl[\rho^2 d\phi^2 + e^{2\gamma} \bigl(d\rho^2 + dz^2\bigr)\Bigr] - f \bigl(dt - \omega d \phi\bigr)^2 \,, \label{electric-metric} \\
A &= A_t dt + A_\phi d\phi \,, \label{electric-Maxwell}
\end{align}
\end{subequations}
where the dependence on the coordinates of the functions $f$, $\omega$, $\gamma$,  $A_t$ and $A_\phi$ is only on $(\rho,z)$.

Since the formalism is formulated in cylindrical coordinates $(\rho,z,\phi)$, and the functions depend only on $(\rho, z)$, it will be convenient to express the equations in a three-dimensional Euclidean space notation. This involves introducing quantities like the axis unit vectors $(\vec{e}_{\rho},\vec{e}_{z},\vec{e}_{\phi})$, the gradient, the curl, the divergence, and the Laplacian in these coordinates, which are reported in Appendix~\ref{A-Cyl}.

\vfill
\clearpage
\subsection{Rewriting the Einstein-Maxwell Equations} \label{subsec:rewriting-einstein-maxwell}
By defining the Maxwell equations as
\beq
M^\nu \coloneq \nabla_\mu F^{\mu\nu} \,,
\eeq
it can be shown that the $\phi-$component of these equations can be written as
\beq
\label{m-phi}
\begin{split}
M^\phi & =  \frac{e^{-2\gamma} f}{\rho ^2}  \biggl[ \, \vec{\nabla} f \cdot \bigl( \vec{\nabla} A_\phi + \omega \vec{\nabla} A_t \bigr) 
-  \frac{2 f}{\rho }  \biggl( \frac{\partial A_\phi}{\partial\rho} + \omega \frac{\partial A_t}{\partial\rho} \biggr) \\
&\quad +  f  \Bigl( \nabla^2 A_\phi + \omega \nabla^2 A_t + \vec{\nabla}\omega \cdot \vec{\nabla} A_t \Bigr) \biggr] \equiv 0 \,,
\end{split}
\eeq
that can be proved to be equivalent to
\beq
\label{ernst-max1}
\vec{\nabla} \cdot \biggl[\, \frac{f}{\rho ^2}\bigl( \vec{\nabla} A_\phi + \omega\vec{\nabla} A_t \bigr) \biggr] = 0 \,.
\eeq
Similarly, it is possible to manipulate the $t-$component of the Maxwell equations as
\beq
\begin{split}
M^t & = e^{-2\gamma} \biggl[
\frac{\omega f^2}{\rho^2} \Bigl( \vec{\nabla} f \cdot \bigl( \vec{\nabla} A_\phi + \omega \vec{\nabla} A_t \bigr) + \nabla^2 A_\phi + \omega \nabla^2 A_t \Bigr) + \frac{1}{f} \vec{\nabla} f \cdot \vec{\nabla} A_t   \\
&\quad + \frac{f^2}{\rho^2} \vec{\nabla} \omega \cdot \bigl( \vec{\nabla} A_\phi + 2\,\omega \vec{\nabla} A_t \bigr)
- \frac{2\, \omega f^2}{\rho^3} \biggl( \frac{\partial A_\phi}{\partial\rho} + \omega \frac{\partial A_t}{\partial\rho} \biggr) - \nabla^2 A_t  \biggr] \equiv 0 \,,
\end{split}
\eeq
which is equivalent to
\beq
\label{ernst-max2}
\vec{\nabla} \cdot \biggl[ \,\frac{1}{f} \vec{\nabla} A_t - \frac{\omega f}{\rho^2} \bigl( \vec{\nabla} A_\phi + \omega\vec{\nabla} A_t \bigr) \biggr] = 0 \,.
\eeq
Analogously, the Einstein equations in the trace-reversed form~\eqref{Ricci-Einstein-equations-Maxwell} can be defined as
\beq
E_{\mu\nu} \coloneq
R_{\mu\nu} - 8\pi T_{\mu\nu} \,.
\eeq
Hence, the $tt-$component of these equations is then given by
\beq
\begin{split}
\label{ernst-einstein-tt}
E_{tt} & = -\frac{e^{-2\gamma}}{2\rho^2} \biggl[ \,2 f \bigl(\vec{\nabla} A_t\bigr)^2 + \frac{2 f^3}{\rho^2} \Bigl(  \omega^2 \bigl(\vec{\nabla} A_t\bigr)^2 + \bigl(\vec{\nabla} A_\phi\bigr)^2 + 2\ \omega \vec{\nabla} A_\phi \cdot \vec{\nabla} A_t \Bigr) \\
&\quad + \bigl(\vec{\nabla} f\bigr)^2 -\frac{f^4}{\rho^2} \bigl(\vec{\nabla}\omega\bigr)^2 - f \nabla^2 f \biggr] \equiv 0 \,,
\end{split}
\eeq
that can also be written as
\beq
\label{ernst-einstein1}
f \nabla^2 f = \bigl(\vec{\nabla} f \bigr)^2 - \frac{f^4}{\rho^2}  \bigl(\vec{\nabla}\omega \bigr)^2 + 2 f  \bigl(\vec{\nabla} A_t \bigr)^2 + \frac{2 f^3}{\rho^2} \bigl( \vec{\nabla} A_\phi + \omega \vec{\nabla} A_t \bigr)^2 \,.
\eeq
Likewise, the $t\phi-$component gives
\beq
\label{ernst-einstein-tphi}
\begin{split}
E_{t\phi} & =
-\frac{e^{-2\gamma}}{2\rho^2} \biggl[ \rho^2 \omega \bigl(\vec{\nabla} f\bigr)^2 + 2\, \omega f^3 \Bigl(  \omega^2 \bigl(\vec{\nabla} A_t\bigr)^2 + \bigl(\vec{\nabla} A_\phi\bigr)^2 + 2\,\omega \vec{\nabla} A_\phi \cdot \vec{\nabla} A_t \Bigr) \\
&\quad - 2 \rho^2 f \Bigl(   \vec{\nabla} f \cdot \vec{\nabla} \omega + 2\vec{\nabla} A_\phi \cdot \vec{\nabla} A_t + \omega (\vec{\nabla} A_t)^2 + \frac{\omega}{2} \nabla^2 f \Bigr) \\
&\quad - \rho f^2 \biggl( \rho \nabla^2\omega - 2\frac{\partial\omega}{\partial\rho} \biggl) - \omega f^4 \bigl(\vec{\nabla}\omega\bigr)^2 \biggr]
\equiv 0 \,.
\end{split}
\eeq
As a result, the combination of Eq.~\eqref{ernst-einstein-tphi} and Eq.~\eqref{ernst-einstein-tphi} as
\beq
\begin{split}
E_{t\phi} + \omega E_{tt} & =
-\frac{fe^{-2\gamma}}{2\rho^4} \biggl[ 4 \vec{\nabla} A_\phi \cdot\vec{\nabla} A_t + 4\, \omega \bigl(\vec{\nabla} A_t\bigr)^2 \\ & \quad + 2 \vec{\nabla} f \cdot\vec{\nabla}\omega + f \vec{\nabla}^2 \omega - \frac{2 f}{\rho} \frac{\partial\omega}{\partial\rho} \biggr]
\equiv 0 \,,
\end{split}
\eeq
can be shown to be equivalent to
\beq
\label{ernst-einstein2}
\vec{\nabla} \cdot \biggl[ \
\frac{f^2}{\rho^2} \vec{\nabla}\omega + \frac{4 f}{\rho^2} \bigl( \vec{\nabla} A_\phi + \omega \vec{\nabla} A_t \bigr) A_{t} \biggl] = 0 \,.
\eeq
The other non-trivial Einstein equations define $\gamma$ by quadratures. Indeed, the equations for $E_{\rho z}$ and $E_{\rho\rho} - E_{zz}$ result in
\begin{subequations}
\label{gamma-quadratures}
\begin{align}
\begin{split}
& \rho \, \frac{\partial\gamma}{\partial z}
+ \frac{\rho^2}{f}\biggl(\frac{\partial A_t}{\partial z}\biggr) \biggl(\frac{\partial A_t}{\partial \rho} \biggr)
- 2f \biggl(\omega\frac{\partial A_t}{\partial z}  - \frac{\partial A_\phi}{\partial z} \biggr)\biggl(\omega\frac{\partial A_t}{\partial \rho}-\frac{\partial A_\phi}{\partial \rho}\biggr) \\
&\quad - \frac{\rho^2}{2 f^2}\biggl(\frac{\partial f}{\partial z}\biggr) \biggl(\frac{\partial f}{\partial \rho} \biggr)
+ \frac{f^2}{2}\biggl(\frac{\partial \omega}{\partial z}\biggr) \biggl(\frac{\partial \omega}{\partial \rho} \biggr) = 0 \,,
\end{split} \\
\begin{split}
& \rho \, \frac{\partial\gamma}{\partial \rho} + \frac{\rho^2}{f}\biggl(\frac{\partial A_t}{\partial \rho}\biggr)^2  \!\!- \biggl(\frac{\partial A_t}{\partial z} \biggr)^2
\!\!+  \frac{\rho^2}{f^2}\biggl[\biggl(\frac{\partial f}{\partial z} \biggr)^2 \!\!- \biggl(\frac{\partial f}{\partial \rho} \biggr)^2\,\biggr]
+ f^2 \biggl[\biggl(\frac{\partial \omega}{\partial z} \biggr)^2 \!\!- \biggl(\frac{\partial \omega}{\partial \rho} \biggr)^2\,\biggr]\\
&\quad + f \biggl[ \biggl(\frac{\partial A_\phi}{\partial z}\biggr)^2 \!\!- \biggl(\frac{\partial A_\phi}{\partial \rho}\biggr)^2 \,\biggl] +\, \omega^2\biggl[\biggl(\frac{\partial A_t}{\partial z}\biggr)^2 \!\!- \biggl(\frac{\partial A_t}{\partial \rho}\biggr)^2 \,\biggr] \\
&\quad + 2\,\omega\biggl[\biggl(\frac{\partial A_t}{\partial \rho}\biggr)\biggl(\frac{\partial A_\phi}{\partial \rho}\biggr) - \biggl(\frac{\partial A_t}{\partial z}\biggr)\biggl(\frac{\partial A_\phi}{\partial z}\biggr)\biggr]  = 0 \,.
\end{split}
\end{align}
\end{subequations}
Thus, $\gamma$ can be found by integrating these latter equations~\eqref{gamma-quadratures} once the functions present in the metric are known.
\vfill
\clearpage
\subsection{Ernst Potentials and Ernst Equations} \label{subsec:Ernst-Potentials}
\vspace{-0.05cm}
The rewriting of the Maxwell equation for the $\phi-$component~\eqref{ernst-max1} results in a total divergence. Hence, for this equation, it is possible to introduce a potential for which the divergence of the corresponding vector is equal to zero. Indeed, it can be proved that for any function “sufficiently well-behaved” $\chi(\rho,z)$, it holds
\beq
\label{potential-regular}
\vec{\nabla} \cdot \biggl[ \,\frac{1}{\rho}\bigl(\vec{e}_{\phi}  \times \vec{\nabla} \chi\bigr)\biggr] = 0 \,.
\eeq
Thus, after defining the so-called {\bfseries electromagnetic twisted potential} $\tilde{A}_\phi$ through
\beq
\label{max-potential}
\vec{e}_{\phi} \times\vec{\nabla} \tilde{A}_\phi \coloneqq
\frac{f}{\rho} \bigl( \vec{\nabla} A_\phi + \omega \vec{\nabla} A_t \bigr) \,,
\eeq
it is straightforward, thanks to~\eqref{potential-regular}, that the equation~\eqref{ernst-max1} resulting from the $\phi-$component of the Maxwell equations, can be rewritten as
\beq
\label{potential-twisted-divergence}
\vec{\nabla}\cdot  \biggl[\,\frac{1}{\rho}\bigl(\vec{e}_{\phi} \times \vec{\nabla} \tilde{A}_{\phi} \bigr) \biggr] = 0 \,.
\eeq
Moreover, applying $\displaystyle \vec{e}_{\phi}\, \times$ to the definition of the electromagnetic twisted potential~\eqref{max-potential} yields
\beq
\frac{1}{\rho}\bigl(\vec{e}_{\phi} \times \vec{\nabla} A_{\phi} \bigr)= - \frac{1}{f} \vec{\nabla} \tilde{A}_\phi - \frac{\omega}{\rho} \bigl(\vec{e}_{\phi}\times \vec{\nabla} A_t\bigr) \,,
\eeq
which, when composed with the operator $\displaystyle \vec{\nabla}\, \cdot$, results in
\beq
\vec{\nabla}\cdot \biggl[\, \frac{1}{f} \vec{\nabla} \tilde{A}_\phi + \frac{\omega}{\rho} \bigl(\vec{e}_{\phi} \times \vec{\nabla} A_t\bigr) \biggr]
= - \vec{\nabla} \cdot \biggl[ \,\frac{1}{\rho}\bigl(\vec{e}_{\phi} \times \vec{\nabla} A_{\phi}\bigr)\biggr] \overset{\eqref{potential-regular}}{=} 0 \,,
\eeq
that substitutes Eq.~\eqref{ernst-max1} as an equation of motion.

In a similar manner, the equation~\eqref{ernst-max2} arising from the $t-$component of the Maxwell equations can be written as
\beq
\vec{\nabla} \cdot \biggl[ \,\frac{1}{f} \vec{\nabla} A_t  - \frac{\omega}{\rho}\bigl( \vec{e}_{\phi} \times \vec{\nabla} \tilde{A}_\phi \bigr)\biggr] = 0 \,.
\eeq
In summary, after implementing the definition of the potential $\tilde{A}_\phi$~\eqref{max-potential}, the Maxwell equations are equivalent to the following system:
\begin{subequations}
\label{ernst-system1}
\begin{align}
\vec{\nabla}\cdot \biggl[\, \frac{1}{f} \vec{\nabla} \tilde{A}_\phi + \frac{\omega}{\rho} \bigl(\vec{e}_{\phi} \times \vec{\nabla} A_t\bigr) \biggr] & = 0 \,, \\
\vec{\nabla} \cdot \biggl[\,\frac{1}{f} \vec{\nabla} A_t  - \frac{\omega}{\rho}\bigl( \vec{e}_{\phi} \times \vec{\nabla} \tilde{A}_\phi \bigr)\biggr] &= 0 \,.
\end{align}
\end{subequations}
Furthermore, the two potentials $A_t$ and $\tilde{A}_\phi$ can be effectively packed in what is known as the {\bfseries electromagnetic complex Ernst potential}
\beq
\label{em-complex-ernst-potential}
\Phi \coloneqq A_t + i \tilde{A}_\phi \,,
\eeq
so that Eqs.~\eqref{ernst-system1} can be expressed as the following single complex equation
\beq
\vec{\nabla} \cdot \biggr[ \, \frac{1}{f} \vec{\nabla} \Phi + \frac{i \, \omega}{\rho} \bigl( \vec{e}_{\phi} \times \vec{\nabla} \Phi \bigr) \biggr] = 0 \,.
\eeq
It is useful to note that Eq.~\eqref{potential-regular}, applied to the product of the two functions $\chi=A_t\tilde{A}_\phi$, gives
\beq
\vec{\nabla} \cdot \biggl[ \, \frac{1}{\rho}\bigl( \vec{e}_{\phi} \times A_t \vec{\nabla}\tilde{A}_\phi \bigr)\biggr] = -\vec{\nabla} \cdot \biggl[ \frac{1}{\rho}\bigl( \vec{e}_{\phi} \times \tilde{A}_\phi \vec{\nabla} A_t \bigr)\biggr] \,.
\eeq
With this result, the Einstein equation Eq.~\eqref{ernst-einstein2} can be written as
\beq
\label{ernst-einstein2-twist}
\vec{\nabla}  \cdot \biggl[ \, \frac{f^2}{\rho^2} \vec{\nabla}\omega + \frac{2}{\rho} \Bigl( \vec{e}_{\phi}\times \Im\bigl( \Phi^* \vec{\nabla}\Phi \bigr) \Bigr) \biggr]= 0 \,,
\eeq
as can be explicitly verified by using the expansion of the product
\beq
\Im \bigl (\Phi^{*} \vec{\nabla}\Phi \bigr) = A_{t} \vec{\nabla} \tilde{A}_{\phi} - \tilde{A}_{\phi} \vec{\nabla} A_{t} \,.
\eeq
Consequently, as done for the electromagnetic twisted potential $\tilde{A}_{\phi}$~\eqref{max-potential}, it is also possible to introduce a {\bfseries twisted gravitational potential} $h$ from Eq.~\eqref{ernst-einstein2-twist}, in such a way that
\beq
\label{grav-potential}
\vec{e}_{\phi} \times \vec{\nabla} h \coloneqq - \frac{f^2}{\rho} \vec{\nabla}\omega - 2\, \vec{e}_{\phi} \times \Im\bigl( \Phi^* \vec{\nabla}\Phi \bigr) \,.
\eeq
Hence, applying $\displaystyle \vec{e}_{\phi}\,\times$ to the gravitational twisted potential~\eqref{grav-potential} results in
\beq
\frac{1}{\rho}\bigl( \vec{e}_{\phi} \times\vec{\nabla}\omega \bigr) = \frac{1}{f^2}\vec{\nabla} h + \frac{2}{f^2}\, \Im\bigl( \Phi^* \vec{\nabla}\Phi \bigr) \,,
\eeq
which, when composed with the operator $\displaystyle \vec{\nabla}\,\cdot$, yields
\beq
\vec{\nabla}\cdot \biggl[ \, \frac{1}{f^2} \vec{\nabla} h + \frac{2}{f^2} \Im\bigl( \Phi^* \vec{\nabla}\Phi \bigr) \biggr]
= - \vec{\nabla} \cdot \biggl[ \,\frac{1}{\rho}\bigl(\vec{e}_{\phi} \times \vec{\nabla} \omega\bigr)\biggr] \overset{\eqref{potential-regular}}{=} 0 \,,
\eeq
that substitutes Eq.~\eqref{ernst-einstein2}.

Finally, the last Einstein equation~\eqref{ernst-einstein1} with the implementation of the twisted potential $h$ becomes
\beq
f \nabla^2 f = (\vec{\nabla} f)^2 - \Bigl[ \vec{\nabla} h + 2\Im\bigl( \Phi^* \vec{\nabla}\Phi \bigr)\Bigr]^2 + 2 f \bigl(\vec{\nabla}\Phi^* \cdot \vec{\nabla}\Phi \bigr) \,.
\eeq
\vfill
\clearpage
\noindent Summarising, the Einstein equations provided by Eq.~\eqref{ernst-einstein1} and Eq.~\eqref{ernst-einstein2} are replaced by the equations
\begin{subequations}
\label{ernst-system2}
\begin{gather}
\vec{\nabla}\cdot \biggl[ \, \frac{1}{f^2} \vec{\nabla} h + \frac{2}{f^2} \Im\bigl( \Phi^* \vec{\nabla}\Phi \bigr) \biggr] = 0 \,, \\
f \nabla^2 f = (\vec{\nabla} f)^2 - \Bigl[ \vec{\nabla} h + 2\Im\bigl( \Phi^* \vec{\nabla}\Phi \bigr)\Bigr]^2 + 2 f \bigl(\vec{\nabla}\Phi^* \cdot \vec{\nabla}\Phi \bigr) \,.
\end{gather}
\end{subequations}
By defining the {\bfseries gravitational complex Ernst potential} as
\beq
\label{grav-complex-ernst-potential}
\ernst \coloneqq f - \Phi^* \Phi + i h \,,
\eeq
together with the electromagnetic complex Ernst potential $\Phi$~\eqref{em-complex-ernst-potential}, the equations of motion represented by the Eqs.~\eqref{ernst-system1} and Eqs.~\eqref{ernst-system2} can be expressed as the following two complex equations:
\begin{subequations}
\label{ernst-eqs}
\begin{align}
\bigl( \Re\ernst + \Phi^* \Phi \bigr) \nabla^2\ernst & =
\vec{\nabla}\ernst\cdot \bigl( \vec{\nabla}\ernst + 2\Phi^* \vec{\nabla}\Phi \bigr) \,, \\
\bigl( \Re\ernst + \Phi^* \Phi \bigr) \nabla^2\Phi & = \vec{\nabla}\Phi\cdot \bigl( \vec{\nabla}\ernst + 2\Phi^*\vec{\nabla}\Phi \bigr) \,,
\end{align}
\end{subequations}
called the {\bfseries Ernst equations}, which thus formally reduces the problem of solving the Einstein-Maxwell equations for a stationary and axisymmetric spacetime to the problem of solving these equations.

Similarly, the equations for $\gamma$, in terms of the Ernst potentials $\Phi$ and $\ernst$, are given by
\begin{subequations}
\begin{align}
\begin{split}
\partial_{\rho}\gamma & = \frac{\rho}{4\bigl(\Re\ernst + \Phi^* \Phi\bigr)^2} \biggl[ \bigl(\partial_{\rho}\ernst + 2\Phi^{*}\partial_{\rho}\Phi\bigr) \bigl(\partial_{\rho}\ernst^{*} + 2\Phi\partial_{\rho}\Phi^{*}\bigr) \\
&\quad - \bigl(\partial_{z}\ernst + 2\Phi^{*}\partial_{z}\Phi\bigr) \big(\partial_{z}\ernst^{*} + 2\Phi\partial_{z}\Phi^{*}\bigr) \biggr] \\
&\quad - \frac{\rho}{\bigl(\Re\ernst + \Phi^* \Phi\bigr)} \big(\partial_{\rho}\Phi^{*}\partial_{\rho}\Phi - \partial_{z}\Phi^{*}\partial_{z}\Phi\big) \,,
\end{split} \\
\begin{split}
\partial_{z}\gamma & = \frac{\rho}{4\bigl(\Re\ernst + \Phi^* \Phi\bigr)^2} \biggl[ \big(\partial_{\rho}\ernst + 2\Phi^{*}\partial_{\rho}\Phi\big) \big(\partial_{z}\ernst^{*} + 2\Phi\partial_{z}\Phi^{*}\big) \\
&\quad + \big(\partial_{z}\ernst + 2\Phi^{*}\partial_{z}\Phi\big) \big(\partial_{\rho}\ernst^{*} + 2\Phi\partial_{\rho}\Phi^{*}\big) \biggr] \\
&\quad - \frac{\rho}{\bigl(\Re\ernst + \Phi^* \Phi\bigr)} \big(\partial_{\rho}\Phi^{*}\partial_{z}\Phi + \partial_{z}\Phi^{*}\partial_{\rho}\Phi\big) \,,
\end{split}
\end{align}
\end{subequations}
that, as already mentioned, can be solved by quadratures.
\vfill
\clearpage
\subsection{Symmetries of the Ernst Equations}  \label{subsec:Ernst-symmetries}

It can be noted that Ernst equations~\eqref{ernst-eqs} represents an effective three-dimensional problem, thus allowing one to overlook the four-dimensional origin of the problem.
Furthermore, these equations also enable an efficient way to study the symmetries of the Einstein-Maxwell equations and subsequently make use of such symmetries to construct new solutions from an already known one. Indeed, it can be proved that the Ernst equations~\eqref{ernst-eqs} can be derived from the effective three-dimensional action
\beq
\label{ernst-action}
S = \int d^3x \,
\biggl[\frac{\bigl(\vec{\nabla}\ernst + 2\Phi^*\vec{\nabla}\Phi\bigr)\cdot\bigl(\vec{\nabla}\ernst^* + 2\Phi\vec{\nabla}\Phi^*\bigr)}{\bigl(\ernst + \ernst^* + 2\Phi^* \Phi\bigr)^2} - \frac{2\vec{\nabla}\Phi^*\cdot\vec{\nabla}\Phi}{\bigl(\ernst + \ernst^* + 2\Phi^* \Phi\bigr)}\biggr] \,.
\eeq
Therefore, the symmetries of the Ernst equations~\eqref{ernst-eqs} can be found just by studying the symmetries of the Ernst action~\eqref{ernst-action}\footnote{To be more precise, not all the symmetries of the equations of motion correspond to the symmetries of the action. Indeed, the equations of motion might possess more symmetries than the action does. However, the symmetries of the action are always symmetries of the equations of motion, and for the Ernst equations they also coincide.}.

A smart way to study the symmetries of the Ernst action~\eqref{ernst-action} is to analyze the quadratic form associated with this action, also called the associated metric~\cite{Stephani}.
Hence, after considering $\ernst$ and $\Phi$ as complex coordinates with real coordinates $(x,y,u,v)$ as
\begin{subequations}
\begin{align}
\ernst &= x + i \,y \,, \\
\Phi &= u + i \,v \,,
\end{align}
\end{subequations}
the metric associated with the action~\eqref{ernst-action} is given by
\begin{align}
\label{ernst-metric}
ds^2 & = \frac{\bigl(d\ernst + 2\Phi^*d\Phi\bigr)\bigl(d\ernst^* + 2\Phi d\Phi^*\bigr)}{\bigl(\ernst + \ernst^* + 2\Phi^* \Phi\bigr)^2} - \frac{2\,d\Phi^* d\Phi}{\bigl(\ernst + \ernst^* + 2\Phi^* \Phi\bigr)} \, \\
        & = \frac{dx^2 + dy^2 - 4\,x \bigl( du^2 + dv^2 \bigr) + 4\, dy \bigl( \,u\, dv - v\, du \bigr) + 4\, dx\, \bigl(v\, dv + u \,du \bigr) }{4\, \bigl(x+u^2+v^2\bigr)^2} \,.
\end{align}
Solving the Killing equation~\eqref{Killing-equation} for this four-dimensional metric yields a total of eight Killing vectors
\begin{subequations}
\label{ernst-killing}
\begin{align}
\xi_1 & = 4\,x\,y\, \partial_x + 2\bigl(y^2-x^2\bigr) \partial_y + 2\bigl(x\,v+y\,u\bigr) \partial_u + 2\bigl(y\,v-x\,u\bigr) \partial_v \,, \\
\xi_2 & = 2\bigl(x\,v+y\,u\bigr) \partial_x + 2\bigl(y\,v-x\,u\bigr) \partial_y + \bigl(4\,u\,v-y\bigr) \partial_u + \bigl(2\,v^2-2\,u^2+x\bigr) \partial_v \,, \\
\xi_3 & = 2\bigl(x\,u-y\,v\bigr) \partial_x + 2\bigl(x\,v+y\,u\bigr) \partial_y + \bigl(2\,u^2-2\,v^2+x\bigr) \partial_u + \bigl(4\,u\,v+y\bigr) \partial_v \,, \\
\xi_4 & = 4\,x\, \partial_x + 4\,y\, \partial_y + 2\,u\, \partial_u + 2\,v\, \partial_v \,, \\
\xi_5 & = -v\, \partial_u + u\, \partial_v \,, \\
\xi_6 & = -2\,v\, \partial_x + 2\,u\, \partial_y + \partial_v \,, \\
\xi_7 & = 2\,u\, \partial_x + 2\,v\, \partial_y - \partial_u \,, \\
\xi_8 & = 4\, \partial_u \,,
\end{align}
\end{subequations}
which are then equivalent to the eight infinitesimal generators of the symmetries of the Ernst action~\eqref{ernst-action}.

In order to find the finite transformations generated by these Killing vectors~\eqref{ernst-killing}, it is necessary to integrate the flow generated by such vectors. The equations that define the flow are:
\beq
\frac{\partial x^i}{\partial \epsilon} = \xi^i\bigl(x^i\bigr) \,,
\eeq
where $x^i=(x,y,u,v)$ are the coordinates, and $\epsilon$ is the flow parameter of the finite transformation.

Integrating the infinitesimal transformations provided by Eqs.~\eqref{ernst-killing} yields the so-called finite Kinnersley transformations~\cite{Stephani}:
\begin{subequations}
\label{ernst-group}
\begin{align}
\label{gauge1}
\ernst' & = \lambda^* \lambda\, \ernst \,, \qquad\qquad\qquad\quad\;\,\hspace{0.75pt}
\Phi' = \lambda \Phi \,, \\
\label{gauge2}
\ernst' & = \ernst + i\,b \,, \qquad\qquad\quad\quad\;\;\;\;
\Phi' = \Phi \,, \\
\label{Ehlers}
\ernst' & = \frac{\ernst}{1 + i\, \jmath\, \ernst} \,, \qquad\qquad\qquad\;
\Phi' = \frac{\Phi}{1 + i\,\jmath\,\ernst} \,, \\
\label{gauge3}
\ernst' & = \ernst - 2\beta^*\Phi - \beta^* \beta \,, \qquad\quad\!\hspace{0.35pt}
\Phi' = \Phi + \beta \,, \\
\label{Harrison}
\ernst' & = \frac{\ernst}{1 - 2\,\alpha^*\Phi - \alpha^* \alpha \, \ernst} \,, \quad\;\;\,
\Phi' = \frac{\alpha\,\ernst + \Phi}{1 - 2\,\alpha^*\Phi - \alpha^* \alpha \, \ernst} \,,
\end{align}
\end{subequations}
where $\alpha$, $\beta$ and $\lambda$ are complex parameters, while $b$ and $\jmath$ are real.

It can be proved~\cite{Stephani, Martelli-Thesis} that the transformations given by Eq.~\eqref{gauge1}, Eq.~\eqref{gauge2} and Eq.~\eqref{gauge3} are actually gauge transformations, which thus do not produce any new solutions, in the sense that these transformations are diffeomorphisms which do not alter the nature of the solutions. For this reason, these maps are not particularly interesting. In contrast, the transformations provided by Eq.~\eqref{Ehlers} and Eq.~\eqref{Harrison}, respectively known as {\bfseries Ehlers transformations}~\cite{Ehlers} and {\bfseries Harrison transformations}~\cite{Harrison}, usually act non-trivially on a seed solution of the Einstein-Maxwell equations, therefore generating a new and independent solution\footnote{This is always true for black holes, however, an Ehlers or Harrison transformation always results in a change of coordinates when applied to the Minkowski background in the electric ansatz.}. Moreover, the Kinnersley transformations~\eqref{ernst-group} form a SU$(2,1)$ group~\cite{Neugebauer, Kinnersley}, which is consistent with the fact that SU$(2,1)$ is also the symmetry group of the Einstein-Maxwell equations.

\subsubsection{Symmetries of the Ernst Equations in Pure Gravity}  \label{subsec:Ernst-symmetries-NON-electro}
In the vacuum case, $\Phi = 0$, the Ernst equations reduce to the single complex equation
\beq
\label{ernst-eqs-pure-gr}
\bigl( \Re\ernst\bigr) \nabla^2\ernst = \vec{\nabla}\ernst\cdot \vec{\nabla}\ernst \,, \\
\eeq
whose corresponding action is then
\beq
S = \int d^3x \,\frac{\vec{\nabla}\ernst^*\!\cdot \vec{\nabla}\ernst}{\bigl(\ernst + \ernst^*\bigr)^2} \,, \label{ernst-action-pure-gr}
\eeq
which, after considering the gravitational Ernst potential $\ernst$ as a complex coordinate
\beq
\ernst = x + i \,y \,,
\eeq
gives rise to the following associated metric
\beq
ds^2  = \frac{d\ernst^* d\ernst}{\bigl(\ernst + \ernst^*\bigr)^2} = \frac{dx^2 + dy^2}{4\, x^2} \,. \label{ernst-metric-pure-gr}
\eeq
As previously seen, the number of Killing vectors in the electrovacuum case is eight in total~\eqref{ernst-killing}. On the other hand, the associated metric when $\Phi = 0$~\eqref{ernst-metric-pure-gr} has only the following three Killing vectors
\begin{subequations}
\label{ernst-killing-pure-gr}
\begin{align}
\xi_1 & = 4\,x\,y\, \partial_x + 2\bigl(y^2-x^2\bigr) \partial_y  \,, \\
\xi_2 & = 4\,x\, \partial_x + 4\,y\, \partial_y \,, \\
\xi_3 & = 2\, \partial_y \,,
\end{align}
\end{subequations}
which, after integration, results in the generalized Ehlers transformations~\cite{AstorinoMelvinCosmological}
\beq
\label{generalized-Ehlers-pure-GR}
\ernst' = \frac{a\bigl(a\ernst + i\,b\bigr)}{\bigl(1 + b\,\jmath + i\,a\, \jmath\bigr)} \,,
\eeq
where $a$, $b$ and $\jmath$ are real parameters.

As expected, the generalized Ehlers transformations~\eqref{generalized-Ehlers-pure-GR} have the same symmetry group SL$(2,\R)$ (or SU$(1,1)$) as General Relativity. Moreover, these transformations are equivalent to an Ehlers transformation~\eqref{Ehlers} composed with the two gauge transformations provided by Eq.~\eqref{gauge1} and Eq.~\eqref{gauge2}. Therefore, in contrast to the Ehlers transformations, it holds that the Harrison transformations~\eqref{Harrison} are additional independent transformations only for the Einstein-Maxwell theory.

\vfill

\subsection{The Magnetic Ansatz}  \label{subsec:magnetic-ansatz}

As mentioned in Sec.~\ref{subsubsec:LWP-Conjugation}, it is always possible to perform a double-Wick rotation~\eqref{double-Wick} on a metric written in the electric LWP form~\eqref{Electric-LWP} in order to obtain a new solution expressed in the magnetic LWP~\eqref{Magnetic-LWP}, which can subsequently be recast in the electric one just by performing a reshuffling of the terms and a redefinition of the functions.

Therefore, the Ernst formalism can also be applied to what is called the {\bfseries magnetic ansatz}
\vspace{-0.5cm}
\begin{subequations}
\label{magnetic-ansatz}
\begin{align}
ds^2_m & = \frac{1}{f}\Bigl[-\rho^2 d t^2 + e^{2\gamma} \bigl(d\rho^2 + dz^2\bigr)\Bigr] + f \bigl(d\phi - \omega d t\bigr)^2 \,, \label{magnetic-metric} \\
A & = A_t d t + A_\phi d\phi \,, \label{magnetic-maxwell}
\end{align}
\end{subequations}
by using the following “magnetic” Ernst potentials
\begin{subequations}
\label{magnetic-ernst-potentials}
\begin{align}
\ernst & \coloneqq - f - \Phi^*\Phi + i h \,, \label{magnetic-grav-complex-ernst-potential}\\
\Phi & \coloneqq \tilde{A}_{t} - i A_{\phi} \,, \label{magnetic-em-complex-ernst-potential}
\end{align}
\end{subequations}
and “magnetic” twisted potentials
\begin{subequations}
\label{magnetic-twisted-potentials}
\begin{align}
\vec{e}_{\phi} \times \vec{\nabla} \tilde{A}_{t} & = - \frac{f}{\rho} \Bigl( \vec{\nabla} A_{t} + \omega \vec{\nabla} A_{\phi} \Bigr) \,, \label{magnetic-max-potential}\\
\vec{e}_{\phi} \times \vec{\nabla} h & = - \frac{f^2}{\rho} \vec{\nabla} \omega - 2\, \vec{e}_{\phi} \times \Im \bigl (\Phi^{*} \vec{\nabla}\Phi \bigr) \,, \label{magnetic-grav-potential}
\end{align}
\end{subequations}
where then
\beq
\Im \bigl (\Phi^{*} \vec{\nabla}\Phi \bigr) = A_{\phi} \vec{\nabla} \tilde{A}_{t} -  \tilde{A}_{t} \vec{\nabla} A_{\phi} \,.
\eeq
As can be seen, the roles of the time and the azimuthal coordinates are exchanged, as expected from the application of the double-Wick rotation~\eqref{double-Wick}, which also introduces a correction in the signs of the functions defining the “magnetic” Ernst potentials $\ernst$~\eqref{magnetic-ernst-potentials} and $\Phi$~\eqref{magnetic-twisted-potentials}.

Moreover, the magnetic ansatz has the same symmetries~\eqref{ernst-group} as the electric one, with the Ehlers~\eqref{Ehlers} and Harrison transformations~\eqref{Harrison} being the only possible non-trivial transformations.

Therefore, this new ansatz implies that the same Ehlers or Harrison transformation can generate another independent solution depending on which ansatz the seed solution is expressed in, increasing the number of independent and meaningful transformations from two to four. Hence, depending on which ansatz the transformation is applied to, it will be referred to either as an {\bfseries electric Ehlers-Harrison transformation} or as a {\bfseries magnetic Ehlers-Harrison transformation}.
\vfill
\clearpage
\noindent The terminology “electric” and “magnetic” has gained popularity only in recent years~\cite{Vigano-Thesis, Martelli-Thesis} \cite{Illy-thesis, Swirling, BarrientosCisterna}. Nevertheless, the meaning should be straightforward given the dual nature of the double-Wick rotation. In general, these terms are also employed due to the way the Harrison transformation works with the different seeds when the transformation parameter is real.
Indeed, a real Harrison map adds an electric charge when applied to the electric ansatz~\eqref{electric-ansatz}, while it adds an external magnetic field in the case of the magnetic ansatz~\eqref{magnetic-ansatz}.
\subsection{Magnetic Transformations on the Schwarzschild Black Hole}  \label{subsec:magnetic-transformations}
The Schwarzschild metric~\eqref{Schwarzschild-metric}, which is reported here for convenience:
\beqNN
ds^2=-\biggl(1 - \frac{2 M}{r}\biggr) dt^2 + \frac{dr^2}{\Bigl(1 - \frac{2 M}{r}\Bigr)} + r^2\bigl(d\theta^2 + \sin^2\theta\, d\phi^2\bigr)\,,
\eeqNN
can be expressed in the magnetic ansatz~\eqref{magnetic-ansatz} as
\begin{align*}
ds^2_m & = \frac{1}{f}\Bigl[-\rho^2 d t^2 + e^{2\gamma} \bigl(d\rho^2 + dz^2\bigr)\Bigr] + f \bigl(d\phi - \omega d t\bigr)^2 \,, \\
A & = A_t d t + A_\phi d\phi \,,
\end{align*}
by means of the following transformation from cylindrical coordinates
\begin{subequations}
\begin{align}
\label{magnetic-coordinates}
\rho & = \sqrt{r^2-2Mr} \sin \theta \,, \\
z & = \bigl(r-M\bigr) \cos \theta \,,
\end{align}
\end{subequations}
from which the functions present in the magnetic ansatz~\eqref{magnetic-ansatz} are
\begin{subequations}
\begin{align}
\label{magnetic-ernst-functions}
f & = r^2 \sin^2 \theta \,, \\
\gamma & = \frac{1}{2} \log \biggl[\frac{r^4 \sin^2 \theta}{r^2-2 M r+ M^2\sin^2\theta} \biggr] \,, \\
\omega & = A_{t} = A_{\phi} = 0 \,.
\end{align}
\end{subequations}
It has to be noted that the value of $\gamma$ is not fundamental because it is invariant under the Ehlers and Harrison transformations, however, it has been made explicit for completeness.
With these functions, it is possible to obtain the magnetic twisted potentials by solving the differential equations that define them~\eqref{magnetic-twisted-potentials}, which results in these twisted potentials being zero up to an integration constant
\beq
\label{magn-twist-scharz}
h = \tilde{A}_{t} = 0\,.
\eeq
The following step will be applying an Ehlers or Harrison transformation to the Ernst potentials of the magnetic ansatz, that were defined as~\eqref{magnetic-ernst-potentials}
\begin{align*}
\ernst & \coloneqq - f - \Phi^*\Phi + i h \,,\\
\Phi & \coloneqq \tilde{A}_{t} - i A_{\phi} \,,
\end{align*}
which, using Eqs.~\eqref{magnetic-ernst-functions}~\eqref{magn-twist-scharz}, in this case are
\begin{subequations}
\label{ernst-potentials-magnetic-Schwarzschild}
\begin{align}
\ernst & = -r^2 \sin^2 \theta\,, \\
\Phi & = 0 \,.
\end{align}
\end{subequations}
\subsubsection{Magnetic Ehlers: Swirling} \label{subsubsec:magnetic-ehlers}
An Ehlers transformation~\eqref{Ehlers} with real parameter $\jmath$ :
\begin{align*}
\ernst' & = \frac{\ernst}{1 + i\, \jmath\, \ernst} \,, \\
\Phi' & = \frac{\Phi}{1 + i\,\jmath\,\ernst} \,,
\end{align*}
when applied to the Ernst potentials under consideration~\eqref{ernst-potentials-magnetic-Schwarzschild}, yields
\begin{subequations}
\begin{align}
\ernst' & = -\frac{r^2 \sin^2 \theta}{1 - i\, \jmath\, r^2 \sin^2\theta}\,, \\
\Phi' & = 0  \,,
\end{align}
\end{subequations}
which, using Eqs.~\eqref{magnetic-ernst-potentials}, result in
\begin{subequations}
\begin{align}
f' & = - \Re\ernst' - \Phi'^*\Phi' = \frac{r^2 \sin^2 \theta}{1+\jmath^2r^4\sin^4\theta} \,, \\
h' & = \Im \ernst'= - \frac{\jmath \, r^4 \sin^4 \theta}{1+\jmath^2r^4\sin^4\theta} \,, \\
\gamma' & = \gamma \,, \\
A'_{t} & = A'_{\phi} = 0 \,, \label{em-pot-swirl}
\end{align}
\end{subequations}
while $\omega'$ has to be found from the definition of the twisted potential $h'$~\eqref{magnetic-grav-potential}, which gives
\beq
\label{omega-swirling}
\omega' = - 4 \jmath \bigl(r- 2M\bigr) \cos\theta + \omega_0 \,,
\eeq
where $\omega_0$ is an integration constant that can be reabsorbed being related to the frame of reference. 

\noindent Consequently, the result for the metric is
\beq
\label{schwarzschild-swirling}
\pms{ds} = F \biggl[ - \biggl( 1 - \frac{2M}{r}\biggr) dt^2 + \frac{dr^2}{\Bigl(1 - \frac{2M}{r}\Bigr)} + r^2 d\theta^2 \biggr] + \frac{r^2 \sin^2 \theta}{F} \biggl[ d\varphi \,+\, 4\,\jmath\bigl(r-2M\bigr)\cos \theta dt \biggr]^2 ,
\eeq
where the function $F$ has been defined as
\beq
F\coloneqq 1+\jmath^2r^4 \sin^4 \theta \,,
\eeq
and the electromagnetic potential remains null as stated in Eq.~\eqref{em-pot-swirl}, meaning that a magnetic Ehlers transformation maps a vacuum solution into a vacuum solution. Obviously, Eq.~\eqref{schwarzschild-swirling} reduces again to the seed Schwarzschild metric for $\jmath=0$.

This solution thus represents a Schwarzschild black hole~\eqref{Schwarzschild-metric} with mass $M$, embedded in what is called the {\bfseries swirling universe}~\cite{Swirling}, which is a sort of rotating universe, in the sense that, by setting to zero the mass, $M = 0$, it is possible to remove the black hole and recover the background swirling universe:
\beq
\label{swirling-background}
\pms{ds} = F \biggl[ - dt^2 + dr^2+ r^2 d\theta^2 \biggr] + \frac{r^2 \sin^2 \theta}{F} \biggl[ d\varphi + 4\,\jmath\, r\cos \theta dt \biggr]^2 \,,
\eeq
corresponding to a universe with a rotation due to a frame-dragging effect~\eqref{frame-dragging-eq} equal to 
\beq
\Omega = \frac{d \phi}{d t} = -4 \,\jmath\, r \cos\theta \,.
\eeq
Moreover, this rotation is also modified by the black hole presence. Indeed, in the massive case, this parameter has an additional term proportional to the mass\footnote{As we will see in Sec.~\ref{sec:Kerr-Newman-Melvin-Swirling} the rotation is in general modified by all the parameters of the theory.}:
\beq
\Omega = -4 \jmath \bigl(r - 2M\bigr) \cos\theta \,.
\eeq
In particular, using the cylindrical coordinates~\eqref{magnetic-coordinates}, this can be understood as a rotation around the symmetry $z-$axis, with the two hemispheres rotating in the opposite direction from each other, indeed, in these coordinates the rotation becomes:
\beq
\Omega = -4\, \jmath\,z \,,
\eeq
which is an expression that holds both for the background and also for the Schwarzschild black hole embedded in this universe.

\subsubsection{Magnetic Harrison: Electromagnetic Melvin}  \label{subsubsec:magnetic-harrison}

Similarly, a Harrison transformation with a complex parameter $\alpha$ :
\begin{align*}
\ernst' & = \frac{\ernst}{1 - 2\,\alpha^*\Phi - \alpha^* \alpha \, \ernst} \,, \\
\Phi' & = \frac{\alpha\,\ernst + \Phi}{1 - 2\,\alpha^*\Phi - \alpha^* \alpha \, \ernst} \,,
\end{align*}
applied to the Ernst potentials given by Eqs.~\eqref{ernst-potentials-magnetic-Schwarzschild} results in
\begin{subequations}
\begin{align}
\ernst' & = -\frac{r^2 \sin^2\theta}{1+\alpha^*\alpha \, r^2 \sin^2\theta} \,, \\
\Phi' & = -\frac{\alpha\, r^2 \sin^2\theta}{1+\alpha^*\alpha\, r^2 \sin^2\theta} \,,
\end{align}
\end{subequations}
and therefore
\begin{subequations}
\begin{align}
f' & = \frac{r^2 \sin^2\theta}{1+\alpha^*\alpha \, r^2 \sin^2\theta} \,, \\
h' & = \omega' = 0 \,, \\
\gamma' & = \gamma \,.
\end{align}
\end{subequations}
Additionally, the magnetic Harrison transformations also generate an electromagnetic field, which can be obtained by using the definition of the Ernst potential $\Phi'=\tilde{A}'_t - i A'_{\phi}$~\eqref{magnetic-em-complex-ernst-potential} and by integrating the equations for the electromagnetic twisted potential $\tilde{A}'_t$~\eqref{magnetic-max-potential}. Hence, after defining the complex Harrison parameter as $\alpha\coloneqq \frac{1}{2}\bigl(E - i\,B\bigr)$, the metric and the electromagnetic potential are given by
\begin{subequations}
\label{em-Melvin-Schwarzschild}
\begin{align}
\pms{ds} & = F^2 \biggl[ -\biggl(1 - \frac{2M}{r}\biggr) dt^2 + \frac{dr^2}{\Bigl(1 - \frac{2M}{r}\Bigr)}+ r^2 d\theta^2 \biggr] + \biggl[\frac{r^2 \sin^2\theta}{F^2}\biggr]d\phi^2  \,, \\
A' & = E \bigl(r-2 M\bigr) \cos\theta dt + \biggl[\frac{B\, r^2 \sin^2\theta}{ 2\,F}\biggr]  d\phi \,,
\end{align}
\end{subequations}
where
\beq
F \coloneqq 1 + \biggl[\frac{E^2 + B^2}{4}\biggr] r^2 \sin^2\theta \,.
\eeq
The background is again recovered by setting $M=0$:
\begin{align}
\label{em-Melvin-background}
\pms{ds} & = F^2 \bigl( - dt^2 + dr^2 + r^2 d\theta^2 \bigr) + \biggl[\frac{r^2 \sin^2\theta}{F^2}\biggr] d\phi^2 \,, \\
A' & = E\, r \cos\theta dt + \biggl[\frac{B\,r^2 \sin^2\theta}{2\,F}\biggr]  d\phi \,.
\end{align}
\vfill
\clearpage
\noindent This background~\eqref{em-Melvin-background} thus corresponds to a universe filled with a “uniform” electric field $E$ and magnetic field $B$, called the {\bfseries electromagnetic universe}, and that represents a generalization of the purely magnetic sub-case, $E=0$, known as the Melvin universe~\cite{Melvin}. For this reason, Eq.~\eqref{em-Melvin-Schwarzschild} represents a Schwarzschild black hole embedded in an electromagnetic universe~\cite{ErnstMelvin}.

In particular, the purely {\bfseries magnetic Melvin universe} is
\begin{subequations}
\label{Melvin-background}
\begin{align}
\pms{ds} & = \bigl(1 + \frac{1}{4} B^2\, r^2 \sin^2\theta\bigr)^2\biggl[ - dt^2 + dr^2 + r^2 d\theta^2 \biggr] + \frac{r^2 \sin^2\theta}{\bigl(1 + \frac{1}{4} B^2\, r^2 \sin^2\theta\bigr)^2}d\phi^2  \,, \\
A' & = \frac{B\, r^2 \sin^2\theta}{2 \bigl(1 + \frac{1}{4}B^2\, r^2 \sin^2\theta\bigr)}\ d\phi \,.
\end{align}
\end{subequations}
\subsection{Electric Transformations on the Schwarzschild Black Hole}  \label{subsec:electric-transformations}
In a similar way to what was done for the magnetic ansatz in Sec.~\ref{subsec:magnetic-transformations}, it holds that the Schwarzschild spacetime~\eqref{Schwarzschild-metric}
\beqNN
ds^2 = -\biggl(1 - \frac{2M}{r}\biggr) dt^2 + \frac{dr^2}{\Bigl(1 - \frac{2M}{r}\Bigr)} + r^2 \bigl( d\theta^2 + \sin^2\theta d\phi^2 \bigr) \,,
\eeqNN
can be written in the electric ansatz~\eqref{electric-ansatz}
\begin{align*}
ds_{e}^2 & = \frac{1}{f}\Bigl[\rho^2 d\phi^2 + e^{2\gamma} \bigl(d\rho^2 + dz^2\bigr)\Bigr] - f \bigl(dt - \omega d \phi\bigr)^2 \,,\\
A &= A_t dt + A_\phi d\phi \,,
\end{align*}
using the same cylindrical coordinates as the magnetic ansatz
\begin{subequations}
\begin{align}
\label{electric-coordinates}
\rho = \sqrt{r^2 - 2Mr} \sin \theta \,, \\
z = \bigl(r - M\bigr) \cos \theta \,,
\end{align}
\end{subequations}
from which the functions present in the electric ansatz~\eqref{electric-ansatz} correspond to
\begin{subequations}
\begin{align}
\label{electric-ernst-functions}
f & = 1 - \frac{2M}{r} \,, \\
\gamma & = \frac{1}{2} \log \biggl[\frac{r^2-2 M r}{r^2-2 M r+ M^2\sin^2\theta} \biggr] \,, \\
\omega &  = A_{t} = A_{\phi} = 0 \,.
\end{align}
\end{subequations}
\vfill
\clearpage
\noindent With these functions, it is possible to obtain the twisted potentials, using their definitions as Eqs.~\eqref{max-potential}~\eqref{grav-potential}, which can be proved to be zero up to an integration constant,
\beq
h = \tilde{A}_{\phi} = 0\,,
\eeq
while the Ernst potentials of the electric ansatz~\eqref{grav-complex-ernst-potential}~\eqref{em-complex-ernst-potential}
\begin{align*}
\ernst & = f - \Phi^*\Phi + i\,h \,, \\
\Phi & = A_t + i \tilde{A}_\phi \,,
\end{align*}
in this case, are given by
\begin{subequations}
\label{ernst-potentials-electric-Schwarzschild}
\begin{align}
\ernst & = 1 - \frac{2M}{r} \,, \\
\Phi & = 0 \,.
\end{align}
\end{subequations}
\vspace{-1.05cm}
\subsubsection{Electric Ehlers: Taub-NUT}  \label{subsubsec:electric-ehlers}
\vspace{-0.1cm}
Therefore, an Ehlers transformation~\eqref{Ehlers}
\begin{align*}
\ernst' & = \frac{\ernst}{1 + i\, \jmath\, \ernst} \,, \\
\Phi' & = \frac{\Phi}{1 + i\,\jmath\,\ernst} \,,
\end{align*}
applied to the seed Ernst potentials~\eqref{ernst-potentials-electric-Schwarzschild}, results in
\begin{subequations}
\begin{align}
\ernst' &= \frac{r^2 - 2Mr}{r^2 + \jmath^2 \bigl(r-2M\bigr)^2}  -i \,\frac{\jmath \bigl(r - 2M\bigr)^2}{r^2 + \jmath^2 \bigl(r-2M\bigr)^2} \,, \\
\Phi' &= 0 \,.
\end{align}
\end{subequations}
This implies that an electric Ehlers transformation does not add an electromagnetic field to a seed solution that does not possess one from the beginning, thus mapping a vacuum solution into a vacuum solution, as was the case for the magnetic Ehlers transformation.

Consequently, the functions needed to construct the new metric are
\begin{subequations}
\begin{align}
f' & = \frac{r^2 - 2Mr}{r^2 + \jmath^2 \bigl(r-2M\bigr)^2} \,, \\
h' & = - \frac{\jmath\bigl(r - 2M\bigr)^2}{r^2 + \jmath^2 \bigl(r-2M\bigr)^2} \,, \\
\omega' & = 4\,\jmath\, M \cos\theta \,.
\end{align}
\end{subequations}
Therefore, the new metric after the electric Ehlers transformation is:
\beq
\begin{split}
\pms{ds} & = -\frac{r^2-2Mr}{r^2+\jmath^2\bigl(r-2M\bigr)^2} \bigl(dt - 4\jmath M \cos\theta d\phi\bigr)^2 + \frac{r^2+\jmath^2\bigl(r-2M\bigr)^2}{r^2-2Mr} dr^2 \\
&\quad + \Bigl[r^2+\jmath^2\bigl(r-2M\bigr)^2\Bigr] \bigl(d\theta^2 + \sin^2\theta d\phi^2 \bigr) \,. \label{taub-nut-0}
\end{split}
\eeq
In order to identify the physical meaning of this metric~\eqref{taub-nut-0}, it is useful to perform the following change of coordinates\footnote{The transformation of the radial coordinate is suggested by the $S^2$ element in the metric.}:
\begin{subequations}
\begin{align}
R & = \frac{r + \jmath^2 \bigl(r - 2M\bigr)}{\sqrt{1+\jmath^2}} \,, \\
T & = \frac{t}{\sqrt{1+\jmath^2}} \,,
\end{align}
\end{subequations}
that, together with a redefinition of the parameters as
\begin{subequations}
\begin{align}
M & = -\frac{n}{2\jmath} \sqrt{1+\jmath^2} \,, \\
\jmath & = \frac{m - \sqrt{m^2 + n^2}}{n} \,,
\end{align}
\end{subequations}
allows rewriting the obtained solution~\eqref{taub-nut-0} as the well-known {\bfseries Taub-NUT} metric~\cite{Taub,NUT}:
\beq
\begin{split}
\pms{ds} &= -\frac{R^2-2mR-n^2}{R^2+n^2} \bigl(dT - 2n\cos\theta d\phi\bigr)^2 + \frac{R^2+n^2}{R^2-2mR-n^2} dR^2 \\
&\quad + \bigl(R^2+n^2\bigr) \bigl(d\theta^2 + \sin^2\theta d\phi^2 \bigr) \,,
\end{split}
\eeq
where $m$ is the mass and $n$ is the so-called NUT parameter.

An important consideration is that if the seed is the flat Minkowski spacetime, i.e.~the massless case, it holds that, after an electric Ehlers transformation, the result is again the Minkowski spacetime, thus without the addition of any new parameter. Indeed, for $M=0$ the metric in Eq.~\eqref{taub-nut-0} reduces to
\beq
\pms{ds} = -\frac{dt^2}{\bigl(1+\jmath^2\bigr)} + \bigl(1+\jmath^2\bigr) dr^2 + r^2\bigl(1+\jmath^2\bigr)\bigl(d\theta^2 + \sin^2\theta d\phi^2 \bigr) \,, \label{taub-nut-minkowski}
\eeq
that is exactly the Minkowski background modulo a rescaling of the coordinates.
\vfill
\clearpage

\subsubsection{Electric Harrison: Dyonic Reissner-Nordstr\"om}  \label{subsubsec:electric-harrison}

Finally, applying a Harrison transformation
\begin{align*}
\ernst' & = \frac{\ernst}{1 - 2\,\alpha^*\Phi - \alpha^* \alpha \, \ernst} \,, \\
\Phi' & = \frac{\alpha\,\ernst + \Phi}{1 - 2\,\alpha^*\Phi - \alpha^* \alpha \, \ernst} \,,
\end{align*}
to the considered seed~\eqref{ernst-potentials-electric-Schwarzschild}, leads to the following Ernst potentials
\begin{subequations}
\begin{align}
\ernst' & = \frac{r-2M}{r-\alpha^*\alpha \bigl(r-2M\bigr)} \,, \\
\Phi' & =  \frac{\alpha\bigl(r-2M\bigr)}{r-\alpha^*\alpha \bigl(r-2M\bigr)} \,,
\end{align}
\end{subequations}
from which
\begin{subequations}
\begin{align}
\label{functions-primed-rn}
f' & = \frac{r\bigl(r-2M\bigr)}{\Bigl[\, r-\alpha^*\alpha \bigl(r-2M\bigr) \Bigr]^2} \,, \\
h' & = \omega' = 0 \,, \\
\gamma' & = \gamma \,, \\
A' & = \frac{\alpha_{r}\bigl(r-2M\bigr)}{r-\alpha^*\alpha \bigl(r-2M\bigr)} dt + 2\,\alpha_I \,M \cos\theta d\phi \,,
\end{align}
\end{subequations}
where the electromagnetic twisted potential $\tilde{A}_\phi$ has been obtained using its definition as in Eq.~\eqref{max-potential}, and the complex parameter of the Harrison transformation has been defined as $\alpha \coloneqq \alpha_R + i\,\alpha_I$.

Thus, using the functions obtained after performing the Harrison transformation~\eqref{functions-primed-rn}, the resulting metric is
\beq
\begin{split}
\label{rn-0}
\pms{ds} & = -\frac{r\bigl(r-2M\bigr)}{\Bigl[\, r-\alpha^*\alpha \bigl(r-2M\bigr) \Bigr]^2} dt^2 + \frac{\Bigl[\, r-\alpha^*\alpha \bigl(r-2M\bigr) \Bigr]^2}{r\bigl(r-2M\bigr)} dr^2 \\
&\quad + \Bigl[\, r-\alpha^*\alpha \bigl(r-2M\bigr) \Bigr]^2\bigl( d\theta^2 + \sin^2\theta d\phi^2 \bigr) \,,
\end{split}
\eeq
which, after performing the following change of coordinates
\begin{subequations}
\begin{align}
R & = r\bigl(1-\alpha^*\alpha\bigr) + 2M\alpha^*\alpha \,, \\
T & = \frac{t}{1-\alpha^*\alpha} \,,
\end{align}
\end{subequations}
together with the reparametrizations 
\vspace{-0.1cm}
\begin{subequations}
\begin{align}
p & = 2\,\alpha_I M \,, \\
e & = -\frac{2\,\alpha_R M}{1-\alpha^*\alpha} \,, \\
m & = M \bigl(1+\alpha^* \alpha\bigr) \,,
\end{align}
\end{subequations}
and by defining the dyon $q^2 \coloneqq e^2+p^2$, it holds that the final result for the electromagnetic potential and the metric is
\begin{align}
\label{dyonic-rn}
\pms{ds} & = - \biggl(1 - \frac{2m}{R} + \frac{q^2}{R^2}\biggr) dt^2 + \frac{dR^2}{\Bigl(1 - \frac{2m}{R} + \frac{q^2}{R^2}\Bigr)} + R^2 \bigl( d\theta^2 + \sin^2\theta d\phi^2 \bigr) \,, \\
A' & = \frac{e}{R} dt + p \cos\theta d\phi \,,
\end{align}
known as the {\bfseries dyonic Reissner-Nordstr\"om black hole}~\cite{Reissner,Nordstrom}, i.e.~the non-rotating, $a=0$, sub-case of the dyonic Kerr-Newman black hole~\eqref{kn-magnetic}, thus representing an electromagnetically charged black hole, with mass $m$, electric charge $e$ and magnetic charge $p$.

As in the case of the electric Ehlers transformation, an electric Harrison transformation also does \emph{not} introduce any new parameters when applied to the Minkowski spacetime. Indeed, the metric given by Eq.~\eqref{taub-nut-0} also reduces to the Minkowski background modulo a rescaling of the coordinates for $M=0$:
\beq
\pms{ds} = -\frac{dt^2}{\bigl(1+\alpha^* \alpha)} + \bigl(1+\alpha^* \alpha) dr^2 + r^2\bigl(1+\alpha^* \alpha\bigr)\bigl(d\theta^2 + \sin^2\theta d\phi^2 \bigr) \,. \label{rn-minkowski}
\eeq
\subsection{Physical Meaning of the Ehlers-Harrison Transformations} \label{subsec:physical-ehlers-harrison}

The physical meaning of the parameters added by the electric and magnetic Ehlers-Harrison transformations can therefore be understood from the metrics obtained in Sec.~\ref{subsec:magnetic-transformations} and Sec.~\ref{subsec:electric-transformations}, through the application of these transformations to the simple case of a Schwarzschild black hole~\eqref{Schwarzschild-metric}.

Summarizing, a Harrison transformation applied to an electric ansatz has the effect of adding a dyon, i.e. it adds both an electric and a magnetic charge. Thus generating the so-called {\bfseries dyonic Reissner-Nordstr\"{o}m black hole} when starting from the Schwarzschild one. Similarly, on the magnetic ansatz, the Harrison transformation embeds the seed spacetime in the {\bfseries electromagnetic universe}, which is a spacetime filled with a “uniform” electromagnetic field. In particular, when the parameter of the Harrison transformation is real and the mass is set to zero, the resulting background is a universe filled just with a  \emph{uniform magnetic field}, known as the {\bfseries Melvin universe}.

\noindent On the other hand, an Ehlers transformation applied to an electric ansatz has the effect of adding the NUT parameter~\cite{Reina}, therefore generating what is referred to as the {\bfseries Taub-NUT black hole} when the seed is the Schwarzschild black hole. Analogously, on the magnetic ansatz, an Ehlers transformation embeds the given spacetime in the {\bfseries swirling universe}, which is a sort of \emph{rotating universe}~\cite{Swirling}, as explained in Sec.~\ref{subsubsec:magnetic-ehlers}.
{\begin{table}[H]
\begin{center}
\doublespacing
\begin{tabular}{ |p{4.5cm}||p{4.25cm}|p{4cm}|  }
\hline
\multicolumn{3}{|c|}{Ehlers-Harrison on the Schwarzschild Black Hole} \\
\hline
                                                                   & Harrison                                                                            & Ehlers                                  \\
\hline
Electric\,\,\,\,\,\,Transformation                   & Reissner-Nordstr\"{o}m  BH                                            	& Taub-NUT BH                              \\
\hline
Magnetic\,\,Transformation                        & Schwarzschild BH in an                                                    & Schwarzschild BH  in a        \\
\,                                                                 & electromagnetic universe                                                  & swirling universe        \\
\hline
\end{tabular}
\caption{\small Table summarizing the effect of an electric or magnetic Ehlers-Harrison transformation applied to the Schwarzschild black hole.}
\end{center}
\label{table:Ehlers-Harrison}
\end{table}}
\vspace{-0.2cm}
\noindent Moreover, an Ehlers transformation never generates an electromagnetic field if not already present in the seed solution, thus always mapping a vacuum solution into a vacuum solution. In contrast, a Harrison transformation always generates an additional electromagnetic field, except when applied to the Minkowski spacetime in the electric ansatz.

Indeed, as pointed out in Sec.~\ref{subsec:electric-transformations}, while any \emph{magnetic} transformation always generates a new independent solution when applied to the Minkowski background, this is not the case for an \emph{electric} transformation, since the application of any electric transformation to the Minkowski spacetimes always results in a mere change of coordinates. The reason behind this result relies upon the fact that the magnetic transformations add new parameters characterizing the background universe, whereas the electric transformations add new parameters to the black hole, which the Minkowski spacetime obviously does \emph{not} contain.

Another important consideration is that the Elhers and Harrison transformations are Lie point symmetry of the Ernst equations, which basically means that, after one of these transformations, the result will always be in the same ansatz as the seed solution, thus implying that it is always possible to compose an arbitrary number of these transformations. Moreover, these transformations also commute, hence, for example, a Melvin and swirling universe can be obtained either through the application of a magnetic Ehlers transformation followed by a magnetic Harrison, or equivalently using the opposite order of compositions. Since it will be relevant in the following sections, the explicit metric and electromagnetic potential of the Melvin-swirling universe are reported in Appendix~\ref{sub:melvin-swirling-universe}. 

Furthermore, In Sec.~\ref{sec:Kerr-Newman-Melvin-Swirling}, the magnetic transformations will be used to obtain, for the first time, the dyonic Kerr-Newman black hole~\eqref{kn-magnetic} embedded in a Melvin-swirling universe.

\clearpage
\section{The Melvin-swirling Universe}
\label{sec:Melvin-Swirling-Universe}
There is a profound relation between the parameters that can be added using the Ernst formalism. Indeed, the swirling background~\eqref{swirling-background} can also be obtained in a particular way starting from the Taub-NUT spacetime with a flat base manifold~\cite{Swirling}, which in turn can be generated via an Ehlers transformation~\eqref{Ehlers}, from the Schwarzschild metric~\eqref{Schwarzschild-metric}, previously composed with a double-Wick rotation~\eqref{double-Wick}\footnote{While this is true for a generic sign of the constant curvature of the seed base manifold, only the metrics with positive curvature can be interpreted as black holes in General Relativity.}.

In fact, the flat Taub-NUT spacetime, with mass $m$ and NUT parameter $n$ is given by the metric
\beq
ds^2 = -\frac{2 m r}{r^2 + n^2}\bigl(d\tau-2n\theta d\varphi\bigr)^2 + \frac{r^2 + n^2}{2mr}dr^2 + \bigl(r^2 + n^2\bigr) \bigl(d\theta^2+d\varphi^2\bigr) \,,\label{Taub-NUT-flat}
\eeq
which, after the following double-Wick rotation and reparametrizations
\begin{subequations}
\label{rep-nut-swirl}
\begin{align}
n &= \Bigl(\,\frac{m}{2\jmath}\,\Bigr)^{\frac{1}{3}} \,,\\
\varphi &= -\frac{i \, t}{n} = -i\Bigl(\,\frac{2\jmath}{m}\,\Bigr)^{\frac{1}{3}}t \,,\\
\tau &= -\frac{i\, n^2}{m}\phi = -i\Bigl(\,\frac{m}{2\jmath}\,\Bigr)^{\frac{2}{3}}\frac{\phi}{m} \,,\\
\theta &= -\frac{z}{n} = -\Bigl(\,\frac{2\jmath}{m}\,\Bigr)^{\frac{1}{3}}z \,,\\
r &= \frac{m}{2n^2}\rho^2 = \frac{m}{2}\Bigl(\,\frac{2\jmath}{m}\,\Bigr)^{\frac{2}{3}}\rho^2 \,,
\end{align}
\end{subequations}
yields
\beq
ds^2 = \bigl(1+\jmath^2\rho^4\bigr)\bigl(-dt^2+d\rho^2+dz^2\bigr)+\frac{\rho^2}{\bigl(1+\jmath^2\rho^4\bigr)}\bigl(d\phi+4\jmath z\,dt\bigr)^2 \,,\label{swirling-background-cylindrical} \\
\eeq
that is exactly the swirling background, with swirling parament $\jmath$, in cylindrical coordinates
\begin{subequations}
\label{cyl-mv-sec}
\begin{align}
\rho &= r \sin \theta \,, \\
 z & = r \cos \theta \, \,.
\end{align}
\end{subequations}
\clearpage
\noindent Analogously, the Melvin background can be also obtained starting from the magnetic Reissner-Nordstr\"om spacetime with a flat base manifold~\cite{GibbonsWiltshire, GibbonsHerdeiro}, which is given by
\begin{subequations}
\begin{align}
ds^2 & = - \frac{2 m r + p^2}{r^2} dt^2 + \frac{r^2}{2mr + p^2}dr^2 + r^2 \bigl( d\theta^2 + d\varphi^2 \bigr) \,, \\
A& = p\, \theta\, d\varphi \,,
\end{align}
\end{subequations}
where $m$ is the mass and $p$ is the magnetic charge. On the other hand, this magnetic Reissner-Nordstr\"om with a flat base manifold can also be generated from the Schwarzschild metric via a Harrison transformation~\eqref{Harrison} previously composed with a double-Wick rotation.

Moreover, both the Melvin universe and the swirling universe are obtained respectively by applying a Harrison transformation and an Ehlers transformation on a seed expressed in the magnetic ansatz. Thus, it is possible to summarize the correspondence between these backgrounds by the following proportion:
\beq
\boxed{\text{\small Melvin Universe : Harrison transformations = Swirling Universe : Ehlers transformations}} \label{proportion}
\eeq
Furthermore, the Melvin universe can be demonstrated to be equivalent to the spacetime generated by two magnetic Reissner-Nordstr\"om black holes with opposite charges that are infinitely distant along the symmetry axis~\cite{Emparan}. Similarly, as explained in~\cite{Swirling}, the swirling background should have an analogous interpretation, with the Reissner-Nordstr\"om black holes replaced by two Taub-NUT black holes with opposite NUT parameters. Although this interpretation has not been yet demonstrated, it seems to be a valid hypothesis given the many analogies between the Melvin and the swirling universes, where, in particular, we have that the swirling universe can be obtained as a double-Wick rotation of the flat Taub-NUT black hole in the same way as the Melvin universe can be obtained as a double-Wick rotation of the flat magnetic Reissner-Nordstr\"om black hole. Moreover, the swirling background can also be thought of, with a certain degree of approximation, as the gravitational setting generated by the interplay of two sources counter-rotating around the same axis, such as two counter-rotating galaxies or black holes~\cite{AstorinoRemoval}. 

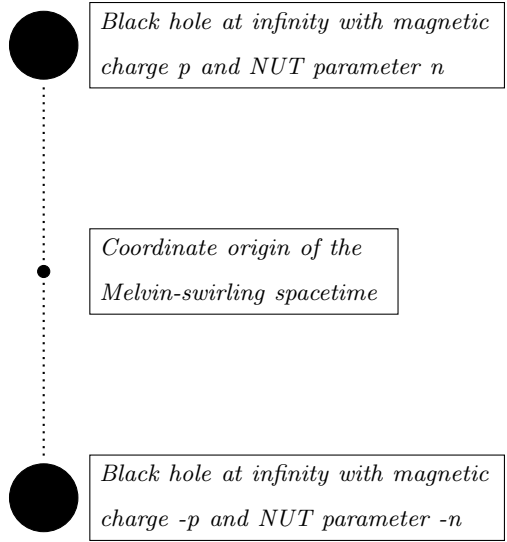
\begin{figure}[H]
\begin{center}
\begin{tikzpicture}[trim right=5.57cm]
\filldraw[color = black, fill = black] (0,3) circle (0.45cm);
\filldraw[color = black , fill = black] (0,-3) circle (0.45cm);
\filldraw[color = black , fill = black] (0,0) circle (0.08cm);
\path [draw = black, thick, dotted] (0,0.15) -- (0,2.5);
\path [draw = black, thick, dotted] (0,-0.15) -- (0,-2.5);
\node[draw] at (0.6,3) [anchor=west, text width = 52 mm] {\footnotesize \it Black hole at infinity with magnetic \\ charge p and NUT parameter n};
\node[draw] at (0.6,0) [anchor=west, text width = 38 mm] {\footnotesize \it Coordinate origin of the \\ Melvin-swirling spacetime};
\node[draw] at (0.6,-3) [anchor=west, text width = 52 mm] {\footnotesize \it Black hole at infinity with magnetic \\ charge -p and NUT parameter -n};
\end{tikzpicture}
\caption{\small Schematic representation of the Melvin-swirling background as two infinitely distant, along the symmetry axis, magnetic Reissner-Nordstr\"om-Taub-NUT black holes with opposite charges.} \label{fig:representation} 
\end{center}
\end{figure}

Combining these interpretations, the Melvin-swirling universe\footnote{The metric and the electromagnetic potential for this solution are reported in Sec.~\ref{sub:melvin-swirling-universe}.}, which is obtained by applying a magnetic Ehlers transformation to the Melvin universe, or equivalently, by applying a real magnetic Harrison transformation to the swirling universe, can be thought of as the spacetime generated by two magnetic Reissner-Nordstr\"om-Taub-NUT black holes pushed at infinity along the symmetry axis and with opposite magnetic and NUT charges, as sketched in Figure~\ref{fig:representation}. Therefore, performing a double-Wick rotation on the flat magnetic Reissner-Nordstr\"om-Taub-NUT black hole should reproduce the Melvin-swirling background.
\vfill
\clearpage
\null \vspace{0.5cm}
\noindent In addition, as we will prove in Sec.~\ref{subsec:curvature-dknms}, the Melvin-swirling universe is a spacetime locally asymptotically flat
\beq
R_{\mu \nu \rho \sigma}R^{\mu \nu \rho \sigma} \overset{r\to \infty}{\approx} \frac{49152}{\bigl[B^4 + 16 \jmath^2\bigr]^2\sin^{12}\theta\, r^{12}} \,,\eeq
with a constant curvature on the symmetry axis
\beq
R_{\mu \nu \rho \sigma}R^{\mu \nu \rho \sigma}\big\rvert_{\theta = 0, \pi} = 4 \bigl(5 B^4 - 48 \jmath^2\bigl) \,,
\eeq
where we recall that $B$ and $\jmath$ are, respectively, the Melvin and swirling parameters.
\vfill
\vspace{-0.5cm}
\subsection{The Electromagnetic-swirling Universe with a Cosmological Constant}

There are two other considerations to address before testing the claim that the Melvin-swirling background should result from a double-Wick rotation on the flat magnetic Reissner-Nordstr\"om-Taub-NUT black hole. 

\clearpage 
\noindent The first consideration is that it is also possible to add a cosmological constant with the method aforementioned.  Indeed, the flat Taub-Nut black hole with a cosmological constant is given by
\beq
ds^2 = -\frac{\pazocal{R}}{r^2+n^2}\bigl(d\tau-2n\theta d\varphi\bigr)^2+\frac{r^2+n^2}{\pazocal{R}}dr^2+\bigl(r^2+n^2\bigr)\bigl(d\theta^2+d\varphi^2\bigr) \label{Taub-Nut-Cosmological-Flat} \,,
\eeq
where
\beq
\pazocal{R}=2mr + \Lambda \, n^4 - 2\, \Lambda \, n^2 r^2 - \frac{\Lambda}{3} r^4  \,,
\eeq
$m$ is the mass, $n$ is the nut parameter, and $\Lambda$ is the cosmological constant.

\noindent Thus, using the double-Wick rotation and the reparametrizations given by Eqs.~\eqref{rep-nut-swirl}, this solution becomes:
\beq
ds^2 = \bigl(1+\jmath^2\rho^4\bigr)\biggl[-dt^2+\frac{\rho^2d\rho^2}{\pazocal{S}}+dz^2\biggr]+\frac{{\pazocal{S}}}{\bigl(1+\jmath^2\rho^4\bigr)}\bigl(d\phi+4\jmath z\,dt\bigr)^2 \,,\label{swirling-background-cylindrical-cosmological} \\
\eeq
where
\beq
\pazocal{S} = \rho^2 + \frac{\Lambda}{4\jmath^2}  - \frac{\Lambda}{2}\rho^4 - \frac{\jmath^2 \Lambda}{12}\rho^8\,.
\eeq
As can be verified, the metric provided by Eq.~\eqref{swirling-background-cylindrical-cosmological}, is a solution of the Einstein equations with a cosmological constant. Therefore, it is the correct generalization, in cylindrical coordinates~\eqref{cyl-mv-sec}, of the swirling universe with a cosmological constant. Moreover, the same can be done for the Melvin universe, starting from the flat magnetic Reissner-Nordstr\"om black hole with a cosmological constant~\cite{AstorinoMelvinCosmological}.

The second consideration is that the electromagnetic generalization of the Melvin universe, potentially with a cosmological constant, can also be obtained in the same way starting from the flat \emph{dyonic} Reissner-Nordstr\"om black hole. 

Collecting all these results, we will now obtain the electromagnetic-swirling universe with a cosmological constant, starting from the flat dyonic Reissner-Nordstr\"om-Taub-NUT black hole with a cosmological constant, whose metric and potential are
\begin{subequations}
\begin{align}
ds^2& = -\frac{\pazocal{R}}{r^2+n^2}\bigl(d\tau-\,2n\theta d\varphi\bigr)^2+\frac{r^2+n^2}{\pazocal{R}}dr^2+\bigl(r^2+n^2\bigr)\bigl(d\theta^2+d\varphi^2\bigr) \,,\label{Dyonic-RN-Taub-Nut-Cosmological-Flat} \\
A& = -\frac{e\, r}{\bigl(r^2+n^2\bigr)}\bigl(d\tau-2n\theta d\varphi \bigr) + \frac{p}{r^2+n^2}\Bigl[n\,d\tau+ \bigl(r^2-n^2\bigr) \theta d\varphi \Bigr] \,,
\end{align}
\end{subequations}
where
\beq
\pazocal{R} = 2mr + e^2 + p^2 + \Lambda \, n^4 - 2\, \Lambda \, n^2 r^2 - \frac{\Lambda}{3} r^4 \,,\\
\eeq
$m$ is the mass, $e$ is the electric charge, $p$ is the magnetic charge, $n$ is the NUT parameter and $\Lambda$ is the cosmological constant.
\vfill
\clearpage
\noindent In addition to the double-Wick rotation and reparametrizations given by Eqs.~\eqref{rep-nut-swirl}, which were used to obtain the swirling spacetime with a cosmological constant, we can also define a reparametrization for the charges as:
\begin{subequations}
\label{rep-charges-cosmo}
\begin{align}
e & = \frac{i \, m}{n} E = i \bigl(2 m^2  \jmath\bigr)^{\frac{1}{3}}E \,,\\
p &= \frac{i\, m}{n} B = i \bigl(2 m^2  \jmath\bigr)^{\frac{1}{3}}B \,.
\end{align}
\end{subequations}
Therefore, by means of Eqs.~\eqref{rep-nut-swirl} and Eqs.~\eqref{rep-charges-cosmo}, the flat dyonic Reissner-Nordstr\"om-Taub-NUT black hole with a cosmological constant~\eqref{Dyonic-RN-Taub-Nut-Cosmological-Flat} becomes:
\begin{subequations}
\label{Dyonic-Background-Swirling-Cosmological-Cylindrical} 
\begin{align}
ds^2& = \bigl(1+\jmath^2\rho^4\bigr)\biggl[-dt^2+\frac{\rho^2d\rho^2}{\pazocal{S}}+dz^2\biggr]+\frac{{\pazocal{S}}}{\bigl(1+\jmath^2\rho^4\bigr)}\bigl(d\phi+4 \jmath z dt\bigr)^2 \label{metric-patologic-dnb} \,,\\
A& = \biggl[\frac{1}{1+\jmath^2 \rho^4}\biggr]\biggl[\bigl(B-\jmath E \rho^2)d\phi + 2\jmath z \bigl(B - 2\jmath E \rho^2 - \jmath^2 B \rho^4\bigr) dt\biggr]\,, \\
\pazocal{S} & = \rho^2-\bigl(E^2+B^2\bigr) + \frac{\Lambda}{4\jmath^2}  - \frac{\Lambda}{2}\rho^4 - \frac{\jmath^2 \Lambda}{12}\rho^8 \,.
\end{align}
\end{subequations}
However, this new solution~\eqref{Dyonic-Background-Swirling-Cosmological-Cylindrical} presents some odd behaviors. Indeed, if the cosmological constant is zero, $\Lambda = 0$, there is a singularity for $\rho^2=\bigl(E^2+B^2\bigr)$, nonetheless, this is a \emph{coordinate} singularity, that can be fixed by the following coordinate transformation 
\beq
\pms{\rho} = \rho^2 + \bigl(E^2+B^2\bigr) \,.
\eeq
In order to better understand the coordinate nature of this singularity, it can be useful to install the same coordinate singularity in the Minkowski metric expressed in cylindrical coordinates. Indeed, after the opposite shift, $\pms{\rho} = \rho^2 - c^2$, we have that the Minkowski metric results in
\beq
ds^2 = -dt^2 + \frac{\pms{\rho} d\pms{\rho}}{\pms{\rho} - c^2}  + dz^2 + \bigl(\pms{\rho} - c^2\bigr) d\phi^2 \,.
\eeq
Moreover, there is also the problem that in the non-swirling case, $\jmath = 0$, the electromagnetic potential~\eqref{metric-patologic-dnb} becomes zero, up to a gauge transformation, meaning that setting the swirling parameter to zero also removes the electromagnetic background fields. Problem that can be solved by performing a rescaling of the electric and magnetic fields, $E' = \frac{E}{\jmath}$, $B' = \frac{B}{\jmath}$.

When we reached this point, during the mid-stages of this thesis, an article~\cite{BarrientosCisterna} was published where they obtained all the possible backgrounds constructed by composing two different Ehlers-Harrison transformations. In particular, they also found the electromagnetic Melvin-swirling universe with a cosmological constant, using the same double-Wick rotation as we were doing. For this reason, we did not proceed any further in searching for a better possible coordinate transformation.
\vfill
\clearpage

\noindent Nevertheless, we can express this result in the following form:
\begin{subequations}
\label{em-melvin-swirling-cosmological}
\begin{align}
ds^2 & = F \biggl[-dt^2+\frac{\rho^2 d\rho^2}{\pazocal{S}}+dz^2\biggr]+\frac{{\pazocal{S}}}{F}\bigl(d\phi - \Omega \,dt\bigr)^2 \,, \\
A & = A_{0} \,dt + \frac{A_{3}}{F}\bigl(d\phi - \Omega\, dt\bigr) \,,
\end{align}
\end{subequations}
where
\begin{subequations}
\begin{align}
A_{0} & = E z \,, \\
A_{3} & = \frac{1}{2}\Bigl[B \rho^2 - \jmath E \rho^4 + \frac{B X^2}{4} \rho^4\Bigr] \,, \\
\Omega & = - 4 \jmath z \,, \\
F & = 1 + \frac{X^2}{2} \rho^2 + \biggl[\jmath^2 + \frac{X^4}{16}\biggr] \rho^4 \,, \\
\begin{split}
\pazocal{S} &  = -\frac{4\Bigl[\bigr(48\jmath^2 + X^4\bigr)^2 - 3072\jmath^4 \Bigr]\Lambda}{3\bigl(16\jmath^2 + X^4\bigr)^3} + \biggl[1 - \frac{4X^2\bigl(48\jmath^2 + X^4\bigr) \Lambda}{3\bigl(16\jmath^2 + X^4\bigr)^2}\biggr]\rho^2  \\
& \quad - \frac{\Lambda}{2}\rho^4 - \frac{X^2 \Lambda}{12}\rho^6 - \frac{\bigl(16\jmath^2 + X^4\bigr)  \Lambda}{192}\rho^8 \,,
\end{split} \\
X^2 & = B^2 + E^2 \,,
\end{align}
\end{subequations}
with $\jmath$ the swirling parameter, $\Lambda$ the cosmological constant, $E$ the uniform electric field and $B$ the uniform magnetic field. This solution~\eqref{em-melvin-swirling-cosmological} thus represents an {\bfseries electromagnetic-Melvin-swirling universe with a cosmological constant}. 

\begin{figure}[H]
\begin{center}
\begin{tikzpicture}[trim right=5.57cm]
\filldraw[color = black, fill = black] (0,2.8) circle (0.4cm);
\filldraw[color = black , fill = black] (0,-2.8) circle (0.4cm);
\filldraw[color = black , fill = black] (0,0) circle (0.08cm);
\path [draw = black, thick, dotted] (0,0.15) -- (0,2.5);
\path [draw = black, thick, dotted] (0,-0.15) -- (0,-2.5);
\node[draw] at (0.6,2.8) [anchor=west, text width = 65 mm] {\footnotesize \it Black hole at infinity with electric charge e, magnetic charge p and NUT parameter n};
\node[draw] at (0.6,0) [anchor=west, text width = 60 mm] {\footnotesize \it Coordinate origin of the electromagnetic Melvin-swirling spacetime};
\node[draw] at (0.6,-2.8) [anchor=west, text width = 65 mm] {\footnotesize \it Black hole at infinity with electric charge -e, magnetic charge -p and NUT parameter -n};
\end{tikzpicture}
\caption{\small Schematic representation of the electromagnetic Melvin-swirling background as two infinitely distant, along the symmetry axis, dyonic Reissner-Nordstr\"om-Taub-NUT black holes with opposite charges.} \label{fig:representation-2} 
\end{center}
\end{figure}
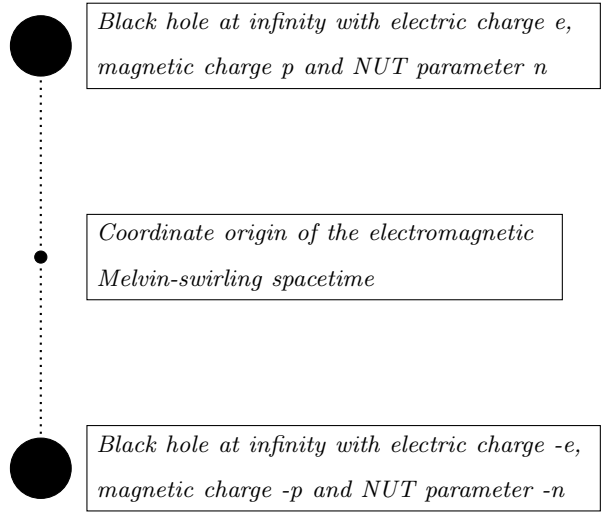

In particular, in the case of a negative cosmological constant $\Lambda = - \frac{3}{\ell^2}$, we find that the function $\pazocal{S}$ becomes:
\begin{align}
\begin{split}
\pazocal{S} &  = \frac{4\Bigl[\bigr(48\jmath^2 + X^4\bigr)^2 - 3072\jmath^4 \Bigr]}{\ell^2\bigl(16\jmath^2 + X^4\bigr)^3} + \biggl[1 + \frac{4X^2\bigl(48\jmath^2 + X^4\bigr) }{\ell^2\bigl(16\jmath^2 + X^4\bigr)^2}\biggr]\rho^2  \\
& \quad + \frac{3}{2\ell^2}\rho^4 + \frac{X^2}{4\ell^2}\rho^6 + \frac{\bigl(16\jmath^2 + X^4\bigr) }{64\ell^2}\rho^8 \,.
\end{split}
\end{align}
Similarly, if the cosmological constant is set to zero, $\Lambda = 0$, we recover the {\bfseries electromagnetic-Melvin-swirling universe}, whose function $\pazocal{S}$ is simply
\beq
\pazocal{S}  = \rho^2 \,.
\eeq
Therefore, as sketched in Figure~\ref{fig:representation-2}, we can conclude that the electromagnetic Melvin-swirling background can be thought of as the spacetime generated by two dyonic Reissner-Nordstr\"om-Taub-NUT black holes pushed at infinity along the symmetry axis and with opposite charges.
\vfill
\clearpage
\noindent Moreover, similarly to the magnetic Melvin-swirling universe, we have that, by means of the following coordinate transformation:
\begin{subequations}
\label{cyl-coor-emswirl}
\begin{align}
\rho & = r \cos\theta \,, \\
z & = r \sin\theta \,,
\end{align}
\end{subequations}
the electromagnetic Melvin-swirling universe is also locally asymptotically flat
\beq
R_{\mu \nu \rho \sigma}R^{\mu \nu \rho \sigma} \overset{r\to \infty}{\approx} \frac{49152}{\bigl[X^4 + 16 \jmath^2\bigr]^2\sin^{12}\theta\, r^{12}} \,,
\eeq
with a constant curvature on the symmetry axis
\beq
R_{\mu \nu \rho \sigma}R^{\mu \nu \rho \sigma}\big\rvert_{\theta = 0,\pi} =  4 \bigl(5 X^4 - 48 \jmath^2\bigl) \,.
\eeq
Analogously, if the cosmological constant $\Lambda$ is not zero, using the same coordinate transformation~\eqref{cyl-coor-emswirl}, we find:
\begin{subequations}
\begin{align}
R_{\mu \nu \rho \sigma}R^{\mu \nu \rho \sigma}\big\rvert_{\theta \neq 0,\pi} \,&\xrightarrow{r\to \infty}\,  \frac{8 \Lambda^2}{3} \,, \\
\begin{split}
R_{\mu \nu \rho \sigma}R^{\mu \nu \rho \sigma}\big\rvert_{\theta = 0,\pi} & = \frac{4}{3}\biggl[15 X^4 - 144 \jmath^2 + 2\Lambda^2 - \frac{16384 \jmath^4 \Lambda}{ \bigl(16 \jmath^2 + X^4\bigr)^6}\Bigl[ 3 X^2 \bigl(16 \jmath^2 + X^4\bigr)^4\\
& \quad  + 16 \jmath^2 \bigl(\Lambda - 6X^2\bigr)\bigl(16\jmath^2+X^4\bigr)^3 - 512 \jmath^4 \Lambda \bigr(16\jmath^2-3X^4\bigr)^2\Bigr]  \biggr] \,.
\end{split}
\end{align}
\end{subequations}
\vfill

\clearpage
\section{New Charged and Rotating Black Holes in a Melvin-swirling Universe}
\label{sec:Kerr-Newman-Melvin-Swirling}
\vspace{0.5cm}
In this chapter, we will analytically obtain the new solution of the Einstein-Maxwell equations representing a {\bfseries dyonic Kerr-Newman black hole in a Melvin-swirling universe}, whose physical interpretation is that of an electromagnetically charged and rotating black hole (dyonic Kerr-Newman~\eqref{kn-magnetic}) embedded in a rotating universe (swirling~\eqref{swirling-background}) permeated by a uniform magnetic field (Melvin~\eqref{Melvin-background}).

Moreover, we will also obtain some other new solutions as sub-cases of this new spacetime, such as the \emph{Kerr black hole}~\eqref{kerr-magnetic} (i.e.~a rotating but not charged black hole) and the \emph{dyonic Reissner-Nordstr\"om black hole}~\eqref{dyonic-rn} (i.e.~a charged but not rotating black hole) embedded in a Melvin-swirling universe.

In principle, this new solution can be obtained starting from the simple case of a Kerr black hole without any additional background. Indeed, as explained in Sec.~\ref{sec:Ernst}, a complex electric Harrison transformation can be used to add the electromagnetic charges, thus obtaining the dyonic Kerr-Newman black hole. Then, a real magnetic Harrison transformation can add the uniform magnetic-Melvin field. Finally, the swirling background can be added by means of a magnetic Ehlers transformation. Or, equivalently, in any other order since these transformations commute.

However, since the dyonic Kerr-Newman black hole in a Melvin universe has already been obtained~\cite{Gibbons}, we will start directly from this spacetime, which we will generalize by adding the swirling parameter through a magnetic Ehlers transformation.
\vfill
\clearpage
\subsection{Generating the New Solutions} \label{subsec:generation}
\vspace{-0.15cm}
Therefore, we begin with the dyonic Kerr-Newman-Melvin metric~\cite{Gibbons}, which we have found can be expressed in the following particular form that ties the functions present in the metric and in the electromagnetic potential:
\begin{subequations}
\label{DKNM}
\begin{align}
ds^2 & = F \biggl[ -\frac{\Delta}{\Sigma}\,dt^2 + \frac{dr^2}{\Delta} + d\theta^2\biggr] + \frac{\Sigma\sin^2\theta}{F}\biggl[ d\phi - \frac{\Omega}{\Sigma} \,dt \biggr]^2 \,, \\
A & = \frac{A_0}{\Sigma} \,dt + \frac{A_{3}}{F}\biggl[ d\phi - \frac{\Omega}{\Sigma}\,dt \biggr] \,,
\end{align}
\end{subequations}
where the various functions have been defined as:
\begin{subequations}
\begin{align}
A_0 &= \chi_{(0)} + \frac{B}{2}\,\chi_{(1)} + \frac{3B^2}{4}\,\chi_{(2)} + \frac{B^3}{8}\,\chi_{(3)}  \,, \\
A_3 &= \varphi_{(0)} +  \frac{B}{2}\,\varphi_{(1)} + \frac{3B^2}{4}\,\varphi_{(2)} + \frac{B^3}{8}\,\varphi_{(3)}  \,, \\
\Omega &= a\,\lambda + 2B\,\chi_{(0)} +  \frac{B^2}{2}\,\chi_{(1)} + \frac{B^3}{2}\,\chi_{(2)} + \frac{B^4}{16} \chi_{(3)}  \,, \\
F & = R^2 + 2B\,\varphi_{(0)} +  \frac{B^2}{2}\,\varphi_{(1)} + \frac{B^3}{2}\,\varphi_{(2)} + \frac{B^4}{16} \varphi_{(3)} \,, \\
\Sigma & = \bigl(r^2 + a^2\bigr)^2 - \Delta a^2 \sin^2\theta \,, \\
\Delta & = r^2 - 2Mr + Z^2 + a^2 \,, \\
Z^2 & = Q^2 + H^2 \,,
\end{align}
\end{subequations}
which in turn are expansions of the following functions:
\begin{subequations}
\begin{align}
\Xi & = \bigl(r^2 + a^2\bigr )\sin^2\theta + Z^2\cos^2\theta \,, \\
\lambda & = r^2 + a^2 - \Delta = 2Mr - Z^2 \,, \\
R^2 & = r^2 + a^2\cos^2\theta \,, \\
\chi_{(0)} & = a H\Delta \cos \theta - Q r \bigl(r^2+a^2\bigr) \,, \\
\chi_{(1)} & = -3 a Z^2 \Bigl[ \lambda + \Delta \bigl(1+\cos^2\theta \bigr) \Bigr] \,, \\
\begin{split}
\chi_{(2)} & = Q \biggl[ r^3 \Bigl( \lambda + \Delta \bigl(1+\cos^2\theta \bigr) \Bigr) + a^2 \Bigl( \Delta \cos^2\theta \bigl( 3r - 4 M \bigr) -  r \bigl( Z^2 + \Delta \bigr)   \Bigr) - 2 M a^4 \biggr] \\
&\quad + a H \Delta \cos\theta  \bigl( \, \Xi + 2 R^2 \bigr)  \,,
\end{split}
\\\begin{split}
\chi_{(3)} & = -a \biggl[  a^2 \Delta \cos^2\theta \Bigl ( \bigl( Z^2 + 4 M^2 - 6 M r \bigr)\cos^2\theta + Z^2 + 12 M^2 - 12 M r - 6 r^2 \Bigr) \\
&\quad + 2 a^4 M \bigl( 2 M + r \bigr) + a^2 Z^2\Delta - 4 a^2 M r \bigl( r^2 - 2 Z^2 + 3 M r \bigr) - 6 M r^5 \\
& \quad + \Delta \cos^2\theta \Bigl( 6 r^2 \bigl(\Delta - r^2 \bigr) + \bigl( Z^4 + 2 M r^3 - 3 Z^2 r^2 \bigr) \cos^2\theta \Bigr)\biggr] \,,
\end{split}\\
\nb \\
\varphi_{(0)} & = a Q r \sin^2\theta - H \bigl( r^2 + a^2 \bigr) \cos\theta \,, \\
\varphi_{(1)} & = \Sigma \sin^2 \theta + 3 Z^2 \bigl( r ^2\cos^2\theta + a^2 \bigr) \,, \\
\begin{split}
\varphi_{(2)} & = a Q \biggl[\bigl( 1 + \cos^2\theta \bigr) \Bigl( r^3 +  \bigl (2 M + r \bigr) a^2 \Bigr) + r \cos^2\theta \Bigl(2 Z^2 - \Delta \bigl(3 - \cos^2\theta \bigr) \Bigr)\biggr] \\
&\quad + H \cos\theta \Bigl[ 2 a^2 \lambda \sin^2\theta - \bigl(r^2 +a^2) \, \Xi \Bigr]  \,,
\end{split}
\\
\begin{split}
\varphi_{(3)} & = Z^2 \bigg[2 a^4 \bigl( 1 + \cos^2\theta \bigr)^2 +  r^2 \cos^2\theta \bigl( \, \Xi + R^2 \sin^2\theta \bigr) \\
&\quad + a^2 \cos^2\theta \Bigl( 2\, \Xi + 3 Z^2 + r^2 \bigl( 5 + 6 \sin^2\theta + 3 \cos^4\theta \bigr) - 8 \Delta \Bigr) \biggr] \\
&\quad + a^6 \sin^6\theta + a^2 \Bigl[ \lambda^2 \cos^2\theta \bigl( 3 - \cos^2\theta \bigr)^2  + r^3 \sin^6\theta ( 4 M - r )\Bigr] \\
&\quad +2 a^4 \Bigl[ 2 M^2 \bigl( 1 + \cos^2\theta \bigr)^2 - \Delta \sin^6\theta \Bigr]  + \bigl( r^2 + a^2 \bigr)^3 \sin^4\theta \,,
\end{split}
\end{align}
\end{subequations}
$M$ is the mass, $a$ is the rotation parameter, $Q$ and $H$ are respectively the electric and magnetic charges, while $B$ is the parameter representing the intensity of the uniform magnetic field.

This spacetime can be expressed in the magnetic ansatz~\eqref{magnetic-ansatz}
\begin{align*}
ds^2_m & = \frac{1}{f}\Bigl[-\rho^2 d t^2 + e^{2\gamma} \bigl(d\rho^2 + dz^2\bigr)\Bigr] + f \bigl(d\phi - \omega d t\bigr)^2 \,,\\
A & = A_t d t + A_\phi d\phi \,,
\end{align*}
by means of the Weyl coordinates
\begin{subequations}
\begin{align}
\rho &= \sqrt{\Delta}\sin\theta \,, \\
z &= \bigl(r - M \bigl) \cos\theta \,,
\end{align}
\end{subequations}
and the functions
\begin{subequations}
\begin{align}
\omega &= \frac{\Omega}{\Sigma} \,,\\
f & = \frac{\Sigma \sin^2\theta}{F} \,,\\
\gamma & = \frac{1}{2} \log\biggl[ \frac{ \Sigma \sin^2\theta }{ \Delta + \bigl( M^2 - a^2 - Z^2\bigr)\sin^2\theta }\biggr] \,, \\
A_{\phi} & =  \frac{A_{3}}{F} \,, \\
A_{t} & = \frac{A_0}{\Sigma} - \omega A_{\phi} \,.
\end{align}
\end{subequations}
\vfill
\clearpage
\noindent Using these functions, we have that the magnetic Ernst potentials~\eqref{magnetic-ernst-potentials}
\begin{align*}
\ernst & = - f - \Phi^*\Phi + i h \,, \\
\Phi & = \tilde{A}_{t} - i A_{\phi} \,,
\end{align*}
and the twisted potentials~\eqref{magnetic-twisted-potentials} $\tilde{A}_{t}$ and $h$, can be obtained through integrations of the equations that define these twisted potentials:
\begin{align*}
\vec{e}_{\phi} \times  \vec{\nabla} \tilde{A}_{t} & = - \frac{f}{\rho} \Bigl( \vec{\nabla} A_{t} + \omega \vec{\nabla} A_{\phi} \Bigr) \,, \\
\vec{e}_{\phi} \times \vec{\nabla} h & = - \frac{f^2}{\rho} \vec{\nabla} \omega - 2\, \vec{e}_{\phi} \times \text{Im} \bigl (\Phi^{*} \vec{\nabla}\Phi \bigr) \,, \\
\Im \bigl (\Phi^{*} \vec{\nabla}\Phi \bigr) & = A_{\phi} \vec{\nabla} \tilde{A}_{t} -  \tilde{A}_{t} \vec{\nabla} A_{\phi} \,,
\end{align*}
which result in
\begin{subequations}
\begin{align}
\tilde{A}_{t} & = \frac{1}{F}\biggl[ \tilde{\chi}_{(0)} + B \tilde{\chi}_{(1)} + \frac{B^2}{4} \, \tilde{\chi}_{(2)}\biggr] \,, \\
h & = -\frac{1}{F}\biggl[ 2 \tilde{\chi}_{(1)} + B\, \tilde{\chi}_{(2)} \biggr] \,,
\end{align}
\end{subequations}
where we have defined the following ancillary functions:
\begin{subequations}
\label{ancillary-seed}
\begin{align}
\tilde{\chi}_{(0)} & = - Q \cos\theta \bigl( r^2 + a^2 \bigr) - a H r \sin^2\theta \,, \\
\tilde{\chi}_{(1)} & = - a \cos\theta \biggl[ M \Bigl(\bigl( 3 r^2 + a^2 \bigr) - \bigl( r^2 - a^2\bigr)\cos^2\theta \Bigr) - Z^2 r \sin^2\theta \biggr] \,, \\
\begin{split}
\tilde{\chi}_{(2)} & = a H \biggl[ 2 M \Bigl( a^2 + \cos^2\theta \bigl( 2 r^2 + a^2 \bigr) \Bigr) + r \sin^2\theta \bigl( \lambda + \Delta \sin^2\theta \bigr)\biggr] \\
&\quad+ Q \cos\theta \biggl[\, \Xi  \bigl( r^2 + a^2 \bigr) - 2 a^2 \lambda  \sin^2\theta \biggr] \,,
\end{split}
\end{align}
\end{subequations}
Therefore, by applying an Ehlers transformation~\eqref{Ehlers} with real parameter $\jmath$ :
\begin{align*}
\ernst' & = \frac{\ernst}{1 + i\, \jmath\, \ernst} \,, \\
\Phi' & = \frac{\Phi}{1 + i\,\jmath\,\ernst} \,,
\end{align*}
\clearpage
\noindent we obtain that the new functions are given by
\begin{subequations}
\label{new-function-kn}
\begin{align}
f' & = - \Re\ernst' - \Phi'^* \Phi' = \frac{\Sigma \sin^2\theta}{F'} \,, \\
h' & = \Im\ernst' \,,\\
\tilde{A}'_{t} & = \Re \Phi' \,, \\
A'_{\phi} & = - \Im \Phi' =  \frac{A'_{3}}{F'}\,, \\
\gamma' & = \gamma \,, \\
\omega' & = \frac{\,\,\Omega'}{\,\!\Sigma} \,, \\
A'_{t} & = \frac{A'_0}{\Sigma} - \omega' A'_{\phi} \,,
\end{align}
\end{subequations}
where the functions $\omega'$ and $A'_{t}$ have been obtained by using again the definition of the twisted potentials~\eqref{magnetic-twisted-potentials} with the primed quantities. 

Thus, in order to finalize the derivation of the new spacetime, we should now explicitly state the full expressions for the new functions listed in Eqs.~\eqref{new-function-kn}. However, to avoid repetition, we only present here the result for the two twisted potentials $h'$ and $\tilde{A}'_{t}$, while the full expressions for the remaining functions will be provided in the next section along with the final result.
\begin{subequations}
\label{new-twisted}
\begin{align}
\tilde{A}_{t'} & = \frac{1}{F'}\biggl[ \tilde{\chi}_{(0)} + B \tilde{\chi}_{(1)} + \frac{B^2}{4} \, \tilde{\chi}_{(2)} + \jmath \, \tilde{\chi}_{(3)} + \frac{\jmath B}{2}\, \tilde{\chi}_{(4)} \biggr] \,, \\
h' & = -\frac{1}{F'}\biggl[ 2 \tilde{\chi}_{(1)} + B\, \tilde{\chi}_{(2)} + \jmath \, \tilde{\chi}_{(4)} \biggr] \,,
\end{align}
\end{subequations}
where the new ancillary functions are given by
\begin{subequations}
\begin{align}
\begin{split}
\tilde{\chi}_{(3)} & = a Q \biggl[ 2 M \Bigl( a^2 + \cos^2\theta \bigl( 2 r^2 + a^2 \bigr) \Bigr) + r \sin^2\theta  \bigl( \lambda + \Delta \sin^2\theta \bigr)\biggr] \\
&\quad - H \cos\theta \biggl[ \Xi  \bigl( r^2 + a^2 \bigr) - 2 a^2 \lambda  \sin^2\theta \biggr]\,,
\end{split}\\
\begin{split}
\tilde{\chi}_{(4)} & = 2 a^4 \biggl[2 M^2 \bigl( 1 + \cos^2\theta \bigr)^2 - \Delta  \bigl( 2 - \cos^2\theta \bigr) \cos^2\theta + 2 M r \sin^2\theta \biggr]  \\
& \quad + a^2 \cos^2\theta \biggl[ \Delta^2 \bigl( 2-\cos^2\theta \bigr)^2  - 4 M r \Bigl(Z^2 + 2 \cos^2\theta \bigl( M r - Z^2 \bigr) \Bigr) \\
& \quad + 2 r^2 \Bigl( 10 M^2 - \Delta \sin^2\theta - Z^2 \Bigr) - r^4 \biggr] + a^6 \cos^2\theta + r^2 \Xi ^2+ 4 M a^2 r^3  \,,
\end{split}
\end{align}
\end{subequations}
while $\tilde{\chi}_{(0)}$, $\tilde{\chi}_{(1)}$, and $\tilde{\chi}_{(2)}$ are the same as in Eq.~\eqref{ancillary-seed}. 
\vfill
\clearpage
\noindent Therefore, in an equivalently way, the new twisted potentials~\eqref{new-twisted} can be expressed as
\begin{subequations}
\begin{align}
\tilde{A}'_{t} & = \frac{F}{F'}\biggl[ \tilde{A}_{t} + \jmath \, \tilde{\chi}_{(3)} + \frac{\jmath B}{2}\, \tilde{\chi}_{(4)} \biggr] \,, \\
h' & = -\frac{F}{F'}\biggl[ h + \jmath \, \tilde{\chi}_{(4)} \biggr] \,.
\end{align}
\end{subequations}
\subsection{Dyonic Kerr-Newman in a Melvin-swirling Universe} \label{subsec:DKNMS}
Finally, the new metric and electromagnetic Maxwell field, resulting from the application of a magnetic Ehlers transformation~\eqref{Ehlers} to the dyonic Kerr-Newman-Melvin spacetime~\eqref{DKNM}, are respectively given by\footnote{In the following all the quantities will be expressed without the prime superscript in order to lighten the notation.}:
\begin{subequations}
\label{DKNMS}
\begin{align}
ds^2 & = F \biggl[ -\frac{\Delta}{\Sigma}\,dt^2 + \frac{dr^2}{\Delta} + d\theta^2\biggr] + \frac{\Sigma\sin^2\theta}{F}\biggl[ d\phi - \frac{\Omega}{\Sigma} \,dt \biggr]^2 \,, \label{DKNMS-metric} \\
A & = \frac{A_0}{\Sigma} \,dt + \frac{A_{3}}{F}\biggl[ d\phi - \frac{\Omega}{\Sigma}\,dt \biggr] \,, \label{DKNMS-potential}
\end{align}
\end{subequations}
where the functions have been defined as:
\begin{subequations}
\begin{align}
A_0 &= \chi_{(0)} + \frac{B}{2}\,\chi_{(1)} + \frac{3B^2}{4}\,\chi_{(2)} + \frac{B^3}{8}\,\chi_{(3)} + \jmath\,\chi_{(4)} \,, \\
A_3 &= \phi_{(0)} +  \frac{B}{2}\,\varphi_{(1)} + \frac{3B^2}{4}\,\varphi_{(2)} + \frac{B^3}{8}\,\varphi_{(3)} + \jmath\,\phi_{(4)} \,, \\
\Omega &= a\,\lambda + 2B\,\chi_{(0)} +  \frac{B^2}{2}\,\chi_{(1)} + \frac{B^3}{2}\,\chi_{(2)}  + \biggr[ \jmath^2 + \frac{B^4}{16} \biggr]\chi_{(3)} + 2 \jmath B \,\chi_{(4)} + \jmath \, \Omega_{(1)} \,, \\
F & = R^2 + 2B\,\varphi_{(0)} +  \frac{B^2}{2}\,\varphi_{(1)} + \frac{B^3}{2}\,\varphi_{(2)} + \biggr[ \jmath^2 + \frac{B^4}{16} \biggr]\varphi_{(3)} + 2 \jmath B \, \varphi_{(4)} + \jmath\, F_{(1)} \,, \label{F-DKNMS}\\
\Sigma & = \bigl(r^2 + a^2\bigr)^2 - \Delta a^2 \sin^2\theta \,, \label{Sigma-DKNMS} \\
\Delta & = r^2 - 2Mr + Z^2 + a^2 \,, \\
Z^2 & = Q^2 + H^2 \,, \\
& \nb
\end{align}
\end{subequations}
and where
\begin{subequations}
\begin{align}
\Xi & = \bigl(r^2 + a^2\bigr )\sin^2\theta + Z^2\cos^2\theta \,, \\
\lambda & = r^2 + a^2 - \Delta = 2Mr - Z^2 \,, \\
R^2 & = r^2 + a^2\cos^2\theta \,, \\
\chi_{(0)} & = a H\Delta \cos \theta - Q r \bigl(r^2+a^2\bigr) \,, \\
\chi_{(1)} & = -3 a Z^2 \Bigl[ \lambda + \Delta \bigl(1+\cos^2\theta \bigr) \Bigr] \,, \\
\begin{split}
\chi_{(2)} & = Q \biggl[ r^3 \Bigl( \lambda + \Delta \bigl(1+\cos^2\theta \bigr) \Bigr) + a^2 \Bigl( \Delta \cos^2\theta \bigl( 3r - 4 M \bigr) -  r \bigl( Z^2 + \Delta \bigr)   \Bigr) - 2 M a^4 \biggr] \\
&\quad + a H \Delta \cos\theta  \bigl( \, \Xi + 2 R^2 \bigr)  \,,
\end{split}
\\
\begin{split}
\chi_{(3)} & = -a \biggl[  a^2 \Delta \cos^2\theta \Bigl ( \bigl( Z^2 + 4 M^2 - 6 M r \bigr)\cos^2\theta + Z^2 + 12 M^2 - 12 M r - 6 r^2 \Bigr) \\
&\quad + 2 a^4 M \bigl( 2 M + r \bigr) + a^2 Z^2\Delta - 4 a^2 M r \bigl( r^2 - 2 Z^2 + 3 M r \bigr) - 6 M r^5 \\
& \quad + \Delta \cos^2\theta \Bigl( 6 r^2 \bigl(\Delta - r^2 \bigr) + \bigl( Z^4 + 2 M r^3 - 3 Z^2 r^2 \bigr) \cos^2\theta \Bigr)\biggr] \,,
\end{split}
\\
\begin{split}
\chi_{(4)} & = H \biggl[ - 2 a^4 M  - a^2 \Bigl(r \bigl( Z^2 - r^2 \bigr) + 4 M \Delta \cos^2\theta + r \Delta \bigl( 1 - 3 \cos^2\theta \bigr) \Bigr) \\
& \quad + r^3 \bigl( \Delta \cos^2\theta + r^2 \bigr) \biggr] - a Q \Delta \cos\theta \bigl( \, \Xi + 2 R^2 \bigr) \,,
\end{split}
\\
\varphi_{(0)} & = a Q r \sin^2\theta - H \bigl( r^2 + a^2 \bigr) \cos\theta \,, \\
\varphi_{(1)} & = \Sigma \sin^2 \theta + 3 Z^2 \bigl( r ^2\cos^2\theta + a^2 \bigr) \,, \\
\begin{split}
\varphi_{(2)} & = a Q \biggl[\bigl( 1 + \cos^2\theta \bigr) \Bigl( r^3 +  \bigl (2 M + r \bigr) a^2 \Bigr) + r \cos^2\theta \Bigl(2 Z^2 - \Delta \bigl(3 - \cos^2\theta \bigr) \Bigr)\biggr] \\
&\quad + H \cos\theta \Bigl[ 2 a^2 \lambda \sin^2\theta - \bigl(r^2 +a^2) \, \Xi \Bigr]  \,,
\end{split}
\\
\begin{split}
\varphi_{(3)} & = Z^2 \bigg[2 a^4 \bigl( 1 + \cos^2\theta \bigr)^2 +  r^2 \cos^2\theta \bigl( \, \Xi + R^2 \sin^2\theta \bigr) \\
&\quad + a^2 \cos^2\theta \Bigl( 2\, \Xi + 3 Z^2 + r^2 \bigl( 5 + 6 \sin^2\theta + 3 \cos^4\theta \bigr) - 8 \Delta \Bigr) \biggr] \\
&\quad + a^6 \sin^6\theta + a^2 \Bigl[ \lambda^2 \cos^2\theta \bigl( 3 - \cos^2\theta \bigr)^2  + r^3 \sin^6\theta ( 4 M - r )\Bigr] \\
&\quad +2 a^4 \Bigl[ 2 M^2 \bigl( 1 + \cos^2\theta \bigr)^2 - \Delta \sin^6\theta \Bigr]  + \bigl( r^2 + a^2 \bigr)^3 \sin^4\theta \,,
\end{split}
\\
\begin{split}
\varphi_{(4)} & = a H \biggl[2 M \Bigl(a^2 +  \cos^2\theta \bigl(2 r^2 + a^2 \bigr) \Bigr) + r \sin^2\theta \bigl( \lambda + \Delta \sin^2\theta \bigr)\biggr] \\
&\quad + Q \cos\theta \Bigl[ \bigl( r^2+ a^2 \bigr)\, \Xi - 2 \lambda a^2 \sin^2\theta \Bigr] \,,
\end{split}\\
\Omega_{(1)} & = -4 \Delta \cos \theta  \biggl[r^3+a^2 \Bigl(\bigl(r-M\bigr) \cos^2\theta - M \Bigr)\biggr] \,, \\
F_{(1)} & = -4 a \cos\theta \Bigl[M \bigl( 1 + \cos^2\theta \bigr) \bigl( r^2 + a^2 \bigr)+\lambda\, r \sin^2\theta \Bigr] \,,
\end{align}
\end{subequations}
$M$ is the mass, $a$ is the rotation parameter, $Q$ and $H$ are respectively the electric and magnetic charges, $B$ is the parameter representing the intensity of the uniform magnetic field, and $\jmath$ is the swirling parameter.\\

\noindent This new spacetime~\eqref{DKNMS} thus represents a {\bfseries dyonic Kerr-Newman black hole in a Melvin-swirling universe}, i.e.~an electromagnetically charged and rotating black hole~\eqref{kn-magnetic}, embedded in a swirling~\eqref{swirling-background} and magnetic Melvin universe~\eqref{Melvin-background}. \newline

\noindent Furthermore, as sub-cases of this new solution~\eqref{DKNMS}, we also derived several new black holes that were not yet obtained. These new spacetimes, some of which are reported in detail in Appendix~\ref{B-Subcases}, correspond to the following black holes:
\begin{itemize}
\item {\bfseries Electric Kerr-Newman in a Melvin-swirling universe} $\bigl(H=0\bigr)$,
\item {\bfseries Magnetic Kerr-Newman in a Melvin-swirling universe} $\bigl(Q=0\bigr)$,
\item {\bfseries Kerr in a Melvin-swirling universe} $\bigl(H=Q=0\bigr)$,
\item {\bfseries Dyonic Reissner-Nordstr\"om in a Melvin-swirling universe} $\bigl(a=0\bigr)$,
\item {\bfseries Magnetic Reissner-Nordstr\"om in a Melvin-swirling universe} $\bigl(a=Q=0\bigr)$.
\end{itemize}
For the other remaining sub-cases, we have that the electric Reissner-Nordstr\"om black hole in a Melvin-swirling universe was already obtained in~\cite{Illy-thesis}, while the dyonic Kerr-Newman black hole in a swirling universe was obtained in~\cite{AstorinoRemoval}.

\vfill
\clearpage
\section{Analysis of the New Solutions}
\label{sec:Properties}
In this chapter, we will analyze the properties of the new solution~\eqref{DKNMS} and all its sub-cases. We will encompass all the singularities that are usually found in spacetimes, i.e.~conical singularities, Dirac and Misner strings, and curvature singularities. Moreover, we will check the presence of closed timelike curves (CTCs). Afterward, we will study the shape of the event horizons and of the ergoregions, and at the end we will obtain the so-called near-horizon extremal limit of these spacetimes.

In particular, we will show that the new spacetime we found~\eqref{DKNMS} is \emph{completely} regular outside the event horizon according to an appropriate choice of the parameters: a constraint on the parameters will allow us to remove both the conical singularities and the Dirac string (which naturally appear in the presence of a magnetic charge). In addition, this choice of parameters will also make the new spacetime free of curvature singularities, while CTCs will still be present, but “protected” by the event horizon. In a similar manner, we will also find that its non-rotating and non-Melvin sub-case~\eqref{DRNS} is also completely regular.
\subsection{Coordinate Singularities} \label{subsec:coordinate-sing-dknms}
It is straightforward to notice that the metric of the new solution~\eqref{DKNMS} is in the same form as the seed metric~\eqref{DKNM}. This implies that the new solution is also singular whenever $F=0$, $\theta = 0$, $\Delta = 0$, or $\Sigma = 0$. Following Sec.~\ref{subsec:Black-Holes}, by analyzing the Kretschmann scalar~\eqref{Kretschmann-scalar}, we have that the only possible curvature singularity is for $F=0$, which will be discussed in detail in Sec.~\ref{subsec:curvature-dknms}, while the others are all coordinate singularities\footnote{To be more precise, $\theta = 0$ can also be a curvature singularity in three very peculiar sub-cases, which will be analyzed in Sec.~\ref{subsubsec:curv-along-axis-dknms}. However, this singularity also requires $F=0$.}. In particular, the functions corresponding to these coordinate singularities are \emph{not} modified by the swirling or Melvin parameters.
 
From this last consideration we obtain the event horizons are found in the same locations of the respective black hole embedded in the Minkowski background~\eqref{KN-horizon}:
\beq
\label{horizon-DKNMS}
\Delta = 0 \quad \Rightarrow \quad r_{\pm} = M\pm\sqrt{M^2-a^2-Z^2} \,,
\eeq
which also implies that the extremality condition is also \emph{not} modified,  $M_{ext}^2 = a^2 + Z^2$.
\vfill
\clearpage
\subsection{Conical Singularities}  \label{subsec:conical-sing-dknms}
Axial conical singularities are a type of spacetime pathology due to a defect of the azimuthal angle, which results in a non-regularity of the symmetry axis. Furthermore, these singularities always initially appear in a pair, one for each half of the symmetry axis. The reason for their name is that, at the location where a conical singularity is present, the spacetime locally resembles the pointed tip of a cone.

To compute the angular defects, we consider the ratio between the perimeter of a small circle around the symmetry axis and its radius, for both halves of the symmetry axis, $\theta=0$ and $\theta=\pi$:
\begin{subequations}
\begin{align}
\label{con-sing-0}
\delta_0 & = \lim_{\theta \to 0} \frac{1}{\theta} \int_{0}^{2 \pi} \sqrt{\frac{g_{\phi\phi}}{g_{\theta\theta}}}d\phi \,, \\
\delta_\pi & =  \lim_{\theta \to \pi} \frac{1}{\pi-\theta} \int_{0}^{2 \pi} \sqrt{\frac{g_{\phi\phi}}{g_{\theta\theta}}}d\phi \,.
\end{align}
\end{subequations}
Therefore, the condition for a spacetime to \emph{not} possess conical singularities is
\beq
\label{conical-condition}
\delta_0 = \delta_\pi = 2\pi \,.
\eeq
From this condition, we can see that it is always possible to fix at least one conical defect by choosing a new periodicity $\frac{2 \pi}{\delta \phi}$ for the azimuthal angle $\phi$, or equivalently, by rescaling this angle as $\phi\mapsto\frac{2 \pi}{\delta \phi}\phi$ while maintaining the usual periodicity of $2 \pi$. Indeed, with this method, it is always possible to set at least one between $\delta_0$ and $\delta_\pi$ equal to $2 \pi$.

After this procedure, the other conical singularity, if still present, could be fixed by constraining the parameters of the solution, however, such a constraint may not exist or it might be unphysical and thus unacceptable.

For this reason, conical singularities are said to be \emph{never present} if the condition given by Eq.~\eqref{conical-condition} is trivially satisfied. Similarly, conical singularities are said to be \emph{removable} if there is a physical relation among the parameters and/or a rescaling of the azimuthal angle $\phi\mapsto\frac{2 \pi}{\delta \phi}\phi$ that satisfies the condition provided by Eq.~\eqref{conical-condition}. Otherwise, if this requirement can never be satisfied, conical singularities are said to be \emph{non-removable}. Moreover, we say that the initial conical singularities have \emph{different conicity} if a rescaling of the azimuthal coordinate $\phi\mapsto\frac{2 \pi}{\delta \phi}\phi$ is not sufficient to remove both of these singularities.

Furthermore, as explained in~\cite{Swirling}, the presence of a non-removable conical singularity implies that a “cosmic string” or a strut has to be postulated to compensate for the “force” effect induced by the spin-spin interaction between the black hole and the background, which would otherwise tend to add an acceleration to the black hole.
\vfill
\clearpage
\noindent In the most general case~\eqref{DKNMS} we obtain:
\begin{subequations}
\label{con-sing-1}
\begin{align}
\delta_0 & = \frac{32 \pi}{D_{(0)} + 16 \bigl(1 - 4 \jmath a M \bigr)^2 + 32 B\bigl( \jmath Q Z^2 - H\bigr) - 8 B^3 H Z^2} \,, \\
\delta_\pi & = \frac{32 \pi}{D_{(0)} + 16 \bigl(1 + 4 \jmath a M \bigr)^2 - 32 B\bigl(  \jmath Q Z^2 - H\bigr) + 8 B^3 H Z^2} \,,
\end{align}
\end{subequations}
where
\beq
D_{(0)} \coloneqq 24 B^2 Z^2 + 32 B^3 a M Q + B^4 \bigl(Z^4 + 16 a^2 M^2 \bigr) +  128 \jmath B a M H+ 16 \jmath ^2 Z^4 \,.
\eeq
This means that there are sub-cases for which the conical singularities are non-removable, others for which they are never present, and some particular cases for which these singularities are removable, as summarized in the following Table~\ref{table:Conical-singularities}:
{\begin{table}[H]
\begin{center}
\onehalfspacing
\small
\begin{tabular}{ |p{4.65cm}||p{1.9cm}|p{1.9cm}|p{1.9cm}|p{2.5cm}|  }
\hline
\multicolumn{5}{|c|}{Conical Singularities} \\
\hline
Spacetime                                                  & Minkowski   & Melvin          & Swirling        & Melvin-swirling \\
\hline
Background                                                & No               & No                & No                & No              \\
Schwarzschild                                            & No               & No                & No                & No               \\
Electric\,\,\,\,\,\,Reissner-Nordstr\"{o}m      & No               & Removable   & Removable  & Yes               \\
Magnetic\,\,Reissner-Nordstr\"{o}m           & No               & Yes               & Removable   & Yes               \\
Dyonic\,\,\,\,\,\,\,\,Reissner-Nordstr\"{o}m   & No               & Yes               & Removable   & Removable    \\
Kerr                                                             & No              & Removable   & Yes               & Yes                 \\
Electric\,\,\,\,\,\,Kerr-Newman                      & No              & Removable   & Yes               & Removable     \\
Magnetic\,\,Kerr-Newman                           & No              & Yes                & Yes              & Removable      \\
Dyonic\,\,\,\,\,\,\,\,Kerr-Newman                   & No              & Yes                & Yes             & Removable      \\
\hline
\end{tabular}
\caption{\small Table summarizing for all the sub-cases whether the conical singularities are never present (No), non-removable (Yes), or removable by a certain choice of the parameters and/or a rescaling of the azimuthal angle (Removable).}\label{table:Conical-singularities}
\end{center}
\end{table}}
\noindent Thus, if both the Melvin and swirling parameters are present, the conical singularities are removable only if both of the charges are non-null, $Q, H\neq0$, or if the rotation parameter is non-zero, $a\neq0$. Additionally, for these Melvin-swirling sub-cases, the conical singularities are never present only if the charges and the rotation parameters are all null, $a = Q= H = 0$.

Moreover, for the swirling sub-cases, $B=0$, the conical singularities are never present or removable only if the rotation parameter is equal to zero, $a=0$. Similarly, for the Melvin sub-cases, $\jmath=0$, these singularities are removable or never present only if the magnetic charge is null, $H=0$.

\subsubsection{Sub-cases with Non-Removable Conical Singularities}
In particular, we list here some of the sub-cases with non-removable conical singularities.
\vspace{-0.25cm}
\paragraph{Kerr in a Melvin-swirling universe ($Q=H=0$):}
\label{kerr-melvin-swirling-conical}
\begin{subequations}
\begin{align}
\delta_0 & = \frac{32\pi}{16  - 8 a \jmath M + a^2\bigl(B^4 + 16 \jmath^2\bigr)M^2} \,, \\
\delta_\pi & = \frac{32\pi}{16 + 8 a \jmath M + a^2\bigl(B^4 + 16 \jmath^2\bigr)M^2} \,.
\end{align}
\end{subequations}
We can see that, for this sub-case, the non-removability of the conical singularities exactly arises from the spin-spin interaction between the black hole $(aM)$ and the background $(\jmath)$. For its non-Melvin sub-case $B=0$, the respective condition was already discussed in~\cite{Swirling}. Nevertheless, we can say that the addition of the uniform magnetic field $B$ is insufficient to remove the conical singularities of the rotating black hole in a Melvin-swirling universe.
\vspace{-0.20cm}
\paragraph{Magnetic Reissner-Nordstr\"{o}m in a Melvin-swirling universe ($a=0$):}
\begin{subequations}
\begin{align}
\delta_0 & = \frac{32\pi}{16 \jmath^2 H^4  + \bigl(2 - BH\bigr)^4} \,, \\
\delta_\pi & = \frac{32\pi}{16 \jmath^2 H^4 + \bigl(2 + BH\bigr)^4} \,.
\end{align}
\end{subequations}
In a similar manner, for this sub-case, the non-removable conical singularities are due to the coupling between the magnetic Melvin background $B$ and the black hole magnetic charge $H$.  \newline
\vspace{-0.05cm}
\noindent Finally, the other sub-cases for which it is not possible to remove both conical singularities have already been studied, as discussed at the end of Sec.~\ref{subsec:DKNMS}.
\vspace{-0.10cm}
\subsubsection{Sub-cases with Removable Conical Singularities}
Analogously, we present here all the sub-cases with the respective conditions under which the conical singularities are removable, also including, for the sake of completeness, the already-known sub-cases for which the swirling parameter is absent, $\jmath = 0$.
\vspace{-0.3cm}
\paragraph{Dyonic Kerr-Newman in a Melvin-swirling universe:}
\begin{subequations}
\label{free-con-sing}
\begin{align}
\jmath & = - \frac{ B H \bigl( 4 + B^2 Z^2 \bigr)}{16 a M - 4 B Q Z^2} \,, \\
\delta\phi & = \frac{32 \pi \bigl( B Q Z^2 - 4 a M \bigr)^2\Bigl[ 16 a^2 M^2 - 8 a B M Q Z^2 + B^2 Z^6\Bigr] ^{-1}}{\Bigl[16 + 8 B^2 \bigl( H^2 + 3 Q^2 \bigr) + 32 a B^3 M Q + B^4 \bigl(16 a^2 M^2 + Z^4 \bigr) \Bigr]} \,.
\end{align}
\end{subequations}
\paragraph{Magnetic Kerr-Newman in a Melvin-swirling universe ($Q=0$):}
\begin{subequations}
 \label{conical-MKNMS}
\begin{align}
\jmath & = - \frac{ B H \bigl( 4 + B^2 H^2 \bigr)}{16 a M} \,,  \\
\delta\phi & = \frac{512 \pi a^2 M^2}{\Bigl[ 16 a^2 M^2 + B^2 H^6\Bigr] \Bigl[16 a^2 M^2 B^4 + \bigl(4 + B^2 H^2 \bigr)^2\Bigr]} \,.
\end{align}
\end{subequations}
\paragraph{Electric Kerr-Newman in a Melvin-swirling universe ($H=0$):}
\begin{subequations}
\label{conical-EKNMS}
\begin{align}
a & = \frac{ B Q^3}{4 M} \,, \\
\delta\phi & = \frac{32 \pi}{\Bigl[ 1 + B^2 Q^2 \Bigr] \Bigl[16 + 8 B^2 Q^2 + \bigl( B^4 + 16 \jmath^2\bigr) Q^4 \Bigr]} \,.
\end{align}
\end{subequations}
\paragraph{Dyonic Reissner-Nordstr\"{o}m in a Melvin-swirling universe ($a=0$):}
\begin{subequations}
\label{conical-DRNMS}
\begin{align}
\jmath & = \frac{ H \bigl( 4 + B^2 Z^2 \bigr)}{4 Q Z^2} \,, \\
\delta\phi & = \frac{32\pi Q^2 }{Z^2 \Bigl[16 + 8 B^2 \bigl(H^2 +3 Q^2 \bigr) + B^4 Z^4 \Bigr]} \,.
\end{align}
\end{subequations}
\paragraph{Reissner-Nordstr\"{o}m in a swirling universe ($B=a=0$):}
\beq
\label{conical-DRNS}
\delta \phi = \frac{2 \pi}{1 + \jmath^2 Z^4} \,,
\eeq
where it is not specified if the black hole is dyonic, magnetic, or electric because this condition is the same for all three sub-cases.
\paragraph{Electric Kerr-Newman in a Melvin universe ($\,\jmath=H=0$):}
\beq
\delta\phi = \frac{32 \pi}{16 + 24 B^2 	Q^2 + 32 a B^3 M Q + B^4\bigl(Q^4 + 16 a^2 M^2\bigr)} \,. \label{conical-EKNM}
\eeq
\paragraph{Kerr in a Melvin universe ($\,\jmath=Q=H=0$):}
\beq
\delta\phi = \frac{2 \pi}{1 + a^2 B^4 M^2} \,. \label{conical-KM}
\eeq
\paragraph{Electric Reissner-Nordstr\"{o}m in a Melvin universe ($\,\jmath=a=H=0$):}
\beq
\delta\phi = \frac{32 \pi}{16 + 24 B^2 	Q^2 + B^4 Q^4} \,. \label{conical-ERNM}
\eeq
\subsection{Dirac Strings}  \label{subsec:dirac-strings-dknms}
Dirac strings are a type of pathology that arises in the presence of magnetic charges~\cite{DiracString, DiracStringII}. Indeed, the presence of the magnetic charge cannot be described by a smooth magnetic potential and its discontinuity generates a line singularity, which is the Dirac string.  These strings can be seen as one-dimensional curves in space that connect two Dirac monopoles with opposite magnetic charges, or similarly from one magnetic monopole out to infinity.

The condition for a Maxwell potential $A_{\mu}$ to \emph{not} present Dirac strings is 
\beq
\label{dirac-condition}
\lim_{\theta \to 0} A_{\phi} = \lim_{\theta \to \pi} A_{\phi} = 0\,.
\eeq
Thus, in a similar manner to the conical singularities explained in Sec.~\ref{subsec:conical-sing-dknms}, we have that Dirac strings are said to be \emph{never present} if the condition provided by Eq.~\eqref{dirac-condition} is trivially satisfied. Analogously, Dirac strings are said to be \emph{removable} if there is a physical relation among the parameters and/or a gauge transformation $A_{\phi}\mapsto A_{\phi}-\delta A_{\phi}$ that satisfies the requirement given by Eq.~\eqref{dirac-condition}. Otherwise, if this condition can never be satisfied, we say that Dirac strings are \emph{non-removable}.

Moreover, if one wants to remove both the conical singularities and the Dirac strings, it is more convenient to simultaneously consider the rescaling of the azimuthal coordinate $\phi\mapsto\frac{2 \pi}{\delta \phi}\phi$ with the gauge transformation as $A_{\phi} \mapsto \frac{2 \pi}{\delta \phi} A_{\phi} - \delta A_{\phi}$. \newline
\vspace{-0.4cm}

\noindent For the most general case found~\eqref{DKNMS}, we have:
\begin{subequations}
\begin{align}
\label{Dirac-sing-1}
\lim_{\theta \to 0} A_{\phi} & = \biggl[\frac{\delta_0}{16 \pi}\biggl]\Bigl[ A_{(0)} + A_{(1)}\Bigr] \,, \\
\lim_{\theta \to \pi} A_{\phi}  & = \biggl[\frac{\delta_\pi}{16 \pi}\biggl]\Bigl[ A_{(0)} - A_{(1)}\Bigr] \,,
\end{align}
\end{subequations}
where $\delta_{0}$ and $\delta_{1}$ are those which define the condition for conical singularities~\eqref{con-sing-1}, and
\begin{subequations}
\begin{align}
A_{(0)} & \coloneqq 32 \jmath a H M + 12 B Z^2 + 24 B^2 a M Q + B^3 \bigl(16 a^2 M^2 + Z^4\bigr)  \,,\\
A_{(1)} & \coloneqq - 8 H + 8 \jmath Q Z^2 - 6 B^2 H Z^2 \,.
\end{align}
\end{subequations}
Therefore, collecting the results of Table~\ref{table:Conical-singularities} and Table~\ref{table:Dirac-strings}, the only two possible sub-cases for which it is possible to remove both the Dirac strings and the conical singularities\footnote{With this statement we are considering only the spacetimes which initially have both of these singularities, otherwise there are other sub-cases that can result in a spacetime without any of these singularities.} are the new dyonic Kerr-Newman black hole in a Mevin-swirling universe~\eqref{DKNMS}, and its non-rotating and non-Melvin sub-case, the dyonic Reissner-Nordstr\"om black hole in a swirling universe~\eqref{DRNS}, whose constraints needed to achieve this result will be reported in the following pages.
\vfill
\clearpage

\noindent For the other sub-cases, we have that if both the Melvin and swirling parameters are present, the Dirac strings are always removable or never present, except if the black hole is rotating but not charged, which is quite a surprising result since the Dirac strings usually arise in black holes that do carry a magnetic charge.

On the other hand, if only the swirling parameter is present, it is possible to have Dirac strings if any type of charge is present, which however is an expected result since, in a certain sense, the addition of the swirling rotation generates a magnetic charge from an electric one, and vice-versa\footnote{In Appendix~\ref{sub:drn-swirling} is reported the dyonic Reissner-Nordstr\"om black hole in a swirling universe with a clarification on this topic.}.

Finally, for the Melvin sub-cases, we have that the addition of the uniform magnetic field to the Minkowski background has the effect of making the Dirac strings always removable except for the non-rotating magnetic black hole.

{\begin{table}[H]
\begin{center}
\onehalfspacing
\small
\begin{tabular}{ |p{4.65cm}||p{1.9cm}|p{1.9cm}|p{1.9cm}|p{2.5cm}|  }
\hline
\multicolumn{5}{|c|}{Dirac Strings} \\
\hline
Spacetime                                                  & Minkowski    & Melvin                  & Swirling                        & Melvin-swirling \\
\hline
Background                                                & No                & No                         & No                               & No                           \\
Schwarzschild                                            & No                & No                         & No                               & No                            \\
Electric\,\,\,\,\,\,Reissner-Nordstr\"{o}m      & No                & No                         & Yes                              & Removable$^{*}$     \\
Magnetic\,\,Reissner-Nordstr\"{o}m           & Yes               & Yes                        & Yes                              & Removable$^{*}$     \\
Dyonic\,\,\,\,\,\,\,\,Reissner-Nordstr\"{o}m   & Yes               & Removable            & Removable$^{**}$      & Removable$^{***}$    \\
Kerr                                                             & No                & No                         & No                               & Yes                            \\
Electric\,\,\,\,\,\,Kerr-Newman                      & No                & No                         & Yes                              & Removable$^{***}$    \\
Magnetic\,\,Kerr-Newman                           & Yes               & Removable$^{*}$  & Removable                  & Removable$^{***}$    \\
Dyonic\,\,\,\,\,\,\,\,Kerr-Newman                   & Yes               & Removable$^{*}$  & Removable$^{*}$        & Removable$^{**}$      \\
\hline
\end{tabular}
\caption{\small Table summarizing for all the sub-cases whether the Dirac strings are never present (No), non-removable (Yes), or removable by a certain choice of the parameters and/or a gauge transformation.} \label{table:Dirac-strings}
\end{center}
\end{table}}
\vspace{-0.5cm}
{\small \noindent $*$ In these cases, there are additional constraints in addition to the relations between the parameters necessary to remove the Dirac strings. For example, for the magnetic Reissner–Nordstr\"{o}m black hole in a Melvin-swirling universe it is also required that $|B H| > 2$. \\
\noindent $**$ In these cases, the Dirac Strings are removable with constraints on the parameters that are compatible with those necessary to remove the conical singularities. \\
\noindent $***$ In these cases, the conditions to remove the Dirac Strings are \emph{not} compatible with the respective conditions needed in order to remove the conical singularities.}
\vfill
\clearpage
\subsubsection{Sub-cases with Non-Removable Dirac Strings} 
As summarized in Table~\ref{table:Dirac-strings}, not in every sub-case the constraints necessary to remove the conical singularities are compatible with those to remove the Dirac string. For this reason, in this part, we list all the sub-cases with their respective reasons for which this is not possible, corresponding to the difference in the value of the $\phi-$component of the electromagnetic potential on the two halves of the symmetry axis.

\paragraph{Magnetic Kerr-Newman in a Melvin-swirling universe ($Q=0$):}
\beq
\frac{2 \pi}{\delta \phi}\Bigl[A_{\phi}\big\rvert_{\theta=0} - A_{\phi}\big\rvert_{\theta=\pi}\Bigr]\bigg\rvert_{\eqref{conical-MKNMS}} = - \frac{H \bigl(4 + 3 B^2 H^2\bigr)}{2} \,.
\eeq

\paragraph{Electric Kerr-Newman in a Melvin-swirling universe ($H=0$):}
\beq
\frac{2 \pi}{\delta \phi}\Bigl[A_{\phi}\big\rvert_{\theta=0} - A_{\phi}\big\rvert_{\theta=\pi}\Bigr]\bigg\rvert_{\eqref{conical-EKNMS}} = 2 \jmath Q^3 \,.
\eeq

\paragraph{Dyonic Reissner-Nordstr\"{o}m in a Melvin-swirling universe ($a=0$):}
\beq
\frac{2 \pi}{\delta \phi}\Bigl[A_{\phi}\big\rvert_{\theta=0} - A_{\phi}\big\rvert_{\theta=\pi}\Bigr]\bigg\rvert_{\eqref{conical-DRNMS}} = - B^2 H Z^2 \,. \label{not-rem-dirac-DRNMS}
\eeq

\myparagraph{Magnetic and Electric Reissner-Nordstr\"{o}m in a swirling universe ($B=a=0$):}
Using for convenience the result of the dyonic case, we have that
\beq
\frac{2 \pi}{\delta \phi}\Bigl[A_{\phi}\big\rvert_{\theta=0} - A_{\phi}\big\rvert_{\theta=\pi}\Bigr]\bigg\rvert_{\eqref{conical-DRNS}} = 2 \bigl(\jmath Q Z^2 - H\bigr) \,, \label{dirac-DRNS}
\eeq
which is clearly never zero if only one charge is present. Moreover, it also strengthens the interpretation that a charged black hole embedded in a swirling universe acquires a magnetic charge proportional to the electric one, and vice-versa.
\myparagraph{Kerr in a Melvin-swirling universe ($Q=H=0$):}
Moreover, it is also interesting to study the only Melvin-swirling sub-case for which both singularities are not removable. Indeed, as we have already briefly commented, the Kerr black hole in a Melvin-swirling universe also presents non-removable Dirac strings despite not being a charged black hole. 

\vfill
\newpage
\noindent This Dirac string can be interpreted as a singularity that arises from the interaction between the swirling parameter $\jmath$ and a charge of magnitude $p=a\,MB$, as can be seen from the following result:
\beq
A_{\phi}\big\rvert_{\theta=0} - A_{\phi}\big\rvert_{\theta=\pi} =\frac{32 \jmath \bigl( a M B \bigr)^3}{1+2a^2\big(B^4 - 16 \jmath^2\bigr)M^2 + a^4 \bigl(B^4 +16 \jmath^2\bigr)^2M^4} \,.
\eeq

\subsubsection{Sub-cases with Removable Dirac Strings} 
As already pointed out, there are only two possible sub-cases for which the conical singularities and the Dirac strings are both removable. Thus, in the following, we list the two spacetimes with the respective values of the rescaling of the azimuthal coordinate $\phi\mapsto\frac{2 \pi}{\delta \phi}\phi$ and gauge transformation $A_{\phi}\mapsto\frac{2 \pi}{\delta \phi}A_{\phi}-\delta A_{\phi}$, in addition with the constraints on the parameters, needed in order to remove both of these pathologies.
\paragraph{Dyonic Kerr-Newman in a Melvin-swirling universe:}
\begin{subequations}
\label{reg-DKNMS}
\begin{align}
\jmath & = \frac{ H \bigl(4 + 3 B^2 Z^2\bigr)}{4 Q Z^2} \,, \\
a & = \frac{ Q B^3 Z^4}{2 M \bigl(4 + 3 B^2 Z^2\bigr)} \,, \label{reg-DKNMS-a}\\
\delta \phi & = \frac{32 \pi Q^2 \bigl(4 + 3 B^2 Z^2\bigr)^2 \Bigl[ 16 + 8 B^2 \bigl( 3 H^2 + Q^2 \bigr) + B^4 Z^2 \bigl( 9 H^2 + Q^2 ) \Bigr]^{-1}}{Z^2  \Bigl(16 + 8 B^2 \bigl( 3 H^2 + 5 Q^2 \bigr) + B^4 Z^2 \bigl( 9 H^2 + 25 Q^2 \bigr) + 4 B^6 Q^2 Z^4 \Bigr)} \,, \\
\delta A_{\phi} & = \frac{B Z^2 \Bigl[ 192 + 16 B^2 \bigl(23 H^2 + 19 Q^2 \bigr) + 12 B^4 Z^2 \bigl(19 H^2 + 15 Q^2 \bigr) + 45 B^6 Z^6 + 4 B^8 Q^2 Z^6 \Bigr]}{8 \bigl(4 + 3 B^2 Z^2 \bigr)^2} \,.
\end{align}
\end{subequations}
We can notice that the expressions in Eqs.~\eqref{reg-DKNMS} are physically well-posed only for
\beq
\label{physical-req}
a \,Q\, B > 0 \,.
\eeq
Indeed, the mass only appears in the constraint regarding the rotation parameter~\eqref{reg-DKNMS-a}, from which we can always guarantee that the mass is positive, $M>0$, only when the product of the parameters $a$, $Q$ and $B$ is also positive.
Equivalently, if the mass $M$ is taken as a free positive parameter, this physical requirement~\eqref{physical-req} directly determines the direction of the black hole rotation through the constraint provided by Eq.~\eqref{reg-DKNMS-a}.

\myparagraph{Dyonic Reissner-Nordstr\"{o}m in a swirling universe ($B=a=0$):}
Since, for this sub-case, the condition necessary to remove the conical singularities is just a rescaling of the azimuthal coordinate~\eqref{conical-DRNS} we have that solving Eq.~\eqref{dirac-DRNS} exactly corresponds to removing the Dirac strings, independently from the fact the conical singularities are still present or not.

Therefore, the constraints necessary to remove both the Dirac strings and the conical singularities are given by
\begin{subequations}
\label{reg-DRNS}
\begin{align}
\jmath = \frac{H}{Q Z^2} \,, \label{dirac-DRNS-cond}\\
\delta \phi = \frac{2 \pi Q^2}{Z^2} \,,
\end{align}
\end{subequations}
where the condition on the parameters~\eqref{dirac-DRNS-cond} is the only one needed in order to remove the Dirac strings.
\subsection{Closed Timelike Curves} \label{subsec:CTCs-dknms}
Closed timelike curves, also called CTCs, are timelike curves that loop back on themselves, therefore representing worldlines of particles or observers that return to their own past, hence violating causality. A famous example of a spacetime that allows CTCs is the G\"odel universe~\cite{Godel}.

For an axisymmetric spacetime with Killing vector $m = \partial_{\phi}$, whose orbits are closed, the existence of CTCs is thus determined by the region for which $m$ becomes timelike, i.e.
\beq
m^2 = g_{\mu \nu} m^{\mu}m^{\nu} = g_{\phi \phi} < 0 \,.
\eeq
For the new solution~\eqref{DKNMS} we have that 
\beq
g_{\phi \phi} = \frac{\Sigma \sin^2\theta}{F} \,,
\eeq
with the functions $F$ and $\Sigma$ respectively defined by Eq.~\eqref{F-DKNMS} and Eq.~\eqref{Sigma-DKNMS}.

It can be proved numerically that $F \geq 0$ for any choice of the parameters if $M, r \geq 0$. Thus, we have that the regions in which CTCs are allowed are determined by the condition
\beq
\Sigma = \bigl(r^2 + a^2\bigr)^2 - \bigl(r^2 - 2Mr + Q^2 + H^2 + a^2\bigr)\, a^2 \sin^2\theta < 0 \,,
\eeq
which does \emph{not} depend on either the Melvin parameter $B$ or the swirling parameter $\jmath$. This implies that these regions are the same as in the case without the Melvin-swirling background.

\noindent For this reason, CTCs are allowed whenever the rotation parameter $a$ and at least one of the charges $Q$ or $H$ are non-zero, i.e.~for the Kerr-Newman black hole\footnote{To be more precise, if the radial coordinate $r$ is allowed to be negative, $r < 0$, the same argument also applies for the rotating non-charged Kerr black hole.}, and in the same regions independently of a possible embedding of this black hole in a Melvin, swirling, or Melvin-swirling universe. Therefore, CTCs are allowed only in the finite “doughnut” region near the origin $r = 0$ inside the (inner) event horizon.

In order to obtain a visual representation of this region, it is useful to define the “rectangular” coordinates $(x, y, z)$ such that
\begin{subequations}
\label{rectangular-coordinates}
\begin{align}
r & = \sqrt{x^2 + y^2 + z^2} \,, \\
\cos\theta & = \frac{z^2}{x^2 + y^2 + z^2} \,,
\end{align}
\end{subequations}
that maps the symmetry axis to the $z-$axis.

\begin{figure}[H]
\captionsetup[subfigure]{labelformat=empty}
\centering
\hspace{-0.75cm} 
\subfloat[\hspace{1cm} CTCs Cross-section $y=0$]{{\hspace{0.5cm}\includegraphics[width=0.3\textwidth]{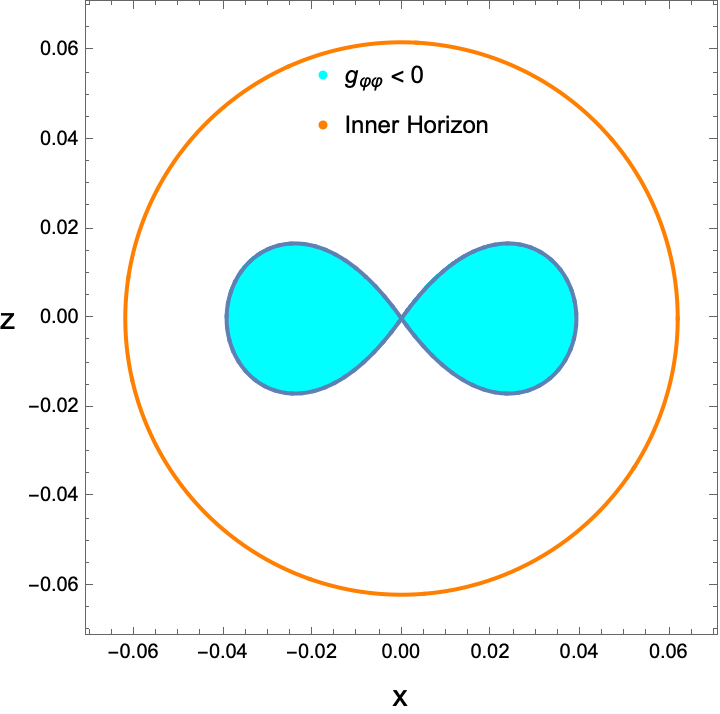}}}
\subfloat[\hspace{1cm} CTCs]{{\hspace{0.5cm}\includegraphics[width=0.3\textwidth]{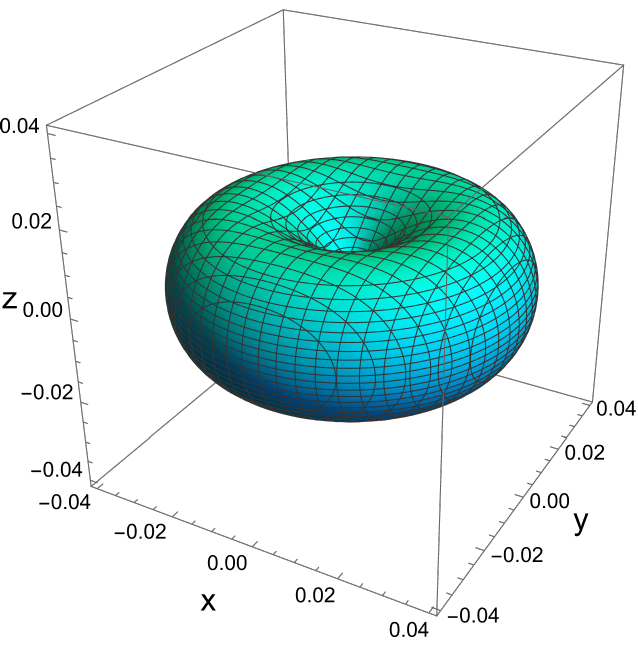}}}
\subfloat[\hspace{1cm} CTCs Cross-section $z=0$]{{\hspace{0.5cm}\includegraphics[width=0.3\textwidth]{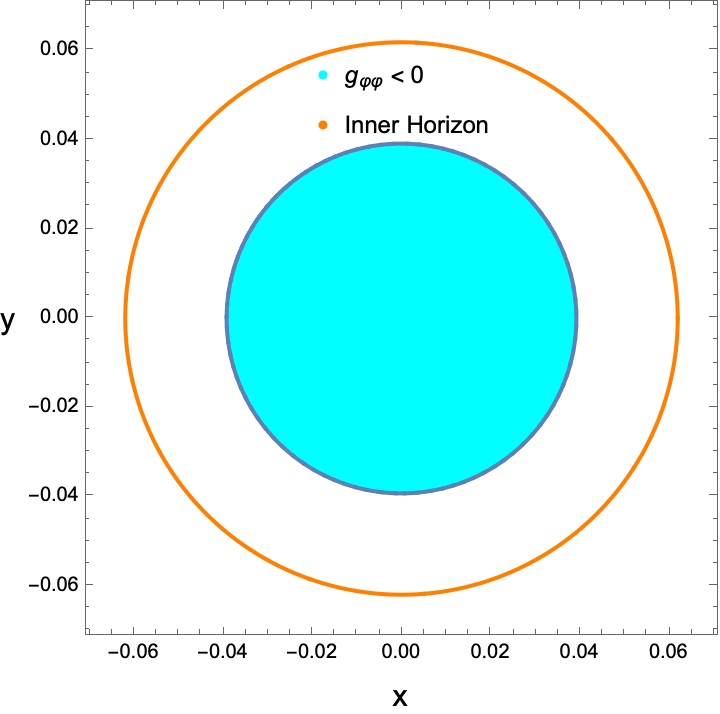}}}
\caption{\small Plots depicting the “doughnut” region in which CTCs are allowed, for the dyonic Kerr-Newman black hole with parameters $M = 1, Q=\frac{1}{5}, H=\frac{1}{5}, a=\frac{1}{5} \Rightarrow \pazocal{I} = 0.12$, independently of the choice of the swirling and Melvin parameters $\jmath$ and $B$. The left and right panels show the cross-sections of this region and of the corresponding inner horizon, respectively for $y=0$ and $z=0$, in order to prove that this region is located inside the inner horizon.} \label{CTC-KN}
\end{figure}

\begin{figure}[H]
\captionsetup[subfigure]{labelformat=empty}
\centering
\hspace{-0.75cm}
\subfloat[\hspace{1cm} CTCs Cross-section $y=0$]{{\hspace{0.5cm}\includegraphics[width=0.3\textwidth]{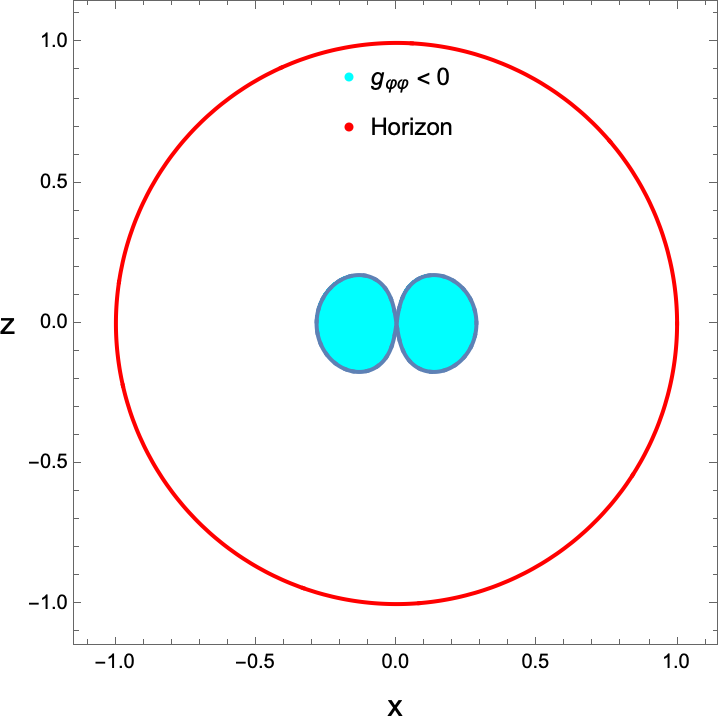}}}
\subfloat[\hspace{0.5cm} CTCs]{{\hspace{0.5cm}\includegraphics[width=0.3\textwidth]{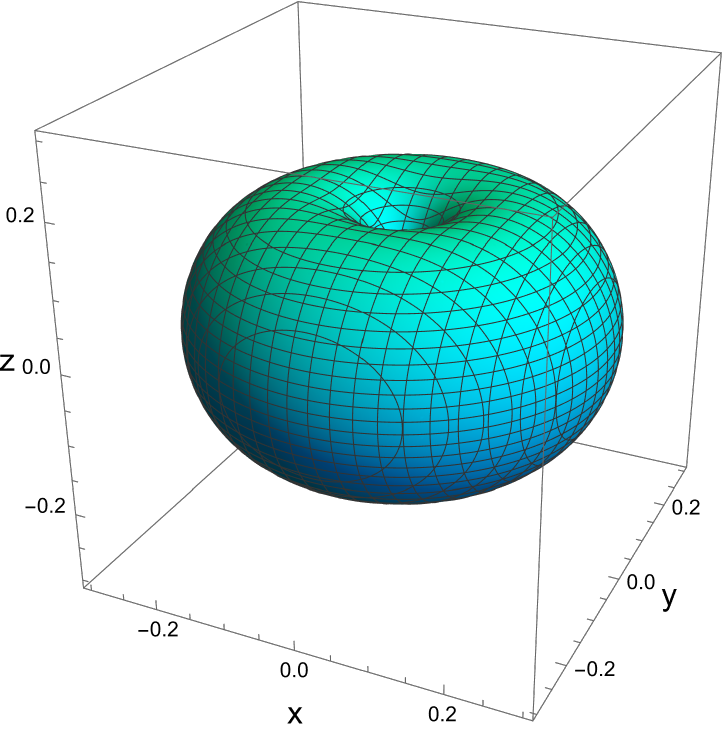}}}
\subfloat[\hspace{1cm} CTCs Cross-section $z=0$]{{\hspace{0.5cm}\includegraphics[width=0.3\textwidth]{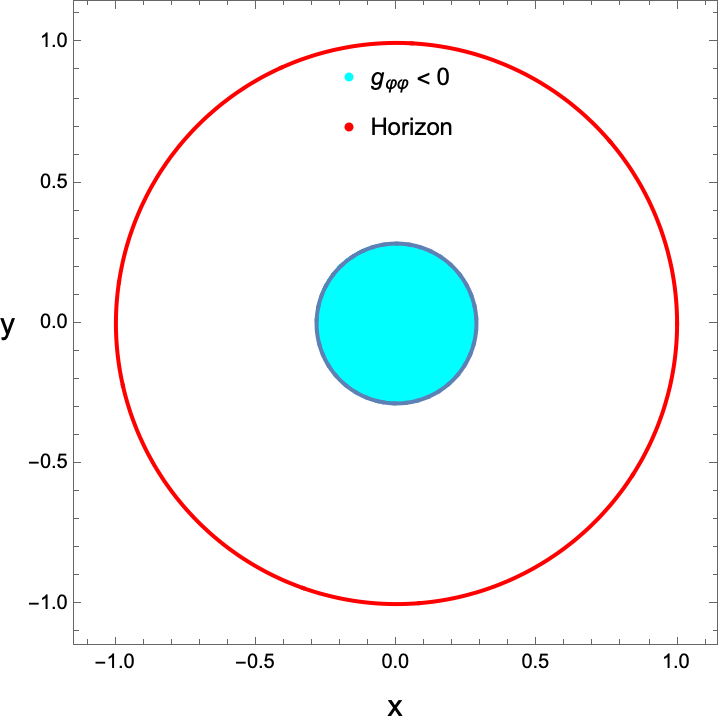}}}
\caption{\small Plots depicting the “doughnut” region in which CTCs are allowed, for the \emph{extremal} dyonic Kerr-Newman black hole with parameters $M = 1, Q=\frac{7}{10}, H=\frac{7}{10}, a=\frac{\sqrt{2}}{10} \Rightarrow \pazocal{I} = 1$, independently of the choice of the swirling and Melvin parameters $\jmath$ and $B$. The left and right panels show the cross-sections of this region and of the corresponding event horizon, respectively for $y=0$ and $z=0$, in order to prove that this region is located inside the event horizon.} \label{CTC-KN-2}
\end{figure}

Additionally, we define the “Extremality Index” $\pazocal{I}$, which represents how close the black hole is to extremality, in the sense that $\pazocal{I} = 1$ if the black hole is extremal:
\beq
\label{extremality-index}
\pazocal{I}=\frac{Q^2+H^2+a^2}{M^2} \,.
\eeq
With these definitions, we can perform a numerical analysis of the function $g_{\phi \phi}$, which is pictured in Figure~\ref{CTC-KN} for the non-extremal case, and in Figure~\ref{CTC-KN-2} for the extremal one.

\clearpage
\null
\vspace{0.5cm}
\noindent However, as we can see from these plots, the use of the “rectangular” coordinates~\eqref{rectangular-coordinates} leads to distorted results since, for example, the horizons are always spheres. Nevertheless, these coordinates are still useful for qualitative analysis. Moreover, as already pointed out, with these coordinates the symmetry axis is entirely mapped to the $z-$axis.

\vspace{0.5cm}
\subsection{Misner Strings} \label{subsec:Misner-strings-dknms}
Misner strings~\cite{MisnerString, Astefanesei}, which are the gravitational analog of the Dirac strings, appear in rotating spacetimes, when the non-diagonal function $\omega = \frac{g_{t\phi}}{g_{tt}}$ is not regular on the symmetry axis, taking different values on the two hemispheres $\theta=0$ and $\theta=\pi$. Furthermore, the presence of a Misner string also leads to CTCs and causality issues~\cite{MisnerString}.

Hence, the condition which guarantees the absence of the singularity is given by
\beq
\lim_{\theta \to 0} \frac{g_{t\phi}}{g_{tt}} = \lim_{\theta \to \pi} \frac{g_{t\phi}}{g_{tt}} \,.
\eeq
This kind of defect is in general associated with the presence of the NUT parameter, and since the latter is absent in our class of spacetimes, we find that, as expected, the new solution~\eqref{DKNMS} and none of its sub-cases are characterized by Misner strings.
\clearpage
\subsection{Curvature} \label{subsec:curvature-dknms}
Finally, we study the curvature of these solutions by analyzing the Kretschmann scalar~\eqref{Kretschmann-scalar}
\beqNN
K=R_{\mu \nu \rho \sigma}R^{\mu \nu \rho \sigma} \,.
\eeqNN 
For large radial distances, $r \to \infty$, the Kretschmann scalar of the dyonic Kerr-Newman metric~\eqref{kn-magnetic} decays as
\beq
K\big\rvert_{\jmath = B = 0} \overset{r\to \infty}{\approx} \frac{48 M^2}{r^6} \,.
\eeq
Conversely, for the Melvin-swirling solutions~\eqref{kn-magnetic}, we have that this scalar invariant decays faster if at least one between the Melvin or swirling parameters is non-zero:
\beq
K \overset{r\to \infty}{\approx} \frac{49152}{\bigl[B^4 + 16 \jmath^2\bigr]^2\sin^{12}\theta\, r^{12}} \,.
\eeq
Therefore, if these parameters are present, the spacetime is locally asymptotically flat.

Moreover, these spacetimes have also an asymptotic constant curvature on the symmetry axis. Indeed, for the most general case we found~\eqref{kn-magnetic}, we have
\begin{subequations}
\label{kretch-inf-poles}
\begin{align}
K\big\rvert_{\theta = 0} \,\xrightarrow{r\to \infty}\,  &  \biggl[\frac{\delta_0}{2\pi}\biggl]^4\Bigl[ K_{(0)} + K_{(1)}\Bigr] \,, \\
K\big\rvert_{\theta = \pi} \,\xrightarrow{r\to \infty}\,&  \biggl[\frac{\delta_\pi}{2\pi}\biggl]^4\Bigl[ K_{(0)} - K_{(1)}\Bigr] \,,
\end{align}
\end{subequations}
where $\delta_{0}$ and $\delta_{1}$ are the same which define the condition for conical singularities~\eqref{con-sing-1}, and
\begin{subequations}
\begin{align}
\begin{split}
K_{(0)} & \coloneqq  2 B^4\Bigr[10 + B^2 \bigl(15 H^2 - Q^2\bigr) - 128 \jmath H Q\Bigr] - 8 \jmath^2 \Bigl[24 + 4 B^2\bigl(H^2 - 15 Q^2\bigr) \Bigr] \\
& \quad  - 24 \Bigl[B^4 + 16 \jmath^2\Bigr] \bigl(B a M \bigr)\bigl( 4 \jmath H + B^2 Q\bigr) + \frac{1}{4}\Bigl[B^4 + 16 \jmath^2\Bigr]^2 \bigl(5 Z^4 - 48 a^2 M^2\bigr) \,,
\end{split}\\
\begin{split}
K_{(1)} & \coloneqq  - 8 B \bigl(5 B^4 H - 32 \jmath B^2 Q - 48 \jmath^2 H \bigl) \\
& \quad + 2 \Bigl[B^4 + 16 \jmath^2\Bigr] \Bigl[48 \jmath a M + 5 B Z^2\big(4\jmath Q - B^2 H\bigr) \Bigr] \,.
\end{split}
\end{align}
\end{subequations}
More precisely, as we can see from Eqs.~\eqref{kretch-inf-poles}, the two hemispheres $\theta = 0$ and $\theta = \pi$ do not asymptotically converge to the same value of constant curvature, except if we remove the conical singularities and possibly impose some additional constraints. Therefore, in the following, we list the cases for which this is achievable, with their respective possible additional constraints and the resulting asymptotic constant curvature $K_{(0)}$.
\vfill
\clearpage
\myparagraph{Dyonic Kerr-Newman in a Melvin-swirling universe:}
Using the conditions necessary to remove the conical singularities~\eqref{free-con-sing}, we obtain, for definition, $\delta_0 = \delta_\pi = 2 \pi$. Moreover, we obtain that $K_{(1)} = 0$ if
\begin{subequations}
\begin{align}
\jmath & = \frac{B^2 H}{4 Q} \,, \\
a & = -\frac{Q}{B M}\,.
\end{align}
\end{subequations}
However, these constraints are \emph{not} compatible with those needed in order to remove the Dirac strings~\eqref{reg-DKNMS}.
\myparagraph{Reissner-Nordstr\"om in a swirling universe ($B=a=0$):}
For the non-rotating charged black hole in a swirling universe, there are no additional conditions other than those necessary to remove the coordinate singularities~\eqref{conical-DRNMS}, and we obtain
\beq
K_{(0)} = - 64 \jmath^2 \bigl( 3 - 5\jmath^2 Z^4\bigr) \,.
\eeq
In particular, in the dyonic case, when both charges $Q$ and $H$ are present, it is also possible to remove the Dirac strings~\eqref{reg-DRNS}, which yields
\beq
K_{(0)} = -\frac{64 H^2 \bigl(3 Q^2 - 5 H^2\bigr)}{Q^4 Z^4} \,,
\eeq
that could be set to zero with the following additional constraint
\beq
\label{cond-asympt-rns}
3 Q^2 = 5 H^2 \,,
\eeq
corresponding to
\beq
\jmath = \pm \frac{\sqrt{15}}{8 Q^2}\,.
\eeq
\myparagraph{Schwarzschild black holes ($a=Q=H=0$):}
For the Schwarzschild black holes we have that
\beq
K_{(0)} = 4 \bigl(5 B^4 - 48 \jmath^2\bigl) \,.
\eeq
Therefore, we can obtain $K_{(0)}=0$ by imposing
\beq
\label{cond-asympt-back}
5 B^4 = 48 \jmath^2 \,.
\eeq
\myparagraph{Backgrounds ($M=a=Q=H=0$):}
For the massless backgrounds, $M=0$, we obtain the same results as for the Schwarzschild black holes. However, in addition, the curvature on the symmetry axis is \emph{exactly} constant for the backgrounds, and not just an asymptotic result for large radial distances $r \to \infty$.
\beq
K\big\rvert_{\theta = 0, \pi} \equiv 4 \bigl(5 B^4 - 48 \jmath^2\bigl) \,.
\eeq
\myparagraph{Other Melvin sub-cases ($\,\jmath=0$):}
Similarly, for the other three Melvin sub-cases which can result in a spacetime free of any singularity, at least outside of the event horizon, i.e.~the Electric Kerr-Newman ($H=0$)~\eqref{conical-EKNM}, Kerr ($Q=H=0$)~\eqref{conical-KM} and Electric Reissner-Nordstr\"om ($a=H=0$)~\eqref{conical-ERNM}, we obtain the following asymptotic constant curvature $K_{(0)}$, without any additional constraint:
\beq
K_{(0)} = 20 B^4 - 2 B^6 Q^2 - 24 B^7 a M Q + \frac{1}{4} \bigl(5 Q^4 - 12 a^2 M^2\bigr) \,,
\eeq
which can also be set to zero in all three sub-cases.
\subsubsection{Curvature Singularities at the “Center” of the Black Hole} \label{subsubsec:curv-center-dknms}
{\begin{table}[H]
\begin{center}
\onehalfspacing
\small
\begin{tabular}{ |p{4.65cm}||p{1.9cm}|p{1.9cm}|p{1.9cm}|p{2.5cm}|  }
\hline
\multicolumn{5}{|c|}{Curvature Singularity for $r = \cos\theta = 0$} \\
\hline
Spacetime                                                  & Minkowski    & Melvin                         & Swirling       & Melvin-swirling \\
\hline
Background                                                & No                & No                               & No               & No                                  \\
Schwarzschild                                            & Yes               & Yes                              & Yes              & Yes                                 \\
Electric\,\,\,\,\,\,Reissner-Nordstr\"{o}m      & Yes               & Yes                              & Yes              & Yes                                  \\
Magnetic\,\,Reissner-Nordstr\"{o}m           & Yes               & Yes                              & Yes              & Yes                                  \\
Dyonic\,\,\,\,\,\,\,\,Reissner-Nordstr\"{o}m   & Yes               & Yes                              & Yes              & Yes                                  \\
Kerr                                                             & Yes               & No                               & No               & No                                   \\
Electric\,\,\,\,\,\,Kerr-Newman                      & Yes               & Condition$^{\star}$     & No               & No                                    \\
Magnetic\,\,Kerr-Newman                           & Yes               & No                               & No               & No                                     \\
Dyonic\,\,\,\,\,\,\,\,Kerr-Newman                   & Yes               & No                               & No               & Condition$^{\star \star}$   \\
\hline
\end{tabular}
\caption{\small Table summarizing for all the sub-cases if the “center” of the black hole $r = \cos\theta = 0$ is a curvature singularity} \label{table:Curvature:singularities}\end{center}
\end{table}}
\vspace{-0.8cm}
{\small \noindent $\star$ In this case, it is possible to have a curvature singularity for $r = \cos\theta = 0$ if the following relation between the parameters holds: $a = -\frac{2 Q}{B M}$. \\
$\star\star$ In this case, there can be a curvature singularity for $r = \cos\theta = 0$ only if the parameters satisfy both of these following constraints: $a = -\frac{2 Q}{B M}, \jmath = \frac{H B^2 }{4 Q}$. However, these conditions are not compatible with those needed in order to remove the other pathologies.}
\vfill

\clearpage 
\noindent As summarized in Table~\eqref{table:Curvature:singularities}, the “center” of the black hole $R = 0$, i.e.~$r = \cos\theta = 0$, is not always a curvature singularity, in contrast to the common notion that these objects always present such a singularity at their center. In general, we have that all the rotating black holes embedded in a Melvin or swirling universe do \emph{not} possess a curvature singularity at their center, which was already known for the black holes embedded in a Melvin universe, but still not verified for the swirling or Melvin-swirling universe. 

To be more precise, the most general case we found~\eqref{DKNMS} could in principle be singular for $R = 0$, if the parameters satisfy both of the following constraints:
\begin{subequations}
\label{curvature-sing-DKNMS}
\begin{align}
a & = -\frac{2 Q}{B M} \,, \label{curvature-sing-DKNMS-a}\\
\jmath & = \frac{H B^2 }{4 Q} \,.
\end{align}
\end{subequations}
However, we can see that the conditions necessary to remove the other pathologies~\eqref{reg-DKNMS}, or just simply those that remove the conical singularities~\eqref{free-con-sing}, are not compatible with the constraints needed for $R = 0$ to be a curvature singularity~\eqref{curvature-sing-DKNMS}. On the other hand, the electric Kerr-Newman in a Melvin universe sub-case $(\jmath=H=0)$ also presents a similar condition, however, for this black hole, there are no other constraints on the parameters needed to remove the other singularities, meaning that it is possible to require that such a black possess a curvature singularity at its center.

In the following, we list the behavior of the Kretschmann scalar $K$~\eqref{Kretschmann-scalar} in the limit $R\to0$, in order to discuss the order of divergence, or non-divergence, of the curvature singularity at that point. \\

\noindent $\bullet$ We begin by recalling the results for the black holes \emph{without} the swirling and Melvin parameters.
\paragraph{Schwarzschild ($a=Q=H=0$) and Kerr ($Q=H=0$):}
\beq
K\big\rvert_{\jmath = B = 0} \overset{R \to 0}{\approx} \frac{48 M^2}{r^6} \,.
\eeq
\paragraph{Reissner-Nordstr\"{o}m ($a=0$) and Kerr-Newman:}
\beq
K\big\rvert_{\jmath = B = 0} \overset{R \to 0}{\approx} \frac{56 Z^4}{r^8} \,.
\eeq \\
\vfill
\clearpage
\noindent $\bullet$ On the other hand, for the black holes in a Melvin-swirling universe we obtain:

\paragraph{Schwarzschild in a Melvin-swirling universe ($a=Q=H=0$):}
\beq
K \overset{R \to 0}{\approx} \frac{48 M^2}{r^6} \,.
\eeq
Thus, this result is not modified by embedding in the Melvin-swirling universe.
\paragraph{Reissner-Nordstr\"{o}m in a Melvin-swirling universe ($a=0$):}
\beq
K \overset{R \to 0}{\approx} \frac{56 Z^4}{r^8} \,.
\eeq
Hence, if the black hole is \emph{not} rotating, the presence of the swirling or Melvin parameters does \emph{not} modify the behavior of the Kretschmann scalar for $R \to 0$.
\paragraph{Kerr in a Melvin-swirling universe ($Q=H=0$):}
\beq
K \overset{R \to 0}{\propto} \frac{1}{a^{8}\Bigl[ \bigl(B^4+16\jmath^2)\, a^2 M^2 \Bigr]^6} \,.
\eeq
\paragraph{Kerr-Newman in a Melvin-swirling universe:}
\beq
K \overset{R \to 0}{\propto} \frac{1}{a^{8} \Bigl[4 B^2 Z^2 + 4 a M B\bigl(4 \jmath H + B^2 Q\bigr) + \bigl(B^4+16\jmath^2) \,a^2 M^2 \Bigr]^6} \,.
\eeq
%
As we can see, if the black hole \emph{is} rotating, the spin-spin interaction between the black hole $(a M)^2$ and the background $\bigl(B^4+16\jmath^2)$ removes the curvature singularity for $R=0$, with additional corrections if the black hole also possesses an electromagnetic charge.
\myparagraph{Singular Dyonic Kerr-Newman in a Melvin-swirling universe:}
However, as we have already pointed out, for the most general solution we found~\eqref{DKNMS}, it is still possible to introduce a curvature singularity for $R=0$ by constraining the parameters as in Eq.~\eqref{curvature-sing-DKNMS}, which indeed yields
\beq
K\big\rvert_{\eqref{curvature-sing-DKNMS}} \overset{R \to 0}{\propto} \frac{1}{r^8} \,.
\eeq
Moreover, the same result is also obtained for the Electric Kerr-Newman black hole in a Melvin universe, using the same conditions~\eqref{curvature-sing-DKNMS} with $H=0$, which in fact results in $\jmath=0$.
\subsubsection{Curvature Singularities Along the Symmetry Axis} \label{subsubsec:curv-along-axis-dknms}
There are some peculiar sub-cases where the origin $R=0$ is not only the possible place for curvature singularities. Indeed, there are three cases for which $F = 0$ represents a curvature singularity on an entire half of the symmetry axis for some special values of the parameters, as sketched in Figure~\ref{fig:line-singularities}. However, these constraints on the parameters are in general not compatible with those necessary to remove the other possible pathologies.
{\begin{figure}[H]
\begin{center}
\begin{tikzpicture}
\filldraw[color = black , fill = black!25] (-2,0) circle (1cm);
\path [draw = dark-red, snake it] (-2,0) -- (-2,2.25);
\path [draw = black, ->] (-2,-2.25) -- (-2, 2.25);
\filldraw[color = black , fill = black!25] (2,0) circle (1cm);
\path [draw = dark-red, snake it] (2,0) -- (2,-2.25);
\path [draw = black, ->] (2,-2.25) -- (2, 2.25);
\end{tikzpicture}
\caption{\small The left figure depicts a curvature singularity along the symmetry axis in the entire “northern” hemisphere $\theta=0$. Similarly, the right figure depicts a curvature singularity in the entire “southern” hemisphere $\theta=\pi$ of the symmetry axis.} \label{fig:line-singularities}
\end{center} 
\end{figure}
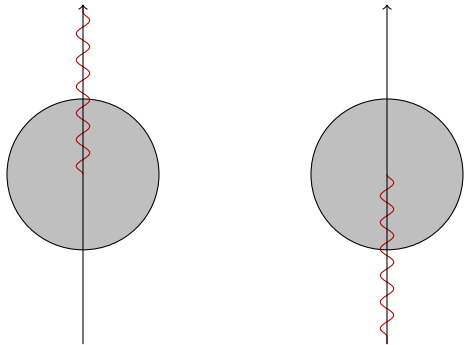 
\noindent Given the peculiarity of these pathologies, we report here the possible sub-cases, with their respective constraints, for which these singularities are present\footnote{The upper sign represents the curvature singularity for $\theta = 0$ while the lower sign is for $\theta = \pi$.}.
\paragraph{Dyonic Kerr-Newman in a Melvin-swirling universe:}
\begin{subequations}
\begin{align}
B & = \pm \frac{2 H}{Z^2} \,, \\
a & = \mp \frac{Q Z^2}{4 H M} \,, \\
\jmath &  = - \frac{HQ}{Z^4} \,.
\end{align}
\end{subequations}
These values of the parameters are \emph{not} compatible with the constraints necessary to remove the conical singularities~\eqref{free-con-sing}. Moreover, these conditions are in general not compatible with the removal of any other pathology.
\vfill
\clearpage
\paragraph{Kerr in a swirling universe ($B=Q=H=0$):}
\beq
\jmath = \pm \frac{1}{4aM} \,.
\eeq
Nonetheless, for this sub-case, these values of $\jmath$ are the same which removes one conical singularity, the one in the same hemisphere of this curvature singularity. As can be verified from Eqs.~\eqref{kerr-melvin-swirling-conical} with $B=0$.
%
\paragraph{Magnetic Reissner-Nordstr\"{o}m in a Melvin universe ($\,\jmath=a=Q=0$):}
\beq
B = \pm \frac{2}{H} \,.
\eeq
For this sub-case, these values of $B$ are the same which removes the conical singularity and the Dirac string located in the same hemisphere of the curvature singularity added along the symmetry axis.%

\subsection{Frame-dragging} \label{subsection:frame-drag-dknms}
Following Sec.~\ref{subsubsec:Frame-dragging}, the frame-dragging~\eqref{frame-dragging-eq} of the whole spacetime~\eqref{DKNMS} is given by
\beq
\omega^{\star} = - \frac{g_{t\phi}}{g_{\phi \phi}} = \frac{ \Omega}{\Sigma} \,.
\eeq
In particular, for large radial distances $r \to \infty$, we find
\begin{align}
\begin{split}
\omega^{\star} & \overset{r\to \infty}{\approx} \Bigl[ - 4 \jmath \cos\theta + 2 \jmath B H \bigl( 1 + \cos^2\theta \bigr) + \frac{Q B^3}{2}  \bigl( 1 + \cos^2\theta \bigr) \\
& \quad + \frac{a M}{8} \bigl( B^4 + 16 \jmath^2 \bigr) \bigl( 3 + 6\cos^2\theta - \cos^4\theta \bigr)  \Bigr] \, r  \,,
\end{split}
\end{align}
which thus receives contributions from the swirling parameter $\jmath$, from a coupling between the Melvin-swirling background and the magnetic charge of the black hole $(\jmath BH)$, from another coupling between the electric charge and the Melvin background $(Q B^3)$, plus a final contribute due to the spin-spin interaction between the angular momentum of the black hole $(a M)$ and the background $( B^4 + 16 \jmath^2)$.

Similarly, we can also obtain the approximation for large radial distances $r \to \infty$ of the angular velocity on the two halves of the symmetry axis
\begin{subequations}
\begin{align}
\omega^{\star} \big\rvert_{\theta = 0} \overset{r\to \infty}{\approx} \Bigl[ 4 \jmath \bigl( B H - 1\bigr) + Q B^3 + a M \bigl( B^4 + 16 \jmath^2 \bigr) \Bigr] \, r \,, \\ 
\omega^{\star}  \big\rvert_{\theta = \pi} \overset{r\to \infty}{\approx} \Bigl[ 4 \jmath \bigl( B H + 1\bigr) + Q B^3 + a M \bigl( B^4 + 16 \jmath^2 \bigr)\Bigr] \, r \,,
\end{align}
\end{subequations}
and on the equatorial plane
\beq
\omega^{\star}  \big\rvert_{\theta = \frac{\pi}{2}} \overset{r\to \infty}{\approx} \Bigl[ 2 \jmath H B  + \frac{Q B^3}{2} + \frac{3 a M}{8} \bigl( B^4 + 16 \jmath^2 \bigr) \Bigr] \, r \,,
\eeq
which, as expected, are not constant.

In principle, since the value of the gravitational frame-dragging grows unbounded, and thus can easily exceed 1, which corresponds to the speed of light in our units, it would be natural to conclude that there exist superluminal observers that violate causality. However, as we proved in Sec.~\ref{subsec:CTCs-dknms}, there is no possible occurrence of closed timelike curves outside the event horizon, meaning that there are no related causality issues. Therefore, the apparent paradox of the superluminal observers can be attributed to a poor choice of coordinates, in the sense that a set of coordinates adapted to timelike observers does not experience an unbounded growth of the frame-dragging, as it happens, for example, for the Alcubierre spacetime~\cite{Alcubierre}.

To be more precise, in order to remove the various pathologies, it is sometimes necessary to constraint some parameters and perform a rescaling of the azimuthal coordinates $\phi \mapsto \frac{2 \pi}{\delta \phi} \phi$. For these cases, the correct frame-dragging to consider is therefore
\beq
\omega  = \biggl[\frac{\delta \phi}{2 \pi}\biggr] \omega^{\star}\bigg\rvert_{\text{“regular”}} \,. \\
\eeq
\subsubsection{Angular Velocity of the Event Horizon} \label{subsubsection:angular-horizon}
Moreover, it is interesting to computer the angular velocity of the horizons $r_{\pm}$~\eqref{horizon-DKNMS}, which is hence given by
\beq
\label{angular-dknms}
\omega^{\star}_{\pm} \coloneqq \omega^{\star}\big\rvert_{r=r_{\pm}} = \frac{2 \Gamma_{\pm} }{\lambda_{\pm}} \,,\\
\eeq
where
\begin{subequations}
\begin{align}
\lambda_{\pm} & = r_{\pm}^2 + a^2 \,, \\
\Gamma_{\pm} &= \frac{a}{2} - BQr_{\pm} -  \frac{3 B^2}{4}  a Z^2 + \frac{B^3}{4} Q\gamma_{\pm} \,+ a M \biggr[ \jmath^2 + \frac{B^4}{16} \biggr]\Bigl[\gamma_{\pm} + 2 r_{\pm} \lambda_{\pm}\Bigr] + \jmath B H \gamma_{\pm} \,, \\
Z^2 & = Q^2 + H^2\,, \\
\gamma_{\pm} & = r_{\pm}^3 - \frac{a^2}{\lambda_{\pm}}\Bigl[2 a^2 M + Z^2 r_{\pm} \Bigr]  \,,
\end{align}
\end{subequations}
which, as expected, is constant.

\subsection{Entropy and Temperature} \label{subsec:entropy-temp-dknms}
Regarding the temperature and the entropy of black holes, which are explained in Sec.~\ref{subsubsec:Black-Holes-Thermodynamics}, it is known that both the Melvin parameter $B$ and the swirling parameter $\jmath$ do \emph{not} modify the area and the surface gravity of the black holes~\cite{AstorinoMagnetised, SiahaanMelvTN, AstorinoRemoval}. Nevertheless, we can prove that this also holds for the new solutions if both parameters are present. 

For the most general solution~\eqref{DKNMS}, we have that the event horizons $r_{\pm}$~\eqref{horizon-DKNMS} are Killing horizons for the Killing vectors
\beq
\xi^{\mu}_{\pm} = \bigl(1, 0 , 0, \omega^{\star}_{\pm} \bigr) \,, \\
\eeq
where $\omega^{\star}_{\pm}$ is the angular velocity of the horizons~\eqref{angular-dknms}.

Therefore, the surface gravity~\eqref{surface-gravity-zero} of these horizons is given by
\beq
\kappa_{\pm}^{\star} = \frac{r_{\pm}-r_{\mp}}{ 2 \bigl( r_{\pm}^2 + a^2 \bigr)} \,, \\
\eeq
which is the same surface gravity of the Kerr-Newman black hole~\eqref{surface-gravity-kn}, and thus results in the same temperature~\eqref{temperature-kn} 
\beq
T^{\star} = \frac{\kappa_{+}}{2 \pi} \,.
\eeq
Moreover, the same happens for the area and the entropy of the black hole
\begin{align}
\pazocal{A}^{\star}_{\pm} & = 4 \pi \bigl( r_{\pm}^2 + a^{2} \bigr) \,,\\
\pazocal{S}^{\star} & = \frac{\pazocal{A}^{\star}_{+}}{4} \,,
\end{align}
which are indeed the same area~\eqref{area-kn} and entropy~\eqref{entropy-kn} of the Kerr-Newman black hole.

However, as discussed in Sec.~\ref{subsection:frame-drag-dknms}, these quantities might depend on the Melvin and swirling parameters after the removal of the various pathologies. Thus, for these cases, the correct physical quantities are
\begin{subequations}
\begin{align}
\kappa_{\pm} & =\kappa_{\pm}^{\star}\Big\rvert_{\text{“regular”}} \,, \\
T & = \frac{\kappa_{+}}{2 \pi} \,, \\
\pazocal{A}_{\pm} & =  \biggl[\frac{2 \pi}{\delta \phi}\biggr]\pazocal{A}^{\star}_{\pm}\bigg\rvert_{\text{“regular”}} \,,\\
\pazocal{S} & = \frac{\pazocal{A}_{+}}{4} \,.
\end{align}
\end{subequations}
\vfill
\clearpage
\subsection{Horizons and Ergoregions} \label{subsec:hor-ergo-dknms}
\vspace{-0.1cm}
We will now analyze the ergoregions, which are explained in Sec.~\ref{subsubsec:Ergoregions}, and the event horizons. Moreover, we will also provide a list of plots, for various values of the parameters, in order to perform a qualitative analysis of these quantities. Specifically, we will start by analyzing the dyonic Kerr-Newman black hole in a Melvin-swirling universe~\eqref{DKNMS}, with the singularities removed through Eqs.~\eqref{reg-DKNMS}.

For the ergoregions, we will perform a numerical analysis of the function $g_{tt}$, utilizing the same “rectangular” coordinates defined for the CTCs~\eqref{rectangular-coordinates}. On the other hand, for the outer event horizon we will use a proper embedding in the Euclidian three-dimensional space $\E^3$, as explained in Appendix~\ref{D-Embedding}, with the differential equations solved numerically. 

In particular, the figures will consist of (at least) one three-dimensional plot for the event horizon and one for the ergoregions, plus additional plots representing the (same) two-dimensional cross-section of the ergoregions for $y=0$, in order to compare the position of the ergoregions with that of the horizons. Since these plots will include both extremal and non-extremal black holes, we will take advantage of the extremality index defined for the CTCs, $\pazocal{I}=\frac{Q^2+H^2+a^2}{M^2}$~\eqref{extremality-index}, in order to indicate how close the black hole is to extremality.

Before showing the plots, it is interesting to notice that the $g_{tt}$ component of these solutions~\eqref{DKNMS} can be written in the following form:
\vspace{-0.05cm}
\beq
\label{gtt}
g_{tt} =  g_{(1)}\bigl(1 - \cos^2\theta\bigr) + g_{(2)}\Delta \,,
\eeq
where $g_{(1)}$ and $g_{(1)}$ are some differentiable functions, and where we recall that and $\Delta = 0 $ are the locations of the horizons~\eqref{horizon-DKNMS}. From this result, we can conclude that the poles of the horizons ($\Delta = 0, \cos^2\theta = 1$) can never be part of the ergoregion, since in these points $g_{tt} = 0$~\eqref{gtt}. Moreover, it can be proved that on the symmetry axis ($\cos^2\theta = 1$) it holds $g_{(2)} < 0$. Therefore, we have that the symmetry axis is never part of the ergoregions outside the outer horizon and also inside the inner horizon, since in these regions $\Delta > 0$. Analogously, the portion of the symmetry axis between the outer and inner horizons is always part of the ergoregions, since between the horizons we have $ \Delta < 0$.

Furthermore, it holds that $g_{(1)} = 0$ if $a=B=0$. Therefore, we can conclude that $g_{tt}=0$ for the entirety of the horizons ($\Delta = 0$) for all the non-rotating black holes in a swirling universe sub-cases, independently from the specific value of the parameters. Due to this last consideration, we will also present plots for the dyonic Reissner-Nordstr\"{o}m black hole in a swirling universe alongside those for its Melvin-swirling generalization, using the same values of the parameters, except for the magnetic field $B$, which is fixed by the constraints needed to remove the conical singularities for the Melvin-swirling black hole~\eqref{conical-DRNMS}, while it is obviously zero for the swirling sub-case.
\vfill
\clearpage
\myparagraph{Dyonic Kerr-Newman in a Melvin-swirling universe:}
\vspace{-0.5cm}
\begin{figure}[H]
\captionsetup[subfigure]{labelformat=empty}
\centering
\hspace{2cm} \subfloat[\mbox{\hspace{-0.25cm} Event Horizon}]{{\includegraphics[height=7.5cm]{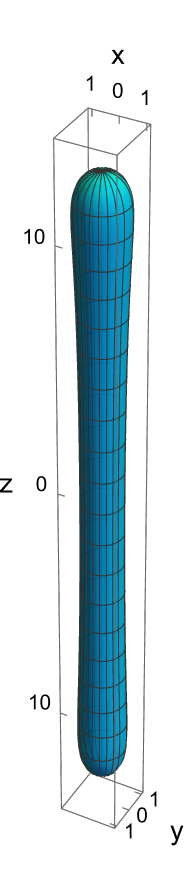}}}
\hspace{2.6cm} \subfloat[\hspace{0.3cm} Ergoregions]{{\hspace{0.5cm}\includegraphics[height=7cm]{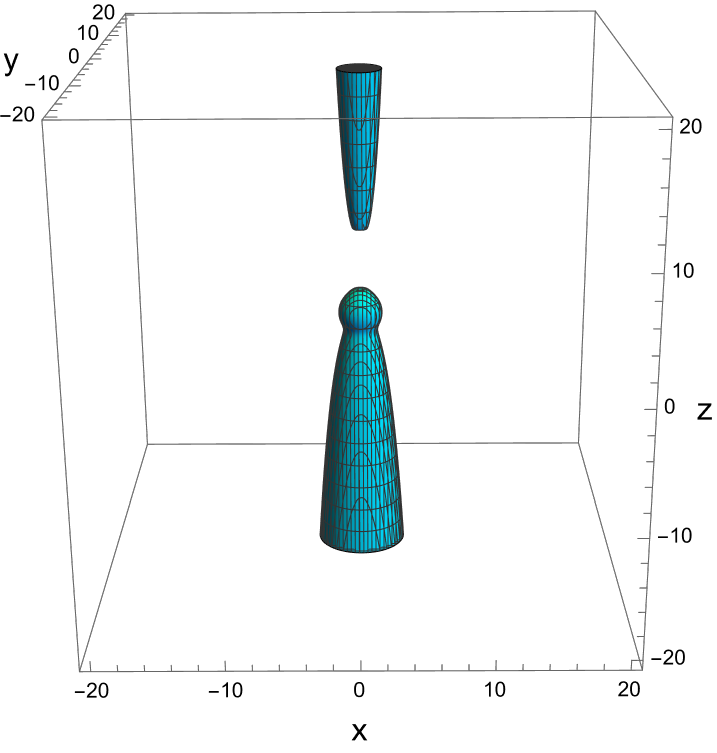}}}\\
\null \, \\
\hspace{-1.5cm} \subfloat[\hspace{1cm} Ergoregions Cross-section $y=0$]{{\hspace{0.5cm}\includegraphics[height=6.5cm]{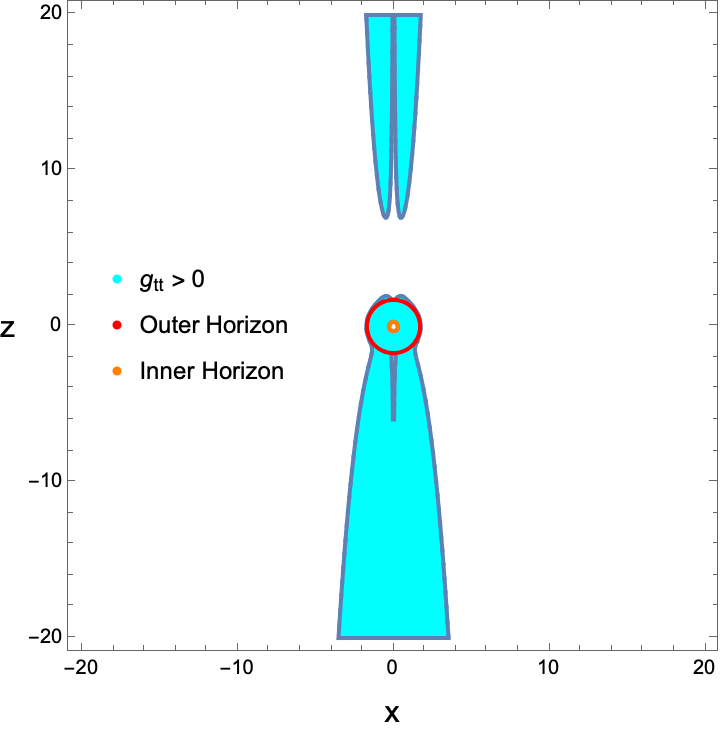}}}
\hspace{1cm} \subfloat[ \hspace{0.5cm}Ergoregions Cross-section $y=0$]{{\includegraphics[height=6.5cm]{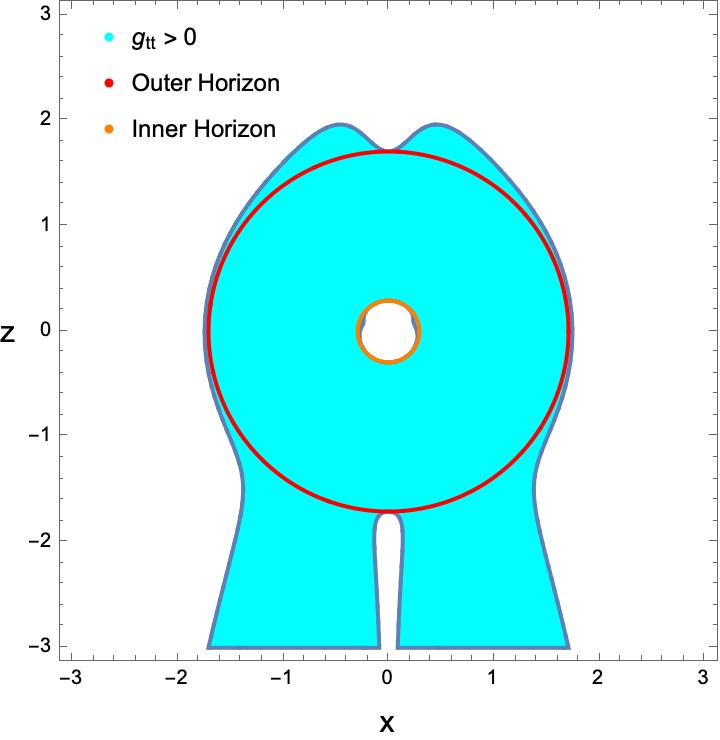}}}
\caption{\small Ergoregions and the event horizon of the dyonic Kerr-Newman black hole in a Melvin-swirling universe, for which both conical singularities and Dirac strings have been removed, with parameters: $M=1, Q=\frac{1}{2}, H=\frac{1}{2}, B=1 \Rightarrow a\simeq 0.01, \jmath = 2.75, \pazocal{I}\simeq 0.5 $.} \label{Plot-DKNMS1}
\end{figure}
\noindent As we can see in Figure~\ref{Plot-DKNMS1}, the presence of the Melvin-swirling background has the effect of making the horizon more prolate, and also of extending through infinity the ergoregions for all the vicinity of the symmetry axis, which is also a common feature of black holes embedded in either just a Melvin or a swirling universe.
\vfill
\begin{figure}[H]
\captionsetup[subfigure]{labelformat=empty}
\centering
\hspace{0.5cm} \subfloat[Event Horizon]{{\includegraphics[height=6cm]{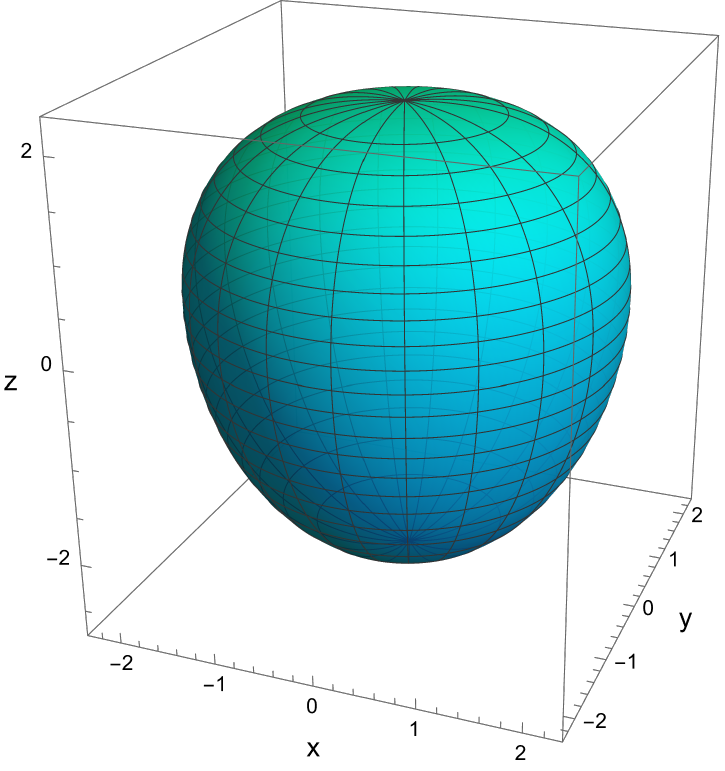}}}
\hspace{0.3cm} \subfloat[\hspace{0.2cm} Ergoregions]{{\hspace{0.5cm}\includegraphics[height=6cm]{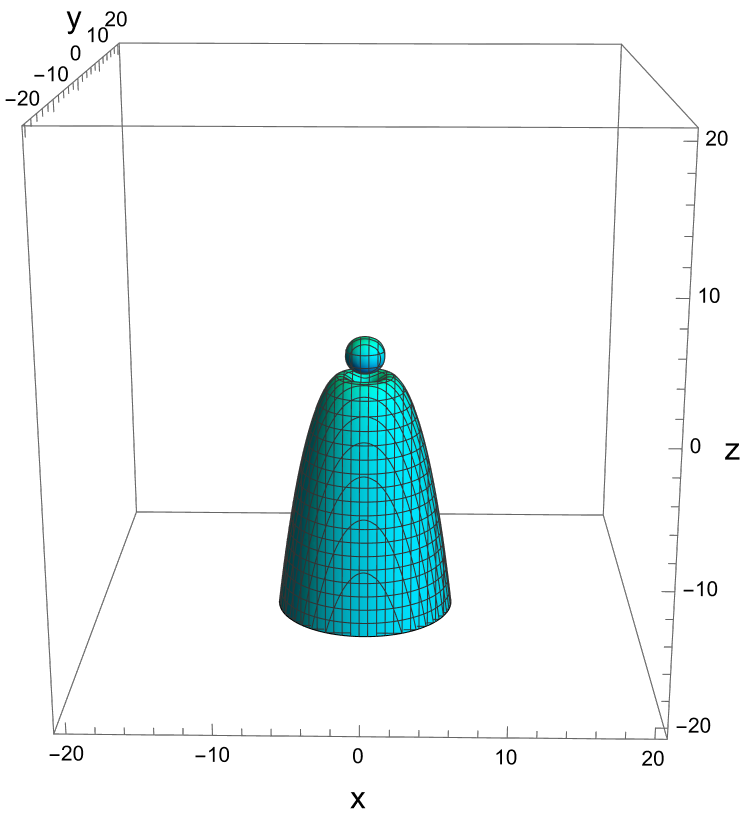}}} \\
\subfloat[\hspace{1cm} Ergoregions Cross-section $y=0$]{{\hspace{0.5cm}\includegraphics[height=6cm]{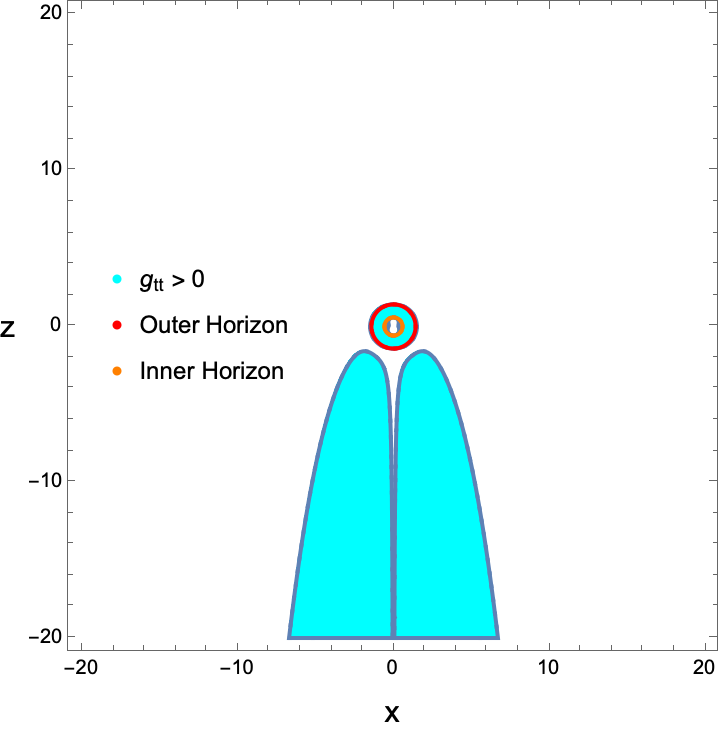}}}
\subfloat[\hspace{1cm} Ergoregions Cross-section $y=0$]{{\hspace{0.5cm}\includegraphics[height=6cm]{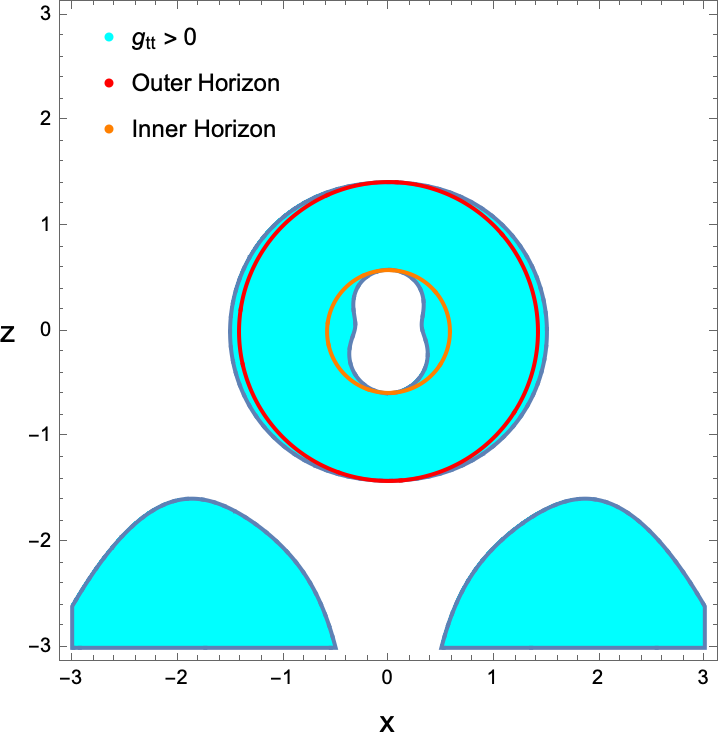}}} \\
\subfloat[\hspace{1cm} Ergoregions Cross-section $y=0$]{{\hspace{0.5cm}\includegraphics[height=6cm]{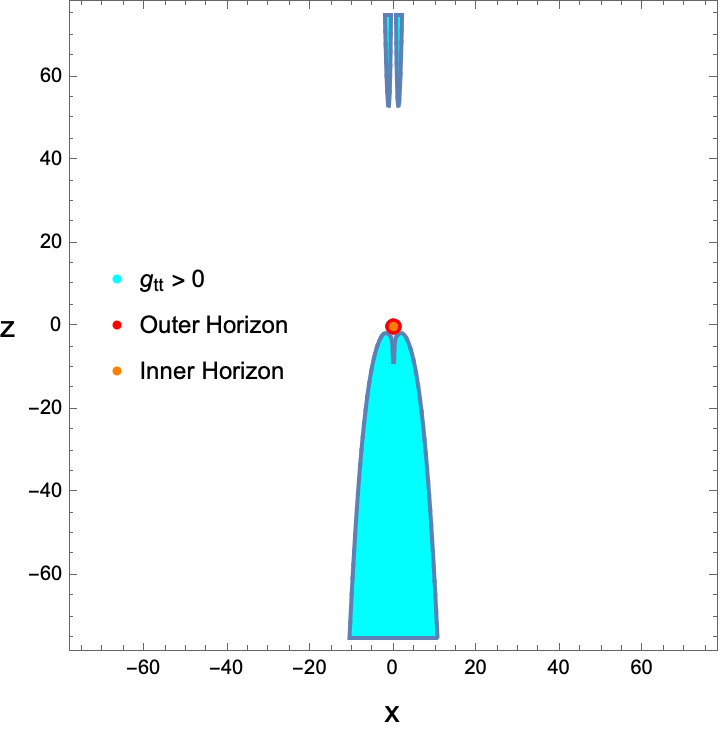}}}
\hspace{0.3cm} \subfloat[\hspace{0.2cm} Ergoregions]{{\hspace{0.5cm}\includegraphics[height=6cm]{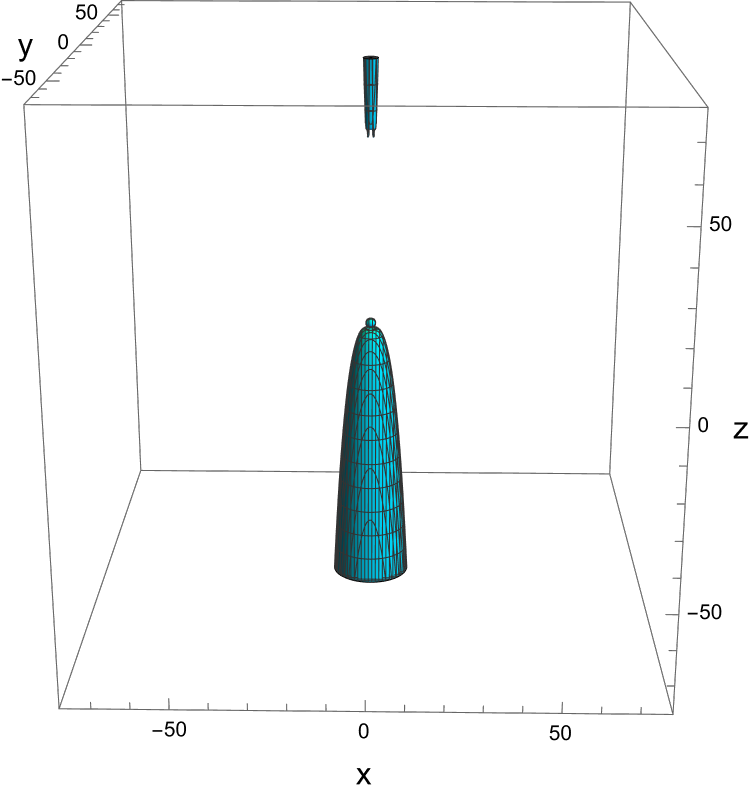}}}
\caption{\small Ergoregions and the event horizon of the dyonic Kerr-Newman black hole in a Melvin-swirling universe, for which both conical singularities and Dirac strings have been removed, with parameters: $M=1, Q=\frac{9}{10}, H=\frac{1}{8}, B=1 \Rightarrow a \simeq 0.047, \jmath \simeq 0.272, \pazocal{I} \simeq 0.828$.} \label{Plot-DKNMS2}
\end{figure}

\clearpage
\begin{figure}[H]
\captionsetup[subfigure]{labelformat=empty}
\centering
\hspace{0.75cm} \subfloat[Event Horizon]{{\includegraphics[height=7.5cm]{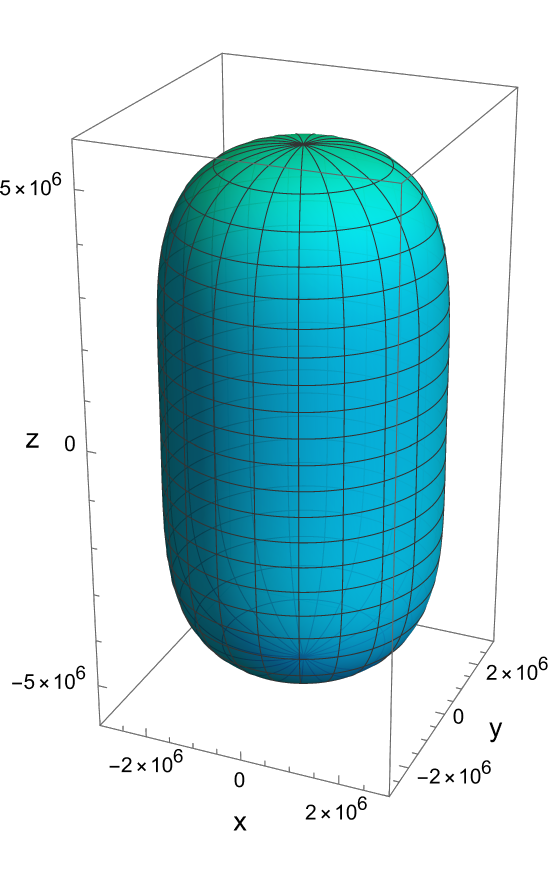}}}
\hspace{1.25cm} \subfloat[\hspace{0.5cm} Ergoregions]{{\hspace{0.5cm}\includegraphics[height=7cm]{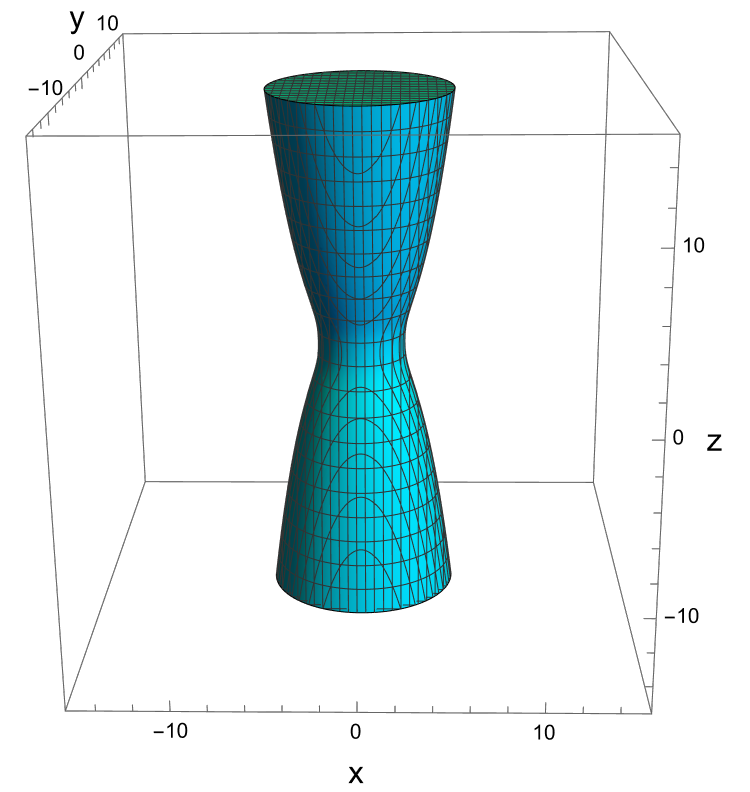}}} \\
\null \,
\hspace{-1.5cm} \subfloat[\hspace{1cm} Ergoregions Cross-section $y=0$]{{\hspace{0.5cm}\includegraphics[height=6.5cm]{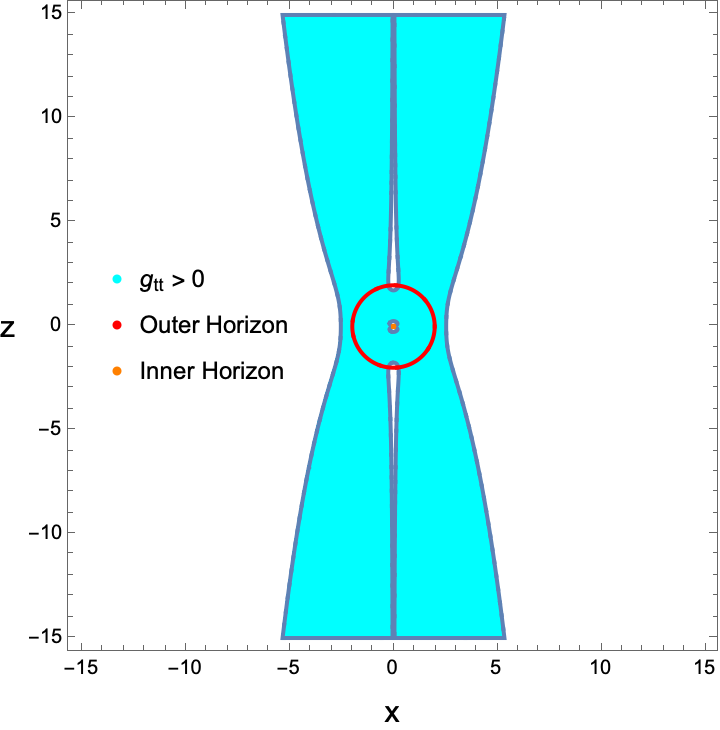}}}
\hspace{1cm} \subfloat[\hspace{0.5cm} Ergoregions Cross-section $y=0$]{{\includegraphics[height=6.5cm]{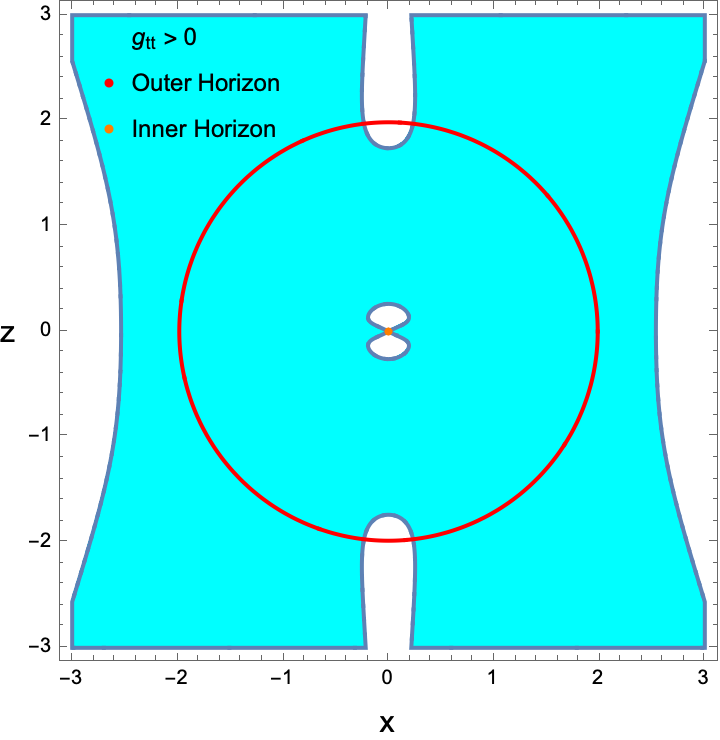}}}
\caption{\small Ergoregions and the event horizon of the dyonic Kerr-Newman black hole in a Melvin-swirling universe, for which both conical singularities and Dirac strings have been removed, with parameters: $M=1, Q=\frac{1}{8}, H=\frac{1}{8}, B=10^3 \Rightarrow a\simeq 0.65, \jmath \simeq 7.5 * 10^5, \pazocal{I}\simeq 0.455$.} \label{Plot-DKNMS3}
\end{figure}
\noindent In the sub-case provided in Figure~\ref{Plot-DKNMS2}, we have that the horizon is not symmetric with respect to the equatorial plane, and that the ergoregion in the upper hemisphere is “pushed away” from the black hole, which are both effects due to coupling between the charges of the black hole and the Melvin background. On the other hand, in Figure~\ref{Plot-DKNMS3}, can see that if the rotation of the black hole is \emph{not} negligible, the shape of the horizon becomes less oblate, despite a possible enormous value of the Melvin or swirling parameters.
%
\clearpage
\myparagraph{Extremal dyonic Kerr-Newman in a Melvin-swirling universe ($M^2=a^2+Z^2$):}
\begin{figure}[H]
\captionsetup[subfigure]{labelformat=empty}
\centering
\hspace{0.125cm} \subfloat[Event Horizon]{{\includegraphics[height=5.8cm]{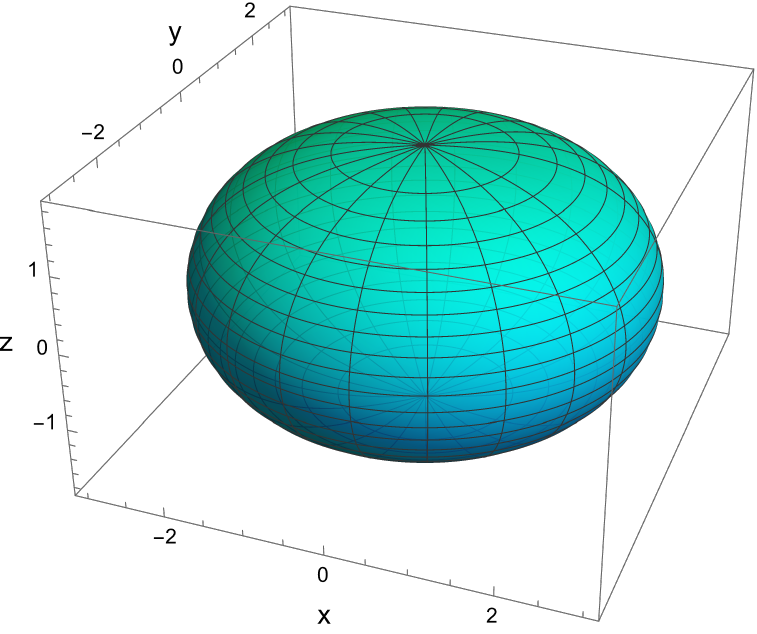}}}
\hspace{0.125cm} \subfloat[\hspace{0.5cm} Ergoregions]{{\hspace{0.5cm}\includegraphics[height=7cm]{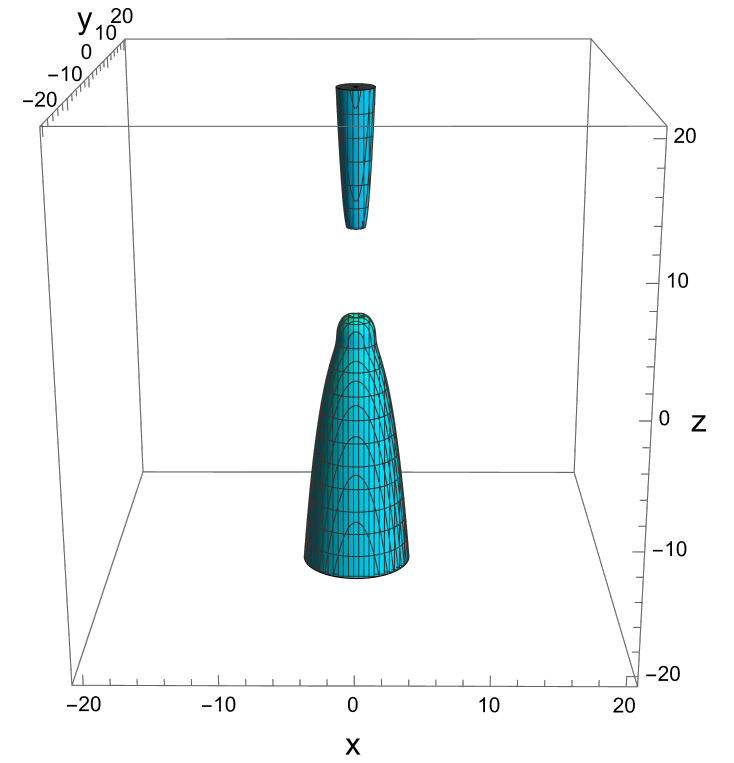}}}\\
\null \, \\
\hspace{-1cm} \subfloat[\hspace{1cm} Ergoregions Cross-section $y=0$]{{\hspace{0.5cm}\includegraphics[height=6.5cm]{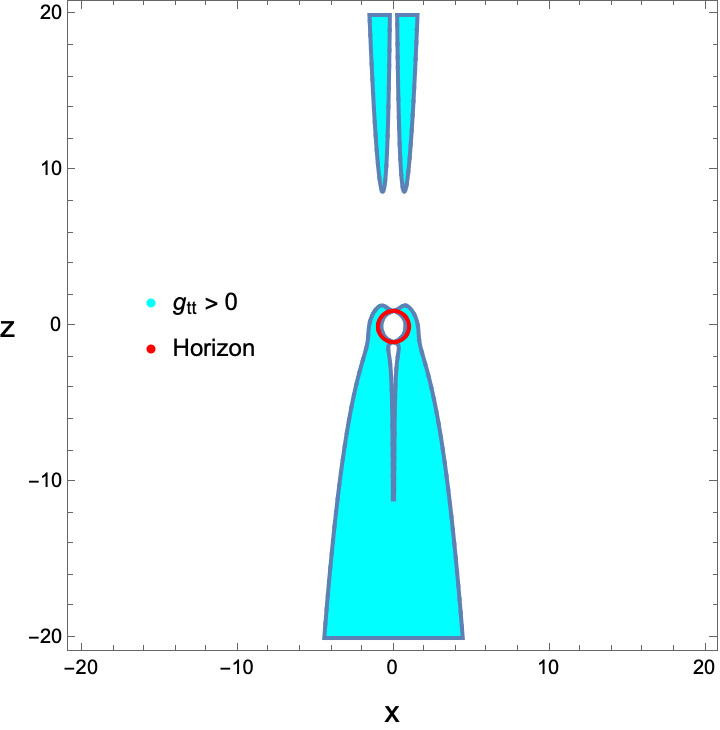}}}
\hspace{0.5cm} \subfloat[\hspace{1cm} Ergoregions Cross-section $y=0$]{{\hspace{0.5cm}\includegraphics[height=6.5cm]{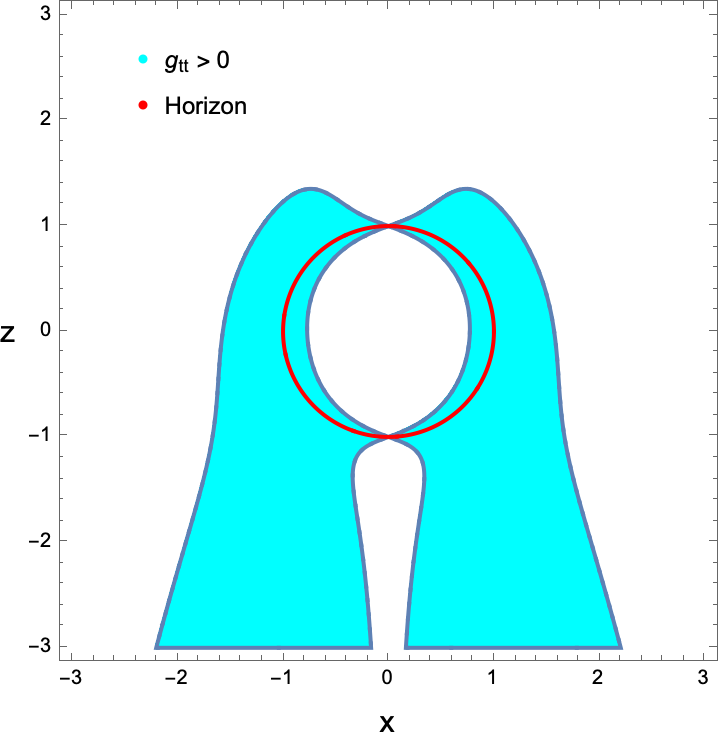}}}
\caption{\small Ergoregions and the event horizon of the \emph{extremal} dyonic Kerr-Newman black hole in a Melvin-swirling universe, for which both conical singularities and Dirac strings have been removed, with parameters:  $M=1, Q=0.706, H=0.706, B=1 \Rightarrow a \simeq 0.05, \jmath \simeq 1.75, \pazocal{I}=1$.} \label{Plot-DKNMS-EXT}
\end{figure}
\noindent As we can see in Figure~\ref{Plot-DKNMS-EXT}, we have that the shape of the horizon in the extremal case tends to return to the shape of the black hole not embedded in the Melvin-swirling universe. However, as we will prove later, the shape of the horizon is not \emph{exactly} the same as that of the non-Melvin-swirling black hole, at least not for all the sub-cases.
%

%
%
\clearpage
\myparagraph{Dyonic Reissner-Nordstr\"om in a (Melvin-)swirling universe ($a=0$):}
\null
\begin{figure}[H]
\captionsetup[subfigure]{labelformat=empty}
\centering
\subfloat[\hspace{-0.25cm} \mbox{Event Horizon}]{{\includegraphics[height=0.4\textwidth]{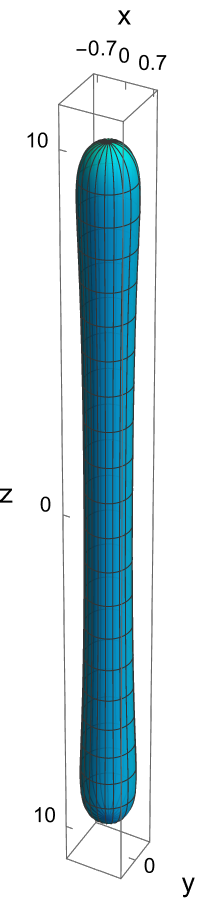}}}
\subfloat[\hspace{0.4cm} Ergoregions]{{\hspace{0.5cm}\includegraphics[width=0.3\textwidth]{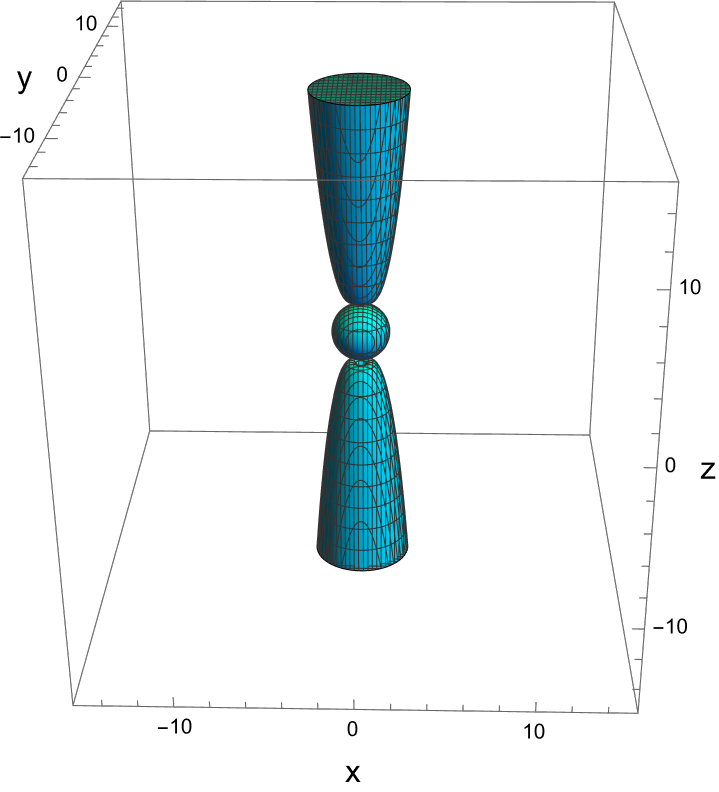}}}
\subfloat[\hspace{0.6cm} Ergoregions Cross-section $y=0$]{{\hspace{0.5cm}\includegraphics[width=0.3\textwidth]{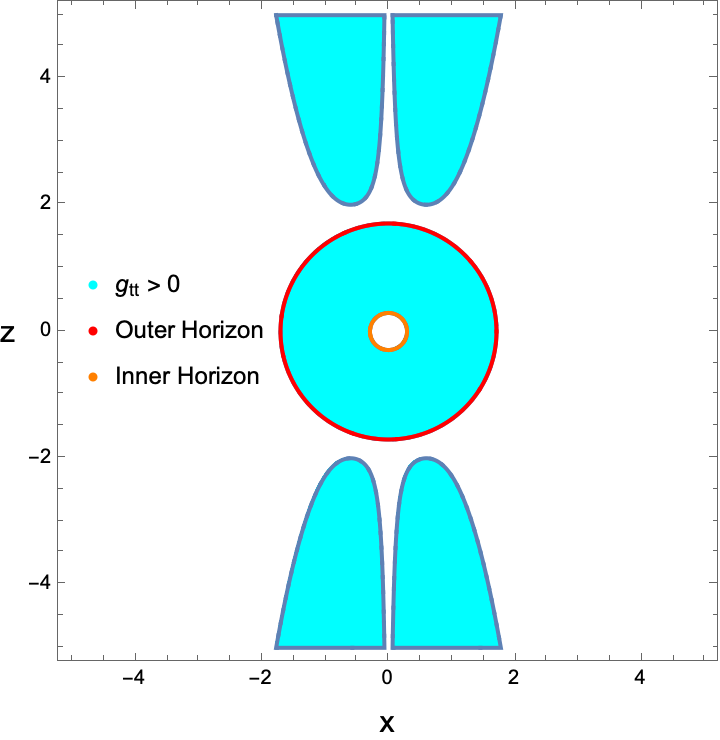}}}
\caption{\small Ergoregions and the event horizon of the dyonic Reissner-Nordstr\"om hole in a swirling universe with parameters: $B = 0, M=1, Q=\frac{1}{2}, H=\frac{1}{2}, \jmath = \frac{54}{25} \Rightarrow \pazocal{I}=\frac{1}{2}$.} \label{Plot-DRNS}
\end{figure}
\begin{figure}[H]
\captionsetup[subfigure]{labelformat=empty}
\centering
\subfloat[\hspace{-0.25cm} \mbox{Event Horizon}]{{\includegraphics[height=0.4\textwidth]{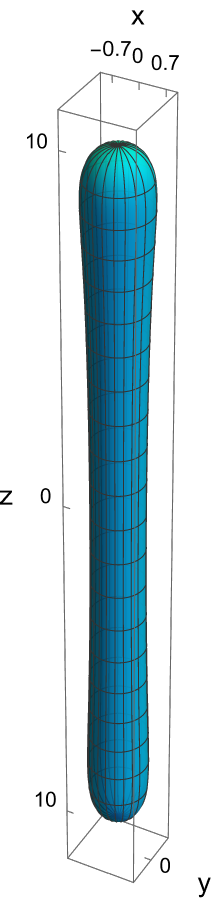}}}
\subfloat[\hspace{0.4cm} Ergoregions]{{\hspace{0.5cm}\includegraphics[width=0.3\textwidth]{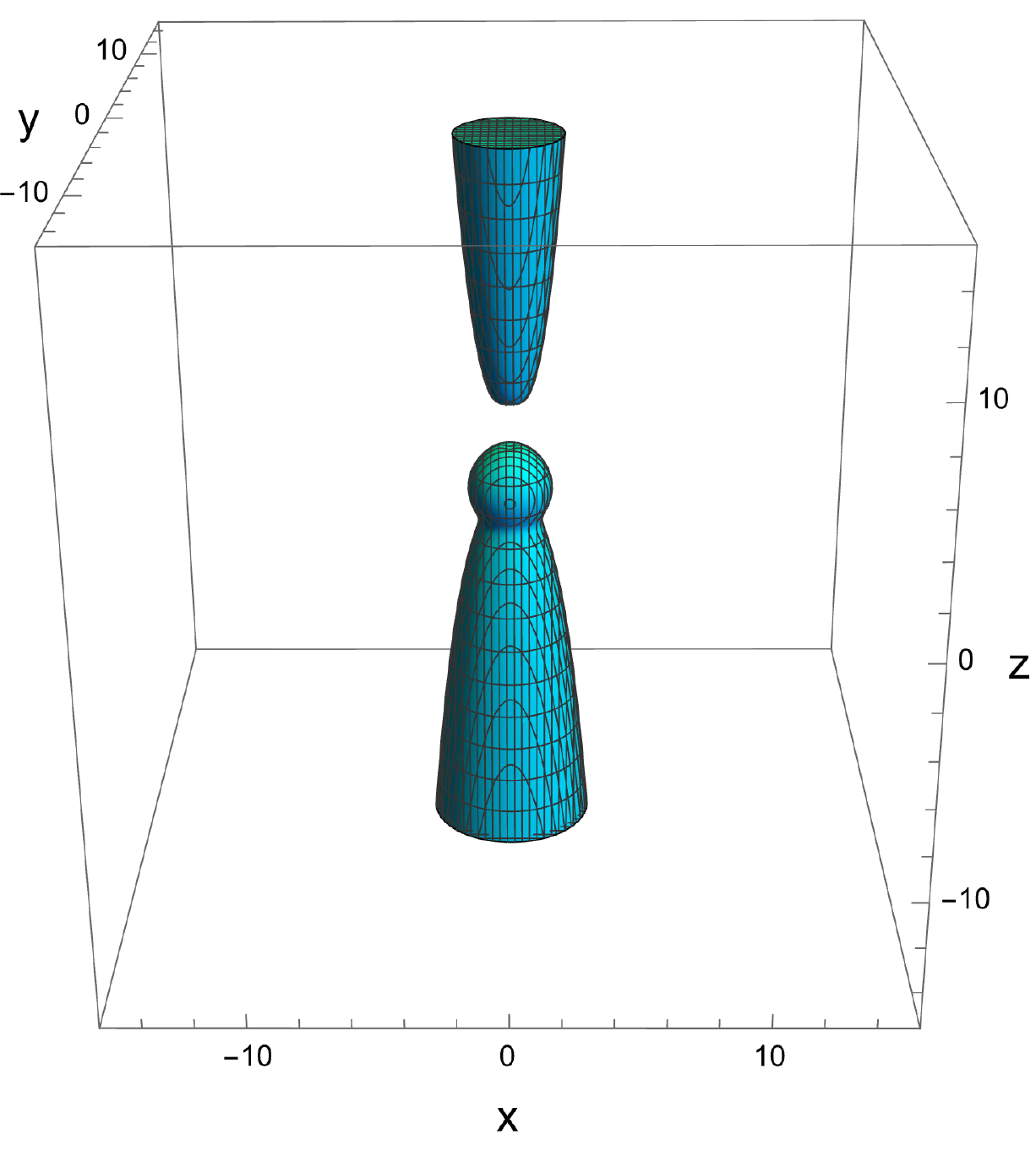}}}
\subfloat[\hspace{0.6cm} Ergoregions Cross-section $y=0$]{{\hspace{0.5cm}\includegraphics[width=0.3\textwidth]{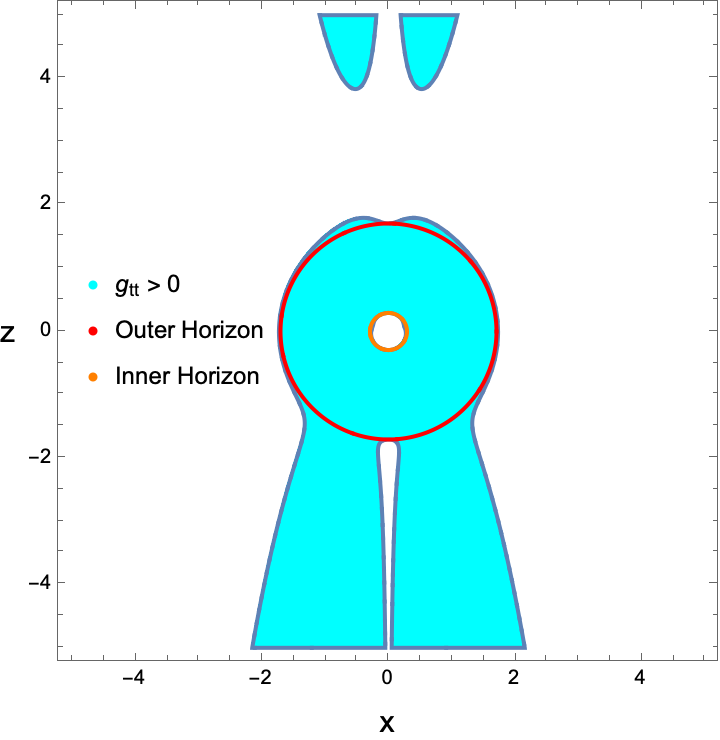}}}
\caption{\small Ergoregions and the event horizon of the dyonic Reissner-Nordstr\"om hole in a Melvin-swirling universe with parameters: $B = \frac{4}{5}, M=1, Q=\frac{1}{2}, H=\frac{1}{2}, \jmath = \frac{54}{25} \Rightarrow \pazocal{I}=\frac{1}{2}$.} \label{Plot-DRNMS}
\end{figure}
\clearpage
\myparagraph{\small Extremal dyonic Reissner-Nordstr\"om in a (Melvin-)swirling universe ($a=0,\,M^2=Z^2$):}

\null

\null

\null
\begin{figure}[H]
\captionsetup[subfigure]{labelformat=empty}
\centering
\subfloat[Event Horizon]{{\includegraphics[width=0.3\textwidth]{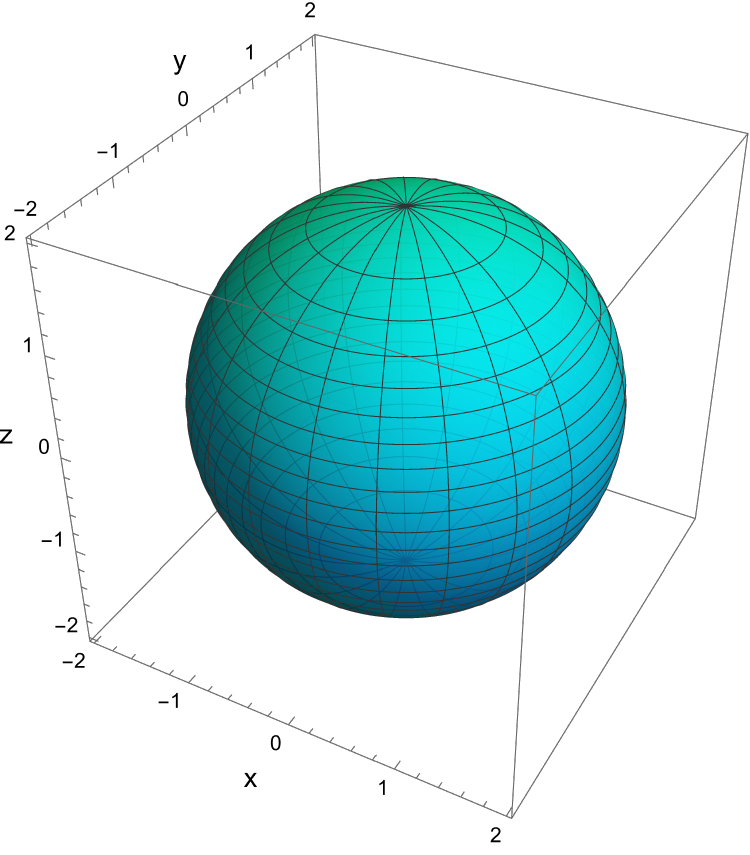}}}
\subfloat[\hspace{0.4cm} Ergoregions]{{\hspace{0.5cm}\includegraphics[width=0.3\textwidth]{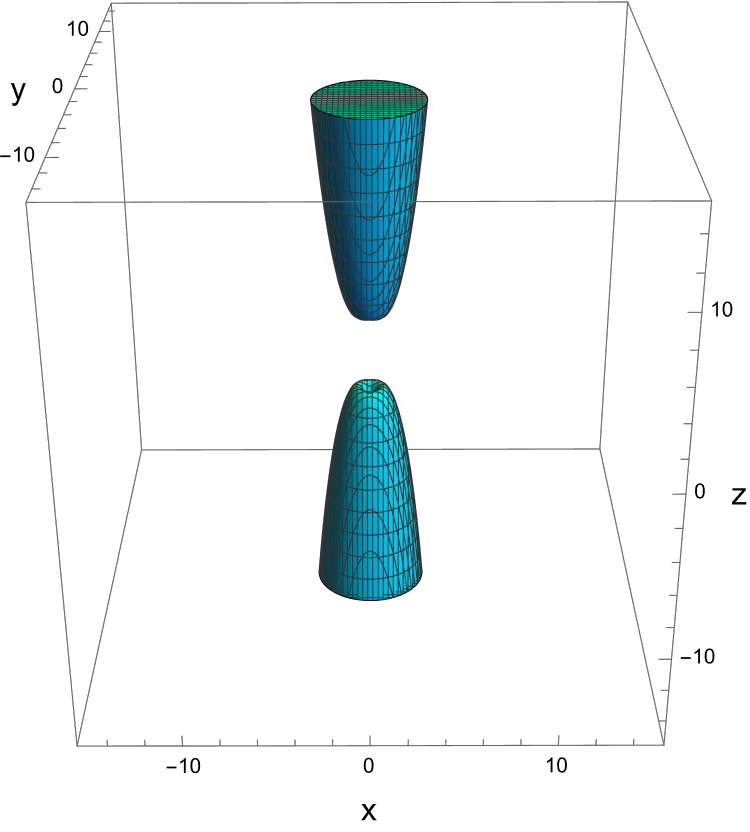}}}
\subfloat[\hspace{0.6cm} Ergoregions Cross-section $y=0$]{{\hspace{0.5cm}\includegraphics[width=0.3\textwidth]{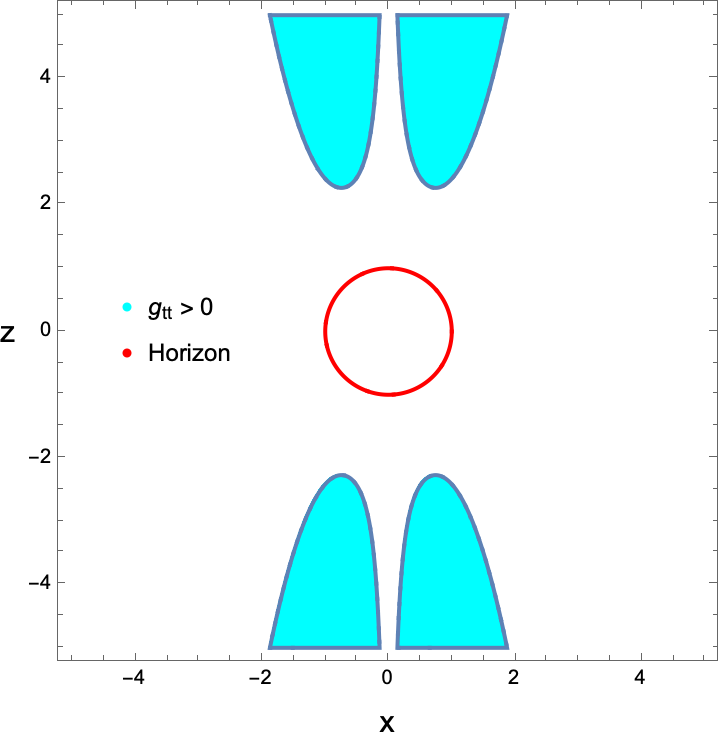}}}
\caption{\small Ergoregions and the event horizon of the \emph{extremal} dyonic Reissner-Nordstr\"om hole in a swirling universe with parameters: $B = 0, M=1, Q=\frac{3}{5}, H=\frac{4}{5}, \jmath = \frac{116}{75} \Rightarrow \pazocal{I}=1$.} \label{Plot-DRNS-EXT-1}
\end{figure}
\begin{figure}[H]
\captionsetup[subfigure]{labelformat=empty}
\centering
\subfloat[Event Horizon]{{\includegraphics[width=0.3\textwidth]{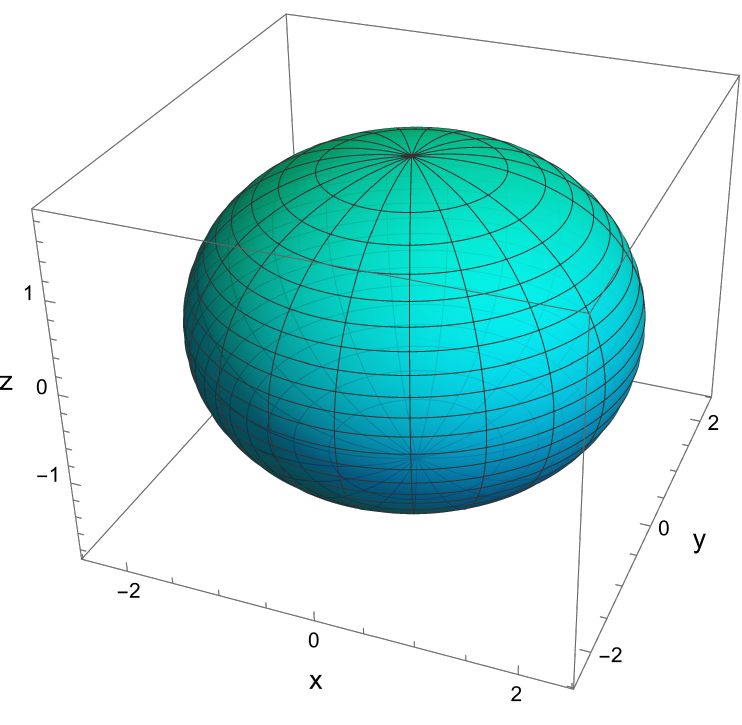}}}
\subfloat[\hspace{0.4cm} Ergoregions]{{\hspace{0.5cm}\includegraphics[width=0.3\textwidth]{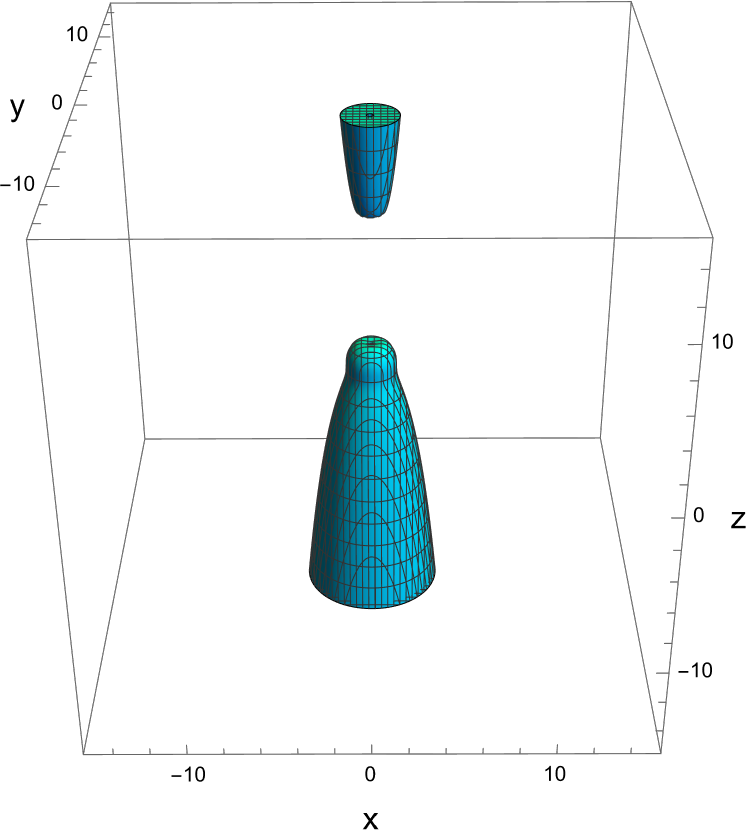}}}
\subfloat[\hspace{0.6cm} Ergoregions Cross-section $y=0$]{{\hspace{0.5cm}\includegraphics[width=0.3\textwidth]{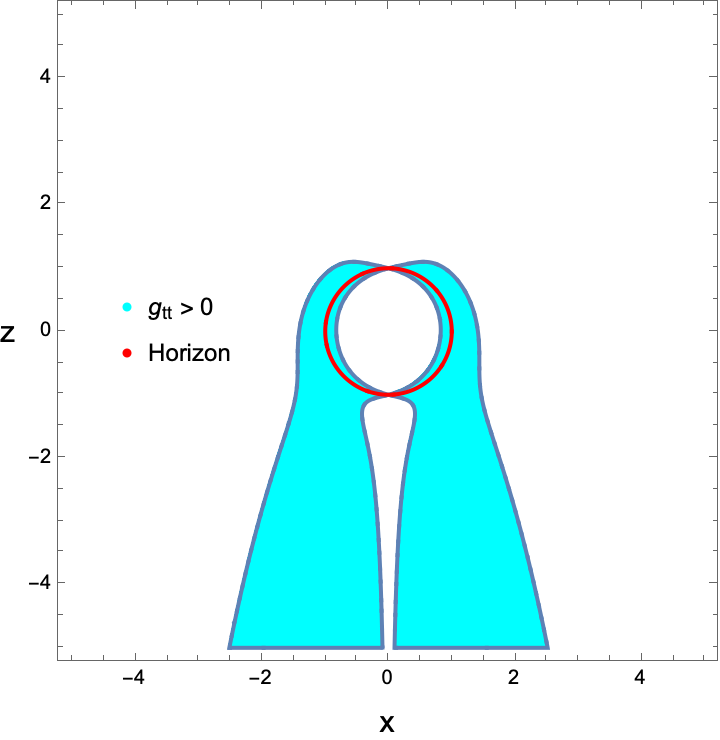}}}
\caption{\small Ergoregions and the event horizon of the \emph{extremal} dyonic Reissner-Nordstr\"om hole in a Melvin-swirling universe with parameters: $B = \frac{4}{5}, M=1, Q=\frac{3}{5}, H=\frac{4}{5}, \jmath = \frac{116}{75} \Rightarrow \pazocal{I}=1$.} \label{Plot-DRNMS-EXT-1}
\end{figure}
\clearpage
\null

\begin{figure}[H]
\captionsetup[subfigure]{labelformat=empty}
\centering
\subfloat[Event Horizon]{{\includegraphics[width=0.3\textwidth]{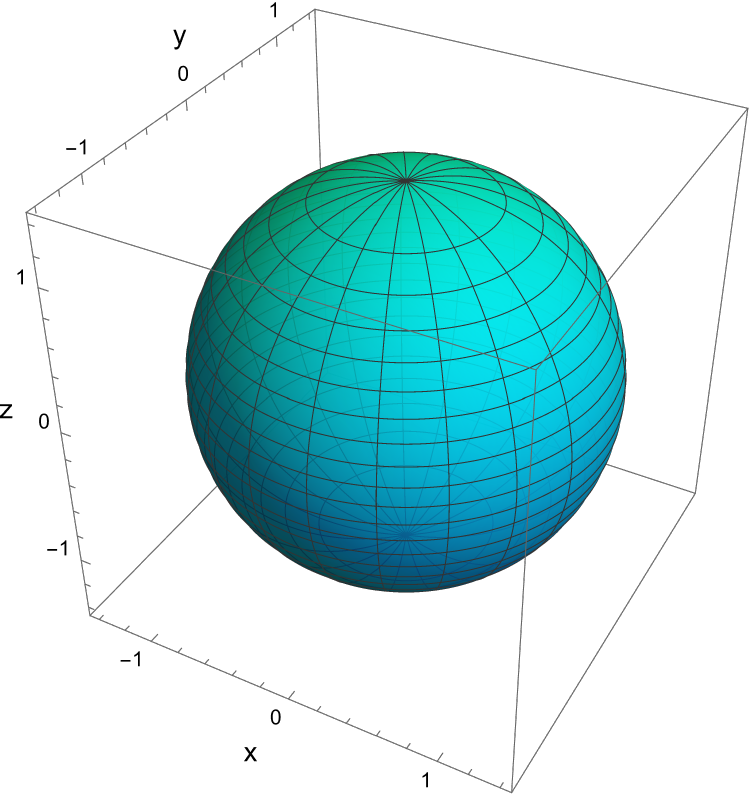}}}
\subfloat[\hspace{0.4cm} Ergoregions]{{\hspace{0.5cm}\includegraphics[width=0.3\textwidth]{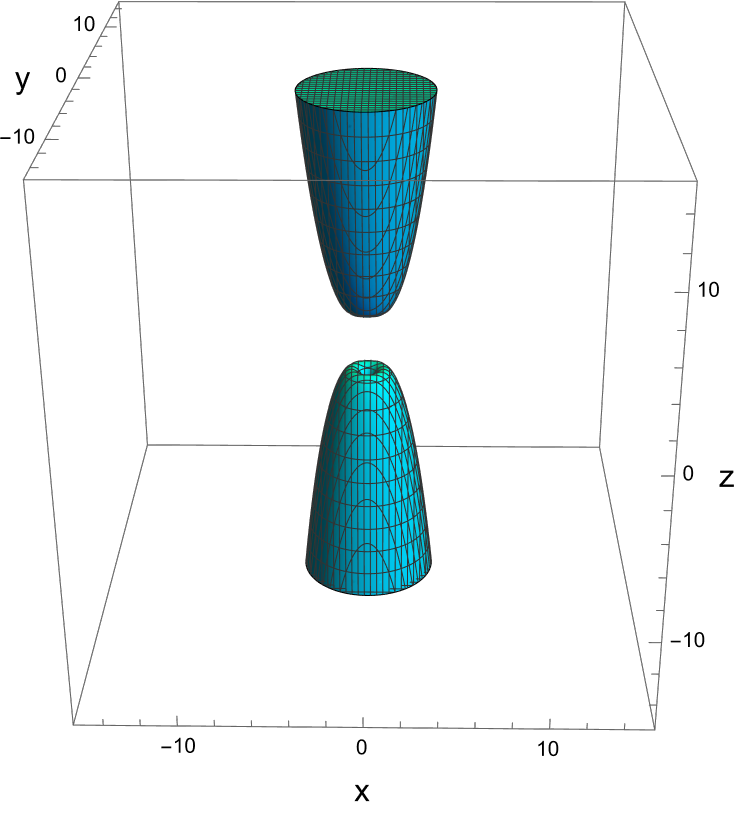}}}
\subfloat[\hspace{0.6cm} Ergoregions Cross-section $y=0$]{{\hspace{0.5cm}\includegraphics[width=0.3\textwidth]{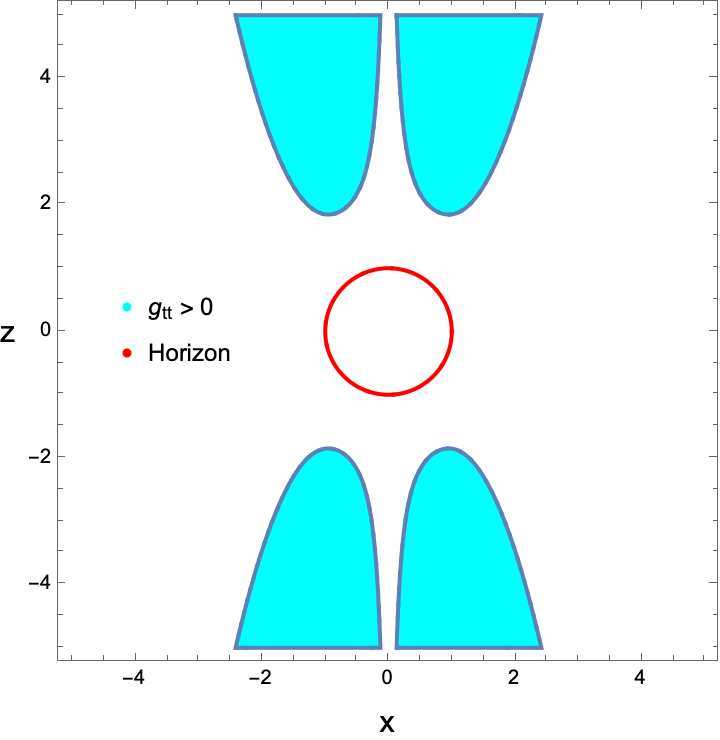}}}
\caption{\small Ergoregions and the event horizon of the \emph{extremal} dyonic Reissner-Nordstr\"om hole in a swirling universe with parameters: $B = 0, M=1, Q=\frac{4}{5}, H=\frac{3}{5}, \jmath = \frac{87}{100} \Rightarrow \pazocal{I}=1$.} \label{Plot-DRNS-EXT-2}
\end{figure}
\begin{figure}[H]
\captionsetup[subfigure]{labelformat=empty}
\centering
\subfloat[Event Horizon]{{\includegraphics[width=0.3\textwidth]{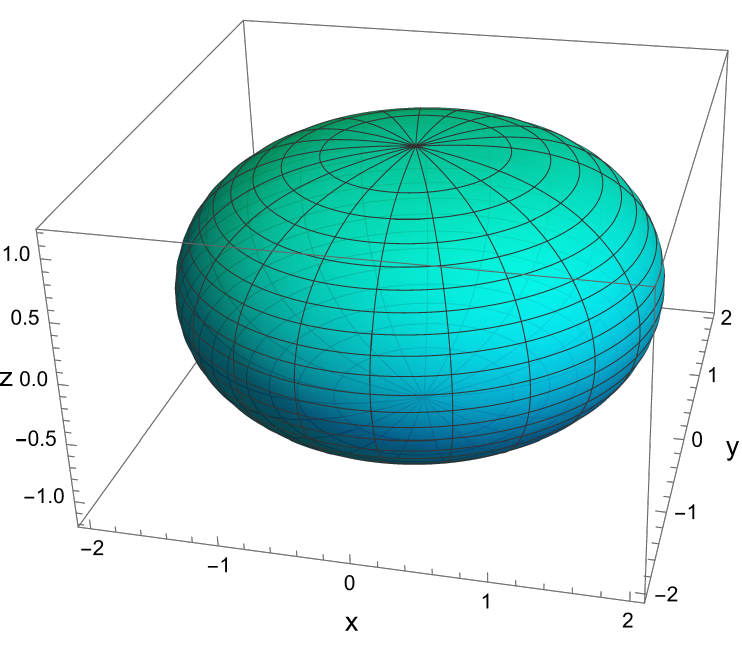}}}
\subfloat[\hspace{0.4cm} Ergoregions]{{\hspace{0.5cm}\includegraphics[width=0.3\textwidth]{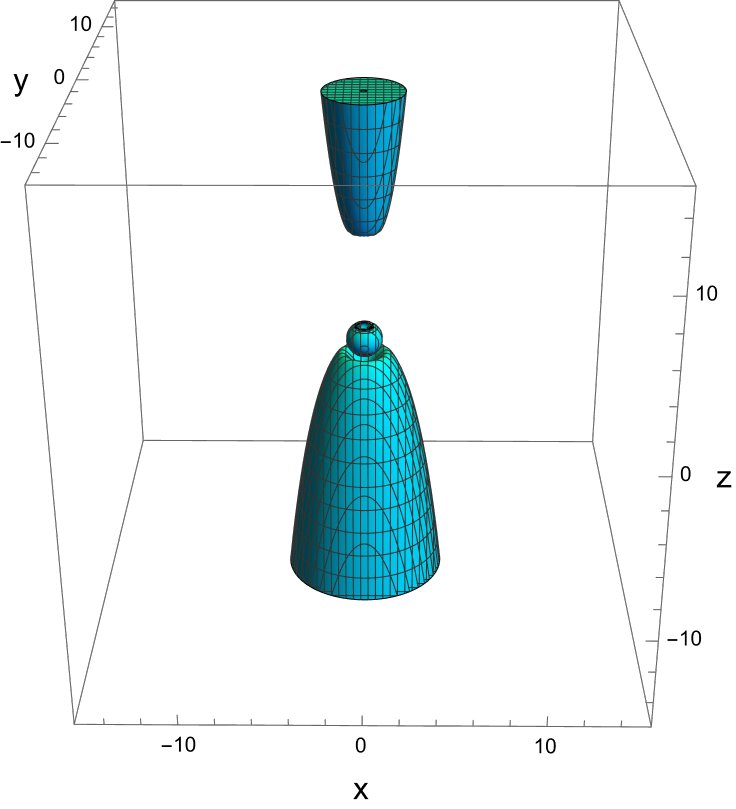}}}
\subfloat[\hspace{0.6cm} Ergoregions Cross-section $y=0$]{{\hspace{0.5cm}\includegraphics[width=0.3\textwidth]{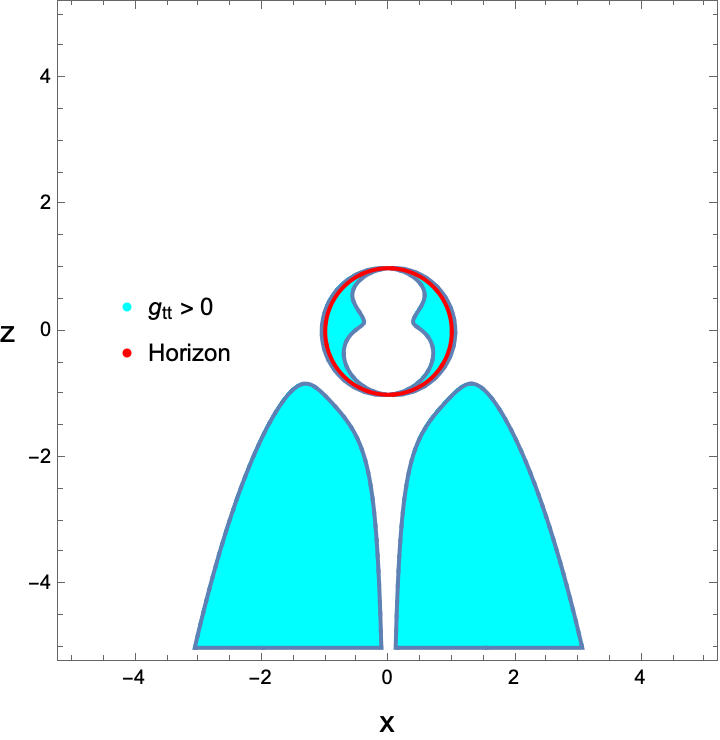}}}
\caption{\small Ergoregions and the event horizon of the \emph{extremal} dyonic Reissner-Nordstr\"om hole in a Melvin-swirling universe with parameters: $ B = \frac{4}{5}, M=1, Q=\frac{4}{5}, H=\frac{3}{5}, \jmath = \frac{87}{100} \Rightarrow \pazocal{I}=1$.} \label{Plot-DRNMS-EXT-2}
\end{figure}
%
\noindent Surprisingly, the plots in Figure~\ref{Plot-DRNS-EXT-1} and Figure~\ref{Plot-DRNS-EXT-2} seem to suggest that the extremal dyonic Reissner-Nordstr\"om black hole in a swirling universe resembles the “classic” Reissner-Nordstr\"om black hole in the proximity of their event horizon. Indeed, the horizon of this swirling black hole seems to be a perfect sphere, with the ergoregions not in the vicinity of the event horizon. In contrast to what happens in the Melvin-swirling case as we can see in Figure~\ref{Plot-DRNMS-EXT-1} and Figure~\ref{Plot-DRNMS-EXT-2}.
\clearpage
\noindent In order to better understand the results of these plots, we can analyze the geometrical proprieties of the horizon itself. In particular, we will determine whether in the extremal case the shape of the event horizon returns to that of the same black hole not embedded in the Melvin-swirling universe.

As we have already seen~\eqref{horizon-DKNMS}, the extremality condition for the black holes we are studying in this thesis is given by:
\begin{align}
M^2 & = Z^2 + a^2 \quad \quad \Rightarrow \quad \quad  r_{\pm} = M \,.
\end{align}
Therefore, we have that the area, the length of the equator, and the length of a meridian of the event horizon in the extremal case are, respectively:
\begin{subequations}
\begin{align}
\pazocal{A}^{\star} & =  \int_{0}^{2 \pi}\!d\phi \int_{0}^{\pi} \!d\theta\, \sqrt{g_{\theta \theta} g_{\phi \phi}} \,\bigg\rvert_{r=M} \,, \\
\pazocal{C}_{equator}^{\star} & =  \int_{0}^{2 \pi}\!d\phi\, \sqrt{g_{\phi \phi}} \,\bigg\rvert_{r=M,\,\, \theta=\frac{\pi}{2}} \,, \\
\pazocal{C}_{meridian}^{\star} & =  \int_{0}^{\pi} \!d\theta\,\sqrt{g_{\theta \theta}} \,\bigg\rvert_{r=M} \,,
\end{align}
\end{subequations}
which after an eventual removal of the conical singularities become:
\begin{subequations}
\begin{align}
\pazocal{A} & = \biggl[\frac{2 \pi}{\delta \phi}\biggr] \pazocal{A}^{\star}\bigg\rvert_{\text{“conical removed”}} \,,\\
\pazocal{C}_{equator} & = \biggl[\frac{2 \pi}{\delta \phi}\biggr] \pazocal{C}_{equator}^{\star} \bigg\rvert_{\text{“conical removed”}} \,,\\
\pazocal{C}_{meridian} & = \pazocal{C}_{meridian}^{\star}  \bigg\rvert_{\text{“conical removed”}} \,.
\end{align}
\end{subequations}
Using these quantities, we can consider the following ratios:
\begin{subequations}
\begin{align}
\pazocal{R}_{1} & \coloneqq \frac{\pazocal{A}}{2\,\pazocal{C}_{equator}} \,, \label{r1} \\
\pazocal{R}_{2} &  \coloneqq \frac{2\, \pazocal{C}_{meridian}}{\pazocal{C}_{equator}}  \label{r2}\,.
\end{align}
\end{subequations}
In particular, $\pazocal{R}_{2}$~\eqref{r2} indicates the degree of “sphericity” of the event horizon. Indeed, it is straightforward to verify that $\pazocal{R}_{2} = 1$ for a perfect sphere, it increases as the horizon becomes more prolate, and it decreases as it becomes more oblate. Similarly, $\pazocal{R}_{1}$~\eqref{r1} is exactly the physical radius when the horizon is a perfect sphere, and in general we have that the volume of the black hole is proportional to $\pazocal{R}_{1}$. Moreover, $\pazocal{R}_{1}$ is also independent from the rescaling of the azimuthal coordinate needed to remove the conical singularities.
\vfill
\clearpage
\myparagraph{\small Extremal dyonic Kerr-Newman in a Melvin swirling universe ($M^2=a^2+Z^2$):}
For the new general solution~\eqref{DKNMS}, in the extremal case, we have:
\vspace{-0.1cm}
\begin{subequations}
\begin{align}
\pazocal{A}^{\star (ext)} & = 4 \pi \bigl( M^2 + a^{2} \bigr) \,, \\
\pazocal{C}_{equator}^{\star (ext)}  & = \frac{ 2 \pi \bigl( M^2 + a^2 \bigr) }{\sqrt{ |\pazocal{C}^{\star}_{den} |}} \,, \\
\pazocal{R}_{1} & = \sqrt{| \pazocal{C}^{\star}_{den} |} \bigg\rvert_{\text{“conical removed”}} \,, \\
\pazocal{R}_{2} & \neq 1 \,,
\end{align}
\end{subequations}
where
\vspace{-0.15cm}
\begin{align}
\begin{split}
\pazocal{C}^{\star}_{den} & =  \frac{M}{16} \bigl( M^2 + 3 a^2\bigr)\biggl[ 8B^2 M + 8 B^3 a Q + \Bigl(B^4 + 16\jmath^2\Bigr) \bigl( M^2 + 3 a^2\bigr) M + 32 \jmath B a H \biggl] \\
&\quad + M^2 + 2 B a M Q + B^2 a^2 Z^2 \,.
\end{split}
\end{align}
Therefore, the volume of the black hole is also proportional to the swirling and Melvin parameters, and not only from the physical quantities characterizing the black hole.
Moreover, from a numerical evaluation, we have that $\pazocal{R}_{2}$ results in a different value if we set the swirling and Melvin parameters to zero. Hence, despite seeming visually similar, the shape of the event horizon in this extremal case is not exactly equal to that of the corresponding extremal dyonic Kerr-Newman black hole~\eqref{kn-magnetic}.
\myparagraph{\small Extremal dyonic Reissner-Nordstr\"{o}m in a swirling universe ($B=a=0,\,M^2=Z^2$):}
For the extremal dyonic Reissner-Nordstr\"{o}m in a swirling universe, we obtain:
\vspace{-0.1cm}
\begin{subequations}
\begin{align}
\pazocal{A} & = 4 \pi Z^2 \bigl(1 + \jmath^2 Z^4\bigr) \,,\\
\pazocal{C}_{equator} & = 2\,\pazocal{C}_{meridian} = 2 \pi |Z| \sqrt {1 + \jmath^2 Z^4 }\,,\\
\pazocal{R}_{1} & = |Z| \sqrt {1 + \jmath^2 Z^4} \,, \\
\pazocal{R}_{2} & = \frac{2\,\pazocal{C}_{meridian}}{\pazocal{C}_{equator}} = 1 \label{r1-DRNS}\,.
\end{align}
\end{subequations}
Hence, for this sub-case, the horizon is a perfect sphere ($\pazocal{R}_{2}=1$) as it is for the non-swirling black hole. On the other hand, the physical radius $\pazocal{R}_{1}$ receives an additional contribution due to the swirling background. Therefore, the effect of the swirling parameter in the extremal case is merely to change the volume, leaving the horizon perfectly spherical. Furthermore, if we also remove the Dirac strings~\eqref{reg-DRNS}, we obtain
\vspace{-0.15cm}
\beq
\pazocal{R}_{1} = \frac{Z^2}{|Q|} \label{R1-DRNS-reg}\,.
\eeq
%
\vfill
\clearpage
\myparagraph{\small Extremal dyonic Reissner-Nordstr\"{o}m in a Melvin-swirling universe ($a=0,\,M^2=Z^2$):}
However, if the Melvin parameter $B$ is not zero, we have:
\begin{subequations}
\begin{align}
\pazocal{A} & = \frac{4 \pi Z^2 \Bigl[| - 4 H^2 + 4 \jmath H Q Z^2| + \jmath^2 Z^6 \Bigr]}{H^2} \,,\\
\pazocal{C}_{equator} & = \frac{2 \pi \Bigl[| - 4 H^2 + 4\jmath H Q Z^2| + \jmath^2 Z^6 \Bigr]}{|\jmath H| Z^2} \,,\\
\pazocal{C}_{meridian} &= \Bigg[\frac{2 Z \sqrt{| - 4 H^2 + 4 \jmath H Q Z^2| + \jmath^2 Z^6}}{|H|}\Biggr] \text{EllipticE}\bigg[\frac{ |-4 H^2 + 4 \jmath H Q Z^2|}{ |- 4 H^2 + 4 \jmath H Q Z^2| + \jmath^2 Z^6 }\bigg] \,, \\
\pazocal{R}_{1} & = \bigg|\frac{\jmath Z^4}{H}\bigg| \,, \label{R1-DRNMS}\\
\pazocal{R}_{2} & \neq 1 \,,
\end{align}
\end{subequations}
where 
\beq
\text{EllipticE}[x] = \int_{0}^{\frac{\pi}{2}}\sqrt{1 - x \sin^2 \theta}\, d\theta \,,
\eeq
is the complete elliptic integral of the second kind.

Thus, in contrast to the previous case, we now find that the horizon is \emph{not} a perfect sphere. Nevertheless, it is interesting to notice that if we use the same value of the swirling parameter $\jmath$ that removes the Dirac strings in the non-Melvin case~\eqref{reg-DRNS}, we obtain that the ratio $\pazocal{R}_{1}$~\eqref{R1-DRNMS} for this black hole becomes exactly equal to that of the non-Melvin case as in Eq.~\eqref{R1-DRNS-reg}. However, it should be noted that this choice of swirling parameter does \emph{not} remove the Dirac strings for this charged Melvin-swirling black hole, as we proved in Eq~\eqref{not-rem-dirac-DRNMS}.
%
\clearpage
%
\subsection{Near-Horizon Extremal Geometry} \label{subsec:geom-dknms}
In Sec.~\ref{subsec:hor-ergo-dknms} we noticed that the shape of the event horizon of the dyonic Reissner-Nordstr\"{o}m black hole in a swirling universe~\eqref{DRNMS} returns to being a perfect sphere in the extremal case~\eqref{r1-DRNS}. Moreover, it is well known~\cite{AstorinoMagnetised, SiahaanMelvTN, Compere} that the geometry near the event horizon in the extremal case has some enhanced symmetry properties. 

For this reason, we will now study the geometry of the spacetime near the event horizon of the extremal black hole, known as the {\bfseries near-horizon extremal (NHE) geometry} and defined as the limit $\varepsilon \rightarrow 0$ of the metric after having performed a coordinate transformation in the following new dimensionless coordinates:
\begin{subequations}
\begin{align}
r & \mapsto M + \varepsilon \, r_{0} \, \hat{r} \,, \\
t & \mapsto  \frac{ r_{0} }{\varepsilon} \hat{t} \,, \\
\phi & \mapsto \hat{\phi} + \frac{ \Omega_{H}^{(ext)} }{ r_{0}^4 } \hat{t} \,,
\end{align}
\end{subequations}
in conjunction with the following gauge transformation of the electromagnetic potential $A_{\mu}$:
\beq
A_{t} \mapsto A_{t} - \frac{ A_{0 H}^{(ext)} }{ r_{0}^4 } \,,\\
\eeq
\vspace{-0.1cm}
where
\begin{subequations}
\begin{align}
M^2 & = Z^2 + a^2 \,, \\
r_{0}^2 & = M^2 + a^2 \,, \\
\begin{split}
A_{0 H}^{(ext)} & = - r_{0}^2 \biggl[ Q M + \frac{3 B}{2} a Z^2 - \frac{3 B^2}{4} M Q Z^2 \\
&\quad - \frac{B^3}{4}\Bigl( 3 Z^4 + 7 a^2 Z^2 + 4 a^4 \Bigr)a - \jmath H M Z^2 \biggr] \,,
\end{split}\\
\begin{split}
\Omega_{H}^{(ext)} & = r_{0}^2 \Biggl[ a - 2 B Q M + 2 \jmath B M H Z^2 - \frac{3 B^2}{2} a Z^2   \\
&\quad + \frac{B^3}{2} M Q Z^2 + \biggr(2 \jmath^2 + \frac{B^4}{8} \biggr) \bigl( 3 Z^4 + 7 a^2 Z^2 + 4 a^4 \bigr)a \Biggr] \,.
\end{split}
\end{align}
\end{subequations}
Using these definitions, we will now analyze the near-horizon extremal geometry of the new solutions we obtained in this thesis. In particular, we will analyze the effect on the geometry of embedding a black hole in the Melvin-swirling universe, checking if these black holes are isometric to their respective counterparts without the Melvin and swirling parameters.
\vfill
\clearpage
\myparagraph{NHE dyonic Kerr-Newman in a Melvin-swirling universe:}
The near-horizon geometry of the most general case we found~\eqref{DKNMS} is then given by
\begin{subequations}
\label{NHEDKNMS}
\begin{align}
d\hat{s}^2 & = \Gamma \Bigl[ -\hat{r}^2 \,d\hat{t}^2 + \frac{d\hat{r}^2}{\hat{r}^2} + d\theta^2 \Bigr] + \frac{r_{0}^{4}\sin^2\theta}{\Gamma} \biggl[ d\hat{\phi} - \frac{\kappa} { r_{0}^{2} } \, \hat{r}\,d\hat{t} \biggr]^2 \,, \\
\hat{A} & = \frac{ \hat{A}_0 }{ r_{0}^2 } \,  \hat{r} \, d\hat{t} + \frac{ \hat{A}_{3} }{ \Gamma }\biggl[ d\hat{\phi} - \frac{\kappa}{r_{0}^2}  \, \hat{r} \,d\hat{t} \biggr] \,,
\end{align}
\end{subequations}
with
\begin{subequations}
\begin{align}
\hat{A}_0 &= \hat{\chi}_{(0)} + \frac{B}{2}\,\hat{\chi}_{(1)} + \frac{3B^2}{4}\,\hat{\chi}_{(2)} + \frac{B^3}{8}\,\hat{\chi}_{(3)} + \jmath\,\hat{\chi}_{(4)} \,, \\
\hat{A}_3 &= \hat{\phi}_{(0)} +  \frac{B}{2}\,\hat{\phi}_{(1)} + \frac{3B^2}{4}\,\hat{\phi}_{(2)} + \frac{B^3}{8}\,\hat{\phi}_{(3)} + \jmath\,\hat{\phi}_{(4)} \,,\\
\kappa   &  = \kappa_{(0)}    + 2B\,\hat{\chi}_{(0)} +  \frac{B^2}{2}\,\hat{\chi}_{(1)} + \frac{B^3}{2}\,\hat{\chi}_{(2)}  + \biggr[ \jmath^2 + \frac{B^4}{16} \biggr]\hat{\chi}_{(3)} + 2 \jmath B \,\hat{\chi}_{(4)} \,,\\
\Gamma & = \Gamma_{(0)} + 2B\,\hat{\phi}_{(0)} +  \frac{B^2}{2}\,\hat{\phi}_{(1)} + \frac{B^3}{2}\,\hat{\phi}_{(2)} + \biggr[ \jmath^2 + \frac{B^4}{16} \biggr]\hat{\phi}_{(3)} + 2 \jmath B \, \hat{\phi}_{(4)} + \jmath\, \Gamma_{(1)} \,,
\end{align}
\end{subequations}
where the auxiliary functions have been defined as
\begin{subequations}
\begin{align}
\hat{\chi}_{(0)} = \,& Q Z^2 \,, \\
\hat{\chi}_{(1)} = \,& 6 a M Z^2 \,, \\
\hat{\chi}_{(2)} = \, & Q \bigl[8 a^2 M^2 + Z^4\bigr] \,, \\
\hat{\chi}_{(3)} = \,& 2 a M \bigl( M^2 + 3 a^2 \bigr) \bigl( 3 M^2 + a^2 \bigr) \,, \\
\hat{\chi}_{(4)} = \,&  H \bigl[8 a^2 M^2 + Z^4\bigr] \,, \\
\hat{\phi}_{(0)} = \,& a Q M \sin^2\theta - H \bigl( M^2 + a^2 \bigr) \cos\theta \,,\\
\hat{\phi}_{(1)} = \,& 4 a^4 \sin^2\theta + a^2 Z^2 \bigl(7 -  \cos^2\theta \bigr) +Z^4 \bigl(1 +  2\cos^2\theta \bigr) \,,\\
\hat{\phi}_{(2)} = \,& a Q M \Bigl[ M^2 + 3 a^2 + \bigl( 3 M^2 + a^2 \bigr)\cos^2\theta \Bigr] - H Z^2 \bigl( M^2 + a^2 \bigr) \cos\theta \,, \\
\hat{\phi}_{(3)} = \, & Z^6 + a^2 \bigl( 3 M^2 + a^2 \bigr)^2 \bigl(1 + \cos^2\theta \bigr) \,,\\
\hat{\phi}_{(4)} = \,& a H M \Bigl[ M^2 + 3 a^2 + \bigl( 3 M^2 + a^2 \bigr)\cos^2\theta \Bigr] + Q Z^2 \bigl( M^2 + a^2 \bigr) \cos\theta\,, \\
\kappa_{(0)} = \, & - 2 a M \,, \\
\Gamma_{(0)} = \, & M^2 + a^2 \cos^2\theta \,, \\
\Gamma_{(1)} = \, & - 8 a M \bigl( M^2 + a^2 \bigr) \cos\theta \,.
\end{align}
\end{subequations}
Remarkably, this near-horizon geometry~\eqref{NHEDKNMS} is in the same form as the near-horizon geometry of the standard Kerr-Newman black hole~\eqref{kn-magnetic}. Therefore, it posses a $SL(2,\R) \times U(1)$ isometry, which is a warped and twisted product of $AdS_{2} \times S^{2}$.
\newpage
\myparagraph{NHE dyonic Reissner-Nordstr\"{o}m in a swirling universe ($B=a=0$):}
For the dyonic Reissner-Nordstr\"{o}m in a swirling universe~\eqref{DRNS} we have $\Omega_{H}^{(ext)} = A_{0 H}^{(ext)} = 0$, and the near-horizon extremal geometry is given by 
\vspace{-0.15cm}
\begin{subequations}
\begin{align}
\label{NHE-DRNS}
d\hat{s}{}^2 & = \Gamma \Bigl[ -\hat{r}^2\,d\hat{t}^2 + \frac{d\hat{r}^2}{\hat{r}^2} + d\theta^2 + \frac{Z^4 \sin^2\theta}{\Gamma^2}\,d\hat{\phi}^2 \Bigr] \,,\\
\hat{A} & = \Bigl[\jmath H Z^2 + Q \Bigr]  \hat{r} \,d\hat{t} + \frac{Z^2}{\Gamma} \Bigl[\jmath Q Z^2 - H \Bigr]  \cos\theta \, d\hat{\phi} \,,\\ %
\Gamma & = Z^2 \bigl( 1 + \jmath^2 Z^4 \bigr) \,.
\end{align}
\end{subequations}
Moreover, if we remove the conical singularities with the conditions given by Eq.~\eqref{conical-DRNS}, we obtain that the geometry becomes \emph{exactly} $AdS_{2} \times S^{2}$:
\vspace{-0.15cm}
\begin{subequations}
\label{NHEDRN}
\begin{align}
d\hat{s}{}^2 & = \Bigl[e^2 + p^2 \Bigr]\Bigl[ -\hat{r}^2\,d\hat{t}^2 + \frac{d\hat{r}^2}{\hat{r}^2} + d\theta^2 + \sin^2\theta d\hat{\phi}^2 \Bigr] \,,\\
\hat{A} & = e \, \hat{r} \, d\hat{t} + p \, \cos\theta \, d\hat{\phi} \,,\\
e & = \jmath H Z^2 + Q \,,\\
p & = \jmath Q Z^2 - H \,.
\end{align}
\end{subequations}
Therefore, the NHE geometry of the dyonic Reissner-Nordstr\"{o}m in a swirling universe is \emph{exactly} the same as that of the dyonic Reissner-Nordstr\"{o}m without the swirling parameter.

Furthermore, this result also reveals another significant implication of embedding a charged black hole in the swirling universe. Indeed, from the electromagnetic potential of this spacetime~\eqref{A-DRNS}, not necessarily in the extremal case, we already knew that a magnetic Ehlers transformation has the effect of mixing the charges, in the sense that if the seed solution is only electrically charged, $H=0$, we have that the resulting solution embedded in a swirling universe will also possess a magnetic charge proportional to the seed electric charge, and vice-versa if the seed solution is only magnetically charged, $Q=0$. 

However, from the NHE limit of this charged swirling black hole~\eqref{NHEDRN}, we can also conclude that, if both the magnetic and electric charge are present, the swirling parameter $\jmath$ also provides a way to remove the effect of one of the charges at the horizon, without actually setting the corresponding black hole charge to zero. 

Indeed, the condition to remove the Dirac strings~\eqref{reg-DRNS}, $ \jmath = \frac{H}{Q Z^2}$, exactly sets $p = 0$ (i.e.~$\hat{A}_{\hat{\phi}} = 0$) and $e  =  \frac{Z^2}{Q}$. Thus, the removal of the Dirac string is equivalent to constraining the swirling parameter $\jmath$ in such a way that is null the effect of the magnetic charge $p$ in the extremal near-horizon limit. Moreover, this value of the charge $e$ is also the same as the physical radius $\pazocal{R}_{1}$ we found for this black hole~\eqref{R1-DRNS-reg}. 

Analogously, if we set $\jmath = - \frac{Q}{H Z^2}$ we obtain $e = 0$ (i.e.~$\hat{A}_{\hat{t}} = 0$) and $p =  - \frac{Z^2}{H}$, thus removing the effect of the electric charge at the horizon.
\vfill
\clearpage

\myparagraph{NHE magnetic Kerr-Newman in Melvin-swirling universe ($Q=0$):}
For the magnetic Kerr-Newman in Melvin-swirling universe sub-case, $Q=0$, after having removed the conical singularities with the constraints given by Eqs.~\eqref{conical-MKNMS}, it is possible to obtain $\frac{r_{0}^{4}}{\Gamma^2}\Bigr[\frac{4 \pi^2}{\delta \hat{\phi}^2}\Bigr] = 1$ for 
\beq 
B = \pm \frac{2}{\sqrt{4 a^2 + 3 H^2}} \,.
\eeq
Moreover, this constraint of the parameters also sets $\kappa = 0$, which implies that also for this black hole the NHE geometry becomes exactly $AdS_{2} \times S^{2}$.

Additionally, it is also possible to set $\hat{A}_{0} = 0$, and therefore $\hat{A}_{\hat{t}}=0$ since $\kappa = 0$, if we also constrain the magnetic charge as
\beq
H = \pm a \sqrt{3 + \sqrt{17}} \,.
\eeq
\myparagraph{NHE dyonic Kerr-Newman in a Melvin-swirling universe:}
Similarly, for the general rotating and dyonic case~\eqref{NHEDKNMS}, it is also possible to set $\Bigr[\frac{4 \pi^2}{\delta \hat{\phi}^2}\Bigr]  = 1$. However, the possible values of $B$ for which this happens are \emph{not} compatible with the removal of the Dirac Strings~\eqref{reg-DKNMS}. Indeed, for example, a possibility is given by: 
\beq
B = \pm \frac{2 a}{\sqrt{4 a^4 + a^2 \bigl ( 3 H^2 + 5 Q^2 \bigr) + 2 Q \biggl [ Q Z^2 + M \sqrt{ Q^2 Z^2 + a^2 \bigl( 3 H^2 + 4 Q^2 + 4 a^2 \bigr)}\, \biggr] }} \,,
\eeq
which is clearly not compatible with constraints given by Eqs.~\eqref{reg-DKNMS}.
\myparagraph{NHE of the others Melvin-swirling sub-cases without conical singularities:}
In all the other sub-cases where the Melvin parameter $B$ and the swirling parameter $\jmath$ are both present, and the conical singularities are removable, we have that the condition
\beq
\frac{r_{0}^{4}}{\Gamma^2}\biggl[\frac{2 \pi}{\delta \hat{\phi}}\biggr]^2 = 1 \,,
\eeq
can be satisfied only for $\theta = 0$ or  $\theta = \pi$, which means that the geometry can never be exactly $AdS_{2} \times S^{2}$. Moreover, since $\sin\theta = 0$ for these values of $\theta$, we have that there is not a single slice of the spacetime for which the geometry is a “warped” $AdS_{3}$.\\

\noindent Therefore, we can conclude that the near-horizon geometry of the extremal dyonic Reissner-Nordstr\"{o}m in a Melvin-swirling universe is \emph{not}} isometric to that of the “classic” dyonic Reissner-Nordstr\"{o}m, in contrast to what happens for the swirling $B=0$ sub-case~\eqref{NHEDRN}.
%
\clearpage
\section{Supersymmetric Extension} 
\label{sec:Sugra}
\subsection{$\pazocal{N}=2$, $d=4$ Gauged Supergravity}

The theory of {\bfseries Supergravity} is the result of supersymmetrizing General Relativity, or equivalently, turning Supersymmetry into a local gauge symmetry, where gravity is described by the {\bfseries graviton} $e_\mu^a$, with the superpartner called the {\bfseries gravitino} $\psi_{\mu}$.

In its $\pazocal{N}=2$, $d=4$ gauged version~\cite{FreedmanN2, FradkinN2} there are four bosonic and four fermionic degrees of freedom, corresponding to a graviton $e_\mu^a$, a Maxwell gauge field $A_\mu$, and two real gravitini $\psi_\mu^i$ $(i=1,2)$, that can be combined into a single complex spinor $\psi_\mu = \psi_\mu^1 + i\psi_\mu^2$.

This theory is obtained starting from the rigid SO${(2)}$ symmetry rotating the two independent Majorana supersymmetries present in the ungauged theory, which is made local by the introduction of a minimal gauge coupling between the photons and the gravitini, with strength $\ell^{-1}$. Moreover, local Supersymmetry requires a gravitini mass term and fixes the (negative) cosmological constant to $\Lambda = -\frac{3}{\ell^{2}}$, which results in the following Lagrangian\footnote{The gamma matrices that we are using are in the (real) Majorana representation, which is explained in Appendix~\ref{E-Gamma}.}:
\begin{align}
\label{lagrange-2}
\begin{split}
e^{-1}L =& -\frac{1}{4}R  - \frac{3}{2\ell^2}  + \frac{1}{4}F_{\mu \nu}F^{\mu \nu} + \frac{1}{2}\bar{\psi}_\mu\Bigl(\gamma^{\mu \nu \rho}\hat{\pazocal{D}}_\nu - \frac{1}{\ell}\gamma^{\mu \rho}\Bigr)\psi_\rho \\
&  + \frac{i}{8}\bigl(F^{\mu \nu} + \hat{F}^{\mu \nu}\bigr) \bar{\psi}_\rho\gamma_{[\mu }\gamma^{\rho \sigma}\gamma_{\nu ]}  \psi_\sigma \,,
\end{split} 
\end{align}
where $\hat{\pazocal{D}}_\mu$ is the {\bfseries gauge- and Lorentz-covariant derivative} acting on spinorial objects
\beq
\hat{\pazocal{D}}_\mu = \pazocal{D}_\mu - \frac{i}{\ell}A_\mu \,, \label{gaugecovder}
\eeq
with $\pazocal{D}_\mu$ the {\bfseries Lorentz-covariant derivative}~\eqref{lorentz-cov-der-spin}
\beq
\pazocal{D}_\mu = \partial_\mu + \frac{1}{4}\omega_\mu{}^{ab}\gamma_{ab} \,,
\eeq
while $\hat{F}_{\mu \nu}$ denotes the {\bfseries supercovariant field strenght}, which is given by
\beq
\hat{F}_{\mu \nu} = F_{\mu \nu} - \Im(\bar{\psi}_\mu\psi_\nu) \,.\label{supcovfs}
\eeq
The theory is given in a first-order formalism for the gravity sector, which means that the scalar curvature $R$ is a function of both the tetrad $e_{\mu}^{a}$ and the spin connection $\omega_{\mu}{}^{a}{}_{b}$. Hence, the spin connection is fixed by its own algebraic equation of
motion following from the Lagrangian~\eqref{lagrange-2}, which are
\beq
\omega_{\mu ab} = \Omega_{\mu ab} - \Omega_{\mu ba} - \Omega_{ab\mu} \,,
\eeq
where
\beq
\Omega_{\mu \nu}{}^{a} \coloneqq \partial^{}_{[\mu} e^{a}_{\nu]} - \frac{1}{2}{\Re}(\bar{\psi}_\mu\gamma^a\psi_\nu) \,.
\eeq
The action corresponding to the Lagrangian~\eqref{lagrange-2} is invariant under the following $\pazocal{N}=2$ local supertransformations
\begin{subequations}
 \label{transfsusy}
\begin{align}
\delta e_\mu^{a} &= \Re(\bar{\epsilon}\gamma^a\psi_\mu)\,,  \label{transfsusy-1} \\
\delta A_\mu &= \Im(\bar{\epsilon}\psi_\mu)\,,  \label{transfsusy-2}\\
\delta \psi_\mu &= \hat{\nabla}_\mu \epsilon \,,  \label{transfsusy-3}
\end{align}
\end{subequations}
where $\epsilon$ is an infinitesimal Dirac spinor, and $\hat{\nabla}_\mu$ is the {\bfseries supercovariant derivative} given by
\beq
\hat{\nabla}_\mu = \hat{\pazocal{D}}_\mu + \frac{1}{2\ell}\gamma_\mu + \frac{i}{4}\hat{F}_{ab}\gamma^{ab} \gamma_\mu \,. \label{supcovder}
\eeq
Furthermore, the supersymmetry algebra of gauged $\pazocal{N}=2$, $d=4$ Supergravity is osp$(4|2)$. This algebra has the ten bosonic generators $M_{ab}, M_{a4}$ $(a=0,1,2,3)$ of the AdS subalgebra so$(3,2)$, two fermionic generators $Q_{\alpha}^i$ $(i=1,2)$, plus one additional bosonic generator of SO${(2)}$ transformations, rotating the two supersymmetries into each other.

The basic anticommutator is then given by:
\beq
\Bigl\{Q^i_{\alpha},Q^j_{\beta}\Bigr\} = \delta^{ij}\Bigl(\bigl(\gamma^aM_{a4} + i\gamma^{ab}M_{ab}\bigr) C\Bigr)_{\alpha\beta} + i \Bigl (C_{\alpha\beta}Q_e + i \bigl(C\gamma^5\bigr)_{\alpha\beta}Q_m \Bigl)\epsilon^{ij} \,,
\eeq
wherein $Q_e$ and $Q_m$ are the central charges, $C$ denotes the charge conjugation matrix, and $\epsilon^{ij}$ is the permutation symbol in two dimensions.\\
\subsubsection{Killing Spinors}
In the bosonic sector of this theory, $\psi_\mu = 0$, the field equations following from Eq.~\eqref{lagrange-2} coincide with the Einstein-Maxwell equations with a negative cosmological constant. Hence, spacetimes that satisfy these field equations represent possible background solutions of gauged $\pazocal{N}=2$ Supergravity.

\noindent Furthermore, being $\psi_\mu = 0$, the invariance of these background solutions under the supertransformations~\eqref{transfsusy} results in the equation
\beq
\hat{\nabla}_\mu \epsilon = 0 \,. \label{killing-spinor}
\eeq
A spinor $\epsilon$ that satisfies the condition given by Eq.~\eqref{killing-spinor} is called a {\bfseries Killing spinor}.

The basic integrability conditions for Eq.~\eqref{killing-spinor} reads
\beq
\hat{R}_{\mu \nu}\,\epsilon = 0 \,, \label{integr-cond}
\eeq
where
\beq
\hat{R}_{\mu \nu} = \bigl[\hat{\nabla}_\mu , \hat{\nabla}_\nu\bigr] \label{supercurv} \,,
\eeq
is referred to as the {\bfseries supercurvature}.

Moreover, the integrability conditions for Killing spinors~\eqref{integr-cond} are also equivalent to
\beq
\det\bigl(\hat{R}_{\mu \nu}\bigr) = 0 \,. \label{integr-cond-2}
\eeq
It is important to note that the basic integrability conditions~\eqref{integr-cond} are necessary, but not sufficient for the existence of Killing spinors. They assure that a Killing spinor exists locally, but there may be topological reasons that prevent their global existence. For this reason, one must either consider a sufficient set of higher integrability conditions~\cite{KillingCon} or return to the original differential equation~\eqref{killing-spinor}.

\subsection{Supersymmetric Extension of the Melvin-swirling Universe}
In the following, we will investigate whether the solutions discussed in this thesis can admit a supersymmetric extension, by examining the possible existence of Killing spinors in these spacetimes, using the conditions given by Eq.~\eqref{killing-spinor} and Eq.~\eqref{integr-cond}.
\newline

\myparagraph{Electromagnetic Melvin-swirling universe:}
We begin with the electromagnetic Melvin-swirling background~\eqref{em-melvin-swirling-cosmological}, where we write here for convenience the form of the metric for its generalization with a cosmological constant:
\beqNN
ds^2  = F \biggl[-dt^2+\frac{\rho^2 d\rho^2}{\pazocal{S}}+dz^2\biggr]+\frac{{\pazocal{S}}}{F}\bigl(d\phi - \Omega \,dt\bigr)^2 \,. \\
\eeqNN
\vfill
\clearpage
\noindent Following Sec.~\ref{subsubsec:Tetrad}, a possible tetrad for this metric is given by
\begin{subequations}
\begin{align}
e^{0} & = \sqrt{F} dt \,, \\
e^{1} & = \sqrt{\frac{F \rho^2}{\pazocal{S}}} d\rho \,, \\
e^{2} & = \sqrt{F} dz \,, \\
e^{3} & = \sqrt{\frac{{\pazocal{S}}}{F}} \bigl(d\phi - \Omega \,dt\bigr) \,.
\end{align}
\end{subequations}
In the case of vanishing gauge coupling $\ell^{-1}$, i.e.~when the cosmological constant is set to zero, $\Lambda = -\frac{3}{\ell^2} = 0$, we are dealing with the ungauged $\pazocal{N} = 2$ theory. Thus, the supercovariant derivative~\eqref{supcovder} in the bosonic sector of this theory, $\psi_{\mu}=0$, simplifies to:
\beq
\hat{\nabla}_{\mu} = \partial_\mu + \Gamma_{\mu} \,,
\eeq
where the “connection coefficients” $\Gamma_{\mu}$ are given by
\beq
\Gamma_{\mu} = \frac{1}{4}\omega_\mu{}^{ab}\gamma_{ab} + \frac{i}{4}F_{ab}\gamma^{ab} \gamma_\mu \,.
\eeq
Hence, in order to consider the electromagnetic Melvin-swirling universe as the background of this theory, the first step consists of verifying the integrability conditions~\eqref{integr-cond-2} for the supercurvature
\beq
\hat{R}_{\mu \nu} = \bigl[\hat{\nabla}_\mu ,\hat{\nabla}_\nu\bigr] = \partial_{\mu} \Gamma_{\nu} - \partial_{\nu} \Gamma_{\mu} +\bigl[\Gamma_{\mu}, \Gamma_{\nu}\bigl] \,.
\eeq
Since these conditions must be verified for all the components of the supercurvature, we can start by setting $\mu = \rho$ and $\nu = z$, thus obtaining
\beq
\label{cond-1}
\det\bigl(\hat{R}_{\rho z}\bigr) = \frac{\bigl(X^4+16\jmath^2\bigr)^2}{\Bigl[16 + 8 X^2 \rho^2 + \bigl(X^4+16\jmath^2 \bigr)\rho^4 \Bigr]^2} \,,
\eeq
where we recall that $X^2 = B^2 + E^2$.

Unfortunately, we can already see that, to satisfy the necessary condition $\det\bigl(\hat{R}_{\rho z}\bigr) = 0$, all the parameters $\jmath$, $B$, and $E$ must be simultaneously zero. However, this choice of the parameters corresponds to the Minkowski spacetime. Moreover, this also precludes a possible supersymmetric extension of the swirling background alone, or of the other sub-cases.
\vfill
\clearpage
\myparagraph{Electromagnetic Melvin-swirling universe with a cosmological constant:}
If the cosmological constant $\Lambda = -\frac{3}{\ell^2}$ is present we return to the gauged theory. Therefore, using the same procedure adopted for the previous ungauged case, but with the correct supercovariant derivative, we can write the determinant of the supercurvature for $\mu = \rho$ and $\nu = z$ as the following expansion
\vspace{-0.3cm}
\beq
\label{cond-2}
\det\bigl(\hat{R}_{\rho z}\bigr) = \frac{\sum_{n = 0} C_{n}\, \rho^{n}}{Den} \,,
\eeq
where $Den$ is the common denominator and $C_{n}$ are the coefficients of the expansion\footnote{To be more precise, the values of the coefficients of the expansion $C_{n}$ depend on the specific common denominator that is used.}.

The usefulness of this expansion is that, in order for the integrability condition~\eqref{integr-cond-2} to be satisfied, this determinant~\eqref{cond-2} must be zero for any value of the coordinates. This implies that all the coefficients of the expansion $C_{n}$ must be simultaneously zero. 

However, since we request the parameters to be real, it is impossible to find a specific value of these parameters for which all the coefficients $C_{n}$ are simultaneously zero.
\vspace{-0.1cm}
\myparagraph{Swirling universe with a cosmological constant:}
In particular, in the sub-case $E=B=0$, corresponding to the swirling universe with a cosmological constant~\eqref{swirling-background-cylindrical}, if we set $\mu = \rho$ and $\nu = \phi$ the basic integrability condition~\eqref{integr-cond-2} becomes
\vspace{-0.2cm}
\beq
\label{cond-3}
\det\bigl(\hat{R}_{\rho \phi}\bigr) = \frac{16 \bigl(4+\jmath^2 \ell^4\bigr)^2 \rho^4}{\ell^8 \bigl(1+\jmath^2 \rho^4\bigr)^6} \,,
\eeq
which is clearly never zero if the parameters are real.
\vspace{-0.2cm}
\myparagraph{Melvin-swirling universe with a cosmological constant:}
For the magnetic Melvin-swirling universe with a cosmological constant~\eqref{magnetic-melvin-swirling-cosmological}, $E=0$, by setting $\mu = \rho$, $\nu = \phi$, and choosing the common denominator as 
\beq
Den = \ell^8\bigl(B^4+16 \jmath^2\bigr)^6\Bigl[16 + 8 B^2\rho^2+\bigl(B^4+16\jmath^2\bigr)\rho^4\Bigl]^{10} \,,
\eeq
we obtain, for example
\vspace{-0.2cm}
\beq
C_{28} = 2^{16} B^4 \ell^4 \bigl(B^4 - 16\jmath^2\bigr)^4 \bigl(B^4+16\jmath^2\bigr)^8  \,,
\eeq
which can be zero only for $B^4 = 16 \jmath^2$. However, this requirement also results in
\vspace{-0.2cm}
\beq
C_{24} = 2^{32} B^{50} \ell^6 \,,
\eeq 
that is always positive.
\myparagraph{Black Holes in a Melvin-swirling universe:}
For the metric of the dyonic Kerr-Newman black hole in a Melvin-swirling universe~\eqref{DKNMS}
\beqNN
ds^2 = F \biggl[ -\frac{\Delta}{\Sigma}\,dt^2 + \frac{dr^2}{\Delta} + d\theta^2\biggr] + \frac{\Sigma\sin^2\theta}{F}\biggl[ d\phi - \frac{\Omega}{\Sigma} \,dt \biggr]^2 \,,
\eeqNN
a possible choice of a tetrad basis is
\vspace{-0.2cm}
\begin{subequations}
\begin{align}
e^{0} & = \sqrt{\frac{F \Delta}{\Sigma}} dt \,, \\
e^{1} & = \sqrt{\frac{F}{\Delta}} dr \,, \\
e^{2} & = \sqrt{F} d\theta \,, \\
e^{3} & = \sqrt{\frac{\Sigma\sin^2\theta}{F}} \biggl[ d\phi - \frac{\Omega}{\Sigma} \,dt \biggr] \,.
\end{align}
\end{subequations}
Moreover, for a computational purpose, it is useful to perform the following transformation of coordinates
\vspace{-0.15cm}
\beq
y = \cos \theta \,.
\eeq
Using these definitions, the integrability condition~\eqref{integr-cond-2}, for $\mu = r$ and $\nu = y$, can be expressed as
\beq
\label{cond-fin}
\det\bigl(\hat{R}_{\,r y}\bigr) = \frac{\sum_{n, m = 0} C_{n, m}\, r^{n} y^{m}}{Den} \,,
\eeq
wherein, similarly to the case given by Eq.~\eqref{cond-2}, $Den$ represents the common denominator, while $C_{n, m}$ are the coefficients of the expansion.

In particular, the coefficient for $n=12$, $m=0$ is
\beq
C_{12, 0} = 16\bigl(B^4+16\jmath^2\bigr)^2 \,,
\eeq
which does \emph{not} depend on the parameters characterizing the black hole, but only on the swirling and Melvin parameters $\jmath$ and $B$.\newline

\noindent From this result, we can conclude that whenever either the swirling parameter $\jmath$ or the Melvin parameter $B$ is present, a supersymmetric extension of any sub-case of the dyonic Kerr-Newman black hole in a Melvin-swirling universe is \emph{never} possible, regardless of the value of the mass, charges, or angular momentum of the black hole under consideration.\newline

\noindent However, the addition of other parameters, such as a cosmological constant, a NUT parameter, an acceleration of the black hole, or a generalization to the electromagnetic universe, may modify this result.
\clearpage
\fancyhead[LE, RO]{\bfseries \large \thepage}
\fancyhead[RE, LO]{\nouppercase{\it Conclusions}}
\addcontentsline{toc}{section}{Conclusions}
\section*{Conclusions}
\label{sec:Conclusions}
\vspace{-0.5cm}
In this thesis, we used the Ernst formalism explained in Sec.~\ref{sec:Ernst} in order to construct new black hole solutions embedded in a Melvin-swirling universe, whose physical interpretation is given in detail in Sec.~\ref{sec:Melvin-Swirling-Universe}. Specifically, in Sec.~\ref{sec:Kerr-Newman-Melvin-Swirling}, we obtained the most general new solution studied in this thesis, namely the dyonic Kerr-Newman metric in a Melvin-swirling universe, which represents an electromagnetically charged and rotating black hole embedded in a rotating-swirling universe permeated by a uniform magnetic-Melvin field. From this new solution, we also obtained all the sub-cases that were not already discovered, corresponding to the non-magnetic, the non-electric, the non-charged, and the non-rotating sub-cases. \\

\noindent For all the possible sub-cases, in Sec.~\ref{sec:Properties}, we studied the most important proprieties, namely the shape and the area of the event horizon; the entropy and the temperature of the black hole; the ergoregions; the frame-dragging; the presence of pathologies, such as curvature and conical singularities, Dirac and Misner strings, and the occurrence of closed timelike curves. In particular, we found that embedding a black hole in a swirling or Melvin-swirling universe has the effect of removing the curvature singularity at its center for all the rotating solutions that we studied, similar to what happened for the already-known Melvin universe. On the other hand, the addition of the swirling parameter sometimes results in introducing conical singularities or Dirac strings that were not present in the seed solution. For example, quite surprisingly, we found that the non-charged Kerr black hole embedded in a Melvin-swirling universe presents Dirac strings, despite not being electromagnetically charged. Conversely, by imposing particular physical constraints on the parameters, we obtained that the most general case and the non-rotating dyonic-swirling sub-case can represent physical black holes, since outside of the event horizon it is possible to remove all the possible singularities present in these spacetimes.  \vspace{-0.05cm}\\

\noindent Afterward, we studied the geometry near the event horizon in the extremal case, where we obtained a new interpretation of the swirling parameter as a possible degree of freedom that can be used, for the non-rotating charged black hole in a swirling universe, in order to remove the effect on the horizon of one of the charges, without setting to zero the corresponding black hole charge. Moreover, for this black hole, which was also one of the two that we found to be completely regular outside the event horizon, we also proved that the shape of the event horizon always becomes a perfect sphere in the extremal case, due to the geometry of the event horizon exactly reducing to that of its non-swirling counterpart. \\

\noindent Finally, in Sec.~\ref{sec:Sugra}, we discussed a possible supersymmetric extension in $\pazocal{N}=2$, $d=4$ Supergravity, which we found is never possible whenever one between the Melvin or the swirling background is present.  \\

\noindent There are a lot of possible developments in this line of research, it would be interesting to use again the Ernst formalism to add the other remaining parameters, namely the NUT charge to the black hole and the electric uniform field to the universe. Then, other formalisms could be used to add different parameters from those already mentioned, such as an acceleration of the black hole and maybe a cosmological constant. Moreover, one could try to add other fields, such as a scalar field for a generalization in Einstein-Maxwell-Scalar theories.  Once these solutions have been found, one should, for all the sub-cases, study the proprieties as we did in this thesis and discuss whether these new parameters allow one to obtain new physical black holes and to finally construct a supersymmetric extension.

%

\clearpage
\fancyhead[LE,RO]{\bfseries \large \thepage}
\fancyhead[RE,LO]{\nouppercase{\rightmark}}

\appendix
\clearpage
\renewcommand{\theequation}{A.\arabic{equation}}
\titleformat{\section}[display]{\normalfont}{\hspace{0.25 em} \Large \it APPENDIX A}{0pt}  {\bfseries \centering \titlerule[0.8pt]\vspace{3mm}\Huge }[{\vspace{3mm}\titlerule[0.8pt]\vspace{10mm}}]
\section{Differential Operators in Cylindrical Coordinates}
\label{A-Cyl}
In cylindrical coordinates $(\rho,z,\phi)$, the gradient, curl, divergence, and Laplacian for any scalar function $f$ and vector $\vec{V}$ are, respectively, given by:
\begin{align}
\vec{\nabla} \cdot \vec{V} & =  \frac{1}{\rho} \frac{\partial (\rho\, V_\rho)}{\partial\rho} + \frac{\partial V_z}{\partial z}  + \frac{1}{\rho} \frac{\partial V_\phi}{\partial\phi} \,, \\
\vec{\nabla} \times \vec{V} & =  \biggl[\, \frac{1}{\rho} \frac{\partial V_z}{\partial\phi} - \frac{\partial V_\phi}{\partial z} \biggr] \vec{e}_{\rho} + \frac{1}{\rho} \biggl[ \frac{\partial (\rho \, V_\phi)}{\partial\rho} - \frac{\partial V_\rho}{\partial\phi} \biggr]  \vec{e}_{z} +\biggl[ \frac{\partial V_\rho}{\partial z} - \frac{\partial V_z}{\partial\rho} \biggr] \vec{e}_{\phi} \,, \\
\vec{\nabla} f & =  \frac{\partial f}{\partial\rho} \vec{e}_{\rho}  + \frac{\partial f}{\partial z} \vec{e}_{z} + \frac{1}{\rho} \frac{\partial f}{\partial\phi}\vec{e}_{\phi}  \,, \label{Gradient-cyl}\\
\nabla^2 f & =  \frac{1}{\rho} \frac{\partial}{\partial\rho} \biggl[\rho \, \frac{\partial f}{{\partial\rho}}\biggr] + \frac{\partial^2 f}{{\partial z}^2} +\frac{1}{\rho^2} \frac{\partial^2 f}{{\partial\phi}^2} \,,
\end{align}
where $(\vec{e}_{\rho},\vec{e}_{z},\vec{e}_{\phi})$ are the axis unit vectors. Moreover, for any coordinate $x$ and vector $\vec{V}$, it holds that:
\beq
\vec{e}_{x} \times \bigl ( \vec{e}_{x} \times \vec{V} \bigr) = -\vec{V} \,.
\eeq
Finally, for the purpose of this thesis, it is useful to express the partial derivates in cylindrical coordinates using the spherical coordinates $(r, \theta, \phi)$ defined by:
\begin{subequations}
\begin{align}
\rho & = \sqrt{\Delta(r)}\sin \theta \,, \\
z & = \bigl(r-M\bigr)\cos \theta \,,
\end{align}
\end{subequations}
which, for a given function $f(r, \theta)$, result in:
\begin{subequations}
\begin{align}
\frac{\partial f(r, \theta)}{\partial \rho} & =  \frac{2\sqrt{\Delta(r)}}{D}\Biggl[ \bigl(r-M\bigr)\sin\theta\biggl(\frac{\partial f(r, \theta)}{\partial r}\biggr) + \cos\theta \biggl(\frac{\partial f(r, \theta)}{\partial \theta}\biggr)\Biggr]\,, \\
\frac{\partial f(r, \theta)}{\partial z} & = \frac{1}{D}\Biggl[ 2\Delta(r) \cos\theta \biggl(\frac{\partial f(r, \theta)}{\partial r}\biggr) - \sin\theta\biggl( \frac{\Delta(r)}{\partial r}\biggr)  \biggl(\frac{\partial f(r, \theta)}{\partial \theta}\biggr)\Biggr] \,,
\end{align}
\end{subequations}
where 
\beq
D = 2\Delta(r) \cos^2\theta + \bigl(r-M\bigr) \sin^2\theta \biggl(\frac{\Delta(r)}{\partial r}\biggr)  \,.
\eeq
\clearpage
\renewcommand{\theequation}{B.\arabic{equation}}
\titleformat{\section}[display]{\normalfont}{\hspace{0.25 em} \Large \it APPENDIX B}{0pt}  {\bfseries \centering \titlerule[0.8pt]\vspace{3mm}\Huge }[{\vspace{3mm}\titlerule[0.8pt]\vspace{10mm}}]
\section{New Black Holes from Sub-cases of the New Solution}
\label{B-Subcases}
In this appendix, we report some of the new solutions that we obtained as sub-cases of the dyonic Kerr-Newman black hole in a Melvin-swirling universe~\eqref{DKNMS}.
\subsection{Kerr in a Melvin-swirling Universe}
\label{Kerr-Melvin-swirling}
The {\bfseries Kerr black hole in a Melvin-swirling universe} corresponds to the \emph{non-charged} sub-case of the new solution~\eqref{DKNMS}. Thus representing a black hole with rotation parameter $a$, mass $M$, embedded in a rotating universe with swirling parameter $\jmath$ and permeated by a uniform magnetic field $B$. This sub-case is obtained by setting $H=Q=0$ in Eq.~\eqref{DKNMS}, which yields
\begin{subequations}
\label{KMS}
\begin{align}
ds^2 & = F \biggl[ -\frac{\Delta}{\Sigma}\,dt^2 + \frac{dr^2}{\Delta} + d\theta^2\biggr] + \frac{\Sigma\sin^2\theta}{F}\biggl[ d\phi - \frac{\Omega}{\Sigma} \,dt \biggr]^2 \,, \label{KMS-metric} \\
A & = \frac{A_0}{\Sigma} \,dt + \frac{A_{3}}{F}\biggl[ d\phi - \frac{\Omega}{\Sigma}\,dt \biggr] \,, \label{KMS-potential}
\end{align}
\end{subequations}
where the functions have been defined as
\begin{subequations}
\begin{align}
A_0 &=  \frac{B^3}{8}\,\chi_{(3)} \,, \\
A_3 &=\frac{B}{2}\,\varphi_{(1)}+ \frac{B^3}{8}\,\varphi_{(3)}\,, \\
\Omega &= a\,\lambda\, +\biggr[ \jmath^2 + \frac{B^4}{16} \biggr]\chi_{(3)}+ \jmath \, \Omega_{(1)} \,, \\
F & = R^2 +  \frac{B^2}{2}\,\varphi_{(1)}+ \biggr[ \jmath^2 + \frac{B^4}{16} \biggr]\varphi_{(3)} + \jmath\, F_{(1)} \,, \\
\Sigma & = \bigl(r^2 + a^2\bigr)^2 - \Delta a^2 \sin^2\theta \,, \\
\Delta & = r^2 - 2Mr + a^2 \,,
\end{align}
\end{subequations}
\clearpage
\noindent and
\vspace{-0.3cm}
\begin{subequations}
\begin{align}
\lambda & = r^2 + a^2 - \Delta = 2Mr \,, \\
R^2 & = r^2 + a^2\cos^2\theta \,, \\
\begin{split}
\chi_{(3)} & = -a \biggl[  a^2 \Delta \cos^2\theta \Bigl ( \bigl( 4 M^2 - 6 M r \bigr)\cos^2\theta + 12 M^2 - 12 M r - 6 r^2 \Bigr) \\
&\quad + 2 a^4 M \bigl( 2 M + r \bigr) - 4 a^2 M r \bigl( r^2+ 3 M r \bigr) - 6 M r^5 \\
& \quad + \Delta \cos^2\theta \Bigl( 6 r^2 \bigl(\Delta - r^2 \bigr) + 2 M r^3\cos^2\theta \Bigr)\biggr] \,,
\end{split}
\\
\varphi_{(1)} & = \Sigma \sin^2 \theta \,, \\
\begin{split}
\varphi_{(3)} & = a^6 \sin^6\theta + a^2 \Bigl[ \lambda^2 \cos^2\theta \bigl( 3 - \cos^2\theta \bigr)^2  + r^3 \sin^6\theta ( 4 M - r )\Bigr] \\
&\quad +2 a^4 \Bigl[ 2 M^2 \bigl( 1 + \cos^2\theta \bigr)^2 - \Delta \sin^6\theta \Bigr]  + \bigl( r^2 + a^2 \bigr)^3 \sin^4\theta \,,
\end{split}
\\
\Omega_{(1)} & = -4 \Delta \cos \theta  \biggl[r^3+a^2 \Bigl(\bigl(r-M\bigr) \cos^2\theta - M \Bigr)\biggr] \,, \\
F_{(1)} & = -4 a \cos\theta \Bigl[M \bigl( 1 + \cos^2\theta \bigr) \bigl( r^2 + a^2 \bigr)+\lambda\, r \sin^2\theta \Bigr] \,.
\end{align}
\end{subequations}

\subsection{Dyonic Reissner-Nordstr\"om in a Melvin-swirling Universe}
\label{Dyonic-Reissner-Nordstrom-Melvin-swirling}
Similarly, the {\bfseries dyonic Reissner-Nordstr\"om black hole in a Melvin-swirling universe} is the \emph{non-rotating} sub-case of the new solution~\eqref{DKNMS}, therefore corresponding to a black hole with mass $M$, electric charge $Q$ and magnetic charge $H$, embedded in a universe with swirling parameter $\jmath$ and permeated by a uniform magnetic field $B$. Thus, setting $a=0$ in Eq.~\eqref{DKNMS} results in
\vspace{-0.2cm}
\begin{subequations}
\label{DRNMS}
\begin{align}
ds^2 & = F \biggl[ -\frac{\Delta}{r^2}\,dt^2 + \frac{r^2dr^2}{\Delta} + r^2 d\theta^2\biggr] + \frac{r^2\sin^2\theta}{F}\biggl[ d\phi - \frac{\Omega}{r^2} \,dt \biggr]^2 \,, \label{DRNMS-metric} \\
A & = \frac{A_0}{r^2} \,dt + \frac{A_{3}}{F}\biggl[ d\phi - \frac{\Omega}{r^2}\,dt \biggr] \,, \label{DRNMS-potential}
\end{align}
\end{subequations}
\vspace{-0.32cm}
where
\vspace{-0.2cm}
\begin{subequations}
\begin{align}
A_0 &= \chi_{(0)} + \frac{3B^2}{4}\,\chi_{(2)} + \jmath\,\chi_{(4)} \,, \\
A_3 &= \varphi_{(0)} +  \frac{B}{2}\,\varphi_{(1)} + \frac{3B^2}{4}\,\varphi_{(2)} + \frac{B^3}{8}\,\varphi_{(3)} + \jmath\,\varphi_{(4)} \,, \\
\Omega &= 2B\,\chi_{(0)} + \frac{B^3}{2}\,\chi_{(2)}  + 2 \jmath B \,\chi_{(4)} + \jmath \, \Omega_{(1)} \,, \\
F & = 1 + 2B\,\varphi_{(0)} +  \frac{B^2}{2}\,\varphi_{(1)} + \frac{B^3}{2}\,\varphi_{(2)} + \biggr[ \jmath^2 + \frac{B^4}{16} \biggr]\varphi_{(3)} + 2 \jmath B \, \varphi_{(4)} \,, \\
\Delta & = r^2 - 2Mr + Z^2 \,, \\
Z^2 & = Q^2 + H^2 \,,
\end{align}
\end{subequations}
and
\begin{subequations}
\begin{align}
\Xi & = r^2\sin^2\theta + Z^2\cos^2\theta \,, \\
\lambda & = r^2 - \Delta = 2Mr - Z^2 \,, \\
\chi_{(0)} & = - Q r\,, \\
\chi_{(2)} & = Q r \Bigl[ \lambda + \Delta \bigl(1+\cos^2\theta \bigr) \Bigr] \,,\\
\chi_{(4)} & = H r \bigl( \Delta \cos^2\theta + r^2 \bigr) \,, \\
\varphi_{(0)} & =- H \cos\theta \,, \\
\varphi_{(1)} & = r^2 \sin^2 \theta + 3 Z^2 \cos^2\theta \,, \\
\varphi_{(2)} & =  - H \Xi \, \cos\theta  \,,
\\
\varphi_{(3)} & = Z^2 \cos^2\theta \bigl( \, \Xi + r^2 \sin^2\theta \bigr) \,,
\\
\varphi_{(4)} & = Q\Xi\, \cos\theta \,, \\
\Omega_{(1)} & = -4 r \Delta \cos \theta \,.
\end{align}
\end{subequations}
\subsection{Electric Kerr-Newman in a Melvin-swirling Universe}
\label{Electric-Kerr-Newman-Melvin-swirling}
Moreover, the {\bfseries electric Kerr-Newman in a Melvin-swirling universe} is the sub-case of~\eqref{DKNMS} \emph{without} a magnetic charge $H=0$, which is then given by 
\begin{subequations}
\label{EKNMS}
\begin{align}
ds^2 & = F \biggl[ -\frac{\Delta}{\Sigma}\,dt^2 + \frac{dr^2}{\Delta} + d\theta^2\biggr] + \frac{\Sigma\sin^2\theta}{F}\biggl[ d\phi - \frac{\Omega}{\Sigma} \,dt \biggr]^2 \,, \label{EKNMS-metric} \\
A & = \frac{A_0}{\Sigma} \,dt + \frac{A_{3}}{F}\biggl[ d\phi - \frac{\Omega}{\Sigma}\,dt \biggr] \,, \label{EKNMS-potential}
\end{align}
\end{subequations}
where
\begin{subequations}
\begin{align}
A_0 &= \chi_{(0)} + \frac{B}{2}\,\chi_{(1)} + \frac{3B^2}{4}\,\chi_{(2)} + \frac{B^3}{8}\,\chi_{(3)} + \jmath\,\chi_{(4)} \,, \\
A_3 &= \varphi_{(0)} +  \frac{B}{2}\,\varphi_{(1)} + \frac{3B^2}{4}\,\varphi_{(2)} + \frac{B^3}{8}\,\varphi_{(3)} + \jmath\,\varphi_{(4)} \,, \\
\Omega &= a\,\lambda + 2B\,\chi_{(0)} +  \frac{B^2}{2}\,\chi_{(1)} + \frac{B^3}{2}\,\chi_{(2)}  + \biggr[ \jmath^2 + \frac{B^4}{16} \biggr]\chi_{(3)} + 2 \jmath B \,\chi_{(4)} + \jmath \, \Omega_{(1)} \,, \\
F & = R^2 + 2B\,\varphi_{(0)} +  \frac{B^2}{2}\,\varphi_{(1)} + \frac{B^3}{2}\,\varphi_{(2)} + \biggr[ \jmath^2 + \frac{B^4}{16} \biggr]\varphi_{(3)} + 2 \jmath B \, \varphi_{(4)} + \jmath\, F_{(1)} \,, \\
\Sigma & = \bigl(r^2 + a^2\bigr)^2 - \Delta a^2 \sin^2\theta \,, \\
\Delta & = r^2 - 2Mr + Q^2 + a^2 \,,
\end{align}
\end{subequations}
and
\begin{subequations}
\begin{align}
\Xi & = \bigl(r^2 + a^2\bigr )\sin^2\theta + Q^2\cos^2\theta \,, \\
\lambda & = r^2 + a^2 - \Delta = 2Mr - Q^2 \,, \\
R^2 & = r^2 + a^2\cos^2\theta \,, \\
\chi_{(0)} & = - Q r \bigl(r^2+a^2\bigr) \,, \\
\chi_{(1)} & = -3 a Q^2 \Bigl[ \lambda + \Delta \bigl(1+\cos^2\theta \bigr)  \Bigr] \,, \\
\begin{split}
\chi_{(2)} & = Q \biggl[ a^2 \Bigl( \Delta \cos^2\theta \bigl( 3r - 4 M \bigr) -  r \bigl( Q^2 + \Delta \bigr) -2 M a^2  \Bigr)  \\
& \quad  + r^3 \Bigl(  \lambda + \Delta \bigl(1+\cos^2\theta \bigr)  \Bigr)\biggr] \,,
\end{split}
\\
\begin{split}
\chi_{(3)} & = -a \Biggl[  a^2 \Delta \cos^2\theta \Bigl ( \bigl( Q^2 + 4 M^2 - 6 M r \bigr)\cos^2\theta + Q^2 + 12 M^2 - 12 M r - 6 r^2 \Bigr) \\
&\quad + 2 a^4 M \bigl( 2 M + r \bigr) + a^2 Q^2\Delta - 4 a^2 M r \bigl( r^2 - 2 Q^2 + 3 M r \bigr) - 6 M r^5 \\
& \quad + \Delta \cos^2\theta \Bigl( 6 r^2 \bigl(\Delta - r^2 \bigr) + \bigl( Q^4 + 2 M r^3 - 3 Q^2 r^2 \bigr) \cos^2\theta \Bigr)\Biggr] \,,
\end{split}
\\
\chi_{(4)} & = - a Q \Delta \cos\theta \bigl( \, \Xi + 2 R^2 \bigr) \,,
\\
\varphi_{(0)} & = a Q r \sin^2\theta \,, \\
\varphi_{(1)} & = \Sigma \sin^2 \theta + 3 Q^2 \bigl( r ^2\cos^2\theta + a^2 \bigr) \,, \\
\varphi_{(2)} & = a Q \biggl[\bigl( 1 + \cos^2\theta \bigr) \Bigl( r^3 +  \bigl (2 M + r \bigr) a^2 \Bigr) + r \cos^2\theta \Bigl(2 Q^2 - \Delta \bigl(3 - \cos^2\theta \bigr) \Bigr)\biggr] \,,
\\
\begin{split}
\varphi_{(3)} & = Q^2 \bigg[2 a^4 \bigl( 1 + \cos^2\theta \bigr)^2 +  r^2 \cos^2\theta \bigl( \, \Xi + R^2 \sin^2\theta \bigr) \\
&\quad + a^2 \cos^2\theta \Bigl( 2\, \Xi + 3 Q^2 + r^2 \bigl( 5 + 6 \sin^2\theta + 3 \cos^4\theta \bigr) - 8 \Delta \Bigr) \biggr] \\
&\quad + a^6 \sin^6\theta + a^2 \Bigl[ \lambda^2 \cos^2\theta \bigl( 3 - \cos^2\theta \bigr)^2  + r^3 \sin^6\theta ( 4 M - r )\Bigr] \\
&\quad +2 a^4 \Bigl[ 2 M^2 \bigl( 1 + \cos^2\theta \bigr)^2 - \Delta \sin^6\theta \Bigr]  + \bigl( r^2 + a^2 \bigr)^3 \sin^4\theta \,,
\end{split}
\\
\varphi_{(4)} & =  Q \cos\theta \Bigl[ \bigl( r^2+ a^2 \bigr)\, \Xi - 2 \lambda a^2 \sin^2\theta \Bigr] \,,
 \\
\Omega_{(1)} & = -4 \Delta \cos \theta  \biggl[r^3+a^2 \Bigl(\bigl(r-M\bigr) \cos^2\theta - M \Bigr)\biggr] \,, \\
F_{(1)} & = -4 a \cos\theta \Bigl[M \bigl( 1 + \cos^2\theta \bigr) \bigl( r^2 + a^2 \bigr)+\lambda\, r \sin^2\theta \Bigr] \,.
\end{align}
\end{subequations}
\subsection{Magnetic Kerr-Newman in a Melvin-swirling Universe}
\label{Magnetic-Kerr-Newman-Melvin-swirling}
Finally, setting the electric charge to zero, $Q=0$, in the new solution~\eqref{DKNMS}, we obtain the {\bfseries magnetic Kerr-Newman black hole in a Melvin-swirling universe}:
\begin{subequations}
\label{MKNMS}
\begin{align}
ds^2 & = F \biggl[ -\frac{\Delta}{\Sigma}\,dt^2 + \frac{dr^2}{\Delta} + d\theta^2\biggr] + \frac{\Sigma\sin^2\theta}{F}\biggl[ d\phi - \frac{\Omega}{\Sigma} \,dt \biggr]^2 \,, \label{MKNMS-metric} \\
A & = \frac{A_0}{\Sigma} \,dt + \frac{A_{3}}{F}\biggl[ d\phi - \frac{\Omega}{\Sigma}\,dt \biggr] \,, \label{MKNMS-potential}
\end{align}
\end{subequations}
where
\begin{subequations}
\begin{align}
A_0 &= \chi_{(0)} + \frac{B}{2}\,\chi_{(1)} + \frac{3B^2}{4}\,\chi_{(2)} + \frac{B^3}{8}\,\chi_{(3)} + \jmath\,\chi_{(4)} \,, \\
A_3 &= \varphi_{(0)} +  \frac{B}{2}\,\varphi_{(1)} + \frac{3B^2}{4}\,\varphi_{(2)} + \frac{B^3}{8}\,\varphi_{(3)} + \jmath\,\varphi_{(4)} \,, \\
\Omega &= a\,\lambda + 2B\,\chi_{(0)} +  \frac{B^2}{2}\,\chi_{(1)} + \frac{B^3}{2}\,\chi_{(2)}  + \biggr[ \jmath^2 + \frac{B^4}{16} \biggr]\chi_{(3)} + 2 \jmath B \,\chi_{(4)} + \jmath \, \Omega_{(1)} \,, \\
F & = R^2 + 2B\,\varphi_{(0)} +  \frac{B^2}{2}\,\varphi_{(1)} + \frac{B^3}{2}\,\varphi_{(2)} + \biggr[ \jmath^2 + \frac{B^4}{16} \biggr]\varphi_{(3)} + 2 \jmath B \, \varphi_{(4)} + \jmath\, F_{(1)} \,, \\
\Sigma & = \bigl(r^2 + a^2\bigr)^2 - \Delta a^2 \sin^2\theta \,, \\
\Delta & = r^2 - 2Mr + H^2 + a^2 \,, \\
\end{align}
\end{subequations}
and
\begin{subequations}
\begin{align}
\Xi & = \bigl(r^2 + a^2\bigr )\sin^2\theta + H^2\cos^2\theta \,, \\
\lambda & = r^2 + a^2 - \Delta = 2Mr - H^2 \,, \\
R^2 & = r^2 + a^2\cos^2\theta \,, \\
\chi_{(0)} & = a H\Delta \cos \theta \,, \\
\chi_{(1)} & = -3 a H^2 \Bigl[ \lambda + \Delta \bigl(1+\cos^2\theta \bigr) \Bigr] \,, \\
\chi_{(2)} & = a H \Delta \cos\theta  \bigl( \, \Xi + 2 R^2 \bigr)  \,, \\
\begin{split}
\chi_{(3)} & = -a \Biggl[  a^2 \Delta \cos^2\theta \Bigl ( \bigl( H^2 + 4 M^2 - 6 M r \bigr)\cos^2\theta + H^2 + 12 M^2 - 12 M r - 6 r^2 \Bigr) \\
&\quad + 2 a^4 M \bigl( 2 M + r \bigr) + a^2 H^2\Delta - 4 a^2 M r \bigl( r^2 - 2 H^2 + 3 M r \bigr) - 6 M r^5 \\
& \quad + \Delta \cos^2\theta \Bigl( 6 r^2 \bigl(\Delta - r^2 \bigr) + \bigl( H^4 + 2 M r^3 - 3 H^2 r^2 \bigr) \cos^2\theta \Bigr)\Biggr] \,,
\end{split}
\\
\begin{split}
\chi_{(4)} & = -H \biggl[ a^2 \Bigl(r \bigl( H^2 - r^2 \bigr) + 4 M \Delta \cos^2\theta + r \Delta \bigl( 1 - 3 \cos^2\theta \bigr) \Bigr) \\
& \quad + 2 a^4 M - r^3 \bigl( \Delta \cos^2\theta + r^2 \bigr) \biggr]  \,,
\end{split}
\\
\varphi_{(0)} & =- H \bigl( r^2 + a^2 \bigr) \cos\theta \,, \\
\varphi_{(1)} & = \Sigma \sin^2 \theta + 3 H^2 \bigl( r ^2\cos^2\theta + a^2 \bigr) \,, \\
\varphi_{(2)} & =H \cos\theta \Bigl[ 2 a^2 \lambda \sin^2\theta - \bigl(r^2 +a^2) \, \Xi \Bigr]  \,,
\\
\begin{split}
\varphi_{(3)} & = H^2 \bigg[2 a^4 \bigl( 1 + \cos^2\theta \bigr)^2 +  r^2 \cos^2\theta \bigl( \, \Xi + R^2 \sin^2\theta \bigr) \\
&\quad + a^2 \cos^2\theta \Bigl( 2\, \Xi + 3 H^2 + r^2 \bigl( 5 + 6 \sin^2\theta + 3 \cos^4\theta \bigr) - 8 \Delta \Bigr) \biggr] \\
&\quad + a^6 \sin^6\theta + a^2 \Bigl[ \lambda^2 \cos^2\theta \bigl( 3 - \cos^2\theta \bigr)^2  + r^3 \sin^6\theta ( 4 M - r )\Bigr] \\
&\quad +2 a^4 \Bigl[ 2 M^2 \bigl( 1 + \cos^2\theta \bigr)^2 - \Delta \sin^6\theta \Bigr]  + \bigl( r^2 + a^2 \bigr)^3 \sin^4\theta \,,
\end{split}
\\
\varphi_{(4)} & = a H \biggl[2 M \Bigl(a^2 +  \cos^2\theta \bigl(2 r^2 + a^2 \bigr) \Bigr) + r \sin^2\theta \bigl( \lambda + \Delta \sin^2\theta \bigr)\biggr] \,, \\
\Omega_{(1)} & = -4 \Delta \cos \theta  \biggl[r^3+a^2 \Bigl(\bigl(r-M\bigr) \cos^2\theta - M \Bigr)\biggr] \,, \\
F_{(1)} & = -4 a \cos\theta \Bigl[M \bigl( 1 + \cos^2\theta \bigr) \bigl( r^2 + a^2 \bigr)+\lambda\, r \sin^2\theta \Bigr] \,.
\end{align}
\end{subequations}
\clearpage
\renewcommand{\theequation}{C.\arabic{equation}}
\titleformat{\section}[display]{\normalfont}{\hspace{0.25 em} \Large \it APPENDIX C}{0pt}  {\bfseries \centering \titlerule[0.8pt]\vspace{3mm}\Huge }[{\vspace{3mm}\titlerule[0.8pt]\vspace{10mm}}]
\section{Other Useful Melvin-swirling Metrics}
\label{C-OtherBH}
In this appendix, we provide other useful sub-cases in a Melvin-swirling universe.
\subsection{Melvin-swirling Universe with a Cosmological Constant}  \label{sub:melvin-swirling-universe-cosmological}
As explained in Sec.~\ref{sec:Melvin-Swirling-Universe}, the {\bfseries Melvin-swirling universe with a cosmological constant} can be obtained by performing a double-Wick rotation onto the flat magnetic Reissner-Nordstr\"om-Taub-NUT black hole with a cosmological constant, or equivalently, by setting the uniform electric field to zero, $E=0$, in the electromagnetic-swirling universe with a cosmological constant~\eqref{em-melvin-swirling-cosmological}, which yields:
\begin{subequations}
\label{magnetic-melvin-swirling-cosmological}
\begin{align}
ds^2 & = F \biggl[-dt^2+\frac{\rho^2 d\rho^2}{\pazocal{S}}+dz^2\biggr]+\frac{{\pazocal{S}}}{F}\bigl(d\phi - \Omega \,dt\bigr)^2 \,, \\
A & = \frac{A_{3}}{F}\bigl(d\phi - \Omega\, dt\bigr) \,,
\end{align}
\end{subequations}
where
\begin{subequations}
\begin{align}
A_{3} & = \frac{B}{2}\Bigl[1 + \frac{B^2}{4} \rho^2\Bigr] \rho^2 \,, \\
\Omega & = - 4 \jmath z \,, \\
F & = 1 + \frac{B^2}{2} \rho^2 + \biggl[\jmath^2 + \frac{B^4}{16}\biggr] \rho^4 \,, \\
\begin{split}
\pazocal{S} &  = -\frac{4\Bigl[\bigr(48\jmath^2 + B^4\bigr)^2 - 3072\jmath^4 \Bigr]\Lambda}{3\bigl(16\jmath^2 + B^4\bigr)^3} + \biggl[1 - \frac{4B^2\bigl(48\jmath^2 + B^4\bigr) \Lambda}{3\bigl(16\jmath^2 + B^4\bigr)^2}\biggr]\rho^2  \\
& \quad - \frac{\Lambda}{2}\rho^4 - \frac{B^2 \Lambda}{12}\rho^6 - \frac{\bigl(16\jmath^2 + B^4\bigr)  \Lambda}{192}\rho^8 \,.
\end{split}
\end{align}
\end{subequations}
In particular, in the case of a negative cosmological constant $\Lambda = - \frac{3}{\ell^2}$ we obtain:
\begin{align}
\begin{split}
\pazocal{S} &  = \frac{4\Bigl[\bigr(48\jmath^2 + B^4\bigr)^2 - 3072\jmath^4 \Bigr]}{\ell^2\bigl(16\jmath^2 + B^4\bigr)^3} + \biggl[1 + \frac{4B^2\bigl(48\jmath^2 + B^4\bigr) }{\ell^2\bigl(16\jmath^2 + B^4\bigr)^2}\biggr]\rho^2  \\
& \quad + \frac{3}{2\ell^2}\rho^4 + \frac{B^2}{4\ell^2}\rho^6 + \frac{\bigl(16\jmath^2 + B^4\bigr) }{64\ell^2}\rho^8 \,.
\end{split}
\end{align}
\subsection{Melvin-swirling Universe} \label{sub:melvin-swirling-universe}
\vspace{-0.15cm}
The {\bfseries Melvin-swirling universe} (without the cosmological constant) can be obtained by adding the uniform magnetic field $B$ to the swirling universe~\eqref{swirling-background} using a real magnetic Harrison transformation, or equivalently, by applying a magnetic Ehlers transformation to the magnetic Melvin universe~\eqref{Melvin-background} in order to add the swirling parameter $\jmath$. 

Moreover, it can also be trivially obtained by setting $\Lambda = 0$ in the case with a cosmological constant~\eqref{magnetic-melvin-swirling-cosmological}, which, using the spherical coordinates defined by $\rho = r\sin\theta$, $z  = r \cos\theta$, results in:
\vspace{-0.5cm}
\begin{subequations}
\label{magnetic-melvin-swirling-no-cosmological}
\begin{align}
ds^2 & = F \Bigl[-dt^2+ dr^2 + r^2 d \theta^2 \Bigr]+\frac{r^2 \sin^2 \theta}{F}\bigl(d\phi - \Omega \,dt\bigr)^2 \,, \\
A & = \frac{A_{3}}{F}\bigl(d\phi - \Omega\, dt\bigr) \,,
\end{align}
\end{subequations}
where
\vspace{-0.45cm}
\begin{subequations}
\begin{align}
A_{3} & = \frac{B}{2} \Bigl[1 + \frac{B^2}{4} r^2 \sin^2\theta \Bigr] r^2 \sin^2\theta\,, \\
\Omega & = - 4 \jmath \, r \cos\theta \,, \\
F & = 1 + \biggl[\frac{B^2}{2}\biggr] r^2\sin^2\theta + \biggl[\jmath^2 + \frac{B^4}{16}\biggr] r^4 \sin^4\theta \,.
\end{align}
\end{subequations}
\vspace{-0.8cm}
\subsection{Dyonic Reissner-Nordstr\"om in a Swirling Universe}  \label{sub:drn-swirling}
\vspace{-0.15cm}
Setting $B=0$ in Eq.~\eqref{DRNMS}, we recover the {\bfseries dyonic Reissner-Nordstr\"om black hole in swirling universe}, which is then given by:
\begin{subequations}
\label{DRNS}
\begin{align}
ds^2 &= F\Big[-h\,dt^2+\frac{dr^2}{h}+ r^2 d\theta^2 \Big]+\frac{r^2\sin^2\theta}{F}\bigl(d\phi  - \Omega\, dt\bigr)^2 \label{dyonic-rn-swirling} \,,\\
A & = A_{0} \, dt + \frac{A_{3}}{F}\bigl(d\phi - \Omega\, dt\bigr) \,, \label{A-DRNS}
\end{align}
\end{subequations}
where
\vspace{-0.35cm}
\begin{subequations}
\begin{align}
h& = 1 -\frac{2 M}{ r}+\frac{Q^2}{r^2}+\frac{H^2}{r^2} \,, \\
F& = 1+\jmath^2 \Bigl[r^2\sin^2\theta + \bigl(Q^2+H^2\bigr)\cos^2\theta\Bigr]^2 \,,\\ 
\Omega & = -4\,\jmath\, h\,r\cos\theta \,, \\
A_{0}& = - \frac{Q}{r} - \jmath H r \Bigl[1 + h \cos^2\theta\Bigr] \label{A0-DRNS}\,,\\
A_{3} & = \cos\theta\Bigl[H + Q\sqrt{F-1} \Bigr] \label{A3-DRNS}\,.
\end{align}
\end{subequations}
From the components of the electromagnetic potential, we can see that, in a certain sense, the swirling parameter $\jmath$ generates a magnetic charge from an electric one, and vice-versa.
\clearpage
\renewcommand{\theequation}{D.\arabic{equation}}
\titleformat{\section}[display]{\normalfont}{\hspace{0.25 em} \Large \it APPENDIX D}{0pt}  {\bfseries \centering \titlerule[0.8pt]\vspace{3mm}\Huge }[{\vspace{3mm}\titlerule[0.8pt]\vspace{10mm}}]
\section{Embedding of the Event Horizon in $\E^3$}
\label{D-Embedding}

An event horizon can be embedded in the Euclidian three-dimensional space $\E^3$ as a surface of revolution, by comparing the induced metric on the horizon with that of the flat three-dimensional space in cylindrical coordinates.

Indeed, the induced metric of an event horizon of radius $r_{H}$ is obtained by setting $r = r_{H}$ and $t = constant$ in the complete spacetime metric, which yields
\beq
ds^2_{H} = f_{1} d\theta^2 + f_{2} d\phi^2 \,, \label{metr-hor-emb}
\eeq
where 
\begin{subequations}
\begin{align}
f_{1} & = g_{\theta \theta}(\theta, r=r_{H}) \,, \\
f_{2} & = g_{\phi \phi}(\theta, r=r_{H}) \,.
\end{align}
\end{subequations}
On the other hand, the flat three-dimensional metric in cylindrical coordinates is given by
\beq
ds_{3}^2 = dz^2 + dR^2 + R^2 d\phi^2 \,. \label{metr-cyl-emb}
\eeq
Therefore, considering $z$ and $R$ as functions of $\theta$, and comparing the flat three-dimensional metric in cylindrical coordinates~\eqref{metr-cyl-emb} with that induced on the event horizon~\eqref{metr-hor-emb}, it follows that
\beq
R =  f_{2} \,, \label{eq-R-emb}
\eeq
while $z$ is determined by the following differential equation:
\beq
\biggl[\frac{dR}{d\theta}\biggr]^2 + \biggl[\frac{dz}{d\theta}\biggr]^2 = f_{1} \,. \label{eq-z-emb}
\eeq
In conclusion, the embedding of an event horizon in $\E^3$ is determined once the functions $R = R\bigl(\theta\bigr)$ and $z = z \bigl(\theta\bigr)$ are obtained.
\vfill
\clearpage
\renewcommand{\theequation}{E.\arabic{equation}}
\titleformat{\section}[display]{\normalfont}{\hspace{0.25 em} \Large \it APPENDIX E}{0pt}  {\bfseries \centering \titlerule[0.8pt]\vspace{3mm}\Huge }[{\vspace{3mm}\titlerule[0.8pt]\vspace{10mm}}]
\section{Majorana Representation of the Gamma Matrices}
\label{E-Gamma}
The (real) Majorana representation of the gamma matrices $\gamma_{a} = \bigl(\gamma_{0}, \gamma_{1}, \gamma_{2}, \gamma_{3}\bigr)$ can be obtained by applying the unitary transformation
\beq
U = \frac{1}{\sqrt{2}} \begin{pmatrix} \mathds{1}_{2} & \sigma_2 \\ \sigma_2 & -\mathds{1}_{2} \end{pmatrix} \,,
\eeq
to the gamma matrices $\gamma'_a$ in the standard Dirac representation:
\beq
\gamma_{a} = U^{-1}\gamma'_{a} U \,,
\eeq
which yields
\begin{align}
\gamma_{0}&=\begin{pmatrix} 0 & -i\sigma_2 \\ -i\sigma_2 & 0 \end{pmatrix} \,,\\
\gamma_{1}&=\begin{pmatrix}-\sigma_3 & 0\\ 0& -\sigma_3 \end{pmatrix}\,, \\
\gamma_{2}&=\begin{pmatrix} 0 & -i\sigma_2 \\ i\sigma_2 & 0  \end{pmatrix} \,,\\
\gamma_{3}&=\begin{pmatrix} \sigma_1 & 0\\ 0 & \sigma_1  \end{pmatrix} \,, \\
\gamma_{5}&=\gamma_{0}\gamma_{1}\gamma_{2}\gamma_{3}=\begin{pmatrix}i\sigma_2 & 0\\ 0& -i\sigma_2\ \end{pmatrix} \,.
\end{align}
Where the $\sigma_{j}$ denoted the standard representation of the two-dimensional Pauli matrices
\beq
\sigma_{j} = \begin{pmatrix} \delta_{j3} & \delta_{j1}-i\delta_{j2} \\ \delta_{j1}+i\delta_{j2} & -\delta_{j3} \end{pmatrix} \,.
\eeq
%
\clearpage
\fancyhead[LE, RO]{\bfseries \large \thepage}
\fancyhead[RE, LO]{\nouppercase{\it Bibliography}}
\addcontentsline{toc}{section}{Bibliography}
\renewcommand{\refname}{Bibliography}
\bibliographystyle{unsrturl}
\bibliography{biblio}
\end{document}